# Efficiency-Aware Computational Intelligence for Resource-Constrained Manufacturing Toward Edge-Ready Deployment

Qianyu Zhou, Ph.D.

University of Connecticut, 2025

In industrial cyber–physical systems and data-driven manufacturing environments, a fundamental dissonance persists between the idealized assumptions of current learning paradigms and the non-ideal realities of industrial data ecosystems. Heterogeneous sensing modalities, stochastic operation conditions, and dynamic process parameters often yield data that are incomplete, unlabeled, imbalanced, and domain-shifted. High-fidelity digital replicas and experimental datasets remain limited by cost, confidentiality, and time-to-acquisition, while stringent latency, bandwidth, and energy constraints at the edge further restrict the feasibility of centralized learning architectures. These conditions collectively compromise the scalability of conventional deep networks, hinder the realization of digital-twin frameworks, and exacerbate risks of error escape in safety-critical applications. Therefore, a critical gap exists in developing computational intelligence that is data-lean, physics-consistent, uncertainty-calibrated, and resource-efficient, ensuring trustworthy inference and efficient deployment across multimodal and multiscale manufacturing scenarios.

In this dissertation, six complementary studies collectively advance an efficiency grounded computational framework for data-lean manufacturing. 1) Generative Data Augmentation with Dynamic Filtering is formulated to alleviate sample scarcity and severe data imbalance by producing statistically diverse yet physically consistent samples, improving the generalizability of learning under limited observations. 2) Semi-Supervised Pseudo-Labeling with Adaptive Weighting introduces a self-evolving learning mechanism that fuses limited labeled and abundant unlabeled data while adaptively regulating pseudo-label confidence, enabling robust convergence with reduced simulation and annotation cost. 3) Parallel Physics-Informed Representation Learning integrates domain knowledge and spatially correlated features within a unified network to enhance interpretability and suppress false negatives, establishing a reliable pathway for small-data condition monitoring. 4) Spatially Informed Graph-Neural Surrogate Modeling embeds process physics and spatial dependencies into graph message passing to enable efficient and accurate modeling of complex manufacturing processes under constrained computational budgets. 5) Edge–Cloud Collaborative Compression and Reconstruction Framework provides a scalable approach for real-time signal analytics, balancing compression efficiency, information fidelity, and system throughput for distributed industrial computing environments. 6) Zero-Shot Vision–Language Reasoning via Retrieval-Augmented Generation extends few-shot visual intelligence to unseen manufacturing scenarios, coupling multimodal embeddings with knowledge retrieval to achieve generalizable understanding across modalities.

Together, these efforts form a unified paradigm of data-efficient, physics-guided, and edge-ready intelligence that bridges laboratory-scale learning and industrial deployment. The dissertation contributes both methodological insight and system-level practicality toward scaling insight rather than data, enabling the next generation of reliable and resource-aware manufacturing analytics.

# Efficiency-Aware Computational Intelligence for Resource-Constrained Manufacturing Toward Edge-Ready Deployment

Qianyu Zhou

B.S., Huazhong University of Science and Technology, 2019

A Dissertation

Submitted in Partial Fulfillment of the

Requirements for the Degree of

Doctor of Philosophy

at the

University of Connecticut

2025

I

APROVAL PAGE

Doctor of Philosophy Dissertation

# Efficiency-Aware Computational Intelligence for Resource-Constrained Manufacturing Toward Edge-Ready Deployment

Presented by

Qianyu Zhou, B.S.

Approved by

Major Advisor:_______________________________________________
                         Dr. Jiong Tang

Co-Major Advisor:_____________________________________________
                         Dr. Farhad Imani

Associate Advisor:_____________________________________________
                         Dr. Chao Hu

Associate Advisor:_____________________________________________
                         Dr. Jeongho Kim

Associate Advisor:_____________________________________________
                         Dr. Hongyi Xu

University of Connecticut

2025

III

*This work is dedicated to my beloved parents, my younger brother, and my girlfriend for their love and endless support*

# Acknowledgment

First and foremost, I wish to express my deepest gratitude to my major advisor, Prof. Jiong Tang, for his exceptional mentorship, guidance, and unwavering support throughout the past five years of my doctoral study. His rigorous scholarship, profound insight, and dedication to research have not only shaped my academic development but have also influenced the way I view challenges, persistence, and the pursuit of knowledge. It has been an honor to learn under his supervision, and the lessons I have gained from him will continue to guide me throughout my career and life.

My sincere appreciation also goes to my co-major advisor, Prof. Farhad Imani, for his thoughtful guidance, encouragement, and constant support. His perspective and advice have greatly enriched my research experience and shaped my growth as a scholar. I am deeply thankful to my associate advisors, Prof. Chao Hu, Prof. Jeongho Kim, and Prof. Hongyi Xu, for their valuable feedback, instructive suggestions, and continued guidance throughout the course of my research.

I would also like to thank my lab members and research colleagues, especially Yang Zhang and Dong Xu, with whom I have shared this long journey of doctoral research, the discussion, collaboration, late-night work, mutual encouragement, and countless moments of support have been invaluable. I am equally grateful to my lab mates and friends Ting Wang, David Javadian, and Eric Gillespie, among others, for the camaraderie and help they provided both in and outside the lab.

This work would not have been possible without the generous support of NASA, CMSC, and the NSF. Their funding enabled me to dedicate myself fully to research and academic development at the University of Connecticut. I also extend my appreciation to all project collaborators whose contributions made our teamwork both successful and rewarding.

My heartfelt gratitude goes to my parents, my younger brother, and my girlfriend, whose unconditional love, encouragement, and understanding have given me strength through every challenge. During moments of homesickness and fatigue, their voices, care, and belief in me brought comfort and renewed motivation. I am deeply thankful to my extended family and friends whose support has quietly accompanied me along the way.

As I write this acknowledgement, I reflect on the years spent at UConn, years marked by struggle, growth, perseverance, and fulfillment. Time moves quietly, carrying away countless memories, yet leaving behind gratitude that words cannot fully express. Lastly, I would like to thank myself for not giving up, for choosing to endure, and for continuing to believe in the value of learning. Although this dissertation marks an important milestone, the journey of exploration and growth continues forward.

# Table of Contents

# List of Figures

# List of Tables

# Chapter 1.  Introduction

## 1.1. Background and Problem Statement

Modern manufacturing has embraced data-driven methods for predictive maintenance, fault diagnosis, and quality control, yet the statistical conditions under which real data is generated differ markedly from those assumed by standard machine-learning pipelines. As Figure 1-1 suggests, operational sensor streams are scarce, costly to label, class-imbalanced, and noisy; operating regimes shift with load, speed, tooling wear, and environment; and deployment must often occur on resource-constrained edge devices with strict latency budgets [1-4]. Failures are rare by design, so minority classes are under-represented, amplifying the risk that models optimize for the majority class while ignoring safety-critical outliers [5-7]. Ground-truthing requires expert inspection or teardown and is therefore expensive and slow, which in turn limits the breadth of labeled data and motivates active/semi-supervised strategies [8,9]. In addition, labels can be uncertain or noisy because annotations depend on subjective thresholds or indirect proxies such as vibration severity [10]. All of these factors are compounded by dataset shift, differences between the training and deployment distributions driven by changing tools, materials, fixtures, or sensors, which erodes generalization when it is not explicitly modeled [11].

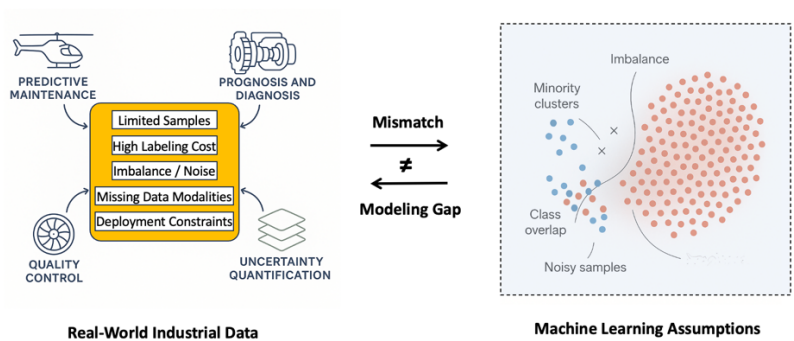

**Figure 1-1.** Mismatch between real industrial PHM data and standard machine learning assumptions. Conventional learning recipes, large, balanced datasets; stationary data-generating processes; clean labels; abundant compute, are codified in widely used textbooks and benchmarks [12,13]. As Figure 1-1 emphasizes, these assumptions rarely hold on the shop floor. When models trained on curated data are deployed in situ, performance degrades because the learned decision boundaries are not calibrated for minority failure modes, drifting conditions, or device-specific noise patterns. Ad-hoc fixes such as oversampling, generic data augmentation, or threshold tuning seldom close the gap at scale [14,15]. The result is a persistent modeling gap between the realities of industrial data and the premises of standard ML.

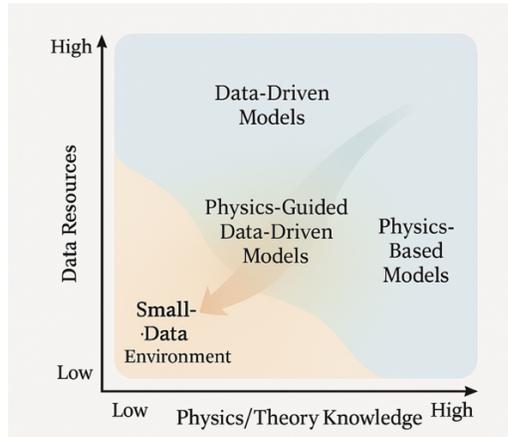

**Figure 1-2.** Positioning of physics-based, data-driven, and physics-guided models under varying data and physics availability.

Figure 1-2 frames this gap in terms of available data and available physics. When data is abundant and representative, purely data-driven models can learn rich, nonlinear mappings; when physics is complete and trusted, mechanistic simulators or analytical models provide reliable predictions and explanations. Manufacturing systems frequently inhabit the small-data quadrant, where neither axis is sufficient on its own: the data are sparse and heterogeneous, and first-principles models omit important nonlinearities, contact phenomena, frictional effects, or stochastic variability. In this regime, purely data-driven models overfit or extrapolate poorly, while purely physics-based models miss critical behaviors [16]. This tension has motivated physics-guided data-driven approaches that embed physical structure, constraints, invariants, inductive biases, into flexible learning architectures [17-19]. By restricting the hypothesis space to solutions consistent with known mechanics while allowing the data to learn residual or unmodeled effects, such hybrids improve sample efficiency, generalization, and interpretability, properties that are essential in safety-critical, small-data manufacturing environments. From a deployment perspective, industrial monitoring also demands edge awareness: encoders must operate under limited CPU, memory, and power

near the sensor, while decoders and analytics may run in the cloud [20,21]. This split architecture imposes an additional constraint on model design: representations must be compact and bandwidth-efficient yet preserve diagnostically salient structure such as harmonics, sidebands, and envelope features known to correlate with incipient faults [22,23]. In short, the central difficulty is not simply achieving high accuracy on a static benchmark, but learning resource-rational, physics-consistent models that remain reliable under label scarcity, class imbalance, distribution shift, and hardware constraints.

This dissertation addresses that difficulty directly. Guided by the perspective in Figure 1-1 and 1-2, it develops algorithms and systems that (i) align learning objectives with the operational characteristics of industrial data (limited labels, imbalance, noise, and shift), (ii) integrate physics knowledge to bias learning toward feasible, interpretable solutions in the small-data regime, and (iii) deliver edge-viable representations that preserve fault-relevant content under strict compute and bandwidth budgets. The subsequent chapters instantiate this agenda across concrete problems in health monitoring and signal compression, demonstrating how physics-guided, edge-aware learning narrows the modeling gap between laboratory ML and real manufacturing practice.

## 1.2. Review of State-of-the-art Approaches

### 1.2.1. Data-Centric Strategies for Small and Imbalanced Industrial Datasets

A central challenge in industrial prognostics and health management (PHM) is the inherent scarcity and imbalance of labeled data. Mechanical components often operate under long periods of healthy conditions, while fault states occur rarely, progressively, or unpredictably, resulting in highly skewed condition distributions [24,25]. Furthermore, annotating degradation states typically requires expert knowledge or controlled destructive testing, making labels expensive and sometimes impractical to obtain [26]. These characteristics stand in contrast to standard machine learning assumptions that rely on abundant, balanced, and independently sampled data, and have motivated a broad family of data-centric strategies that operate directly at the data level to improve model generalization.

One commonly adopted direction is non-generative data augmentation, where the goal is to expand training diversity by applying domain-aware transformations to existing signals. In vibration analysis, this includes geometric manipulations such as time shifting, scaling, jittering, permutation, and window slicing [27,28]. Signal-processing–based transformations, such as adding harmonics, spectral filtering, envelope extraction, and stochastic noise injection, have also been shown to mimic real operating variability while preserving dominant fault signatures [29,30]. These methods are simple to deploy and computationally efficient, but their effectiveness depends heavily on the appropriateness of the chosen transformations, and they may risk introducing artifacts or amplifying class imbalance if not applied carefully.

In contrast, generative augmentation aims to synthesize new samples that reflect realistic fault progression patterns. Earlier approaches used dictionary learning and sparse coding to reconstruct variations around known fault spectra [31,32]. More recent studies have adopted deep generative models, such as Variational Autoencoders (VAE) [33] and Generative Adversarial Networks (GAN) [34], to synthesize fault-bearing waveforms or spectrogram patches under small-sample regimes [35,36]. These models are capable of capturing nonlinear structure in the data distribution but are sensitive to mode collapse, training instability, and the availability of at least minimal seed examples representing each fault type.

Despite their progress, both non-generative and generative augmentation approaches share a common limitation: they act primarily at the data surface level, aiming to increase the apparent data volume without addressing whether the synthesized content meaningfully reflects the underlying physics of machinery behavior. In many practical PHM scenarios, the space of physically realizable signals is constrained by machine dynamics, operational loads, and sensor placement, and synthetic samples that fail to respect these constraints may degrade rather than improve diagnostic robustness [37]. This limitation motivates the later sections of this dissertation, where data augmentation is reconsidered alongside physics-aware representation and learning strategies that explicitly preserve mechanistic relationships between signals and system states.

### 1.2.2. Physics-Guided and Representation Learning Approaches

While data-centric augmentation strategies can help alleviate limited sample size to some extent, they do not fundamentally address the mismatch between physically governed machinery behavior and purely statistical machine learning representations. Mechanical systems such as gearboxes, bearings, and rotating shafts are governed by well-established dynamics, including stiffness–mass relationships, harmonic generation mechanisms, structural mode interactions, and load transfer pathways [38,39]. Corresponding vibration responses exhibit structured spectral patterns, such as gear mesh frequency harmonics, shaft speed sidebands, resonance peaks, and modulation envelopes, which are strongly linked to the underlying health state of the system [40,41]. However, conventional machine learning models often treat signals as arbitrary high-dimensional sequences, learning representations that do not necessarily reflect these mechanistic constraints.

In recent years, physics-guided learning has emerged as a promising direction to integrate mechanistic knowledge with data-driven inference. A common approach is to incorporate domain-informed signal representations that emphasize structural correlates of faults. For example, time–frequency representations such as the short-time Fourier transform (STFT), continuous wavelet transform (CWT), synchrosqueezed wavelet transform, and Hilbert–Huang transform have been used to highlight transient impulses, modulation patterns, and harmonic clusters associated with localized damage [42-44]. These representations serve as structured inputs to learning models, effectively embedding physics into the data domain. However, their performance still depends heavily on the appropriateness of chosen transforms and scale parameters.

A more integrated family of approaches seeks to incorporate physics directly into the model structure or learning objective. Physics-guided neural networks (PGNN) constrain network predictions or intermediate latent variables using governing dynamic relations, modal subspace structure, or spectral priors [45,46]. In rotating machinery diagnostics, spectral energy constraints, modal shape consistency, or envelope periodicity priors have been used to regularize representation learning, improving robustness under noise and limited supervision [47,48]. These methods align learned embeddings with physically meaningful feature spaces, producing diagnostically interpretable outputs and improved generalization. More recently, cross-domain representation learning has been proposed to disentangle shared and condition-specific components in vibration signals [49]. These techniques leverage multi-view or multi-resolution structures, treating machine health signals as composed of low-frequency operational trends and high-frequency structural fault indicators. By explicitly modeling how different frequency bands propagate diagnostic relevance, these methods reduce reliance on raw sample abundance and improve the model's ability to extract fault signatures consistently across operating conditions [50,51].

However, existing physics-guided approaches still face several open challenges. First, many rely on handcrafted constraints derived from simplified physical assumptions, which may not generalize across machine types or evolving fault processes. Second, while the inclusion of physics priors can stabilize training under limited data, overly rigid constraints may restrict the model's expressiveness. Finally, most current formulations treat physics priors as static, whereas degradation processes are dynamic and state-dependent. These challenges motivate the developments in later chapters of this dissertation, where physics guidance is embedded adaptively into representation construction, latent-space compression, and cross-scale feature synthesis, allowing models to preserve fault-relevant signatures while maintaining flexibility across operating conditions

### 1.2.3. Computational-Aware and Edge-Oriented Diagnostic Approaches

While physics-guided representations improve robustness and interpretability, real-world deployment of machinery health monitoring systems introduces additional constraints that are often overlooked in laboratory or benchmark settings. Industrial machinery commonly operates in distributed environments, where sensing nodes are embedded in controllers, edge devices, or maintenance terminals with limited computational resources, memory, and communication bandwidth [52,53]. In such settings, transmitting raw high-frequency vibration data to remote servers or cloud platforms is frequently infeasible due to bandwidth limitations, latency constraints, and energy consumption considerations [54]. Therefore, an important shift in recent research is the development of model-centric approaches that explicitly consider deployment constraints, particularly those involving on-edge processing, compression, online updating, and efficient inference [55].

The central challenge in resource-constrained PHM is to maintain diagnostic fidelity while reducing the computational and communication burden. Traditional digital signal processing (DSP)-based compression schemes, such as PCA, DCT, and wavelet packet decomposition, reduce data size efficiently but may lose local fault signatures and modulation features essential for early-stage diagnostics [56]. More recent autoencoder-based latent compression frameworks enable nonlinear dimensionality reduction while preserving semantic structure. However, these methods typically compress features uniformly, without considering that vibrational fault signatures are spectrally localized and sensitive to specific harmonics, sidebands, or transient bursts. To address this, researchers have explored adaptive and feature-aware compression, where model architectures selectively retain information from frequency regions that hold diagnostic relevance. Examples include attention-based feature selection, spectral-weighted reconstruction losses, and hierarchical token pruning in time–frequency representations. These methods reflect a broader trend in edge-cloud collaborative PHM, where lightweight encoders on the edge extract compact representations, and more computationally expensive processing is executed on cloud servers. Such frameworks preserve the interpretability and reliability emphasized by physics-guided methods, while also satisfying the operational and deployment requirements of industrial settings.

Another important dimension of resource-constrained PHM is the use of multi-task and decision-aware training, where compression is guided not only by numerical reconstruction error but also by downstream diagnostic performance. For example, a compression model may learn to preserve features that maximize classification accuracy or remaining useful life (RUL) prediction, effectively aligning representation learning with the application objective rather than purely signal fidelity. This integrates task-level relevance directly into the model optimization, ensuring that compression does not eliminate faint yet diagnostically critical modulation patterns. Despite these advancements, several limitations remain. First, many existing edge compression frameworks are not explicitly linked to the mechanistic origin of the preserved frequency bands, leaving their interpretability contingent on empirical evaluation rather than physical reasoning. Second, adaptive compression strategies often require temporal consistency constraints to avoid introducing artificial distortion into reconstructed waveforms. Finally, efficient deployment requires not only model compression but scalable tokenization, quantization-aware training, and multi-resolution decoding, especially for systems that exhibit non-stationary degradation behavior. These open challenges motivate the development of the physics-aware and decision-aligned signal compression framework proposed in Chapter 6 of this dissertation.

## 1.3. Dissertation Outline

Building upon the preceding review of data-centric, physics-informed, and computation-aware diagnostic frameworks, this dissertation advances a set of integrated methodologies aimed at enabling reliable condition monitoring and predictive maintenance under practical constraints commonly encountered in manufacturing systems, namely limited labeled data, heterogeneous sensing quality, structural complexity, and resource limitations in edge–cloud deployment. The work presented herein spans sensing, representation, learning, and deployment considerations, forming a progressive investigation into how diagnostic intelligence can be constructed in a manner that is adaptive, interpretable, and deployment-ready. The primary research components are outlined as follows:

- Propose a semi-supervised surface inspection framework that addresses the scarcity and imbalance of labeled defect images in industrial settings. The framework incorporates conditional generative augmentation in tandem with selective pseudo-labeling regulated by a confidence-driven weighting mechanism, thereby enabling the learning process to utilize structurally informative unlabeled samples while mitigating error propagation. This approach establishes an effective paradigm for improving classifier generalization under limited defect observations and heterogeneous imaging conditions.

- Develop a probabilistic sub-labeling method for composite material inspection in which continuous severity estimation is guided through discretized relational constraints embedded in the latent representation space. By inferring intermediate semantic states between annotated damage grades, the framework reduces annotation demands while maintaining interpretability, enabling reliable estimation of degradation progression even when expert labeling resolution is inherently coarse or uncertain.

- Establish a physics-guided dual-branch representation learning architecture for vibration-based gearbox fault diagnosis, where signal morphology is encoded through the fusion of temporal and spectral structural priors. The framework incorporates interpretable harmonic–sideband relations into the representation space, allowing diagnostic distinctions to be amplified even in early-stage fault regimes. This design demonstrates how physical insight can be embedded directly into the representational hierarchy of deep learning models.

- Introduce a graph-based deformation prediction framework for surface quality monitoring in milling operations, wherein structural geometry and spatial correlation among nodes are encoded via dynamic attention modulation. The method integrates physics-derived relationships between local deformation influences and global structural responses, allowing the predictive model to adaptively emphasize salient geometric interactions across machining configurations and process conditions.

- Formulate a resource-constrained signal compression and reconstruction scheme that enables edge-side vibration monitoring under stringent bandwidth and memory limitations. Through harmonics-aware tokenization and variable-resolution quantization regulated by diagnostic relevance, the approach achieves significant compression while preserving prognostic integrity. The decoding is guided by both physical consistency priors and task-level supervisory feedback, enabling reliable reconstruction and diagnosis in cloud environments.

- Demonstrate the scalability and generalizability of the proposed methodologies through comprehensive cross-dataset evaluations, ablation analyses, and deployment-oriented case studies, highlighting the robustness of the developed frameworks under variable sensing conditions, fault types, and operational complexities. The collective investigation confirms the feasibility of establishing actionable intelligence pathways that balance diagnostic fidelity, interpretability, and computational practicality.

# Chapter 2.  GANs Fostering Data-Augmentation for Automated Surface Inspection with Adaptive Learning Bias

In manufacturing, visual inspection of parts' surface is traditionally an important examination before the parts can proceed to the next manufacturing step. For example, the timely detection of minor surface defects, such as dents and scratches, in small-sized airfoils of aircraft engine is typically the final stage of quality assurance before acceptance for assemblage. While such a process is critically important, current practices rely heavily on human operator's judgement, which is subjective and labor-intensive. In this study, we establish an automated, image-based inspection system that utilizes robotic automation to acquire high-resolution images of the parts under inspection and employs a specifically tailored machine learning technique to facilitate decision making of inspection. Leveraging deep learning as the underlying methodology, we address a key challenge in flexible automation of surface inspection, i.e., the scarcity of labeled data during the initial training process. In other words, we tackle the challenge of limited samples with known defects. Specifically, we synthesize an adaptive semi-supervised learning framework, building upon the residual neural network (ResNet) and the deep convolutional generative adversarial network (DCGAN) to extract features from both ground truth and synthetic data. This approach can overcome the shortcomings of the current approaches, leading to more objective and accurate defect detection right from the beginning of implementation with a small labeled dataset. Our results show that the overall classification accuracy on this challenging dataset reaches 92.30%, a 27.79% improvement over the baseline model achieved through optimal use of synthetic and ground truth data. The system also investigates the impact of synthetic data, providing guidelines for integrating it effectively into iterative training. This approach offers a robust solution for surface inspection and quality assurance in diverse manufacturing applications.

## 2.1. Introduction

Part surface inspection is an essential procedure in precision manufacturing, facilitating quality control and safety assessment of the final product. Surface defect detection is critical across industries where product integrity is essential, such as aerospace, automotive, electronics, and heavy manufacturing. For example, in aerospace, even minor surface defects on turbine blades can lead to airflow imbalances and potential failures [57], compromising safety. In automotive manufacturing, surface inspection ensures paint consistency and identifies any structural micro-damage, maintaining both aesthetic and functional standards. Electronics manufacturing uses surface detection for quality control in components like solder joints to prevent circuit failures. Given the high-stakes applications, precise, reliable inspection systems are essential to uphold quality and safety standards. Hence, precise visual inspection is indispensable in the manufacturing process to filter out faulty parts before final acceptance.

The field of surface defect detection has evolved significantly from labor-intensive manual inspections to advanced automated systems that improve precision and efficiency. Early methods primarily relied on visual inspection, where accuracy was dependent on inspectors' expertise, introducing variability and limiting scalability. Automated optical inspection (AOI) and machine vision systems were subsequently developed to reduce human dependency, offering faster and more consistent results. Part surface inspection traditionally involves manually scanning surfaces with a convex lens, where inspectors rely heavily on experience and training for accurate judgment [58,59]. To enhance inspection precision, a range of advanced methods supplements manual techniques. Thermography [60], for example, uses heat patterns to identify surface flaws, making it particularly useful for detecting subsurface irregularities, though it requires controlled conditions for optimal accuracy. Building on visual automation, machine vision systems [61] improve efficiency by automating defect recognition, yet they remain sensitive to lighting and reflective surfaces, requiring precise calibration. Complementing these, laser scanning [62] delivers high-resolution surface mapping, though reflective or glossy materials can cause measurement distortions. Magnetic optical imaging [63] offers an alternative, detecting subsurface flaws in conductive materials; however, it's best suited for materials with favorable magnetic properties and appropriate thickness. For detailed structural imaging, scanning electron microscopy (SEM) [64] achieves micro-level resolution, though it demands significant operator skill and time. Lastly, diffraction-based techniques such as X-ray [65] and fluorescent probes [66] reveal fine structural details, especially valuable in high-stakes applications, but they necessitate rigorous safety protocols and expert interpretation. Collectively, these methods extend inspection capabilities while highlighting the need for skilled oversight and specific environmental conditions, making trained inspectors essential even with advanced technological support. Moreover, even with ample inspection experience, the accuracy of inspector is estimated at approximately 85% [67], given the subjective and fatiguing nature of the task. Obviously, there is a pressing need for a reliable and objective inspection system that can overcome these limitations. Instinctively, an automated robotic system is a technology-driven inspection tool designed to autonomously perform precise, repeatable tasks without human intervention. These systems typically incorporate robotic arms, sensors, machine vision, and AI algorithms, allowing them to inspect complex parts with consistent accuracy. For instance, La et al [68] describes a climbing robot used for inspecting steel structures and bridges; equipped with sensors, it collects real-time data, transmits it to a ground station, and uses image stitching and 3D mapping to detect surface defects. Menendez et al [69] highlights the ROBO-SPECT system, which performs tunnel inspections using a mobile vehicle, a crane, and a robotic arm. This system integrates computer vision and ultrasonic sensors to assess structural integrity by measuring crack width and depth. These examples illustrate the components and functionality of automated robotic systems, which leverage advanced sensors, mobility, and data-processing capabilities to enhance inspection accuracy and efficiency, while minimizing human involvement. Indeed, a number of automated visual defect inspection systems have recently been investigated. Colvalkar et al [70] developed an in-pipe inspection robot equipped with machine vision and image processing capabilities, designed to detect various internal pipe abnormalities, such as blockages and cracks, while navigating autonomously through the inspection area. In manufacturing contexts, such as mobile phone screen glass production [71], an advanced surface detection system equipped with machine vision technology can be employed to identify even minute fissures. When juxtaposed with conventional labor-intensive inspection methods, these automated surface inspection systems [72-74] exhibit markedly enhanced efficiency and superior performance, particularly in the meticulous identification of minor defects in precision components.

The intelligent component of the automated inspection system, e.g., the deep learning framework, necessitates extensive pretraining with ample and high-quality training data to provide accurate and instantaneous predictions on surface images. Nevertheless, obtaining high-resolution images with defects accurately captured from chosen angles requires each individual part to pass through the entire image acquisition system. This process renders the collection of sufficient, pertinent data both arduous and time-consuming, particularly in the context of deep learning, where the foundational dataset must be substantial. Furthermore, the highly imbalanced distribution in the data may exacerbate training performance, as the prevalence of defects on recycled scrap sample blades is largely contingent upon manufacturing equipment and environments. To surmount the issue of limited data size, Huang et al [75] employed a transfer learning-based method, leveraging knowledge from the source task for synthetic aperture radar target classification tasks. Transfer learning has also demonstrated efficacy in fault detection for gear transmission systems when data size is constrained by high data generation costs [76]. However, transfer learning cannot guarantee satisfactory performance when the source and target data exhibit minimal similarities. Recently, deep learning models such as VGG16, Vision Transformers (ViT), and ResNet18 have demonstrated effectiveness in defect detection, although challenges remain with data demands and real-time processing capabilities. VGG16 [77] is effective for complex feature extraction but requires considerable computational resources. In contrast, ViT [78] employs self-attention mechanisms for detailed image segmentation, which can improve model accuracy with large datasets, though it is sensitive to data scarcity. ResNet18 [79], with its residual connections, efficiently manages deeper architectures and mitigates issues like vanishing gradients, though it may face limitations with high-resolution image processing. Despite advancements in deep learning, data scarcity in high-stakes applications like surface inspection presents a significant challenge. Data augmentation has emerged as a critical solution, evolving from traditional non-generative methods to advanced generative techniques. Non-generative augmentation, such as image rotation and scaling, expands datasets but risks overfitting by reusing similar information. Generative approaches, like Generative Adversarial Networks (GANs) [80,81], simulate diverse, realistic defect images, enhancing dataset variety and improving model robustness. Bowles et al [82] demonstrated that GANs could provide a viable and dependable means of extracting additional information from a medical imaging dataset by generating synthetic samples that closely resemble real images.

Incorporating both real and generative data samples into deep learning model training necessitates reliable labeling of unlabeled samples, as this can significantly impact the ultimate performance. Semi-supervised learning techniques [83-85] address data limitations by leveraging both labeled and unlabeled data. Traditional methods assign pseudo-labels to unlabeled data, with approaches such as those introduced by Lee et al [86], which use labeled data to predict labels for unlabeled samples. While effective, such approaches risk error accumulation due to reliance on initial predictions; erroneous predictions could exacerbate self-training [87] outcomes. In response, adaptive frameworks that combine GANs and deep models, like ResNet, have been developed to improve labeling accuracy by incorporating multiple classifiers and adaptive weight strategies. These advancements ensure that even in constrained data scenarios, high model performance is maintained through reliable label assignments. Consequently, we propose an adaptive semi-supervised learning framework that integrates GANs and ResNet [88,89] to simultaneously generate samples and determine the most reliable labels for generative samples using multiple classifiers. The labels assigned to the generative samples will reference decisions made by multiple classifiers, each weighted differently. The pseudo-labeled generative data will then be combined with ground truth labeled data to feed the ResNet, which will perform supervised learning. The proposed semi-supervised learning framework [90-92] is applied to self-collected, image-based surface defect data of engine blades. Results indicate that it surpasses most deep learning models in multi-class classification when faced with limited data sizes. Furthermore, we implemented an adaptive weight strategy to manage learning biases between ground truth data and synthetic data within the custom loss function of the discriminative neural network. This approach facilitates the model's emphasis on the underlying patterns in the ground truth data while concurrently providing a mechanism to fine-tune the deep learning model for diverse scenarios with varying quality control requirements.

The remainder of this paper is structured as follows: Section 2.2 delineates the automated inspection system configuration. Section 2.3 expounds upon the intricacies of the adaptive semi-supervised learning framework designed to address limited data size. Section 2.4 offers an in-depth case investigation supported by experimental analysis. Finally, Section 2.5 delivers concluding observations and remarks.

## 2.2. Automated Intelligent Inspection System Setup

The establishment of an automated intelligent surface inspection system is essential for achieving consistent, accurate, and efficient quality control in manufacturing processes. By integrating automation and intelligent algorithms, we aim to build a robust system capable of real-time analysis, which significantly reduces inspection time and enhances reliability. Our automated inspection system combines a data acquisition component with a collaborative robot, allowing precise control and repeatable inspection movements. This setup also includes a decision-making system that preprocesses raw images and generates reliable predictions through a deep learning model. In this section, we will elaborate on how we addressed the challenges encountered during the setup of the automated data acquisition system. Specifically, we discuss the integration of the robot arm, high-resolution camera, and image preprocessing algorithms, each contributing to an end-to-end solution that is both intelligent and adaptable to various inspection requirements.

### 2.2.1. System Configuration

The full-spectrum configuration of the vision-based apparatus for blade surface inspection is illustrated in Figure 2-1. This system begins with the robotic arm, equipped with grippers, which securely transports each blade into the optical system and precisely positions it in the required orientations for thorough inspection. The robotic arm's accuracy ensures that every critical surface of the blade is appropriately aligned for optimal imaging. Once positioned, an auto-focus system engages, enabling the camera to concentrate on specific subregions with high precision, capturing detailed images essential for accurate analysis. After acquisition, these raw images are transmitted to an image processing workstation. Here, essential information such as the part number, subregion identifiers, and inspection timestamps are systematically recorded and stored in a database, ensuring traceability and facilitating data management for subsequent analysis. Following this data logging, the images proceed to the defect detection system, where they undergo rigorous processing. Advanced algorithms analyze the images, classifying any identified anomalies into distinct categories based on predefined criteria, such as defect type and severity. This structured workflow enhances inspection accuracy, ensures consistent quality control, and provides extensive defect dataset for ongoing process improvements.

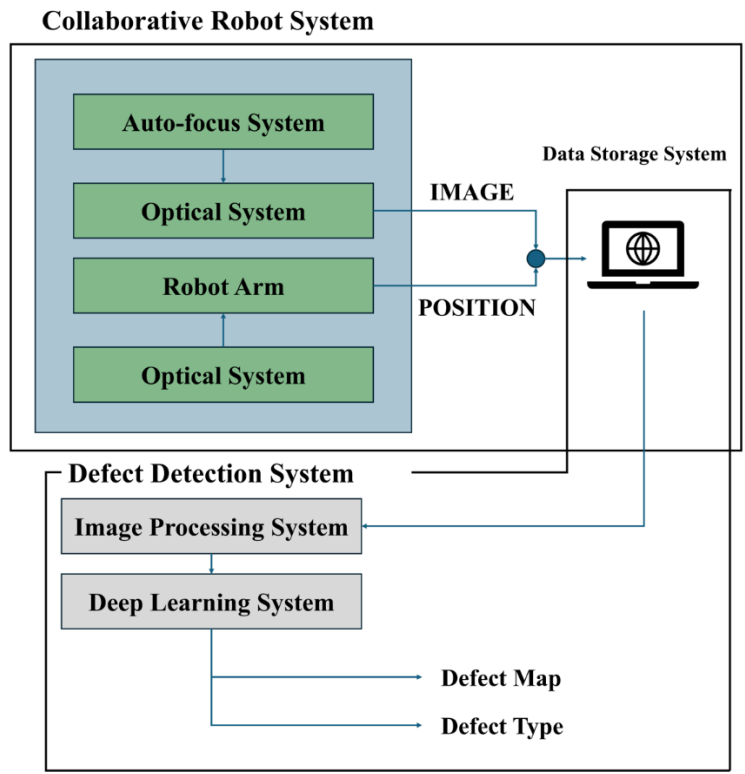

**Figure 2-1.** Architecture of automated intelligent inspection system.

*Auto-Focus System*

Owing to their curvature and semi-specular surface finish, the blades pose significant imaging challenges, particularly in terms of achieving full surface coverage and uniform illumination. The camera hardware's limited depth of field, coupled with the blade's curved geometry [93], diminishes the surface area that can be captured in a single image while maintaining acceptable quality. Moreover, the curved semi-specular blade surfaces non-uniformly scatter light, resulting in uneven lighting distributions across the component. Therefore, the chosen camera is demanded to be capable of capturing high-resolution macro images with sufficient fidelity to be discernible as the defects on the blade surface can have dimensions as minor as 1/64", and importantly it can be compatible with the proposed automated inspection system to implement the automated imaging with auto-focus and subsequent file transfer. For this system, we opt for a Canon EOS 6D Mark II full-frame DSLR camera, combined with a Canon 24-70mm lens featuring macro functionality. This camera lens pairing can satisfy the system requirement and provide a broad range of optical and imaging adjustments. Table 2-1 refers to the key specifications of the camera and lens employed in the system.

**Table 2-1.** Specifications of the camera and lens

| Canon EOS 6D Mark II | |
|---|---|
| Image Resolution | 6240 x 4160 pixels |
| ISO Range | 100 - 4000 |
| Shutter Release Interfaces | External trigger port |
| Flange Focal Length | 44 mm |
| Lens | Canon EF 24-70mm f/4L IS USM |
| Aperture Range | f/4.0 – f/22 |
| Focal Length | 24 – 70 mm |

*Optical System*

Capturing high-quality images of parts requires careful consideration of illumination, as it is a critical factor that directly affects image clarity and accuracy in defect detection [94,95]. Ensuring that parts are evenly illuminated is essential to avoid over- or under-exposure of localized regions, which could lead to the loss of important visual information [96]. Uneven lighting can result in shadows or bright spots that obscure defects, making them harder to detect. Adequate and controlled illumination is also necessary to minimize interference from external lighting conditions, such as ambient light or reflections from nearby surfaces. In addition, the lighting setup must ensure that the exposure level is high enough to provide the camera sensor with sufficient light. This reduces noise caused by slower shutter speeds or increased camera sensitivity (ISO), both of which can degrade image quality and hinder the defect detection process. The specific surface characteristics and complex geometries of the blades introduce additional challenges in achieving uniform illumination. Blades often have semi-specular, reflective surfaces, which are prone to glare and unwanted reflections when exposed to bright-field illumination. These reflections can obscure defects or create artifacts that mimic the appearance of defects, making accurate inspection more difficult. Furthermore, the concave and curved geometries of the blades focus light in certain areas, further complicating the lighting strategy. Bright-field illumination, which typically uses direct light sources, can cause light to concentrate unevenly on these surfaces, while dark-field illumination, which relies on shallow angles of light, can be difficult to implement due to the blade's curved and intricate shapes.

To address the challenges of glare and reflection, the parts are not directly illuminated by lighting sources, as localized light sources would generate significant glare. Instead, the parts are situated within an opaque shooting tent that diffuses light more evenly across the component [97]. Multiple light sources are strategically positioned outside the shooting tent to ensure the most uniform illumination possible across the part surface, as depicted in Figure 2-2. The lights utilized are 60W fluorescent lamps with metal reflectors. Furthermore, a matte black backdrop is placed at the bottom of the shooting tent to maximize the contrast between the part being imaged and the background, a crucial factor when aligning blades in the image processing pipeline. Figure 2-3 presents conforming and non-conforming images that illustrate the challenge we aim to overcome through proper illumination. It is evident that the non-conforming image appears considerably brighter than the conforming one; however, light reflection can obscure defect details, particularly when the surface defect is too minute to detect.

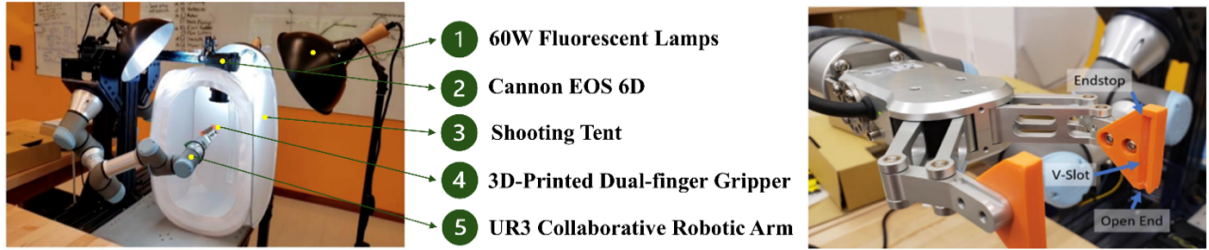

**Figure 2-2.** Collaborative robotic system.

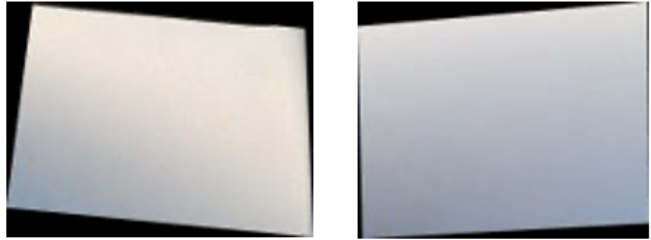

**Figure 2-3.** Illumination illustration (a) non-conforming; (b) conforming.

### 2.2.2. Collaborative Robot

To capture high-quality images of multiple regions along the blade surface, the blade must be accurately positioned in front of the camera for imaging. Blade positioning must be repeatable to ensure that the region of the surface being imaged falls within the focus of imaging system. As mentioned for the auto-focus system, the curvature and semi-specular surface finish render it prohibitively impossible to capture the blade surface in one single image while maintaining the high quality, so the blades will be divided into 12 subregions for front and back sides of surface, corresponding to 24 unique blade positions in front of the camera. In this research, we employ the UR3 collaborative robotic arm [98,99] to offer a swift and repeatable solution for positioning the blade in front of the camera. It can be observed from Figure 2-2 that the robot arm is programmed to position the blade right below the camera and adjust the gestures of the blades to the preset ones at fixed intervals, eventually triggering the camera to capture each subregion in focus.

Additionally, the OnRobotics RG2 Gripper attachment is employed on the end of the UR3 robotic arm to securely hold and manipulate the blades during the inspection process. This gripper is crucial for ensuring stable and repeatable positioning, which is fundamental to achieving consistent inspection results. To further enhance this functionality, custom-designed gripper fingers were developed specifically to interface with the base of the blades. These custom gripper fingers, illustrated in Figure 2-2, were carefully designed to ensure that each blade can be precisely and consistently positioned for inspection. The gripper fingers are 3D printed to allow for rapid prototyping and adjustments based on the unique geometries of the blades. They feature a V-slot design, which is open at one end and closed at the other end with a stop, effectively cradling the base of the blade. This design ensures repeatable and accurate alignment between the blade and the gripper, minimizing the risk of positional shifts during handling. Precision in this alignment is essential because even minor misalignment between the blade and the gripper can lead to misalignment in the object plane during imaging. Such misalignment would result in translational shifts that affect the focus and cropping of the final image, potentially compromising the accuracy of defect detection. By ensuring repeatable positioning, the gripper design contributes to the reliability and consistency of inspection process.

### 2.2.3. Image Preprocessing

The workflow of image processing is presented in Figure 2-4. Owing to the intricate geometry of stator vanes, only specific sections of the blade will be in focus within a single raw image, as presented in Figure 2-4. Correspondingly, these raw images encompass background regions and areas of the part out of focus. As these regions are not pertinent to the inspection task, they have to be eliminated from the image before being input to the deep neural network for defect detection. To circumvent this issue, we employ the bit masking process that utilizes calibrated bitmasks for each distinct part and camera pose, extracting only the relevant, well-focused regions from the raw images. Consequently, we devise 12 poses for capturing in-focus sections, with raw images assigned a corresponding number of bitmasks indicating the sections for which the images are generated. These bitmasks are established during the initial setup and calibration.

In order to systematically extract the pertinent sections from each raw image, the relevant sections within each image must first be defined. Each unique camera pose in the image capture process is calibrated to yield high-quality image data from the multiple predefined subsections on the part surface. The masking process initially identifies sets of pixels within an image from a specific camera pose that correspond to a given subsection on the part. These sets of pixels are defined as binary masks, or bitmasks, which are binary matrices of the same height and width dimensions as the image being masked. Entries within the bitmask matrix determine the set membership of the corresponding pixels in the masked image. Bitmasks simplify the masking process significantly, as a straightforward element-wise multiplication of the bitmask with each color channel of the image leads to the removal of the unimportant sections of the raw image, as demonstrated in Figure 2-4. The calibration part is delineated using blue tapes to demarcate section divisions to create the bitmasks, and subsequently imaged by the robotic cell from pre-established poses to achieve full in-focus surface coverage of all the sections on the blade surface. These calibration images are then imported into image editing software, where the part subregions can be traced, and corresponding bitmasks generated. The area encompassed within the region is filled with white, while the area outside of the region is filled with black, where white regions represent the inspection area and black regions correspond to the non-inspection area. Finally, considering that the inspection regions occupy only a small area within the raw image, the masked image is cropped to minimize the non-useful black area. An association between bitmasks, imaging poses, and subregions is catalogued for reference during data collection and inspection phases. This bit masking approach is readily scalable to accommodate multiple part variations, requiring only an additional mapping for each part variety.

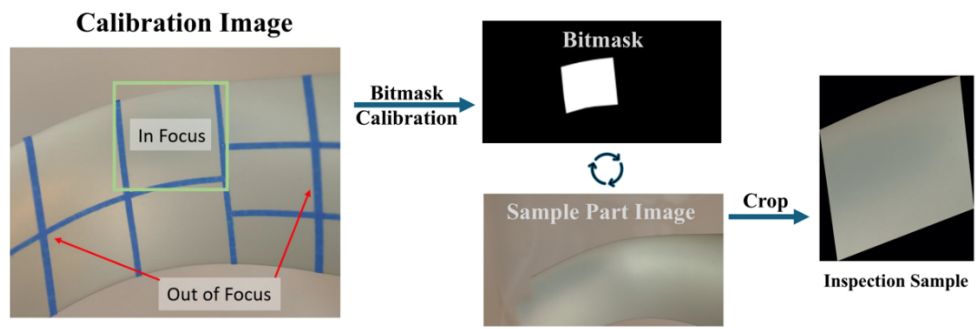

**Figure 2-4.** Workflow of image preprocessing

## 2.3. Adaptive Semi-supervised Learning Framework

The decision-making algorithm built upon an adaptive semi-supervised learning framework is presented in detail in this section.

### 2.3.1. Conditional Deep Convolutional Generative Adversarial Network (cDCGAN)

Humans can innately discern similar patterns from a collection of unlabeled objects; however, deep learning models must identify the most relevant features from an extensive array of samples with the assistance of ground truth labels. Consequently, Generative Adversarial Networks (GANs) are devised to uncover essential knowledge from unlabeled samples and reproduce samples within the same category to achieve unsupervised learning. GANs consist of two distinct models: the Generator model (G) and the Discriminator model (D), which compete against each other during the learning process. The G model endeavors to generate samples as closely resembling real samples as possible, utilizing feedback from the D model. Concurrently, the D model diligently differentiates the generated samples from the real samples. Ultimately, the outputs from the G model are anticipated to become indistinguishable from the real samples, and the D model will converge towards a neutral prediction of 50%, the random prediction accuracy serving as a sanity check performance in binary classification.

In the binary classification of fake or real samples, cross-entropy is favored in the D model to assess the similarity between the sample distribution from the G model and the authentic data distribution, which can be expressed as:

$$H(p,q) = -\sum p_i \log(q_i) \qquad (2.1)$$

where p and q represent the different distributions, we proposed to distinguish. If we substitute the G model and D model into the Equation (2.1), we can have

$$H\left((x_i, y_i)_{i=1}^{N}, D\right) = -\sum_{i=1}^{N} y_i \log(D(x_i)) - \sum_{i=1}^{N}(1-y_i)\log(1-D(G(z))) \qquad (2.2)$$

In Equation (2.2), $x_i$ refers to the input of the real samples and $z$ represents the input samples which are subjected to the a priori noise distribution while $y_i$ indicates the real sample distribution. Correspondingly, the first term on the righthand side of Equation (2.2) can be considered the metric to analyze if all the real samples are classified into the correct class and the second term on the righthand side of Equation (2.2) is presenting if all the generative samples are categorized exactly. This can explain why the absolute value of Equation (2.2) is minimized for G model and maximized for D model, so it can be expressed finally as,

$$\min_{G} \max_{D} V(D,G) = E_{x \sim P_{data}(x)} \left[ \log(D(x)) \right] + E_{z \sim P_{z}(z)} \left[ \log(1 - D(G(z))) \right] \tag{2.3}$$

It is noteworthy that the minimization for the G model or the maximization for the D model will be executed when the counterpart model is frozen, and the models will alternate between being frozen and trained during the learning process. Moreover, Convolutional Neural Networks (CNNs) [100] are incorporated into GANs [101-103] as Deep Convolutional GANs (DCGANs) [104] for unsupervised representation learning, given that convolutional layers have demonstrated success in image pattern recognition by extracting feature maps from the original image inputs, while GANs exclusively employ fully connected layers to discern correlations between real sample distributions and a priori noise distributions. Furthermore, transposed convolutional layers are amalgamated with the G model to upsample data to a larger output feature map, in contrast to the objective of standard convolutional layers, which downsample inputs for a more representative feature map.

It is important to note that GANs or DCGANs are not designed for image-based multi-class classification problems; thus, they cannot generate specific types of data during training, and likewise, the D model is unable to determine which classes the real samples or generative samples should be classified into. However, multi-class classification problems are likely to be one of the most common situations encountered in practice. Consequently, a condition is introduced to the overall architecture of the DCGANs, ensuring that the G model and D model are aware of the type of data they are handling. As a result, labels will be designated for the random inputs before they are mapped to the generative images, enabling the generator to know which types of defects to generate each time. Simultaneously, the labels accompanying the generative samples will inform the discriminator of the defects the sample passed to it represents. Equation (2.3) is subsequently modified as follows:

$$\min_G \max_D V(D,G) = E_{x \sim P_{data}(x)} \left[ \log(D(x|y)) \right] + E_{z \sim P_z(z)} \left[ \log(1 - D(G(z|y))) \right] \quad (2.4)$$

The vector y encoding the label information is added to the input for generator and discriminator, by concatenating it in the model as an additional channel or feature map. The D model will subsequently evaluate whether the samples can be paired with the labels as well as whether the samples are fake or real. The model architecture of the cDCGAN is illustrated in Figure 2-5.

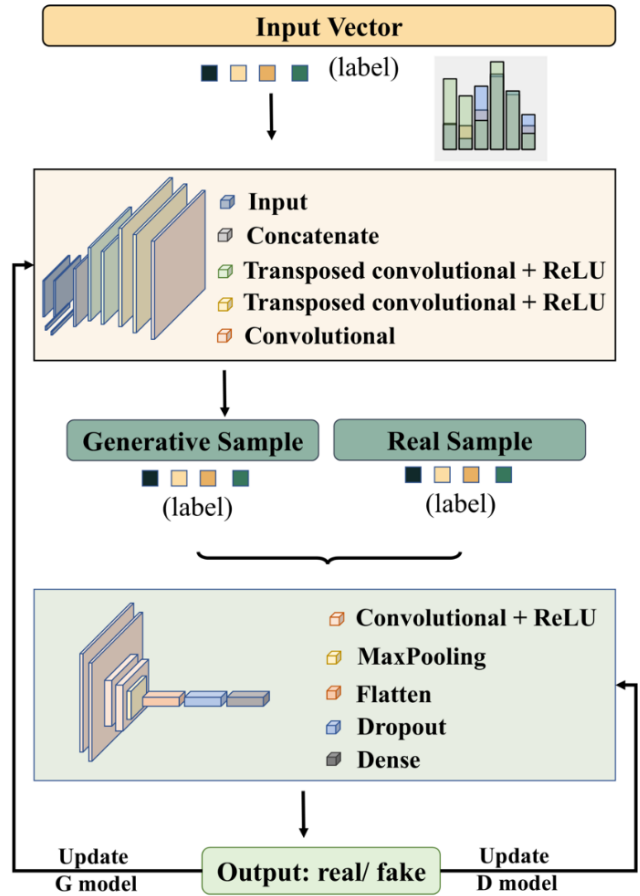

**Figure 2-5.** Model architecture of cDCGAN.

### 2.3.2. Bias-aware ResNet

As CNNs have achieved remarkable success in image-based classification, they are progressively employed in diverse research fields for image-related tasks. Intuitively, the layers learn increasingly complex features as the network becomes deeper, necessitating a deep learning model capable of extracting higher-level features from inputs, particularly in high-dimensional classification problems with an abundance of samples containing intricate and distinct information. However, the gradient vanishing or exploding problem [105] restricts the continuous addition of layers to the neural network architecture, potentially leading to a scenario where weights at the output end are updated rapidly, while those at the input end are adjusted extremely slowly. This implies that the weights of the top layers are barely updated after the total training epochs, and the bottom layers continue learning from the random inputs passed from the top layers. A critical factor causing gradient vanishing or exploding [106,107] is the inevitable exponential increase or decrease of the gradient during backpropagation via chain derivatives, as the gradient is repeatedly multiplied with other gradients. Consequently, the residual block [108,109], as illustrated in Figure 2-6, is introduced to alleviate the problem of training very deep networks.

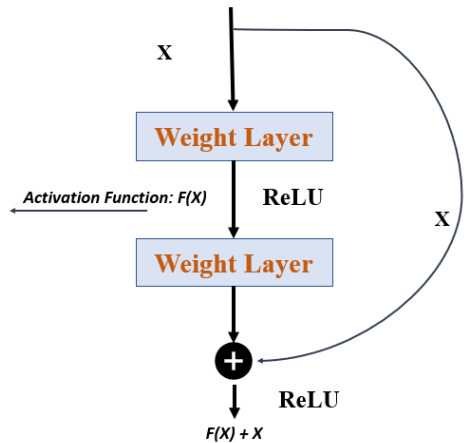

**Figure 2-6.** Residual block.

Within the residual block, a key component is the skip connection, which plays a crucial role in improving the flow of information through the network. This skip connection works by directly adding the output from earlier layers to the input of subsequent layers, allowing raw or less processed information to bypass intermediate layers. This approach ensures that important features are retained and passed on to deeper layers, where they can be further examined or refined. The ability to carry forward information from earlier layers mitigates the risk of vanishing gradients, a common issue in deep networks, where the magnitude of gradients diminishes as they propagate backward, thus hindering effective learning in deeper layers. Moreover, the residual block employs individually assigned weights to the identity mapping (the skip connection) to better align the input information with the output of the residual block. This ensures that the residual, or the difference between the input and output of the block, is properly matched and combined with the output of the main network branch. By making this identity mapping adaptable, the network can dynamically adjust the contribution of the skipped information, ensuring it integrates smoothly with the residual output. This method significantly aids in the training process, enabling the network to be deeper and more powerful without suffering from the degradation problem, where deeper networks tend to perform worse. As a result, the use of residual blocks allows for the construction and efficient training of very deep neural networks capable of capturing more complex patterns and representations in data.

*Bias-aware*

The ResNet [110-112] was proposed for integration into a semi-supervised learning framework, necessitating its ability to concurrently handle both labeled and unlabeled samples. However, even after thousands of training epochs, the cDCGAN cannot guarantee that the generated samples for each class perfectly conform to the real data distribution, let alone the potential uncertainty and instability of prediction by cDCGAN. Consequently, the ResNet is designed to treat real and generated samples differently, mitigating overfitting to inaccurate information contained within the generated samples, which constitutes the confirmation bias that must be addressed when modifying the ResNet. Intuitively, real samples warrant deeper examination, while generated samples, considered less reliable, serve as low-fidelity data for performance enhancement. Therefore, the feedback from real and generated samples for updating layer weights in the learning process should be recalibrated separately to alleviate the confirmation bias [113]. The loss function for classification problems is commonly defined as the categorical cross-entropy,

$$L = -\sum_{i=1}^{N} y_i \log\left(P(x_i)\right) \qquad (2.5)$$

where the labels $y_i$ for each input are multiplied with the logarithm of the prediction probability. In the semi-supervised learning, we mix up the real samples and generated samples, thus assigning different weights for the loss by real samples and generated samples. Correspondingly, the loss function is

$$L^* = \sum_{i=1}^{N} y_i \log(P(x_i)) + \alpha(t) y_i' \log(P(x_i')) \tag{2.6}$$

where $L^*$ represents the overall loss that combines the loss of real samples and generated samples while $x_i'$ refers to the generated samples. The weight $\alpha(t)$ is utilized to control the contribution of generated samples to the overall loss.

*Time-varying weights*

Filtering out low-quality data, which only provides indistinct and confusing features and patterns for the model to classify during the training process, is essential. Instead of discarding all generated samples for which the ResNet yields different predictions from the labels assigned by the G model, we propose retaining all generated samples while fine-tuning the sample rate based on updated predictions during the learning process. This approach counters the assumption that the best performance has been achieved before semi-supervised learning begins, which is clearly not the case. Therefore, we specify the weight as a time-dependent function, which increases gradually during training. This allows the model to acquire baseline classification capabilities for defects when initially trained with real samples only. As the model's performance improves over time, its predictions become more reliable, and the weight increases, placing more emphasis on the generated loss within the overall loss. The weight varies over time as follows:

$$\alpha(t) = \begin{cases} 0, t < T_1 \\ \dfrac{t - T_1}{T_2 - T_1} \alpha_{\max}, T_1 \leq t \leq T_2 \\ \alpha_{\max}, t \leq T_2 \end{cases} \tag{2.7}$$

where $T_1 = 50$, $T_2 = 150$ and $\alpha_{\max} = 2$.

Starting from epoch T1, the trained ResNet begins to play a critical role in refining the sample selection process. After each epoch, ResNet provides predictions for the generated samples, helping to assess their quality and consistency. If there are any samples where the labels predicted by the ResNet are inconsistent with those assigned by the discriminator (D model), these samples are excluded from the subsequent training process. This filtering mechanism ensures that only the most reliable and correctly labeled samples contribute to further training, improving the overall robustness and performance of the model. As training progresses, the predictions made by ResNet are continuously updated to reflect the model's growing accuracy and learning capacity. This dynamic adjustment ensures that the pool of training samples remains relevant and high-quality. The system reselects samples based on these updated predictions, ensuring that only those samples that meet the criteria for consistency and quality are fed into the training loop. This selective process prevents noisy or incorrectly labeled samples from undermining the training process, thereby maintaining the integrity of the learning process. It is important to emphasize that the finalized generated samples produced by the generator (G model) must pass through an additional quality check by the discriminator (D model). This step is crucial because low-quality or suboptimal fake samples can degrade the overall quality of the sample pool if not filtered out. By allowing the discriminator to perform this final verification, the model ensures that only high-quality, realistic samples are used for training, thus safeguarding the model from being influenced by poorly generated data. This strategy contributes to the stability and effectiveness of the training process, enabling the generator to improve its ability to produce realistic and high-quality samples over time.

### 2.3.3. Adaptive Semi-supervised Learning Framework

Employing the modified ResNet designed for joint training with real and generative samples, we establish a semi-supervised learning framework, as depicted in Figure 2-7. In the semi-supervised learning framework detailed in Figure 2-7, a generative adversarial network (works in conjunction with a classifier to enhance the model's ability to detect blade defects. The methodology commences by partitioning the ground truth dataset (GTD) into subsets for training-validation and testing. The classifier undergoes initial training with the training-validation subset, acquiring a foundational ability to recognize blade defects from real, labeled samples. Concurrently, the GAN composed of a generator and a discriminator, undergoes training with the training GTD in an unsupervised manner. The goal is to generate synthetic data (SD) that is qualitatively on par with the GTD.

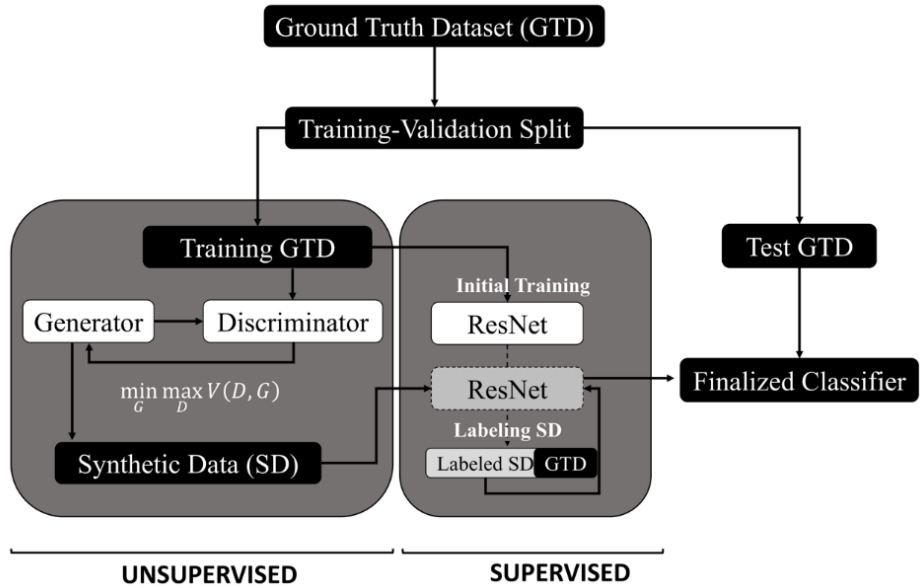

**Figure 2-7.** Adaptive semi-supervised learning framework.

The generator produces new samples while the discriminator critically evaluates them against the real data, incrementally refining the generator's output. This process leads to an iterative cycle where the generated SD is used to make predictions with the classifier. These predictions are evaluated against the SD's original labels, and samples that generate confusion, where predictions deviate significantly from original labels, are culled. The classifier is then retrained with the curated mix of GTD and the high-fidelity SD, enhancing its ability to accurately detect defects. A critical element of this framework is the time-varying weight mechanism that adjusts the synthetic data's influence within the loss function during training. This dynamic weighting allows for the modulation of the synthetic data's significance, depending on the abundance of GTD and the model's evolving proficiency. When a substantial GTD is present, the weight of the synthetic data loss may be reduced to focus on minimizing fault escape. Conversely, when GTD is scarce, an increase in the weight of synthetic data loss—or the creation of additional SD—serves to diversify the training dataset and mitigate the risk of overfitting. The final stage involves validating the retrained classifier against the test GTD to ensure it adheres to the required quality control standards for blade defect inspection. This framework is adaptable, capable of accommodating varying amounts of GTD and different quality control requirements. The incorporation of specialized architectures, such as the proposed cDCGAN and bias-aware ResNet, suggests a customized approach to addressing class imbalances and generating condition-specific synthetic data. Overall, this adaptive framework meticulously merges real and synthetic data to calibrate the classification model's training, aiming for robustness, generalizability, and high accuracy in defect detection to meet the precise demands of quality control tasks. Furthermore, the proposed cDCGAN and bias-aware ResNet architectures are detailed in Table 2-2.

**Table 2-2.** Specifications of the proposed framework

| Sequence | Layer (Size) | Layer (Size) |
|:---:|:---:|:---:|
| | **Generator** | |
| 1 | InputLayer | LabelInputLayer |
| 2 | Dense | Embedding |
| 3 | ReLU | Dense |
| 4 | Reshape | Reshape |
| 5 | Concatenate | |
| 6 | TransposeConv2D (4, 4, 128) + ReLU | |
| 7 | TransposeConv2D (4, 4, 128) + ReLU | |
| 8 | Conv2D (32, 32, 1) + ReLU | |
| | **Discriminator** | |
| 1 | InputLayer | N/A |
| 2 | Embedding | N/A |
| 3 | Dense | N/A |
| 4 | Reshape | LabelInputLayer |
| 5 | Concatenate | |
| 6 | Conv2D (3, 3, 64) + ReLU | |
| 7 | Conv2D (3, 3, 128) + ReLU | |
| 8 | MaxPooling2D (3, 3) | |
| 9 | Flatten | |
| 10 | Dropout | |
| 11 | Dense | |
| | **ResNet** | |
| 1 | Conv2D (7, 7, 64) | |

| | |
|---|---|
| 2 | 2 × Conv2D (3, 3, 64) |
| 3 | 2 × Conv2D (3, 3, 128) |
| 4 | 2 × Conv2D (3, 3, 256) |
| 5 | 2 × Conv2D (3, 3, 512) |
| 6 | AveragePool |

## 2.4. Experimental Studies

In this section, we present the application of our semi-supervised learning framework, combining Generative Adversarial Networks (GANs) and Residual Networks (ResNet), for image-based defect detection on metallic surfaces of engine blades. The framework leverages GANs to generate synthetic samples to augment the limited labeled data, while ResNet performs accurate predictions and classifications. Data acquisition is conducted through an automated intelligent surface defect inspection system, which uses high-resolution cameras and a robotic arm to capture images from various angles. The acquired images form the basis for training and testing the framework. GANs generate additional samples to enrich the dataset, and ResNet handles defect detection on both real and generated images. This approach improves detection accuracy, especially when labeled data is scarce, while also reducing the need for extensive manual labeling. We present both qualitative and quantitative results to showcase the effectiveness.

### 2.4.1. Experimental setup

The automated intelligent surface inspection system meticulously captures images of engine blades affixed to the robotic gripper. As illustrated in Figure 2-4, the blade surface is divided into 12 distinct focus areas, corresponding to the 12 positions adopted for imaging, to guarantee thorough defect inspection coverage. This imaging sequence is methodically executed, collecting sectional images for each blade. Following a rigorously defined protocol not only organizes the workflow but also enhances efficiency. It's imperative to note the requisite technical acumen for precise defect categorization within these images. While undamaged blades typically display no imperfections, the presence of defects on flawed blades may be localized rather than widespread. Consequently, due to the nature of the defects where blades classified as defective often have limited affected areas, resulting in a predominance of healthy sections over defective ones, there arises an inherent unbalance in the dataset. This disparity necessitates each sectional image to undergo thorough scrutiny to ensure accurate defect labeling and underpins the importance of our semi-supervised learning framework, which adeptly generates synthetic data to achieve a more equitable distribution for training purposes. Moreover, to validate the consistency of these labels, each image is meticulously compared with its corresponding physical section on the actual blade. This parallel manual inspection process involves a careful examination to confirm that defects identified in images are intrinsic to the blade's surface and not artifacts of removable debris, dust, or other transient disturbances that could be misinterpreted as flaws. This ensures that the defects cataloged are genuine and not false positives caused by external, non-permanent conditions.

It is crucial to emphasize that the scrap blades utilized in our study are sourced directly from authentic manufacturing processes and have been subjected to expert inspection for precise labeling. Combined with our stringent imaging and labeling protocols, this ensures the data's high quality. However, the rigorous nature of this process and the real-world constraints on the availability of specific defect types render the collection of such high-fidelity ground truth data both costly and limited. This inherent limitation in data quantity, particularly for certain defect categories, further reinforces the significance of our semi-supervised learning approach to judiciously augment the dataset. From a pool of 20 blades, the process yields 480 images, encompassing 203 of unmarred surfaces, 149 with scratches, and 48 with nicks, as demonstrated in Figure 2-8. The images displayed in Figure 8 are captured at an exceptionally high resolution, reflecting the raw output from our sophisticated imaging equipment. These detailed visuals provide an essential reference for discerning the nuanced characteristics of surface defects. However, it's important to clarify that these images undergo downsampling and resizing to a lower resolution before being fed into our model. This step is a practical necessity, as the computational demands of processing such high-resolution images for training are substantial and beyond our resource allocation.

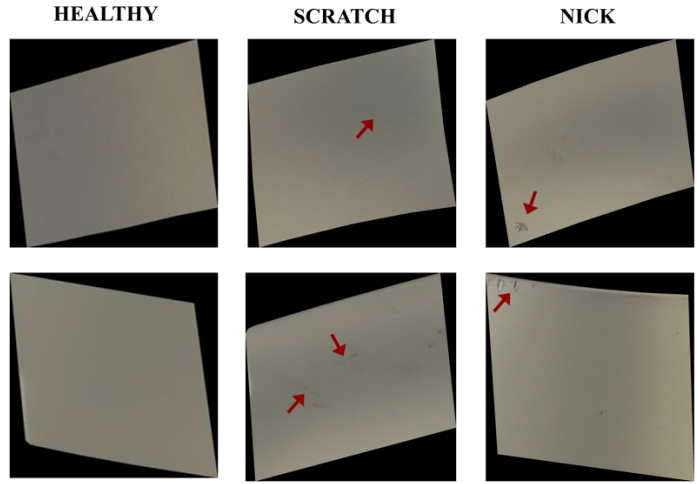

**Figure 2-8.** Sample images for each class.

Even at this high level of detail, certain defects may not be immediately apparent. To aid in identification, arrows have been added to the images, pointing out the defects. This annotation is not just a guide for the observer but also underscores the challenges. From the initial image capture to the final classification, each step demands a meticulous approach to ensure that even the most subtle surface aberrations are correctly labeled and classified. The subtlety of these defects underscores the need for the advanced intelligent inspection system. The entire training regimen for this system is conducted on a solitary NVIDIA GeForce RTX 2070 SUPER GPU within a Windows 10 environment.

### 2.4.2. Result and Discussion

The severely imbalanced distribution of our sample pool, where images labeled as healthy or with scratches vastly outnumber those showing nicks or dents, highlights the necessity of our proposed semi-supervised learning framework. This framework is especially well-suited to handle our dataset's limited size and skewed class distribution, as it effectively leverages both labeled and unlabeled data to improve detection accuracy. However, semi-supervised learning is not the only method to address the challenges posed by data scarcity and imbalance. We also investigate a range of alternative techniques, such as data augmentation, transfer learning, and class rebalancing strategies, which will be evaluated alongside our proposed framework to provide a detailed benchmark comparison using the same dataset.

*Comparative Efficacy in Multi-Class Recognition*

Transfer learning capitalizes on the knowledge gained while solving one problem and applies it to a different but related problem. By harnessing models such as VGG16 [114-117] and ResNet18 [118,119], which have been pretrained on extensive and diverse datasets like ImageNet, we can utilize their deep, feature-rich layers, even with our constrained dataset. This approach is advantageous for small datasets, as it transfers already learned features, which are likely to be beneficial for the new task, thereby reducing the need for learning from scratch. Geometric transformations, a subset of data augmentation techniques, involve modifying images through rotations, flips, and scaling. This method generates additional training data from the existing samples, enhancing the dataset's diversity and size, particularly for underrepresented classes. These alterations maintain the essential characteristics of the defects, allowing the model to learn from varied orientations and scales, thus strengthening its ability to generalize. However, it should be noted that our semi-supervised learning framework diverges from conventional approaches such as transfer learning and geometric transformations. While these methods are commonly adopted to tackle the issues of limited and imbalanced datasets, our framework opts for a different path. Within our framework, we use a generative model to distill more representative feature sets and produce synthetic data that helps balance the dataset. Alongside this, we employ a powerful classifier that dynamically trains on both ground truth and synthetic data with appropriate weighting. This integrated approach ensures that our model not only benefits from the rich representational power of transfer learning but also adapts effectively to our specific dataset challenges. We commence our comparison with these techniques in place, setting the stage for a full-scale assessment of their efficacy within our semi-supervised learning paradigm.

Table 2-3 presents the comparative test outcomes for various modeling techniques within our dataset, which has been divided such that 80% is allocated for training and validation, while the remaining 20% is dedicated to testing. The results shown are derived exclusively from the test data's predictive performance. The transfer learning model utilizing VGG16 as a base reports a baseline accuracy of 0.7179, with a notable instance being its singular correct prediction out of nine test samples for the nick category. To counteract

the class imbalance, we have employed data augmentation strategies and cDCGAN-based data generation, resulting in a curated addition of 54 scratch samples and 155 nick samples, thereby ensuring a balanced class distribution. The synthetic images showcased in Figure 2-9 were generated using a generative model trained on downsampled images and are illustrative of the types of data our model generates. These images serve not as exact duplicates of the original samples but rather as representations that highlight essential features for defect identification within each category. It is crucial to underscore that, akin to the generative process, the ground truth data used in subsequent training phases is also subject to downsampling. Originally captured at a resolution of 1000 × 1000 pixels, these images are resized to 224 × 224 pixels. This downscaling is imperative to achieve an equilibrium between training time and computational resource allocation, ensuring the model's operability without compromising the integrity of feature learning. To assess model performance thoroughly, we use the $F_1$- Score [120] and monitor 'Err Escape' rates. The $F_1$-Score, defined as the harmonic mean of precision and recall, balances accuracy by accounting for both false positives and false negatives, making it valuable for imbalanced datasets. In critical applications, we specifically track 'Err Escape,' or false negatives, where undetected defects could result in serious oversights. Minimizing error escape aligns with our goal of upholding stringent quality control.

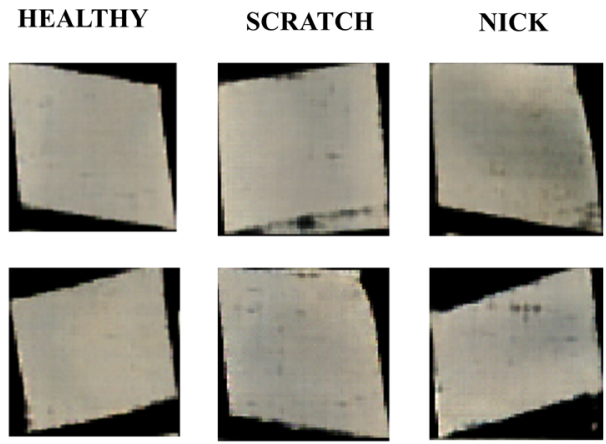

**Figure 2-9.** Generated sample images.

**Table 2-3.** Test results of different model & strategy

| Model | Accuracy | F$_1$- Score (Nick) | Err Escape |
|---|---|---|---|
| VGG16 | 0.7179 | 0.1667 | 0.2368 |
| VGG16* | 0.7692 | 0.4286 | 0.1842 |
| ResNet18* | 0.7820 | 0.4286 | 0.1578 |
| ViT* | 0.8333 | 0.5333 | 0.1315 |
| Proposed Method | 0.9174 | 0.8852 | 0.0263 |

*refers to the selected model with data augmentation*

Employing traditional data augmentation techniques, such as flipping and rotating images, resulted in noticeable improvements in model performance, as indicated by the asterisked models (e.g., VGG16* and ResNet18*). For example, the ResNet18* model's overall accuracy increased from 0.7692 (without augmentation) to 0.7820, reflecting a modest improvement of approximately 1.67%. However, our proposed semi-supervised framework demonstrates a substantial advancement in addressing data imbalance and enhancing defect classification accuracy. Specifically, our method achieved an overall accuracy of 0.9174, representing a 27.79% improvement over the baseline VGG16 model (0.7179) and a 17.30% improvement over the augmented ResNet18* model (0.7820). This significant increase underscores the effectiveness of integrating synthetic data with ground truth data within a semi-supervised learning framework. The F1 score for the "Nick" defect class further highlights the effectiveness of our approach. While models like ResNet18* and ViT* achieved F1 scores of 0.4286 and 0.5333 respectively, our method attained a significantly higher F1 score of 0.8852. This represents an increase of 106.58% over ResNet18* and 66.05% over ViT*. This substantial improvement indicates a robust ability to detect rare defect types accurately, underscoring our framework's capacity to learn critical high-level features from both ground truth and synthetic data. Moreover, the reduction in the Err Escape rate reinforces the reliability of our approach in minimizing false negatives, a critical factor in high-stakes manufacturing applications such as aviation [121,122]. The Err Escape rate of our proposed method is just 0.0263, a substantial decrease of 85.72% compared to VGG16* (0.1842) and 83.33% compared to ResNet18* (0.1578). This significant reduction ensures that far fewer defective parts are mistakenly classified as healthy, enhancing quality control and safety, particularly in fields where undetected defects could lead to severe failures.

Therefore, while traditional data augmentation techniques have provided some improvements, it is our semi-supervised learning framework that more effectively tackles the dual challenges of data imbalance and limited dataset size. Traditional augmentation methods, such as flipping, rotating, or slightly altering existing images, can enhance diversity but often fail to capture the underlying complexities of rare defects like nicks or dents. In contrast, our framework goes beyond these simple transformations by intelligently generating synthetic data that encapsulates key high-level features critical for accurate defect classification. This enables the model to better distinguish between various types of defects, even when faced with significantly imbalanced class distributions. Our approach achieves a commendable accuracy of 0.9174, a significant leap compared to standard methods. More importantly, it substantially reduces the risk of allowing defective parts to pass through quality control, thereby enhancing the reliability of the inspection process. By incorporating both labeled and unlabeled data, the framework ensures that the model learns effectively from limited samples while maintaining high precision. This data-augmented, semi-supervised learning approach represents a meaningful advancement in meeting, and potentially surpassing, the stringent inspection standards required in the manufacturing industry, providing a scalable solution for real-world applications where data scarcity and imbalance are persistent challenges.

Additionally, model performance on each category can be inferred from Figure 2-10 by comparing overall accuracy and $F_1$- Scores for each class label. In multi-class classification, the $F_1$- Score is a critical metric that measures a model's ability to recognize each class accurately. The figure presented uses the darkest shade of red to depict overall accuracy, while three lighter shades represent the $F_1$- Scores for individual classes. This visual distinction allows us to discern the impact of data augmentation and our proposed framework on the model's class-specific recognition capabilities. 'Good' and 'Scratch' are the majority classes, while 'Nick' is the minority class. Data augmentation was employed to enhance the representation of the minority class. The figure reveals that these augmentation techniques, and particularly our proposed method, have significantly elevated the $F_1$- Score of the 'Nick' class. Notably, there is also a marginal increase in the $F_1$- Scores for the majority classes. This suggests that enhancing the model's ability to recognize minority samples not only boosts the true positive rate for that class but also marginally decreases the false negative rate for the majority classes. Therefore, as the proposed semi-supervised learning approach reduces the imbalance in data distribution, we observe a marked improvement in model recognition across all classes, with an especially pronounced effect on minority class recognition.

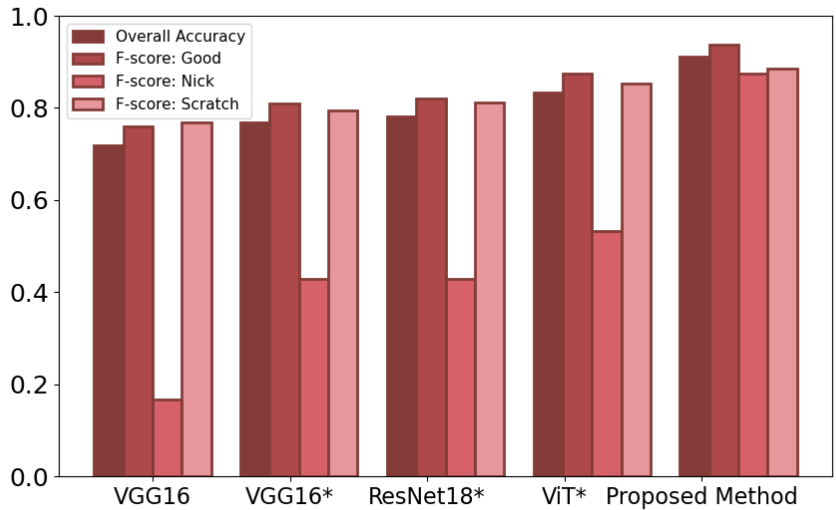

**Figure 2-10.** Overall accuracy & $F_1$- Score of models.

The confusion matrices in Figure 2-11 provide deeper insight into the classification performance of various models on the 'Good', 'Nick', and 'Scratch' classes. By examining the matrices for VGG16, VGG16* (with data augmentation), ViT*, ResNet18*, and the proposed method, we can assess the precision and recall of these models and how they relate to each class. For VGG16, we see that while it correctly classifies 'Good' samples 75% of the time, there is a notable misclassification of 'Nick' and 'Scratch' classes as 'Good', indicating a bias towards the majority class. This is evidenced by 20% of 'Nick' and 13.9% of 'Scratch' being incorrectly labeled as 'Good'. Moving to VGG16* and ResNet18*, both augmented with data techniques, there's an improved recognition of the 'Nick' class, reducing the false positive rate for the 'Good' class, as fewer 'Nick' samples are misclassified as 'Good'. For the 'Scratch' class, a majority, the confusion matrices of augmented models show a consistent high recognition rate, similar to the non-augmented VGG16. This indicates that the increased recognition of the minority class does not adversely affect the recognition of the 'Scratch' majority class. The Vision Transformer (ViT*) shows a more balanced classification, with a lower misclassification rate of 'Nick' samples as 'Good' or 'Scratch', reflecting a more balanced learning from the minority class. However, there is still a significant error rate within the 'Nick' class predictions, with 22.22% being mislabeled as 'Scratch'. The proposed method's confusion matrix presents the most balanced classification across all classes. Notably, it shows no 'Nick' samples misclassified as 'Good', demonstrating a perfect distinction between the minority and majority classes in this regard. Additionally, the misclassification of 'Nick' as 'Scratch' is reduced to 22.22%, similar to the ViT* model, which suggests a substantial improvement in minority class recognition. The evidence from the confusion matrices reinforces the previous conclusion that enhancing the representation of minority classes through data augmentation or the proposed method does indeed enhance the model's recognition ability for all classes. This is particularly evident in the reduction of misclassification rates for the minority 'Nick' class across all models when data augmentation techniques are employed. The proposed method stands out by demonstrating the most substantial reduction in misclassification rates for all classes, emphasizing its effectiveness in dealing with class imbalances.

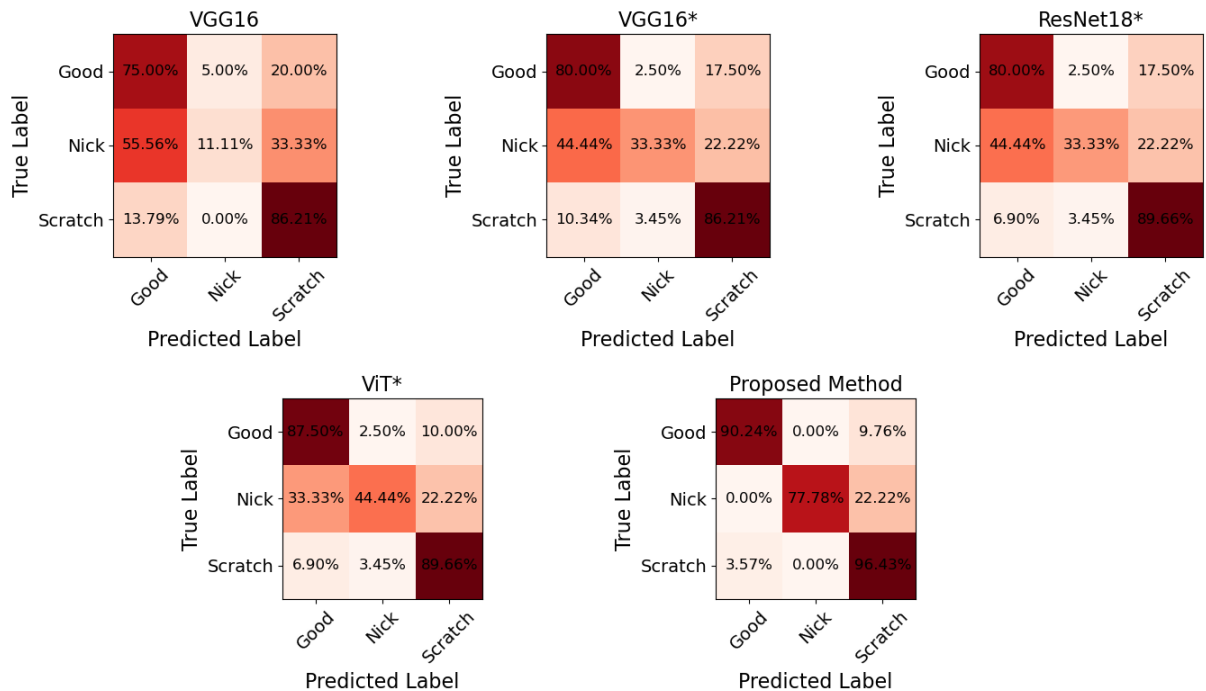

**Figure 2-11.** Confusion matrices of model training.

*Investigation of Performance Enhancement by Synthetic Data*

Upon establishing that our proposed semi-supervised learning framework surpasses other deep learning models and non-generative augmentation techniques, it becomes crucial to assess the impact of integrating synthetic data into the training performance. This investigation is key for verifying the synthetic data's contribution to performance enhancement and for understanding how to optimize the improvement by calibrating the volume of synthetic data used in training. For an impartial evaluation, we preserved a uniform sample count of 48 across classes, supplementing them with varying amounts of synthetic data to discern its influence on model efficacy. This methodology allows us to develop guidelines for the ideal quantity of synthetic samples needed in data augmentation practices. We experimented with augmenting the training set with synthetic data in quantities ranging from one to four times the ground truth data volume, producing results that shed light on the correlation between synthetic data volume and model performance enhancement. These results are summarized in Table 2-4.

**Table 2-4.** Model performance with varying size of synthetic data

| Synthetic / Ground Truth | Accuracy (Ground Truth Only) | Accuracy (Addition of Synthetic) |
|---|---|---|
| 0 | 0.7057 | 0.7057 |
| 1 | 0.7057 | 0.7929 |
| 2 | 0.7057 | 0.8748 |
| 3 | 0.7057 | 0.9230 |
| 4 | 0.7057 | 0.9015 |

Analyzing the results laid out in the Table 4, we observe a clear and consistent upward trend in classification accuracy as synthetic data is incrementally introduced into the training mix. With the initial addition of synthetic data equal to the amount of ground truth data, there's a marked improvement in accuracy from 0.7057 to 0.7929. This increase is indicative of the synthetic data's quality and relevance to the model's learning process. As we double the synthetic data, the accuracy ascends further to 0.8748, reinforcing the notion that additional synthetic data can effectively broaden the training landscape for the model. The most pronounced jump in performance occurs when the synthetic data is tripled, propelling accuracy to a peak of 0.9230. This peak suggests an optimal balance where the synthetic data provides maximum benefit by covering gaps in the ground truth distribution without introducing detrimental noise. However, when the quantity of synthetic data is quadrupled, we note a slight regression in accuracy to 0.9015. Although this accuracy is still significantly higher than the baseline, it indicates that there might be a threshold beyond which the incorporation of synthetic data can start to introduce diminishing returns, potentially overshadowing or distorting the ground truth data's intrinsic patterns. These findings highlight a nuanced understanding of synthetic data's role in enhancing machine learning models: while synthetic data serves as a valuable tool for amplifying the training dataset, especially in scenarios of limited ground truth data, there exists a delicate balance. Beyond a certain point, an excess of synthetic data might lead to overfitting or reduced model generalizability. Therefore, it is paramount to identify the optimal synthetic-to-ground truth data ratio that achieves the best model performance without compromising the integrity of the learning process.

Simultaneously, this inquiry raises the question of whether the available ground truth data points are truly representative of the underlying data distribution. While providing a definitive answer is challenging due to the inherent complexity of real-world data, we can approach this issue from a statistical perspective. Specifically, we can evaluate the model's performance by comparing training accuracies across different volumes of ground truth data. This allows us to examine how effectively the model captures essential patterns as the amount of labeled data increases or decreases. Building on the insights from our previous discussions, we have chosen to increase the volume of synthetic data threefold, which we identified as an optimal augmentation point. To further explore the relationship between ground truth data and model performance, we experiment with varying proportions of ground truth data while maintaining this level of synthetic data. This experimental design enables us to compare the model's performance under different conditions: one scenario where the model is trained exclusively on ground truth data, and another where the training data is supplemented with synthetic examples. Through these comparisons, we aim to uncover how the quantity of ground truth data influences model effectiveness, particularly in relation to its ability to generalize and handle complex defect detection tasks.

The results presented in Table 2-5 reveal the differential impact of ground truth and synthetic data on model accuracy. When relying solely on ground truth data, there's an observable upward trend in model accuracy as the percentage of ground truth data increases: from 37.82% accuracy with 20% of the ground truth data to 72.57% with the full set. This increment, however, exhibits diminishing returns as more ground truth data is included, reflecting a natural plateau in learning from real data alone. Introducing synthetic data into the equation changes the dynamics significantly. With just 20% of ground truth data, the addition of synthetic data marginally improves accuracy to 40.29%. As the ground truth data increases to 40% and 60%, the synthetic data's contribution to accuracy becomes more pronounced, surging to 60.22% and 82.65%, respectively. This trend suggests that while ground truth data is foundational for model training, synthetic data acts as a multiplier of the knowledge base, filling in gaps and providing broader coverage of the feature space. The most substantial gain is observed when the model is trained with 80% of the ground truth data, with synthetic data boosting accuracy to 89.57%. At full ground truth data capacity, the addition of synthetic data achieves the highest accuracy of 92.30%. This pattern indicates that while the base representation of reality provided by ground truth data is essential, the strategic integration of synthetic data can significantly elevate a model's performance by reinforcing and expanding the dataset's diversity.

Table 2-5. Model performance with varying size of ground truth data

| Ground truth data | Accuracy (Ground Truth Only) | Accuracy (Addition of Synthetic) |
| --- | --- | --- |
| 20% | 0.3782 | 0.4029 |
| 40% | 0.5418 | 0.6022 |
| 60% | 0.5964 | 0.8265 |
| 80% | 0.6972 | 0.8957 |
| 100% | 0.7257 | 0.9230 |

From these observations, it can be inferred that a richer, more varied dataset, created through the strategic combination of ground truth and synthetic data, enables the model to develop a more robust and fine-grained understanding of the classes it needs to recognize. The inclusion of diverse synthetic data helps to fill the gaps left by the typically limited and imbalanced real-world datasets, allowing the model to be exposed to a broader range of scenarios and variations that it might not otherwise encounter. This not only improves the model's ability to accurately classify different types of defects but also strengthens its capacity to generalize to new, unseen data. The importance of this balanced approach cannot be overstated. While ground truth data provides the authenticity and precision needed to anchor the model in reality, synthetic data introduces the necessary volume and variation to train the model on rare or underrepresented classes. This synergy between real and synthetic data ensures that the model is not only accurate but also resilient to fluctuations in data quality and distribution, a common issue in practical applications like defect detection in manufacturing. Achieving this balance is critical for developing a model that excels in both accuracy and generalization. Such a model is highly valuable in real-world settings, particularly in automated inspection systems where data scarcity, imbalance, and variability present significant challenges. By leveraging the strengths of both ground truth and synthetic data, we can create a system that performs reliably under a wide range of conditions, making it suitable for integration into high-demand industrial environments or other practical applications where robust, scalable, and consistent performance is essential.

The graph in Figure 2-12 illustrates the relationship between the amount of ground truth data used for training and the resultant classification accuracy, comparing models trained solely on ground truth data with those enhanced by the addition of synthetic data. The stacked bar chart shows two components: the darker section represents the classification accuracy achieved with ground truth data alone, while the lighter section on top indicates the incremental improvement from adding synthetic data. This visualization suggests that the incorporation of synthetic data has a consistently positive effect on the model's performance, especially at lower ratios of ground truth data, where the difference in classification rates between the two approaches is most pronounced. As the amount of ground truth data grows, the incremental benefits of synthetic data diminish slightly, implying a natural cap to the improvement that synthetic data can provide. The convergence of the two lines towards the upper right suggests that the addition of synthetic data is less influential as the amount of ground truth data approaches completeness. In this particular case, where 48 samples per class are utilized, the classification rate has not yet reached its potential maximum, illustrated by neither line touching the perfect classification rate of 1. This indicates that while synthetic data is beneficial, especially when ground truth data is scarce, the ultimate goal of a perfect classification rate likely requires a larger set of ground truth data. While our current framework shows strong performance, further exploration could yield additional improvements. Accumulating more ground-truth data over time could enable retraining that further enhances classification accuracy and adaptability. Optimizing synthetic data generation techniques, especially for more realistic defect samples in underrepresented classes, may also refine model precision. Additionally, experimenting with alternative semi-supervised learning methods, such as adaptive weight adjustments for unbalanced classes, could improve performance. These directions aim to expand model robustness across diverse defect types and broader industrial applications.

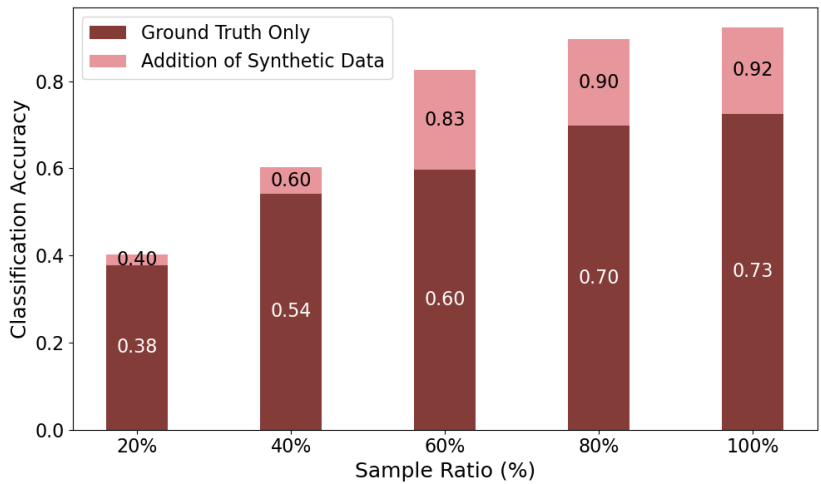

**Figure 2-12**. Model performance with varying size of ground truth data.

## 2.5. Conclusion

In this research, we developed an automated intelligent surface inspection system that achieves consistent data collection through automated protocols and accurately detects defects on convex blade surfaces. The system comprises a data collection module and a defect detection module. The data collection module employs a collaborative robot to facilitate automated image acquisition, ensuring precise positioning and consistent imaging of the blades. The defect detection module utilizes bit masking techniques to focus on specific areas of the blade surface and a deep learning model to identify defect types. Crucially, we incorporated a semi-supervised learning framework into the intelligent surface inspection system, constructed using a conditional Deep Convolutional Generative Adversarial Network (cDCGAN) and a bias-aware Residual Neural Network (ResNet) to address the challenge of limited data size. This framework is designed for scenarios with inherently limited training data sizes, generating additional samples for underrepresented categories. Our approach achieved an overall classification accuracy of 0.9230 representing a significant 27.79% improvement over the baseline benchmark model. Additionally, the error escape rate (false negatives) was substantially reduced from 23.68% to 2.63%, a critical improvement in high-risk industries like aviation manufacturing, where undetected defects can lead to catastrophic failures. These achievements are vital for the flexible automation of blade inspections, especially during initial implementations when data availability is limited. As more ground-truth data accumulates, retraining the network promises further enhancements in performance. We also examined model performance across classes with varying amounts of ground truth and synthetic data, providing deeper insights into our model's adaptability in different scenarios.

# Chapter 3.  Analysis of Microstructure Uncertainty Propagation in Fibrous Composites Empowered by Physics-Informed, Semi-supervised Machine Learning

The advancements in multi-scale computational analysis of fiber reinforced composites have led to the possibility of predicting important material properties based on their microstructure characteristics. Nevertheless, major challenges remain. The fiber distributions feature inherent randomness, which naturally leads to variations in properties such as transverse strength. This in turn undermines the significance of deterministic analysis to guide manufacturing optimization. Direct Monte Carlo simulation for uncertainty analysis is computationally insurmountable, as a single run of finite element simulation is already costly. While several surrogate modeling techniques leveraging supervised learning have been explored, it is commonly recognized that the efficacy of these surrogate models hinges upon the size of training dataset. In this research we establish a semi-supervised learning framework that can produce highly accurate emulation results with much reduced size of labeled training dataset. A random fiber packing algorithm is employed to sample the representative volume element (RVE) images that are subsequently fed to the finite element analysis to generate the ground-truth labeled data used in the training of neural network. To reduce the ground-truth labeling cost while maintaining the deep learning capacity. we employ the pseudo labeling technique where the base model is initially trained on a small set of ground truth labeled data and then used to generate credible pseudo-labels for a larger pool of unlabeled data. The model is subsequently retrained on this augmented dataset with adjusted weights and biases to reflect the varying confidence in the label sources. This framework is successfully employed in the analysis of microstructure uncertainty propagation in fibrous composites. The proposed approach efficiently leverages patterns from both unlabeled and limited labeled samples to predict transverse strength for varied RVE samples, matching the efficacy of a fully supervised model trained with 1,000 ground truth labels while simultaneously slashing labeling efforts by 72%. This framework can be extended to uncertainty propagation analysis using microstructure characteristics of other materials.

## 3.1. Introduction

Carbon fiber reinforced polymer (CFRPs) composites [123] are increasingly used in advanced structural components in aerospace or automobile systems owing to its high stiffness-to-weight and strength-to-weight ratios. Moreover, CFRPs have highly specific mechanical properties and potentially can be tailored for particular needs. Since the employment of CFRPs on those crucial structural components require high standards of safety for material failure and fatigue, the macroscopic properties of fiber reinforced polymer composites (FRCs) [124,125] must be understood thoroughly. However, the ground truth values of the desired mechanical properties are not completely subject to the deterministic predictions, since the randomness is inevitably induced into the CFRPs through the manufacturing processes [126]. In order to identify the direct correlation between the macroscopic behaviors and microstructure design, representative volume elements (RVEs) [127,128] consisting of randomly distributed fibers and a surrounding matrix are generated to simulate the mechanical behavior of composites. RVEs comprising the specified fibers and matrix can be defined with the appropriate boundary conditions to derive the effective properties of heterogeneous material by the knowledge of the constitutive laws and spatial distribution of their constituents. Meanwhile, the random spatial distribution of fibers is generated with specific features to be incorporated into the RVE geometries, predicting the random variation of material properties at the macroscale.

Due to the time-consuming workload and high cost of the direct characterization of the composite microstructure, several algorithms [129,130] have been suggested to numerically generate RVEs, which can allow us to model the uncertainties [131–133] associated with the microstructure. Subsequently, numerical study on those RVE samples can be performed to obtain the corresponding values of mechanical properties. For instance, Stefanou et al [134] propose a framework using the stochastic finite element method to derive the stochastic stiffness matrix from material microstructure. In a separate study, Wang et al [135] detail the establishment of a two-dimensional (2D) finite element model using Representative Volume Elements (RVEs) to explore interphase properties and the damage behavior of composites. Furthermore, digital image correlation techniques [136] are applied to provide experimental characterization of RVEs, demonstrating a practical approach to material analysis. Rigorous derivations of the RVE size and the associated boundary conditions are given in [137]. The critical relation between the stochastic characteristics of the microstructure and the resulting macroscopic properties has been extensively discussed in [138-141]. However, accurately quantifying uncertainties due to fiber spatial distribution requires extensive numerical RVE sampling and time-intensive analysis for each sample, making the process prohibitively computational for certain CFRP designs. Rapid advancements in machine learning algorithms have significantly enhanced computational processes, improving the prediction of mechanical and thermal properties of composite materials [142]. These algorithms excel in both accuracy and efficiency, positioning them as valuable tools for material design and optimization. These algorithms excel in both accuracy and efficiency, positioning them as valuable tools for material design and optimization. Specifically, the Gaussian process [143] has been applied to predict the mechanical properties of carbon fibers, accounting for manufacturing uncertainties and optimizing for specific material attributes. Moreover, while previous studies such as the work by Dey et al [144] have focused on utilizing artificial neural networks for uncertainty quantification in natural frequencies of composite plates, recent advances have significantly broadened the application of machine learning in composite material analysis. For instance, Chen and Gu [145] have elucidated how machine learning, particularly deep learning, can optimize the design and performance of composite materials by accurately predicting their behavior and mechanical properties. Additionally, the comprehensive review by Liu et al [146] discusses the application of machine learning models, specifically ANNs, in nonlinear constitutive modeling, multiscale surrogate modeling, and design optimization of composite materials. Furthermore, the study by Sharma et al [147] highlights the multifunctional properties of polymer composites and explores how machine learning assists in modeling, analysis, and design, focusing on overcoming traditional computational limits and offering recommendations for future research directions in this rapidly evolving field. Despite these advancements, our research presents a fundamentally different and novel approach. Unlike the parameter-based methods typically used in earlier works, our research is entirely image-based, enabling direct predictions from composite microstructure images. This method not only applies to the design phase but also extends to post-manufacturing predictions, accommodating the randomness introduced during the fabrication process. This versatility makes our approach highly relevant in real-world applications where the manufactured product may deviate from initial design specifications due to various uncertainties. Therefore, it's important to acknowledge that traditional machine learning algorithms typically rely on correlations among carefully chosen quantitative indices [148,149] from sample data, which may not fully capture the complexities and uncertainties inherent in RVE images.

Image-based deep learning methods, particularly convolutional neural networks (CNNs), have recently gained prominence, surpassing many traditional surrogate modeling algorithms in image recognition tasks [150–152]. CNNs, with their multiple convolutional filters scanning the entire feature matrix, excel at extracting essential features from the vast amount of parameters in images while retaining critical details. This capability makes CNNs exceptionally suitable for image-based analysis, offering a powerful tool for extracting meaningful insights from complex visual data. Zhou et al [153] exploited the customized CNN and VGG-16 [154,155] to train on approximately 1000 RVE images, aiming to predict the transverse strength of composite materials and to quantify uncertainties introduced during sample generation. This approach enabled the neural network to serve as a predictive model for assessing composite strength through RVE images, achieving an impressive testing mean square error of $7.001 \times 10^{-4}$. However, this method does not overcome the inherent limitations of deep learning, which necessitates substantial data for training to ensure the extraction of generalized features, crucial for the model's performance. Furthermore, the quality and accuracy of labeling each sample are paramount in the training phase, significantly influencing the training direction and effectiveness. To generate labels for this image-based regression challenge, detailed finite element models were constructed from RVE images, with label generation consuming approximately 750 hours for each 45-minute run on advanced workstations. Despite the tailored nature of the trained neural network model, requiring a fresh batch of high-quality data for new composite material design parameters introduces a prohibitive computational burden. This challenge limits the practical application of deep learning's trainability and transferability in scenarios requiring significant computational resources.

The heterogeneity of fiber-reinforced composites introduces significant variability in their mechanical properties, presenting a substantial computational challenge. Traditional approaches to modeling these materials often rely on extensive parameter-based uncertainty quantification, which requires large datasets and significant computational resources. This approach, while comprehensive, often falls short in efficiently handling the microstructural randomness inherent in composite manufacturing processes. To address these challenges, our research introduces a novel surrogate model within a semi-supervised learning (SSL) framework, specifically designed to tackle the complexity of uncertainty propagation in fiber-reinforced composites. This model leverages advanced machine learning techniques to minimize reliance on extensive labeled sample inputs [156–159], focusing instead on efficiently predicting how uncertainties in fiber placement and distribution impact the mechanical properties. By doing so, we provide a more feasible solution for simulating and optimizing composite materials in real-world manufacturing settings. Our surrogate model not only enhances the computational efficiency of simulations but also improves the predictive accuracy under varying microstructural conditions. The initial focus of our study on uncertainty propagation has yielded significant insights into the relationship between microstructural uncertainties and composite performance. Additionally, SSL is particularly suited to our context, where generating a large volume of RVE images is feasible both economically and temporally, yet acquiring a comparable quantity of labeled samples through finite element analysis remains prohibitively expensive. A pivotal challenge is extracting meaningful insights from unlabeled data absent any definitive ground truth. Pseudo-labeling [160] offers a pragmatic solution by using a modest subset of labeled data to augment a larger pool of unlabeled data, thereby improving model efficacy. This is achieved by iteratively assigning pseudo labels to unlabeled data using the trained neural networks, allowing for simultaneous training of both labeled and unlabeled data in each epoch. Furthermore, the pseudo labels and the significance of the unlabeled loss are updated at each epoch's conclusion, reflecting the model's progressive refinement and the increasing reliability of its predictions. Through pseudo-labeling, the necessity for large volumes of labeled data is significantly reduced, enabling extensive use of unlabeled data to uncover the data's intrinsic structure. Consequently, a substantial portion of the ground truth labels for the 1,000 labeled samples are excluded from the training dataset. Instead, pseudo-labeling is applied in this predefined scenario to juxtapose against the supervised learning on the same set of labeled samples. This comparison aims to ascertain whether pseudo-labeling can offer computational cost savings while maintaining or enhancing model performance. After validating the effectiveness of pseudo-labeling, we proceed to leverage all 1,000 labeled samples alongside an additional unlabeled 5,000 samples, applying semi-supervised learning to explore the optimal performance achievable through pseudo-labeling. This comprehensive approach enables us to fully assess the potential of pseudo-labeling in enhancing model accuracy and efficiency in a significantly expanded dataset scenario.

The rest of the paper is organized as follows. In Section 3.2 we outline the RVE generation and the physics-based finite element analysis procedure of composite transverse strength. Section 3.3 introduces our proposed deep learning framework, focusing on achieving robust pseudo labeling via a deep network tailored for regression tasks. Section 3.4 highlights the performance and efficiency improvements over the baseline model and demonstrates streamlined uncertainty quantification with our framework. Concluding remarks are provided in Section 3.5. This research provides a practical technique to use a relatively small amount of labeled data obtained by finite element analysis, with the aid of SSL, to train a reliable neural network as surrogate model that is subsequently employed for highly accurate and robust analysis of uncertainty propagation in fibrous composites.

## 3.2. Physics-based Modeling and Simulation as Foundation

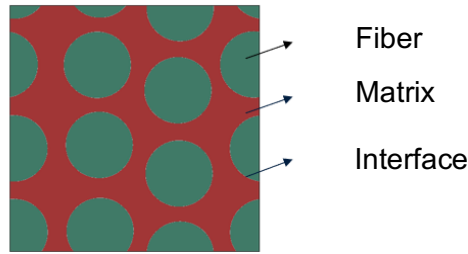

**Figure 3-1.** RVE sample generated by fiber packing algorithm.

The development of a semi-supervised learning framework for predicting the impact of microstructural uncertainties in composites necessitates the use of high-quality training data. This is facilitated through the generation of a substantial number of RVE samples, as depicted in Figure 3-1, employing a fiber packing algorithm [161] to accurately model the spatial randomness inherent in composite microstructures. The focus of this paper is to investigate the transverse tensile strength of a unidirectional composite, which is lower than the strength of the fiber. As a result, the fiber is modeled as an elastic solid, while the matrix is modeled as an elasto-damage solid using a fracture energy based Smeared Crack Approach (SCA) to dictate the progressive damage and failure response. Cohesive elements are assigned at the fiber–matrix interface to capture the debonding behavior. The simulations were performed using the commercial finite element software Abaqus (version 2017), with the SCA algorithms implemented as a user-defined material subroutine, UMAT. Details of the model setup, including the SCA formulation and implementation are given in [153].

The fiber packing algorithm offers several distinct advantages over traditional parameter-based uncertainty quantification methods by more closely mimicking the realistic conditions of composite manufacturing processes. Unlike parameter-based methods, which often require extensive pre-definition of a range for each variable and subsequently test correlations between these ranges and the outputs, our approach integrates uncertainty naturally within the simulation process, reflecting the actual manufacturing variabilities. The fiber packing algorithm incorporates several key assumptions that enhance its realism and effectiveness. While the overall fiber volume fraction and fiber count are meticulously controlled to ensure consistency, the randomness in fiber placement mimics the natural variability observed in manufacturing settings. This variability is reflective of real-world conditions where mechanical vibrations and slight variations in material handling are common. Additionally, the algorithm models fibers as repelling each other to prevent overlap, simulating the physical mechanisms used during the composite lay-up process, where fibers are spread evenly to maintain uniform distribution without excessive clustering. Two critical parameters are central to our simulation's control strategy: the fiber volume fraction and the minimum gap between fibers. The fiber volume fraction is fixed at the onset of each simulation, reflecting standard manufacturing specifications and ensuring that simulations remain consistent across runs for reliable comparative analysis. The minimum gap between fibers, meticulously defined at the start, is crucial for simulating different packing densities and fiber distributions. This parameter influences the final mechanical performance under various loading conditions and is reflective of the touch-and-go nature of fiber placement in real composite layers, where optimal spacing is crucial to avoid material weaknesses. Initially, fibers are positioned according to a Poisson process, suitable for modeling the random scattering of fibers during the initial material lay-up. Adjustments made during the simulation, such as movements due to mechanical interactions, might follow a normal distribution, reflecting minor deviations from mean positions as fibers settle into their final configurations without overlapping. Specifically, users input essential parameters including the total number of fibers, their volume fraction, and the minimum inter-fiber gap. Following this, the system undergoes preparation through the creation of arrays and matrices to store simulation data, encompassing fiber positions, velocities, radii, and masses. Subsequently, fibers are randomly positioned within the simulation domain at reduced radii to minimize the chance of overlap, reflecting an uncompressed fiber tow state. As the simulation progresses, fiber positions and velocities are dynamically updated in response to inter-fiber forces, modeled as repulsive to prevent overlap. Unlike traditional force-based interactions, fiber penetration adjustments are made through direct positional changes, facilitating separation from intersecting fibers based on penetration depth. To maintain the simulation's stability and efficiency, a velocity control mechanism is implemented, monitoring the system's kinetic energy and limiting fiber velocities as necessary. The simulation's time step is also adaptively adjusted according to the velocity of the fastest-moving fiber, ensuring both precision and computational efficiency in modeling the complex interplay of fibers within the RVE samples.

To maintain periodicity in the random-packed fiber simulation, mirror boundary conditions are applied at the computational domain's edges and corners. This ensures that fibers or circles crossing the domain boundaries are mirrored to the opposite sides. This mirroring is achieved by identifying fibers crossing the domain boundaries and calculating their projections across to the opposite edges. These projections form 'images' of the fibers within the periodic domain, which are then utilized for interaction and velocity calculations. Each image receives a unique identifier linked to its original fiber, enabling precise tracking of interactions and force computations between the original fibers and their images. This mechanism of periodicity, domain boundary management, and fiber projection is pivotal, ensuring accurate interaction among fibers and with the domain boundaries while maintaining the desired fiber volume fraction. One of the most challenging aspects of simulating a random packing of fibers is to prevent fiber inter-penetration, while increasing the fiber volume fraction. In the first phase of the simulation, we overcome this challenge through radius expansion. This phase occurs over a specified duration before moving to the next phase, which is attempting to finish the job. The radius of each fiber is allowed to fluctuate during the simulation to help eliminate fiber inter-penetration, while increasing the fiber volume fraction and achieving a random distribution. The algorithm starts by setting the initial radius for the fibers. During the simulation, the radius of the fibers is allowed to expand uniformly to help eliminate inter-penetration. This is done by adding a growth rate to the current radius, and saturating the radius at a maximum value, which is determined automatically. The growth rate is determined by the relationship between the current time and the Phase 1 duration. In the beginning, the fibers grow rapidly according to a saturation growth rate, after which, they are allowed to shrink. The maximum radius, growth rate, and repulsive parameters can be used to control the randomness of the distribution. With a larger maximum radius, the distribution will likely be more uniform, while if the maximum radius is kept near the target radius (determined by desired volume fraction), then it is more likely that distances between fibers will be less uniform, more clumped, and hence more random. Essentially, the uniformity is related to the fiber volume fraction-overshoot during Phase 1. It should be noted that the degree of randomness is also strongly dependent on the desired volume fraction, as the distribution will approach a uniform hexagonal packing array as the volume fraction approaches the maximum theoretical value. Through the radius expansion algorithm, we are able to increase the fiber volume fraction while preventing inter-penetration, resulting in a random packing of fibers that closely approximates the physical system and represents compaction of fibers resulting from manufacturing processes. After detailing the fiber positioning and dynamics within our simulations, it is pertinent to connect these elements with established mathematical and numerical models that deeply explore interface behaviors in composite materials. Kamiński's integrated use [162] of a boundary value problem to derive macroscopic properties through homogenization provides a rigorous framework for understanding how microscale defects influence overall material behavior. Similarly, his application of the Stochastic Finite Element Method, complemented by Monte Carlo simulations, offers valuable insights into handling randomness, crucial for our analysis of fiber dynamics and interactions within composite materials. Complementing these detailed mathematical models, Kim and Mai [163] further enriches our understanding of interface phenomena in composites. While our simulations focus on specific aspects of fiber dynamics and microstructural interactions, the extensive discussions in Kim and Mai's work on engineered interfaces provide broader insights into optimizing the mechanical properties through strategic interface engineering.

As the simulation progresses, it is necessary to check if the system has reached the desired fiber volume fraction and if the fibers have settled into a non-penetrating distribution. This is done in Phase 2 by monitoring the volume fraction, $V_f$, which is calculated as the ratio of the total area occupied by the fibers to the area of the simulation domain. A crucial factor to take into account when working with the finite element meshing of the RVE is the definition of a minimum gap between the fibers. Additionally, the minimum gap serves as another parameter that can be tuned to influence the randomness in the distribution. The code is designed to target a fiber volume fraction that surpasses the user-defined to account for the desired minimum gap between the fibers. In essence, the code selects a target radius that includes half the minimum gap. If there is no overlap between the fibers, then the radius can be reduced to the target value, thus attaining the defined volume fraction and maintaining the minimum gap between the fibers. Following the first phase of the simulation, attempts are made to finish the job. Each increment, it is checked to see if $V_f$ exceeds the desired value and there is no penetration, in which case, the fiber-packing is complete. If it has not, the simulation is augmented with the application of randomized pulses to the system. These pulses cause the fibers to shrink or grow, depending on whether the $V_f$ is above or below the desired value. This step continues until the appropriate $V_f$ is reached while helping fibers to settle in a non-penetrating configuration. The development of the fiber packing algorithm was motivated by the need to create a model that not only accurately represents the microstructure of fiber-reinforced composites but also incorporates both the controlled aspects of manufacturing and the inherent randomness of the process. This approach enables a deeper understanding of how microstructural variations influence the mechanical properties of composites, providing critical insights necessary for the design and optimization of new materials. By employing this sophisticated fiber packing algorithm, we replicate the composite material's microstructure more accurately and gain significant insights into the influence of microstructural randomness on material properties. This method offers a dynamic, integrated approach that closely aligns with actual manufacturing variations and provides a stronger basis for predictive modeling than traditional uncertainty quantification methods.

For each single sample, the fiber volume fraction, fiber number and minimum gap are fixed as 51%, 9 and 40% of the fiber radius to exclude the potential irrelevant factors in this investigation that could have critical impacts on the transverse strength, which is basically reflecting that all the procedures of fabricating the composite samples are strictly controlled in the real scenario of the manufacturing process. On the basis of identical essential features of RVE samples, the uncertainty in the spatial distribution of fibers can be better mapped and identified with the corresponding transverse strength. Eventually, 6000 RVE sample images are generated by implementing the fiber packing algorithm in MATLAB to establish the sample pool. 10 RVE samples generated by the fiber packing algorithm are displayed in Figure 3-2. All of them are given the identical settings of fiber number, minimum gap, and fiber radius, which are demonstrating the discrepancy by randomness of fiber distribution.

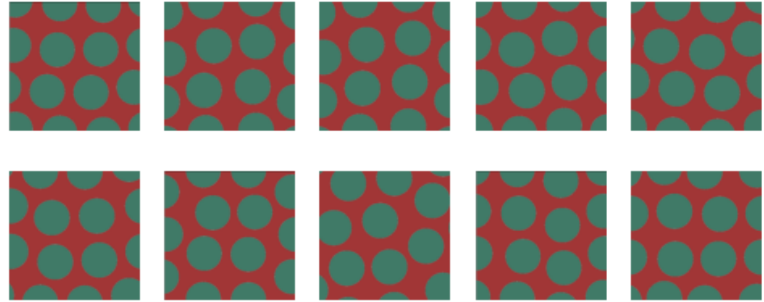

**Figure 3-2.** Uncertainty-induced RVE samples of identical.

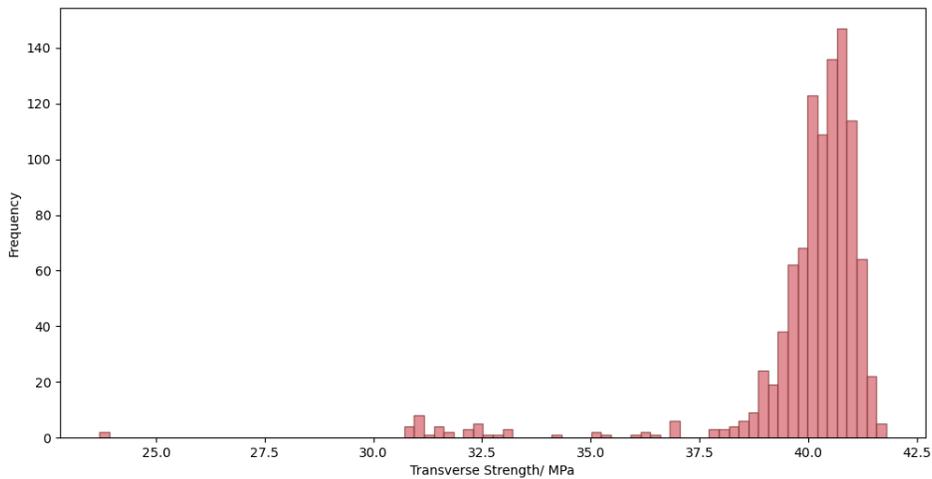

**Figure 3-3.** Transverse strength distribution of labeled sample pool.

Labeling plays a crucial role in creating a high-fidelity dataset for deep learning, serving both as the prediction output for surrogate models and as a key reference for pattern recognition. To this end, 1,000 RVE images are utilized in finite element modeling to conduct progressive damage analysis [164,165], thereby generating computed values of transverse strength. These values are deemed reliable labels for use in the ensuing deep learning analysis. Each single run of the finite element analysis (FEA) on the sample image takes about 45 minutes on a high-end workstation (Intel Xeon W-2155 CPU @3.31 GHz with 20 processors). Following approximately 750 hours of intensive computation, the dataset comprising 1,000 labeled RVE images, derived from finite element analysis, and an additional 5,000 unlabeled RVE images, has been fully prepared for subsequent deep learning analysis. The distribution of transverse strength for the 1,000 labeled samples is depicted in Figure 3-3, where the majority of samples cluster around a strength of 40.0 MPa. However, outliers near the 30.0 MPa margin present a challenge for the deep learning framework to accurately map the full range of uncertainty to transverse strength. While the model is trained to recognize patterns including these outliers, samples with strength values below 25.0 MPa are deemed too divergent from the context of uncertainty considered in this study and are thus excluded from the training dataset to maintain focus on the most relevant patterns.

### 3.3. Robust Pseudo Labeling with Deep Regression Network

The process of generating data has shown that using the finite element method to obtain a single macroscopic mechanical property from images of composite microstructures is computationally costly. To address this, we introduce a deep learning framework as a surrogate model. This model is designed to quickly predict the transverse strength of each Representative Volume Element (RVE) image in less than a second, provided it has been trained with enough data. Gathering and accurately labeling such data, nevertheless, is a time-intensive process, requiring significant computational resources and effort. The necessity for incorporating pseudo-labeling into our deep learning framework arises from the need to streamline the data generation process and reduce the overall time and resources required. Our framework is underpinned by a model built on convolutional layers, enriched with a pseudo-labeling architecture to adeptly uncover and assimilate the patterns concealed within both labeled and unlabeled sample images. CNNs have been consistently lauded for their exceptional ability to identify critical features and differentiate between subtle image patterns across numerous studies. Their proficiency in image-based classification and regression problems is well-documented, often delivering predictions with an accuracy exceeding 90%, based on the learned information. However, one of the main challenges with CNNs is their dependency on large, high-quality datasets for training in a supervised learning context. These datasets provide a wealth of raw pixel information, enabling the model to discern and focus on the most relevant features for accurate decision-making. Pseudo-labeling addresses this challenge by utilizing the available information from unlabeled samples, thus bypassing the need for labor-intensive labeling processes like finite element analysis. This approach significantly enhances the neural network's baseline performance in a semi-supervised learning scenario, making it a strategic complement to our deep learning framework.

### 3.3.1. Convolutional neural network for image-based regression

Representative convolutional neural networks have emerged as a powerful tool and demonstrated remarkable success in various computer vision tasks, which are realized on the publicly available benchmark datasets [166,167]. They can capture and preserve much more low-level local features from images and the accuracy of prediction shades that of the densely connected model which is preferably applied on the training with explicit parameters. Moreover, the majority of research and application work revolves around the image classification problems, training and fine tuning the deep learning models with tons of images classified in different categories, but continuous prediction and estimation based on images are equally demanded in many research and experimental scenarios, so it will be clarified how the specific convolutional neural network is tailored for the image-based regression problem in our case and why they can work out excellent performances on images.

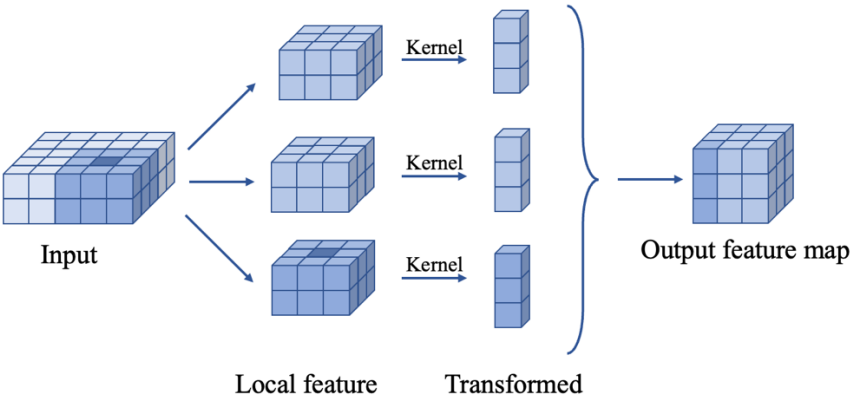

**Figure 3-4.** Workflow of convolutional operation.

The convolution layers, as the essential part of the convolutional neural network, plays a dominant role in extracting the identical and representative features and simultaneously preserve as many local details as possible when looking through all the pixel matrixes of certain size. Instead of learning the global image patterns by multiplying the feature extractor with all the image pixels, the strategy of convolutional layers is to acquire more and more local patterns with certain size of sliding windows performing the convolution operations on one image, which offers more possibilities to obtain miscellaneous local features that may have critical impacts on the final prediction. More importantly, the local features detached from the global patterns can be considered independent indices for identification to determine the desired outputs jointly wherever they are extracted on the images, so image identification will not be restricted by global image patterns. As shown in Figure 3-4 local features from different regions of images are collected by the initial convolutional layers to generate the 3D feature map, sliding the check window of size $3 \times 3$ over the entire image and carrying out the convolution calculation at every stops. Then each 3D local feature map is transformed into a 1D vector of shape, followed by being reassembled into a 3D output feature map. Eventually, the initial feature map can reflect massive original local details and features, including edge, color, textile and so on. Furthermore, increasingly abstract and complex patterns and knowledge will be learned from the collected local features from the previous layers seeing that local features can never give a whole picture for characterizing the images and some certain combinations are expecting to be discovered. The subsequent convolution layers will extract larger patterns made of the local features, which renders the relevant knowledge and features spatially hierarchical. The neural network model thereby can observe and evaluate the sample images from multiple perspectives.

The initial convolution layers generate a feature map teeming with low-level details, yet the sheer volume of parameters and coefficients required for learning can overwhelm the model, leading to significant overfitting. Max-pooling layers address this issue by condensing the feature map, effectively downsampling it. This is achieved by selecting the maximum value from each local region of the feature map across all channels, thereby accentuating the most prominent features. Such a process is crucial for subsequent convolution layers, enabling them to discern more complex, high-level patterns. This pivotal function of max-pooling layers in creating a spatial hierarchy of feature extraction allows the model to analyze input at varying levels of abstraction, ensuring that the most critical information is retained.

Dense layers with selected activation functions embedded are generally served as the last few layers of deep learning models in processing 4D sensor of images for classification problem, since these layers are critical to establish the effective correlations between the extracted features and output categories. In regard to the regression problem, the dense layer without activation function is required to be added in the end of the neural network model, which allows the model can directly investigate how the continuous output values vary as the feature parameters and update the weights with the MSE as the loss.

### 3.3.2. Enhanced VGG-16 for image-based regression analysis

In our research, VGG-16 model [168] is employed to discern the differences among the samples, which principally consists of 13 convolution layers of $3 \times 3$ filter with stride 1 and complemented by 3 fully connected layers culminating in a SoftMax layer for output classification. Illustrated in Figure 3-5, VGG-16 is a substantial model, encompassing approximately 138 million parameters, necessitating a voluminous dataset for effective training. The architectural design, detailed in Figure 3-5, emphasizes the use of 3×3 convolution layers paired with consistent padding and 2×2 max-pooling layers. The model's structure is organized into 5 sequential blocks, each consisting of a set configuration of convolution and max-pooling layers, illustrating a methodical approach to feature extraction and dimensionality reduction.

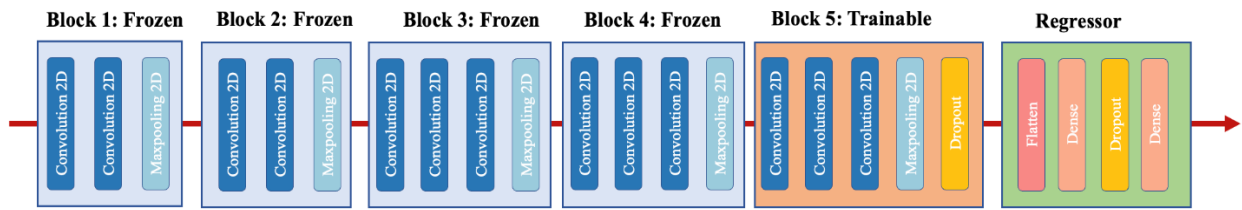

**Figure 3-5.** Architecture of the customized VGG 16.

The decision to employ the VGG16 architecture within our study is underpinned by a careful balance between the risk of overfitting and the necessity for a network of sufficient depth to capture complex patterns, especially given the constraints of our dataset sizes. Initially, our scenario presents us with a limited dataset of only 250 labeled samples for the initial phase of training, a size that poses a significant risk of overfitting if an overly complex network were to be utilized. On the other hand, the availability of 1,000 labeled samples for baseline performance evaluation necessitates a model that is not too shallow, capable of leveraging a larger dataset to achieve a comprehensive understanding of the data's underlying patterns. The VGG-16 model, renowned for its simplicity yet effectiveness, emerges as the ideal candidate under these conditions. In summary, the VGG-16 model provides an optimal balance between depth and simplicity, making it particularly suited to our scenario, where the challenge lies in achieving high performance without succumbing to the pitfalls of overfitting on a modestly sized labeled dataset.

In the quest to enhance the generalization capability of the original deep neural network architecture and to safeguard against overfitting, especially in scenarios with limited training data, dropout has emerged as a pivotal regularization technique. Dropout, as proposed by Srivastava et al [169,170], is a simple yet effective method that randomly "drops" a subset of neurons during the training phase, effectively preventing complex co-adaptations on training data. This stochastic nature ensures that the network maintains a robust feature representation as no single unit becomes too critical to the output, thus encouraging a distributed and sparse representation of features. For our customized VGG-16 architecture, which is tailored for a regression task with a relatively small sample size, the integration of dropout layers is a strategic move to mitigate the risk of overfitting. Given the deep and complex structure of VGG-16, which was originally designed for large-scale image classification tasks, it inherently possesses a high capacity for learning detailed feature hierarchies. While this is advantageous for learning from large datasets, when adapting to a smaller dataset, the model is susceptible to memorizing the noise in the training data rather than learning the underlying data distribution. Therefore, we incorporate the dropout layers into the trainable block and regressor. By positioning a dropout layer at the end of the last trainable block with the convolutional base and in the regressor simultaneously, we aim to reduce the impact of overfitting on the most abstract and complex representations learned by the network. This ensures that the transition from feature extraction to regression does not rely too heavily on any specific activation pattern that might be idiosyncratic to the training set.

In the adapted design of our convolutional network for regression, a novel extension is integrated at the termination of the traditional architecture, incorporating a single-unit dense layer in the regression segment, deliberately devoid of activation functions to accommodate the continuous nature of the target outputs. This approach employs transfer learning to capitalize on the robust, pre-established foundation of the architecture, drawing upon its extensive pre-training on ImageNet [167] for superior image recognition capabilities. To ensure the integrity of the original learned representations, strategic freezing of layers is implemented, specifically preserving the weights within the initial segments of the architecture to prevent undesired adjustments during further training. This measure effectively isolates updates to the concluding segment of the network, alongside the regression extension, allowing for refined adaptation to the dataset in question. Through these modifications, the network emerges as a pivotal component in our deep learning framework, optimized for generating precise regression predictions.

### 3.3.3. Robust pseudo labeling framework

The pseudo labeling technique leverages a pre-trained deep learning model to dynamically assign labels to previously unlabeled samples, refining these labels iteratively throughout the training process. This approach enables the simultaneous training of both originally labeled data and pseudo-labeled data, enhancing the model's understanding of the data's underlying structure. By incorporating both sets of data, the model can achieve a more robust generalization, reducing the uncertainty associated with its predictions. This method effectively expands the dataset, leveraging unlabeled data to improve model performance and reliability.

Assume a set of labeled data is represented by $D_{labeled} = \{(X_1, Y_1), (X_2, Y_2), (X_3, Y_3), ......, (X_n, Y_n)\}$, where $X_i$ and $Y_i$ denote the input images and corresponding ground truth labels respectively, as well as $n$ is the number of labeled samples. For the regression model of our case, $Y_i$ are a series of continuous values in a specified range that represents the mechanical properties of the composite material. Additionally, a set of unlabeled data is demanded for the semi-supervised learning framework, $D_{unlabeled} = \{(X_1^U), (X_2^U), (X_3^U), ......, (X_m^U)\}$, comprising of $m$ unlabeled input images, so it is necessary to utilize the deep learning model trained on $D_{labeled}$ to generate the pseudo labels $Y^*$ for each unlabeled sample. Then, we can have the pseudo-labeled dataset,

$$D_{pseudo} = \{(X_1^U, Y_1^*), (X_2^U, Y_2^*), (X_3^U, Y_3^*), ......, (X_m^U, Y_m^*)\} \quad (3.1)$$

This dataset is to be combined with $D_{labeled}$ to obtain the finalized training dataset,

$$D = \{(X_1, Y_1), (X_2, Y_2), (X_3, Y_3), ......, (X_{(n)}, Y_n), (X_1^U, Y_1^*), (X_2^U, Y_2^*), (X_3^U, Y_3^*), ......, (X_m^U, Y_m^*)\} \quad (3.2)$$

which contains $m+n$ samples in total. During each training epoch, the deep learning model generates predictions for each sample, drawing from the patterns and insights it has previously acquired. Subsequently, the model assesses the precision and relevance of these predictions against the actual ground truth labels, a process encapsulated within the loss functions. For the unlabeled samples, the model initially assesses its predictions against pseudo labels, which are generated from an initial phase of training exclusively with labeled data. Following this initial phase, these pseudo labels can be updated, enabling the calculation of loss for unlabeled samples. Subsequently, these loss functions play a pivotal role in the learning process by providing essential feedback for updating the model's parameters, thereby guiding the model's learning trajectory towards acquiring relevant knowledge. By customizing the loss function, we gain the ability to direct how the model processes and learns from both labeled and pseudo-labeled samples. This adjustment is crucial because the labels derived from pseudo-labeling do not share the same level of reliability as directly labeled data, necessitating differentiated weighting within the model's learning mechanism. The loss function of training samples is assumed to be the mean square error, considered the significant index of the performance of proposed regression model simultaneously, so we have loss function:

$$L_{labeled}(X_i, Y_i) = \frac{1}{n}\sum_i^n (Y_i^p - Y_i)^2, L_{unlabeled}(X_i^U, Y_i^*) = \frac{1}{m}\sum_i^m (Y_i^p - Y_i^*)^2 \quad (3.3)$$

Then the overall weighted loss function will be

$$L = L_{labeled}(X_i, Y_i) + \alpha L_{unlabeled}(X_i^U, Y_i^*) \quad (3.4)$$

where $\alpha$ is the weight assigned to the pseudo labels, which will be updated through the model training and pseudo label revision.

### 3.3.4. Confidence-driven pseudo label refinement

To enhance the pseudo-labeling framework, we incorporate a confidence measure to sift through pseudo labels, prioritizing those with a high confidence level to ensure the integrity of pseudo-labeled data [171], particularly crucial in classification tasks where the model inherently provides a probability for each prediction, directly indicating the uncertainty of these predictions. In regression models, however, a straightforward measure of confidence for predictions is not inherently available, necessitating an alternative approach to ascertain confidence levels for regression tasks. Given the variability in deep learning model outputs where a single sample might yield different predictions across multiple evaluations, including within the pseudo-labeling framework, it is reasonable to infer that a prediction made with high frequency signifies a higher confidence level in that outcome. Consequently, we propose constructing a confidence interval by aggregating output values over multiple predictions, forming a pool of potential labels from which the model can derive a confidence-weighted decision. This method aims to bridge the gap in directly assessing prediction confidence within regression models.

In 50 times of predictions over the single sample, the values are distributed as Figure 3-6. Correspondingly, the confidence interval (CI) [172,173] can be derived from the distribution:

$$CI_{0.05}^{\mu} = \left[ \hat{\mu} - Z \times \frac{\sigma_{\varepsilon}}{\sqrt{n}}, \hat{\mu} + Z \times \frac{\sigma_{\varepsilon}}{\sqrt{n}} \right] \quad (3.5)$$

in which, $\hat{\mu}$ is the average value of the predictions, Z is the Z-score [174], the $\sigma_{\varepsilon}$ is the standard deviation of the sample, and $n$ is the size of the sample. In the context of pseudo-labeling, the confidence interval offers a nuanced perspective by providing both lower and upper bounds, along with a likelihood of prediction precision. A shorter confidence interval is indicative of a higher precision and a smaller margin of error, implying that the model's predictions are consistent and reliable. Contrary to the typical application of confidence intervals where a broad range might be selected to ensure a high likelihood of encompassing the true value, our approach is distinct. We focus on a narrow confidence interval that captures the most consistent predictions across multiple training iterations—reflecting a deep learning model's firm grasp of the underlying patterns in the data. To this end, we employ a 5% confidence interval in Figure 3-6 to delineate the threshold for selecting reliable pseudo labels from a limited number of predictions. This stringent criterion ensures that only predictions that the model consistently makes with high certainty are considered. By doing so, we prioritize precision and consistency in the pseudo labels, which is crucial for the stability and performance of the model as it learns from both labeled and pseudo-labeled data. This approach to utilizing confidence intervals is particularly strategic in ensuring that the incremental learning from pseudo labels is grounded in the most dependable insights the model can generate.

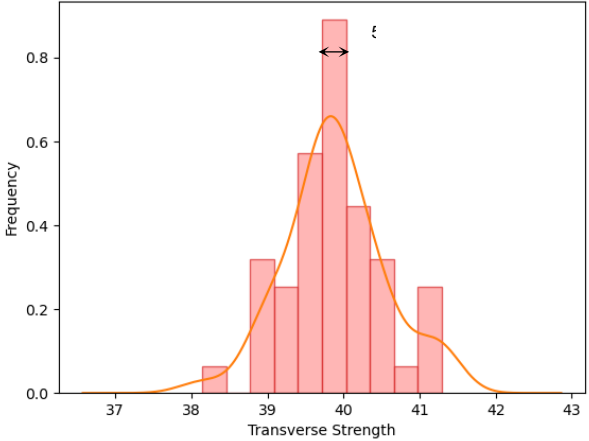

**Figure 3-6.** Distribution of multiple predictions on single sample.

### 3.3.5. End-to-end workflow of confidence-informed pseudo labeling

Having established the importance of confidence measures in our pseudo-labeling approach, we now turn our attention to the end-to-end workflow that encapsulates the entire process. The confidence measures are not standalone components; they are integral to the iterative cycle of model improvement. The accompanying diagram in Figure 3-7 provides a visual depiction of this comprehensive workflow. Starting with an initial model trained on labeled data, we systematically leverage the confidence measures to guide the selection of pseudo-labels from the predictions on unlabeled data. Through consensus voting, the most frequently predicted values are identified and used to create high-confidence pseudo-labels. These pseudo-labels, once weighted by their confidence scores, serve as a novel source of information in subsequent training iterations. This iterative process visualized through the cyclical arrows in the diagram continues until the model achieves a satisfactory level of accuracy and stability. The pseudo-code of the robust pseudo labeling framework in Algorithm 1 provides a step-by-step description of the workflow.

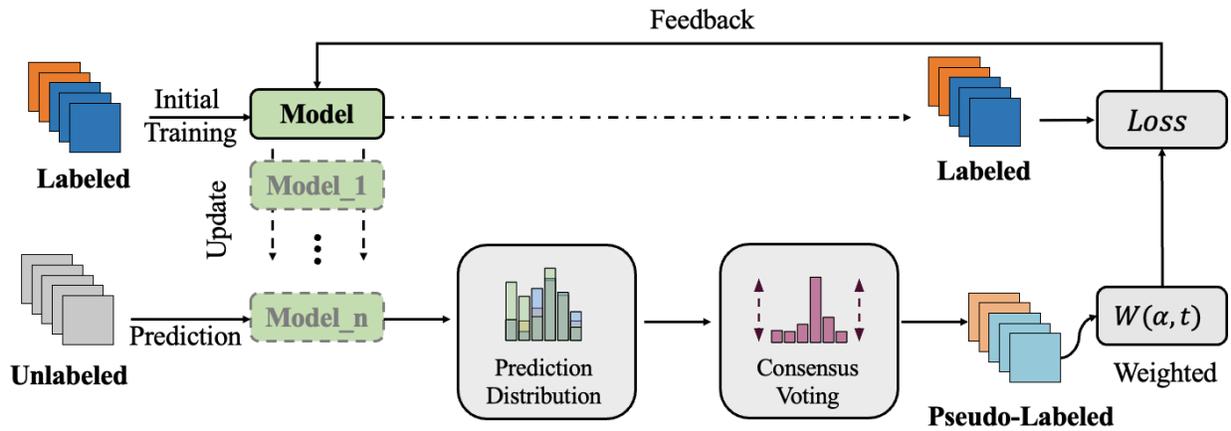

**Figure 3-7.** Architecture of robust pseudo labeling framework. Initial training allows the base model to assign pseudo labels to unlabeled samples. Subsequently, the model is updated with a mix of weighted samples for further improvement.

**Algorithm 1** Confidence-Driven Pseudo-labeling

INPUT:

$D_{labeled} = \{(X_1,Y_1),(X_2,Y_2),(X_3,Y_3),......,(X_n,Y_n)\}$

$D_{unlabeled} = \{(X_1^U),(X_2^U),(X_3^U),......,(X_m^U)\}$

$n_{batch}$ : Batch size of training samples once time of looking

$n_{epochs}$ : Epochs of training

$R_{sample}$ : Sample rate

$\alpha_t$ : Weight of loss

CLASS PSEUDOLABEL:

%% Specify the base deep learning model for training:

$Base = CustomizedVGG16$

%% Assign initial random generated labels for unlabeled data

$$Y_{init}^P = random.uniform(min, max, size)$$

Then, labeled and pseudo-labeled data are combined for training:

$$D = \{(X_1, Y_1), (X_2, Y_2), (X_3, Y_3), \ldots, (X_{(n)}, Y_n), (X_1^U, Y_1^*), (X_2^U, Y_2^*), (X_3^U, Y_3^*), \ldots, (X_m^U, Y_m^*)\}$$

Simultaneously, assign flags 0 or 1 to training data for differentiation:

$$[D, flag] = \{[(X_1, Y_1), 0], [(X_2, Y_2), 0], \ldots, [(X_{(n)}, Y_n), 0], [(X_1^U, Y_1^*), 1], [(X_2^U, Y_2^*), 1], \ldots, [(X_m^U, Y_m^*), 1]\}$$

Shuffle the mixed data and generate training batches:

For $i = 1$ to $n_{batch}$ do:

$$D_{ibatch} = \{(X_{i \cdot n_{batch}}, Y_{i \cdot n_{batch}}), \ldots, (X_{(i+1) \cdot n_{batch}}, Y_{(i+1) \cdot n_{batch}})\}$$

Compute the loss with loss function defined as:

$$Loss = MSE(Y_{true} - Y_{prediction})$$

$$Coefficient = 1.0 - flag + \alpha_t \cdot flag$$

Return $Coefficient * Loss$

For each epoch, $\alpha_t$ varies as:

$$\alpha_t = \begin{cases} 0, n_{epochs} \leq 20 \\ \alpha_f \dfrac{n_{epochs} - 20}{70 - 20}, 20 \leq n_{epochs} \leq 70 \\ \alpha_f, n_{epochs} \geq 70 \end{cases},$$

The prediction on one pseudo-labeled sample will be conducted 50 times

$$Y_{prediction} = \{Y_{m_1}^*, Y_{m_2}^*, Y_{m_3}^*, \ldots, Y_{m_{50}}^*\}$$

Then the confidence interval of prediction is specified to guarantee high-confidence prediction is selected:

$$\bar{y} \pm z \dfrac{\sigma}{\sqrt{50}},$$

Among which, $\bar{y} = \dfrac{1}{50} \sum y_{m_i}^*$, $\sigma$ is the standard deviation, $Z = 0.13$,

$$Y_m^* = Y_{m_i}^* \in [\bar{y} - z \dfrac{\sigma}{\sqrt{50}}, \bar{y} - z \dfrac{\sigma}{\sqrt{50}}]$$

%% The highest frequent prediction values as the representative results to update the pseudo labels

for $i = 1$ to $n_{epochs}$ do:

    repeat the above process

end for

The workflow begins with the collection of input data, consisting of a labeled dataset and an unlabeled dataset. The labeled dataset includes pairs of data points and their corresponding labels, while the unlabeled dataset consists of data points without labels. Essential parameters such as batch size, the number of training epochs, and the sample rate are established. In the initial stage, a deep learning model, specifically an enhanced version of VGG-16, is chosen as the base model for training. This model will learn to predict labels for the unlabeled data. To start, the algorithm assigns random labels to the unlabeled data points, creating an initial set of pseudo-labels. These labels are not accurate and are meant to be refined as the model trains and learns. The labeled and pseudo-labeled datasets are then merged into a single training dataset. To keep track of which data points are truly labeled and which are pseudo-labeled, flags are introduced, marking the true labels with a flag of 1 and the pseudo-labels with a flag of 0. This combined dataset is shuffled to ensure that the model does not learn any particular order of the data points and is then divided into batches according to the specified batch size.

Training proceeds in batches, where the model uses a loss function to evaluate its predictions. The loss function chosen here is the Mean Squared Error (MSE), which measures the average squared difference between the predicted labels and the true labels. The loss for each batch is adjusted by a coefficient that takes into account whether the data points are truly labeled or pseudo-labeled. This coefficient is a combination of a constant term and a variable weight that changes depending on the epoch of training. The purpose of this coefficient is to differentiate the contribution of labeled and pseudo-labeled data to the loss, thus allowing the model to place varying levels of trust in the pseudo-labels as training progresses. After the initial round of training, the model makes predictions on the unlabeled data points multiple times to build a distribution of possible labels. A confidence interval is calculated for these predictions to select the most probable label. A high-confidence prediction is more likely to be accurate. The standard deviation of the predictions is used, along with a constant Z-score, to establish the interval around the mean prediction. Predictions that fall within this interval are considered high confidence.

The pseudo-labels are then updated based on the highest frequency of the predicted values within the confidence interval. This iterative process of updating the pseudo-labels continues for the specified number of epochs, with the weight of the loss for pseudo-labeled data being adjusted as training progresses. The algorithm aims to improve the model's performance by increasingly refining the pseudo-labels with each epoch, resulting in a model that can make more accurate predictions on both the labeled and unlabeled data.

### 3.4. Results and Discussion

This section provides a comprehensive evaluation of our framework's performance, highlighting its efficiency in delivering accurate predictions with a smaller data footprint. We start from a baseline assessment using supervised learning, underscoring the influence of dataset size on model outcomes. Subsequent subsections present the advancements achieved through our semi-supervised learning approach, where we demonstrate that our model maintains high performance levels even when trained on a reduced number of samples. This finding validates our framework's efficiency. We proceed to examine the consistency and refinement of pseudo labels during training, affirming the efficacy of our pseudo-labeling technique. Additionally, we investigate the ideal proportion of unlabeled to labeled data, seeking to optimize performance. This exploration is not only instrumental for future implementations but also illustrates the prowess of our surrogate model in simplifying the quantification of uncertainty. Collectively, these insights accentuate the tangible benefits of adopting our model in practical scenarios.

#### 3.4.1. Performance impact of labeled sample reduction in supervised learning

In this section, we investigate the impact of sample size on the performance of supervised learning within our context. We conduct experiments using 900 and 250 labeled samples, respectively, to assess how reduction in data availability affects model accuracy.

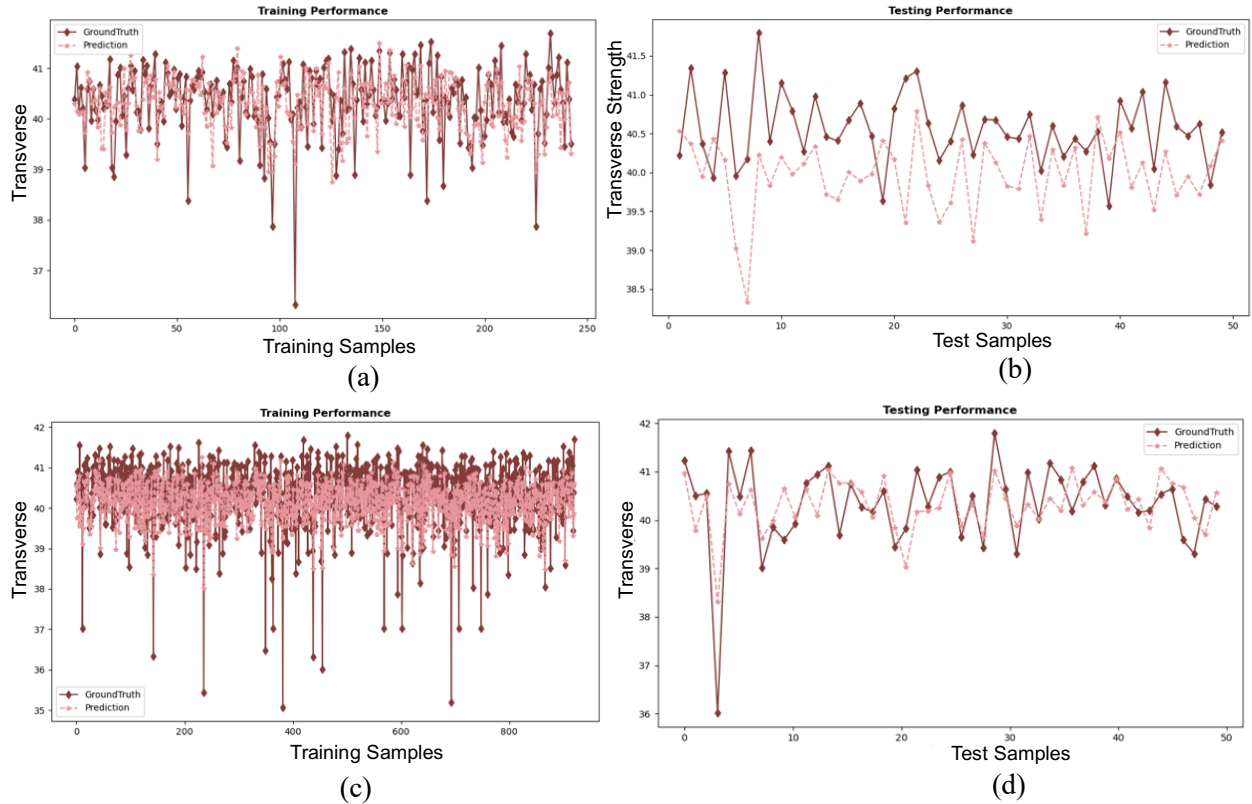

**Figure 3-8.** Model performance of supervised learning (a) training results of 250-sample case; (b) test results of 250-sample case; (c) training results of 900-sample case; (d) test results of 900-sample case.

These experiments employ various convolutional neural networks, with each model undergoing a fine-tuning process to optimize performance irrespective of sample size. Our goal is to determine the highest achievable accuracy for each data scenario, reflecting the neural network's capacity when constrained by the number of labeled samples. Upon testing the models with 50 unseen labeled images, we draw performance comparisons. Figure 3-8(b) illustrates that the model with 250 samples is able to capture the data pattern in the test dataset, albeit with predictions that generally do not coincide precisely with the true values. Conversely, Figure 3-8(d) shows the model trained with 900 samples attaining superior predictions, including a more accurate detection of outliers in the test dataset. This disparity indicates that a decrease in labeled data from 900 to 250 samples considerably degrades the model's performance in supervised learning.

## 3.4.2. Label-efficient training with semi-supervised learning

To rigorously assess the efficacy of semi-supervised learning relative to supervised learning, we maintained an equal count of labeled and unlabeled samples across both models. Specifically, we reclassified 650 out of 900 samples as unlabeled for the semi-supervised model to determine whether it could mirror the performance of its fully supervised counterpart under the same training volume.

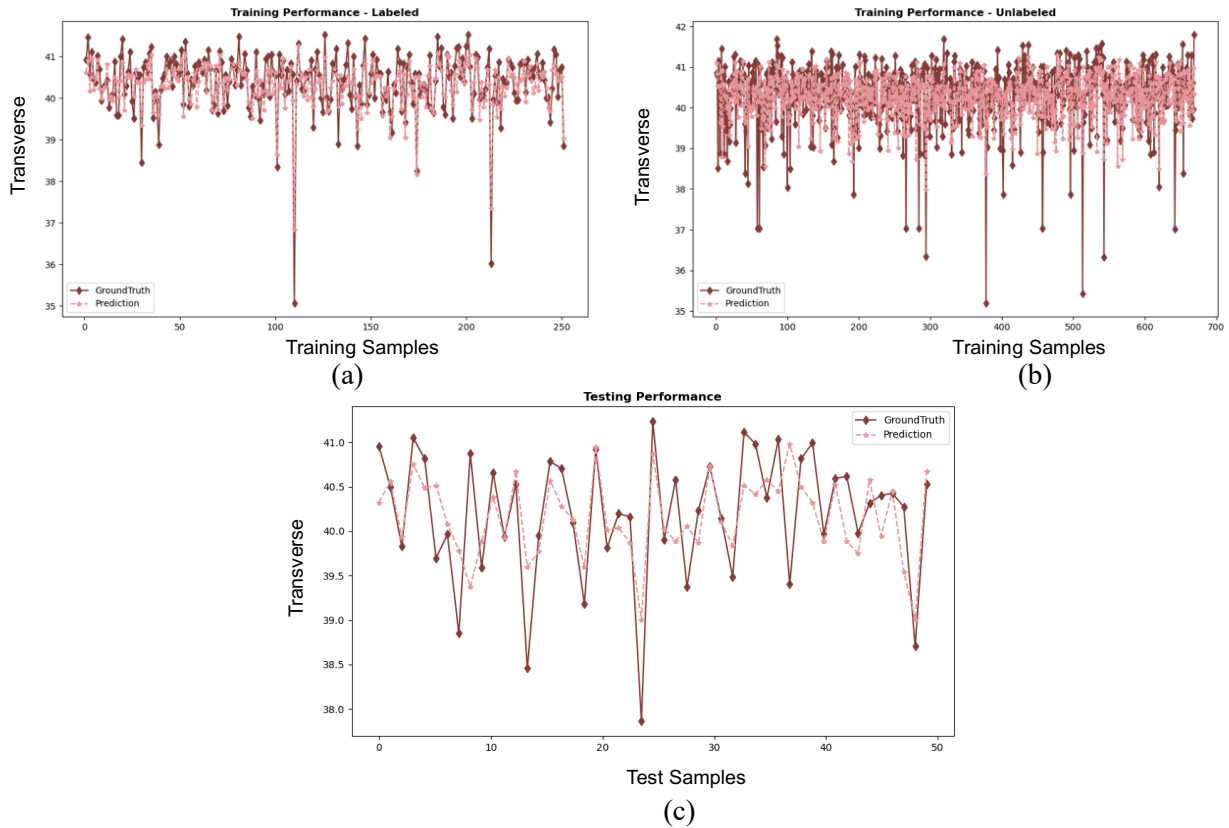

**Figure 3-9.** Model performance of semi-supervised learning (a) training results of labeled samples; (b) training results of pseudo-labeled samples; (c) test results.

In alignment with the pseudo labeling process delineated in previous sections of workflow, the initial model trained with 250 labeled samples assigns pseudo labels to the 650 unlabeled samples. These pseudo labels, alongside the actual labels, are then differentially weighted and fed into the model for further training. This iterative process continues until the loss function stabilizes. The final model's accuracy is subsequently gauged against 50 test samples, as depicted in Figure 3-9(c). It's important to acknowledge that typically, it is not feasible to visualize the training performance on unlabeled samples, as shown in Figure 3-9(b), due to the unavailability of ground truth labels. Nonetheless, for the essential purpose of evaluating our model's accuracy on unlabeled data and thus confirming the validity of the pseudo-labeling method, we have implemented a strategy where the model forecasts these samples' labels. Subsequently, these predictions are meticulously compared against a hidden set of actual labels that were specially set aside for this evaluation.

The mean square error (MSE) achieved through semi-supervised learning closely mirrors the MSE attained in the supervised learning scenario utilizing 900 labeled samples. This parallel suggests that semi-supervised learning is capable of delivering comparable performance with a significantly reduced labeling effort. A comprehensive comparison of the total time invested in the training processes of both supervised and semi-supervised learning, specifically employing the robust pseudo labeling framework, is detailed in Table 3-1. The training durations documented are contingent upon the computing resources used, specifically an Intel Xeon Silver 4210 CPU at 2.20 GHz and an NVIDIA Quadro RTX 5000 GPU.

Table 3-1. Performance comparison between supervised learning & semi-supervised learning

| Method | Labeled/Unlabeled | Testing MSE | Labeling Time | Training Time |
|---|---|---|---|---|
| Supervised | 250/0 | 2.7432e-3 | 187.5 h | 30.64 s |
| Supervised | 900/0 | 2.0693e-4 | 675 h | 120.57 s |
| Semi-supervised | 250/650 | 2.8712e-4 | 187.5 h | 1406.25 s |

Labeling time in semi-supervised learning shows a significant reduction of 72.2% when compared to supervised learning with an equivalent number of samples. This time saving is calculated based on the average duration of 45 minutes for labeling a single representative volume element (RVE) sample via finite element analysis. Therefore, employing 650 unlabeled samples for neural network training can result in considerable time savings, which substantially outweighs the differences in training time costs between the models. Although semi-supervised learning yields a slightly higher mean square error (MSE), indicating certain limitations of pseudo labeling in pattern learning and data structure extraction from labeled samples, this does not detract from the viability of pseudo labeling nor from its substantial time-saving benefits, particularly in applications where time and computational resources are at a premium. Moreover, the notable improvement in MSE from 2.7432e-3 to 2.8712e-4 underscores the positive impact of including unlabeled samples to bolster the training performance through pseudo labeling.

### 3.4.3. Efficacy of pseudo labels and label quantity analysis

Moving deeper into the analysis of our training methodology, we turn our attention to the pseudo-labeling mechanism. The progression of pseudo-label refinement is a key factor in the overall effectiveness of the model. Figure 3-10 provides a visual narrative of how these labels are updated and progressively perfected throughout the training process. The blue curve, aligned with the left vertical axis, represents the MSE of the pseudo-labeled data over the course of training. Despite the usual impracticality of obtaining MSE for pseudo labels due to the absence of ground truth, our unique setup allows for this by utilizing 650 unlabeled samples whose true labels are hidden rather than absent. This approach enables us to validate the effectiveness of our pseudo-labeling framework by comparing pseudo labels against their actual labels. The decreasing trend of the blue curve indicates that the MSE for pseudo labels diminishes as the model training progresses, which signals that the model is learning effectively even from the unlabeled data.

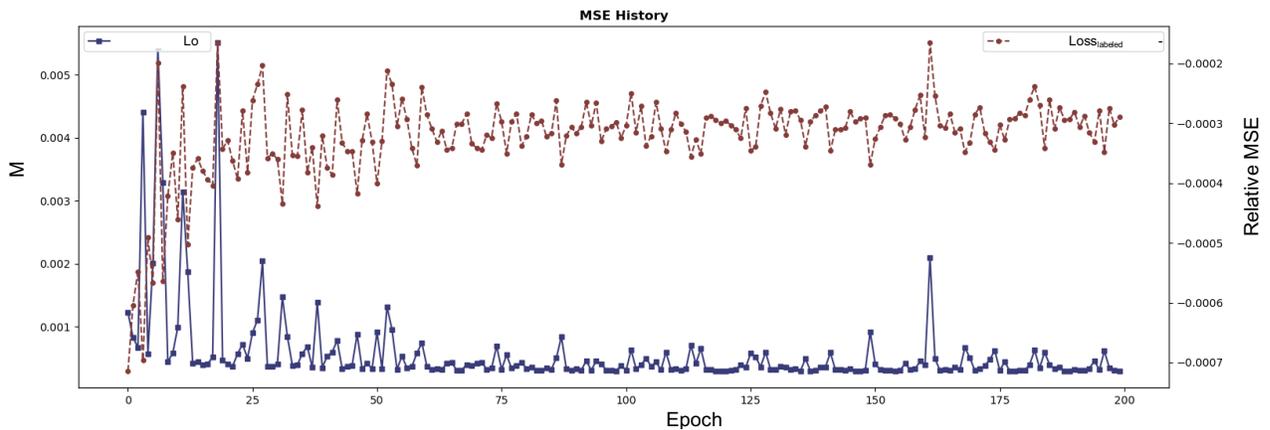

**Figure 3-10.** MSE history of the training process.

On the right vertical axis, we have the relative MSE indicated by the red curve. This measures the difference between the MSE of ground truth labeled samples and the MSE of pseudo-labeled data. The values plotted are negative; therefore, an upward trajectory of the red curve signifies a decreasing gap between the MSE of pseudo labels and the MSE of actual labels. This suggests that the reliability of pseudo labeling is improving throughout the training process, with the pseudo labels becoming increasingly accurate as reflected by the narrowing MSE discrepancy. The trends corroborate that accurate pseudo labels can offer substantial and precise guidance during model refinement.

The efficacy of pseudo-labeling is evident, yet the optimal balance of labeled and pseudo-labeled data remains a key question. Our experiments in Table 3-2 maintained a constant total of 900 samples, divided between labeled and pseudo-labeled samples in varying ratios. Through this method, we determined that diminishing the count of labeled samples below 250 leads to a marked decline in performance, suggesting a threshold for achieving a robust learning structure. Observations from Section 4.2 indicate that supervised learning with only 250 labeled samples yields suboptimal performance. This suggests that any further reduction below this threshold would likely degrade results. Such a scenario forms the initial training basis in semi-supervised learning with fewer than 250 labeled samples. Consequently, we infer that a dataset with less than 250 labeled samples fails to establish a sufficient underlying data structure, which is essential for effective learning using pseudo labels. This is corroborated by the trend observed in semi-supervised learning experiment in Table 3-2, where diminishing the labeled data count consistently correlates with declining performance. Therefore, this threshold must be considered when optimizing the labeled to pseudo-labeled data ratio.

**Table 3-2.** Increasing labeled data count in semi-supervised learning

| Labeled Sample | Unlabeled Sample | Testing MSE | Labeling Time |
|---|---|---|---|
| 50 | 850 | 6.9208e-4 | 37.5 h |
| 100 | 800 | 6.8696e-4 | 75 h |
| 150 | 750 | 6.5282e-4 | 112.5 h |
| 200 | 700 | 3.4634e-4 | 150 h |
| 250 | 650 | 2.8712e-4 | 187.5 h |

### 3.4.4. Unlabeled data exploitation for optimal performance

Our enhanced pseudo labeling framework benefited from an additional 5,000 unlabeled samples sourced from the existing pool. As the model progressed through training, it honed its ability to discern data patterns, leading to improved pseudo label predictions. Consequently, the reliance on pseudo labels grew, bolstered by the increasing accuracy of model predictions. In our study, the VGG-16 model's prior validation under supervised learning, along with the customized CNN model referenced in literature [26], establishes a supervised learning benchmark for performance comparison. With the integration of an extra 5,000 unlabeled samples into the pseudo labeling framework, we anticipate a notable performance boost.

Table 3-3 presents the semi-supervised learning outcomes using 3,000 unlabeled samples, with these findings also depicted in Figure 3-11. These are contrasted with the supervised learning results using 900 labeled samples and the benchmarks from comparable studies [31]. The semi-supervised approach with 3000 unlabeled samples shows only a slight decline in performance compared to these benchmarks, yet it boasts a 27.52% reduction in MSE compared to the same approach with 650 unlabeled samples. These findings highlight the effectiveness of pseudo labeling in harnessing the potential of unlabeled data, though it doesn't guarantee superiority over well-resourced supervised learning. Additionally, the diminishing returns on performance with an increasing volume of unlabeled data suggest a need to determine the optimal ratio of unlabeled to labeled data to optimize training efficiency and minimize data generation costs.

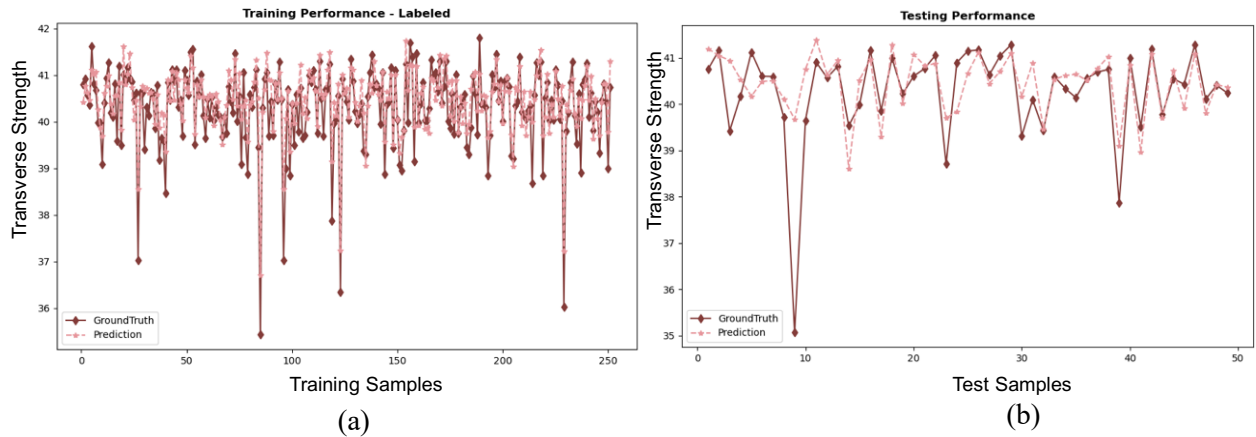

**Figure 3-11.** Model performance of semi-supervised learning with 3,000 unlabeled samples (a) prediction of well-trained model over 250 labeled samples; (b) prediction of well-trained model over test.

**Table 3-3.** Enhancement based on benchmark performance

| Model | Labeled/Unlabeled | Testing MSE | Labeling Time | Training Time |
| --- | --- | --- | --- | --- |
| Supervised (VGG-16) | 900/0 | 2.0693e-4 | 675 h | 120.57s |
| Supervised (CNN) [22] | 900/0 | 2.3420e-4 | 675 h | N/A |
| Semi-supervised | 250/650 | 2.8712e-4 | 187.5 h | 1406.25s |
| Semi-supervised | 250/3000 | 2.0810e-4 | 187.5 h | 7405.62s |

Our experiments explored the impact of varying sample rates on pseudo labeling effectiveness, as documented in Table 3-4. Initially, increasing the sample rate up to 0.6 led to performance improvements. However, further increases to 0.8 and 1.0 resulted in a decline in results. This pattern underscores that adding more unlabeled data doesn't uniformly enhance model performance and may, in fact, diminish it. This decline isn't solely due to the size of the dataset but also factors like data distribution and inherent noise. Notably, the 5,000 unlabeled samples, distinct from the initial 1,000 labeled samples, highlighted that not all augmentations from unlabeled data pools are beneficial. Some results were even less effective than those achieved with 650 unlabeled samples from the original labeled pool. These findings emphasize the importance of adjusting hyperparameters, such as sample rate, based on the specific characteristics of the data, rather than adhering to fixed benchmarks.

The time efficiency of model training, while often overshadowed by the significant 72% reduction in label generation time via finite element analysis, warrants attention. Our examination reveals that the substantial discrepancy in labeling time, amounting to approximately 500 hours, vastly overshadows the combined costs of training and sample generation. Implementing pseudo labeling with 3,000 unlabeled samples on a standard workstation, without the benefit of high-performance GPU acceleration, the training completed in about 33 hours. This observation reinforces our assertion that semi-supervised learning, through pseudo labeling, offers a cost-effective alternative, achieving similar results with considerably lower labeling expenses. The minimal impact of hardware acceleration on overall time savings further supports the suitability of semi-supervised learning for scenarios where labeling is expensive but sample generation is relatively inexpensive.

Table 3-4. Performance investigation with different sample rates

| Labeled Sample | Unlabeled Sample | Sample Rate | Testing MSE | Labeling Time |
|---|---|---|---|---|
| 250 | 1000 | 0.2 | 4.0931e-4 | 187.5 h |
| 250 | 2000 | 0.4 | 3.4300e-4 | 187.5 h |
| 250 | 3000 | 0.6 | 2.0810e-4 | 187.5 h |
| 250 | 4000 | 0.8 | 4.3548e-4 | 187.5 h |
| 250 | 5000 | 1.0 | 4.8779e-4 | 187.5 h |

### 3.4.5. Streamlining uncertainty quantification by surrogate model

The result discussion above on the performance enhancement introduced by the proposed semi-supervised learning framework gradually paints a comprehensive picture of how to effectively apply this method in practice. It suggests that a certain amount of labeled data is necessary for model to acquire the baseline recognition on the sample pattern which yet hinges on what performance we consider as baseline. In our implementation, we are pursuing the minimized mean square errors so that we found out 250 labeled samples could be the balance in the tradeoff between the performance and computational cost in our training as the performance improvement by adding more labeled data vanishes but the relevance and advantage of our proposed method are declining due to the continuously increasing computation cost. However, we are supposed to be informed of the fabrication criteria to determine the baseline performance in the realistic application and subsequently strive for superior performance if computationally affordable. With regard to the unlabeled data involved in the model training, we have observed that significant amount of unlabeled data can facilitate the identification of more precise boundary conditions for model generalization whereas those data points outside the margins of boundary conditions do not help refine the boundary conditions and additionally the outliers can even confuse the model, so it is not surprising that increasing amount of unlabeled data does not ensure performance gains. Given this observation, we could be more cautious in augmenting the unlabeled data in the semi-supervised learning especially considering unlabeled data generation incurs a comparatively low cost, which is outweighed by the label generation cost.

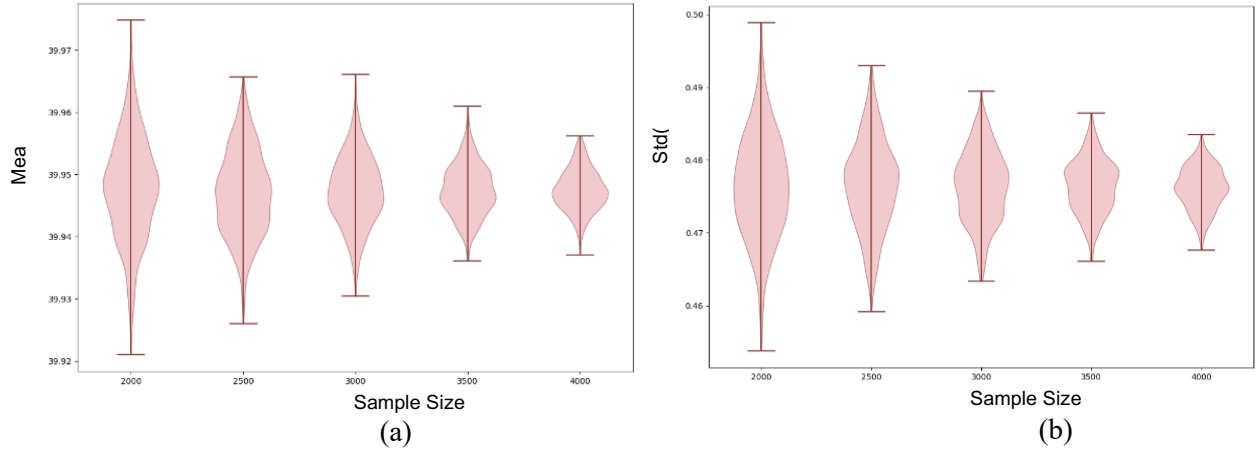

**Figure 3-12.** Sampling Distribution of Prediction (a) Mean Distribution; (b) Std Distribution.

Moreover, we investigate leveraging the well-trained model under the above guidelines to characterize the uncertainty distribution of the unlabeled sample pool, which is the common scenario we would encounter in our application of the model. We sampled 2000, 2500, 3000, 3500, and 4000 specimens from the unlabeled sample pool and computed the mean values and standard deviations of the predicted strength values. Then, the operations of sampling were performed 500 times to unveil the distributions of the mean values and standard deviations. It is noticeable in Figure 3-12 that 3500 samples can statistically guarantee a comprehensive distribution of uncertainty. Ultimately, we investigated and discussed the impact of the labeled data size and unlabeled data size on the model training performance in semi-supervised learning framework and uncertainty analysis convergency. More importantly, the observations we collected and concluded from the experiments and analysis can provide pertinent information and knowledge in applying the proposed method in miscellaneous research.

## 3.5. Conclusion

The semi-supervised deep learning framework proposed in this research represents an advancement in efficiently quantifying uncertainty propagation in composite microstructures. Utilizing Representative

Volume Elements (RVEs) that model the spatial randomness of fiber-packed composites, this approach overcomes the limitations associated with the intensive computational demands of generating accurate ground truth labels through finite element analysis, a process often deemed impractical for supervised learning due to its high demand for labeled data. At the heart of the proposed framework is the innovative use of robust confidence-driven pseudo labeling framework, with the deep regression neural network model laying the foundation. This method adeptly harnesses the potential of unlabeled data, significantly enhancing the model's pattern recognition capabilities and generalization. The result is a level of predictive performance that rivals that of benchmark supervised learning models, achieved with nearly 72% fewer labeled samples. This efficiency translates into a significant reduction of approximately 487.5 hours in label generation time. Moreover, we identify the optimal ratio of unlabeled to labeled samples for better future applications and visually demonstrate the pseudo labeling process to confirm its effectiveness. This work not only showcases the transformative potential of leveraging unlabeled data but also establishes a new standard for conducting efficient and effective uncertainty quantification in the realm of composite microstructures.

# Chapter 4. An Interpretable Parallel Spatial CNN-LSTM Architecture for Fault Diagnosis in Rotating Machinery

In the evolving landscape of Prognostics and Health Management (PHM) enhanced by the Internet of Things (IoT), diagnosing machinery system faults is critical for ensuring operational efficiency and safety across various industries. This research introduces a novel, interpretable deep learning architecture designed to overcome key limitations in existing fault detection methods, such as the high demand for extensive training data and the lack of transparency in feature extraction. Our model uniquely integrates dual branches: one processing raw time-series data through a spatially transformed convolutional neural network, and another incorporating wavelet transform coefficients. This dual approach not only maximizes the effective use of limited data but also significantly enhances model interpretability, eliminating the need for extensive feature engineering and manual feature selection. The significance of this research lies in its innovative methodology, which bridges the gap between advanced deep learning techniques and practical applicability in industrial settings. By leveraging IoT sensors and real-time data processing, our model exemplifies a practical application of IoT in PHM. The proposed algorithm is rigorously evaluated on experimental gearbox data and further validated on a publicly available bearing dataset, demonstrating its generalizability and scalability. Through comprehensive parametric investigations, we elucidate the impact and robustness of the physics-integrated parallel architecture, showcasing its potential to significantly improve fault diagnosis accuracy in diverse operational conditions. This study not only advances the state-of-the-art in fault diagnosis but also provides a framework for developing more interpretable and efficient deep learning models for industrial applications.

## 4.1. Introduction

The application of intelligent fault diagnosis techniques in machinery [175] is a cornerstone for the progression of smart manufacturing as well as transportation systems [176], [177]. Critical components like gearboxes and rolling bearings in these systems demand meticulous attention, as minor faults may escalate into severe, catastrophic failures. Vibration signal analysis, a prominent method for machinery fault detection, has become instrumental. Through the precise monitoring of vibration signals, we can detect various types of faults in bearings, gears, and other vital parts. Early detection is key in averting major failures, thereby minimizing downtime and extending the lifespan of machinery. The integration of IoT technologies in this realm elevates the capabilities of PHM systems, allowing for real-time monitoring and advanced data analytics. Such technological advancements facilitate a more proactive approach to maintenance and ensure the safety and efficiency of the mechanical transmission system [178]-[180]. The focus of this research is on the critical aspect of interpreting the gathered vibration data. While the integration of IoT in PHM provides a robust platform for data acquisition, the crux of fault diagnosis lies in the effective analysis of these vibration signals.

There exist a wide variety of techniques for vibration signal analysis [181], which is categorized based on the characteristics of the vibration signal, such as model-based, knowledge-based, and data-driven approaches. The model-based or knowledge-based methods generally rely on relevant domain expertise to create a mathematical model of the system with known rules, heuristics, or physical laws, which are preferred in fields where ethical considerations demand that decisions be made based on completely transparent and understandable processes. Meanwhile, data-driven methods have gained increasing popularity due to their ability to automate the signal analysis and more importantly the impressive performance without demanding much domain knowledge in many applications. Praveenkumar et al [182] utilized statistical features extracted from mechanical signals and applied them to the support vector machine (SVM), resulting in an average classification efficiency exceeding 95%. Li et al [183] leveraged the sophisticated technique of Fuzzy k-Nearest Neighbor for the nuanced task of fault pattern recognition within gearboxes. This was accomplished through feature extraction via the highly intricate Kernel Independent Component Analysis algorithm. Additionally, the Gaussian Process Classifier, as implemented by Liu et al [184], is capable of furnishing real-time and predictive fault diagnosis of operational signals in wind turbines. These examples of machine learning techniques facilitate the automatic correlation between features and classifications, proving to be efficacious classifiers in the respective case studies. However, the potency of their performance is contingent upon astute feature selection or distribution assumption by kernel functions, whereby physically meaningful and pertinent features are meticulously handpicked for the purpose of classifier training. It is evident that appropriate feature selection or distribution assumption still requires an exhaustive comprehension of the data and subject matter expertise in problem formulation. These prerequisites make such methods inherently case-specific.

As a logical progression in technological advancements, deep learning models are progressively being acknowledged for their capabilities in resolving fault diagnosis via vibration signal analysis. Senanayaka et al [185] adeptly utilized the convolutional neural network (CNN) for classifying gearbox malfunctions. In a noteworthy move, they attempted to decipher the intricacies of the model training by scrutinizing the correlation between characteristic frequencies and the CNN's feature learning filters. In [186], Singh et al proposed the deep learning-based domain adaptation method for gearbox fault diagnosis under different conditions to effectively address the existing shortcomings of manual handcrafted features. In addition, the deep learning based approaches for rotating machinery are summarized together with the existing challenges and possible future research orientations by Tang et al [187].

From these examples, we can observe excellent performances of those deep learning model on the fault diagnosis of rotating machinery, as well as the compelling benefits of requiring no specialized knowledge or engineered features. This offers an advantage by obviating the need for extensive subject matter expertise and intricate, labor-intensive feature creation. However, deep learning models are intrinsically challenging to interpret as they do not readily yield an understanding of the underlying properties that drive their inferential outcomes. This opaque nature necessitates an augmented reliance on both the quality and quantity of the dataset to offset the inherent lack of interpretability when aiming to augment the performance of deep learning model both effectively and efficiently.

Accordingly, a number of physics-informed strategies have been explored to make the models more interpretable. Some researchers have incorporated interpretability by utilizing pre-selected features required by the machine learning algorithm to be physics-informed. This approach aims to make the models more understandable by leveraging domain-specific features that align with physical principles [188]. Others have integrated physically meaningful equations or signal processing techniques with machine learning models, claiming these methods as explainable or physics-informed [189], [90]. These approaches contribute significantly by bridging the gap between machine learning and domain knowledge, making the models more interpretable and relatable to experts in the field. However, these methods come with limitations. The need for meticulously engineered features demands a thorough comprehension of the problem formulation, which can be a challenging and time-consuming task. This requirement can restrict the broader applicability and scalability of the models.

Russel et al [191] advanced the field by integrating generative models with local structure and physics-informed loss terms that incorporate domain knowledge for machine fault diagnosis. They utilized Fast Fourier Transform (FFT) coefficients to provide a distinct perspective of the frequency domain, improving signal reconstruction accuracy. McGowan et al [192] further contributed by directly incorporating physical parameters into the categorical cross-entropy loss function after normalization, enhancing the model's discriminative capabilities. While these methods represent significant advancements in fostering model interpretability, they also have inherent limitations. The immediate integration of normalized singular physical parameters into the loss function may oversimplify complex physical phenomena or inadvertently skew the model towards overfitting a single incorporated physical metric, potentially overlooking other important features or details in the data.

The goal of this research is to address the aforementioned challenges to strike a balance between deep learning and interpretability. Specifically, we propose an interpretable deep learning architecture that utilize binary parallel convolutional branches to accommodate the raw time-series data and the equivalent scalogram representation so that the model can learn the underlying patterns from perspectives of multiple domains. Subsequently, the physics information from the continuous wavelet transform (CWT) [193] can be integrated with the raw information into the merge layer when performing the node-wise weighted addition of the features maps output by the hidden layers of two branches. This approach facilitates a more nuanced representation of the physical properties in the learning process, which can lead to a more generalized and robust model that better captures the underlying complexity of the data. Simultaneously, parallel deep model architecture empowers us to incorporate the empirical physics-informed knowledge into the model training. This approach affords enhanced adaptability and interpretability without necessitating the use of handcrafted feature engineering and oversimplification of physical models.

Parallel architectures have been explored in previous studies, significantly contributing to the field. For instance, Han et al [94] developed a multi-scale dilated CNN-LSTM model for fault diagnosis under noisy environments, demonstrating robustness in complex conditions. Similarly, Wei et al [195] introduced a fault detection method using an oil temperature forecasting model based on deep deterministic policy gradients, highlighting its predictive capabilities. Additionally, Haj Mohamad et al [196] applied a deep CNN-LSTM network to gear fault diagnostics, showcasing efficient data segment processing. Despite these advancements, certain limitations persist. Han et al's approach of reintroducing raw data alongside convolutionally extracted features potentially dilutes the effectiveness of dimensionality reduction achieved by CNNs. Wei et al's static model configuration does not fully leverage the adaptive potential of deep learning frameworks. Furthermore, Haj Mohamad et al process data segments in parallel without enhancing the information content, limiting the depth of analysis possible from each data type.

Our research addresses these limitations by introducing a refined binary parallel convolutional architecture. This architecture integrates raw time-series data and scalogram representations at the merge layer, maximizing data utilization and enhancing model interpretability without extensive feature engineering. Additionally, this parallel deep model architecture incorporates empirical physics-informed knowledge into the training process, further enhancing adaptability and interpretability without relying on handcrafted features or oversimplified physical models.

Moreover, the transformation of 1D time-series data into 2D matrices for image-based training emerges as an astute and efficacious strategy. Given that the CWT of 1D signals yields 2D coefficients, adopting 2D matrices not only ensures the preservation of a larger volume of information but also renders our methodology compatible with the structure of CWT coefficients. At the terminal point of the architecture, the merged feature map, after being reshaped, is interfaced with a self-attention-based Long Short-Term Memory (LSTM) layer [197]. This arrangement is motivated by the hypothesis that convolutional layers excel at discerning localized features which are transposable across various samples, while the LSTM is proposed to excel at recognizing temporal dynamics and distilling long-term information and the self-attention mechanism could enable the LSTM layer to focus on the most salient parts of the sequence and ignore less relevant information. Thus, the fusion of these layers seeks to harness the strengths of both local feature extraction and temporal pattern recognition to provide a robust solution.

In this research, we introduce a dual-branch convolutional neural network architecture that integrates raw vibration data and continuous wavelet transform data to enhance fault diagnosis in rotating machinery. Key contributions include:

- The development of an interpretative model that combines time-series data and image-based features for improved fault classification.
- The demonstration of the model's ability to perform well with limited data through innovative training strategies.
- Extensive validation of the model's effectiveness across varying operational conditions using real-world dataset.

These advancements underline our approach's potential to significantly improve the accuracy and reliability of fault diagnosis in rotating machinery.

The remainder of the paper is structured as follows: Section 4.2 delves into the architecture and characteristics of interpretable parallel deep learning model. Case analysis encompassing a laboratory gearbox is reported in Section 4.3. In Section 4.4, we dissect the training results through a comparative performance analysis. Furthermore, to highlight the methodology generalizability and robustness, a second use case employing publicly accessible bearing dataset is conducted. Finally, Section 4.5 encapsulates our concluding thoughts and observation.

## 4.2. Interpretable Parallel Deep Learning Model

We aim at a practical IoT-based PHM approach. As the primary case example, we have outfitted a gearbox with essential IoT sensors, i.e., accelerometers and tachometers, to capture real-time operational data (Figure 4-1). These sensors are interconnected, forming a network that facilitates continuous data flow to a local workstation. This systematic data collection builds a dataset that is used to train a diagnostic model, employing techniques such as Convolutional Neural Networks (CNN) and Long Short-Term Memory (LSTM) networks, enhanced by self-attention mechanisms for improved performance. After successful training and validation, the model is deployed on the workstation to analyze real-time sensor data for decision making.

The focus of this research is the architecture of the interpretable parallel deep learning model (Figure 4-3) that is predicated upon multiple branches of convolutional neural networks, possessing the ability to simultaneously extract and amalgamate features across various convolutional layers. This unique attribute of parallel extraction facilitates the integration of a more diverse range of empirical and interpretable features into the model training process.

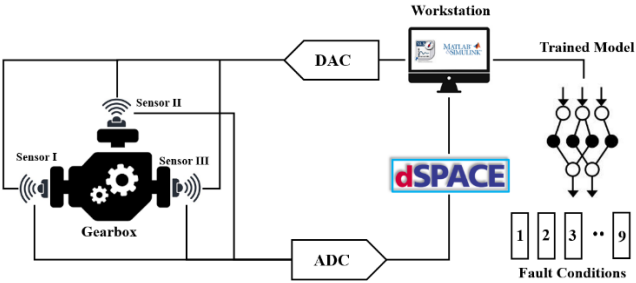

**Figure 4-1.** Gearbox PHM with IoT integration.

### 4.2.1. Convolutional Neural Network

The Convolutional Neural Network (CNN) [198] has demonstrated exceptional performance across a multitude of case implementations, significantly surpassing many other machine learning algorithms, especially in image-based classification tasks. The capability of convolutional layer to perform feature extraction through a sliding observation window allows the model to capture and retain a wealth of low-level local features, identifiable irrespective of their location on the images. Convolutional layers constitute the heart of this architecture, their primary function being the extraction of consistent and indicative features, whilst concurrently preserving as much local detail as possible when examining entire images. The resulting features, disentangled from the broader pattern, compose a feature map that provides a higher-level understanding of pattern recognition. As illustrated in Figure 4-2, local features from various locations of the input image are garnered in the convolutional layers. This is accomplished by sliding a 3×3 inspection window across the image and executing convolution computations at each stride, resulting in a 3D feature map.

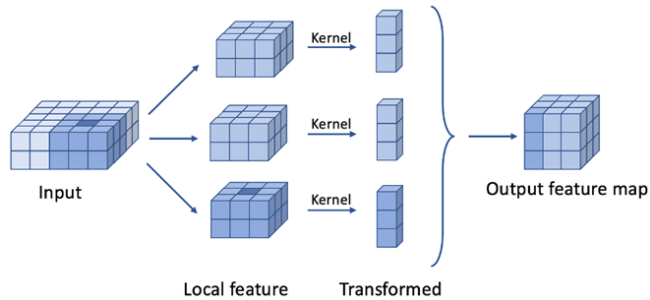

**Figure 4-2.** Workflow of convolutional operation.

Each 3D local feature map is then metamorphosed into a one-dimensional vector of shape, subsequently restructured into a 3D output feature map. This output map encapsulates an array of original local details and features, encompassing aspects such as edges, color, texture, and so on. However, progressively abstract patterns and insights necessitate extraction from the local feature maps, generated by preceding layers, given that local features alone do not present a comprehensive image characterization. Anticipated feature combinations require discovery. In response, the deeper layers engender more global patterns, composed of local features, thus establishing spatially hierarchical knowledge and feature sets. Consequently, CNN is equipped to scrutinize and assess sample images from multiple viewpoints, thereby augmenting its predictive capabilities.

### 4.2.2. Long Short-Term Memory with Self-Attention

The Self-Attention Long Short-Term Memory (LSTM) is a state-of-art approach in the realm of sequence data processing, designed to optimally capture temporal dynamics and long-term dependencies inherent in such data. This technique combines the strengths of the traditional LSTM framework with the power of a self-attention mechanism [199], [200]. LSTM, a particular type of Recurrent Neural Network (RNN) [201], [202], are engineered to avoid the long-term dependency problem encountered by standard RNNs, enabling them to remember and use information from inputs on a large scale. LSTM has gating units to control the flow of information. These gating units are used to solve the vanishing gradient problem of traditional RNNs. The LSTM maintains a cell state and uses three types of gates [203] to control the cell state and hidden states. The first step in LSTM involves determining how much of the past information from the previous cell state $C_{t-1}$ needs to be remembered and how much new information from the current input $x_t$ needs to be stored, as is described below

$$C_t = f_t \cdot C_{t-1} + i_t \cdot \tilde{C}_t \tag{4.1}$$

Here, $f_t$ is the forget gate that decides what to discard from the previous cell state, $i_t$ functions as the input gate determining which values warrant updates, and $\tilde{C}_t$ symbolizes the candidate values proposed for the current state. Having updated the cell state to $C_t$, the LSTM module then determines the output $h_t$, which is a refined version of the current cell state.

$$h_t = o_t \cdot \tanh(C_t) \tag{4.2}$$

In Equation (4.2), $o_t$ acts as the output gate, deciphering which components of the cell state should be transmitted to the output. The functions $f_t$, $i_t$, $\tilde{C}_t$ and $o_t$ are computed based on the input and the prior hidden state. Their respective weights $W$ and bias $b$ parameters in the below equation are the learned values during the training phase, i.e.,

$$f_t = \sigma(W_f \cdot (h_{t-1}, x_t) + b_f)$$
$$i_t = \sigma(W_i \cdot (h_{t-1}, x_t) + b_i) \quad (4.3)$$
$$o_t = \sigma(W_o \cdot (h_{t-1}, x_t) + b_o)$$

The self-attention mechanism, inspired by the transformer model, enables the model to allocate different attention weights to different parts of the sequence, allowing the model to focus more on the significant parts when making predictions. Three matrices are introduced to mathematically represent the self-attention mechanism: the Query matrix (Q), the Key matrix (K), and the Value matrix (V). These matrices are derived from the input data $X$, through learned linear transformations denoted by weight matrices $W$,

$$Q = X \cdot W_q$$
$$K = X \cdot W_k \quad (4.4)$$
$$V = X \cdot W_v$$

The attention weights are calculated as follows,

$$Attention(Q, K, V) = softmax\left(\frac{Q \cdot K^T}{\sqrt{D(K)}}\right) \cdot V \quad (4.5)$$

where $D(K)$ is the dimension of the key matrix, $Attention(Q, K, V)$ [204]–[206] is the output matrix where each element represents an attention score for each part of the sequence. Then we can obtain a weighted sum that indicates which parts of the sequence should be focused on.

In our implementation, the self-attention layer is introduced after the LSTM layer and before the final output layer. The LSTM layer is responsible for learning temporal dependencies in the sequence and it generates a sequence of hidden states. These hidden states carry information about the sequence up to their respective time steps. The self-attention layer then operates on this sequence of hidden states. It calculates attention scores for each hidden state based on its relevance to the other hidden states in the sequence. Incorporating a self-attention mechanism between two LSTM layers can be extremely beneficial for learning complex temporal dependencies across various time steps in a sequence. This approach enables the model to focus on significant parts of the sequence while also considering the temporal relationships between sequence elements, thereby enhancing the model's capacity to understand and learn from the sequence data. By stacking LSTMs and self-attention layers in this manner, the model can effectively learn hierarchical representations of the data, with each LSTM layer and self-attention layer contributing to different levels of abstraction.

### 4.2.3. Proposed Parallel Model Architecture

The Convolutional Neural Network (CNN) is an ingenious design that automates the feature extraction process and identifies the correct mapping between the feature map and the output. This automation, though, paradoxically renders the procedure challenging to interpret when examining visualized feature maps. Consequently, dataset augmentation becomes a conventional and feasible approach for enhancing pattern recognition performance, especially given the constraints of limited dataset sizes. Nevertheless, data augmentation via geometric transformation [207], [208] barely mitigates the problem of overfitting. We posit that overfitting the raw data does not necessarily signify that the model has comprehensively assimilated the relevant information concealed within the data. Instead of geometrically transforming the data sample, our proposition aims to unearth an array of different domain features for the deep learning model to assimilate. Incorporation of such physically meaningful features, obtained from the identical dataset, can ameliorate the challenge posed by limited dataset sizes by integrating more potentially critical features into the model training process. Simultaneously, the significance of these features can be deciphered through the post-hoc results, thereby enhancing the understanding and interpretability of the model training process.

To bring these concepts to fruition, we advocate for a parallel neural network architecture comprising multiple branches, each performing distinct convolutional operations on different inputs. Every branch signifies an independent convolutional extraction process, each with its own specific model parameters. The initial branch of the feature extraction segment, composed of convolutional and max pooling layers [209], can accommodate the image inputs generated by raw data. Although the raw data collected from the gear system via accelerometers represent a one-dimensional time-domain vibration signal, we have transformed this 1D time-series data into 2D images by mapping individual data points to 2D image pixel values. This procedure enables us to create image-based raw inputs for the first convolutional branch. Sequentially, the identical slice of data is subjected to continuous wavelet transform to capture time-frequency features, which are represented by a set of 2D coefficients. To encapsulate this time-frequency information for the second branch of convolutional feature extraction, the coefficients are reshaped into 2D image-based inputs for model parameter updating.

Upon processing the raw data and directing it through the two branches of feature extraction layers, where each branch executes independent convolution operation to produce a high-level feature map for subsequent dense connection, we strategically merge the two feature maps that provide unique perspectives on the underlying structure of the data. We presume that amalgamating the information from both feature maps will yield a more comprehensive representation of the input. In the merging layer, we fuse the two feature maps into a larger one by weighted addition, ensuring that all relevant information is retained for subsequent analysis. The merged feature map will be reshaped into a 1D array before being passed to the LSTM layer. The attention layer follows the first LSTM layer, which provides the model with the ability to focus on specific parts of the sequence while processing the LSTM's output. By assigning different attention to different parts of the sequence, the model can learn to prioritize information that is most relevant for classification. Following the attention layer, the output is sent to another LSTM layer, which provides an additional level of temporal feature learning, enabling the model to capture more complex and potentially hierarchical temporal dependencies. Finally, the output of the second LSTM layer is passed to the dense layer, producing a probability distribution over the classes, indicating the model's predictions.

A detailed illustration of the proposed parallel model structure is depicted in Figure 4-3, and the specifications of the neural network are listed in Table 4-1.

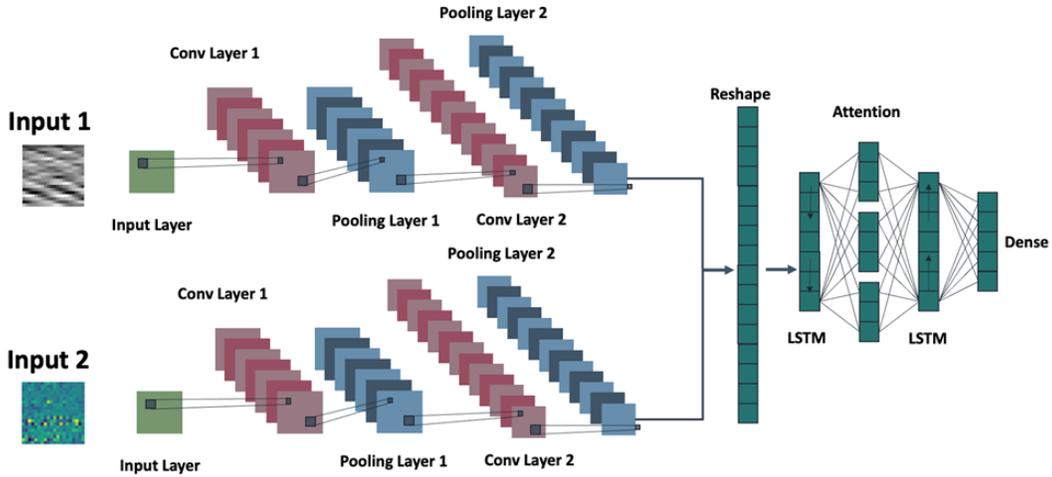

**Figure 4-3.** Parallel model architecture.

**Table 4-1.** Specifications of parallel neural network

| Sequence | Layer (Size) | Layer (Size) |
|---|---|---|
| | **Branch 1** | **Branch 2** |
| 1 | InputLayer (30, 30, 1) | InputLayer (30, 30, 1) |
| 2 | Conv2D + ReLU (32, 3×3) | Conv2D + ReLU (32, 3×3) |
| 3 | MaxPooling | MaxPooling |
| 4 | Conv2D + ReLU (64, 3×3) | Conv2D + ReLU (64, 3×3) |
| 5 | MaxPooling | MaxPooling |
| 6 | Merge Layer | |
| 7 | Conv2D + ReLU (64, 3×3) | |
| 8 | Reshape ( , 1) | |
| 9 | LSTM Layer | |
| 10 | Self-Attention Layer | |
| 11 | LSTM Layer | |
| 12 | Dense (128, ) | |
| 13 | Dense (9, ) | |

### 4.2.4. Physics-informed Merge Layer

The physics-informed merge layer plays a crucial role in integrating the information from both branches of the parallel deep learning model. It amalgamates the output from each branch, taking into account the CWT coefficients. Instead of directly optimizing the model parameters with respect to frequency content by including a loss term generated in the CWT, we opt to consider the frequency components of interest as the supplementary perspective to help the decision in the forward propagation [36]. While a time-domain signal and its CWT are theoretically equivalent, this does not mean that they are functionally identical for deep learning model training. Practically, each domain provides a distinct perspective on the probability distribution at the output, and therefore distinct weight updates.

Before getting into the specifics of the merge layer, it's essential to note that the dimensions of the input data and the output from each of the two branches are consistent with each other. This uniformity is significant because it allows for smooth transition of information and a balanced learning process between two branches, even after the downsampling process. In the merge layer, these matrices undergo a sophisticated integration process, coined as a weighted addition. This operation is governed by the following equation,

$$y^k = \gamma_1 \cdot f_R^{k-1}\left(W_R^{k-1} \cdot y_R^{k-1}\left(\left|X^{(i)}\right|\right) + b_R^{k-1}\right) \\ + \gamma_2 \cdot f_C^{k-1}\left(W_C^{k-1} \cdot y_C^{k-1}\left(\left|X_c^{(i)}\right|\right) + b_C^{k-1}\right) \tag{4.6}$$

where $X^{(i)}$ and $X_C^{(i)}$ denote the raw data input and the corresponding CWT coefficients, $y_R^{k-1}$ and $y_C^{k-1}$ represent the output of the layers before the $(k-1)_{th}$ hidden layer from the two branches to be input into the $(k-1)_{th}$ hidden layer, $W_R^{k-1}$, $W_C^{k-1}$, $b_R^{k-1}$ and $b_c^{k-1}$ are the weights and bias applied to the input for two branches respectively, $f_R^{k-1}$ and $f_C^{k-1}$ are the activation functions used in the $(k-1)_{th}$ hidden layer before they are fused together in the merge layer. More importantly, each output of the $(k-1)_{th}$ hidden layer in the merge layer is computed as a weighted sum of the respective outputs from the two branches, with the weights determined by the merge factors $\gamma_1$ and $\gamma_2$. This operation ensures an effective and balanced integration of information from both branches.

Furthermore, the weighted addition operation facilitates an efficient feedback mechanism, by which the loss from the final output is proportionally allocated back to both branches. In this case, the merge factors play a direct role in how the weight updates are distributed during backpropagation [211], because these factors directly influence the final output of the layer, and so the gradients calculated will be proportional to these factors. In contrast, the merge factors might influence the final output indirectly if used in subsequent layers for weighting the influence of different parts of the concatenated output when performing concatenation in the merge layer, with less direct control over the distribution of weight updates. Such an approach could potentially compromise the interpretability when scrutinizing the performance variations in response to the fine-tuning of merge factors.

### 4.3. Proposed Parallel Model Architecture

The proposed physics-informed parallel deep learning model is executed for fault detection on our experimental vibration data, derived from a laboratory-constructed gearbox system. This section describes our experimental design, data collection process, algorithm implementation, and data processing.

#### 4.3.1. Data Acquisition

Ensuring the healthy operation of gear transmission is crucial in any machinery system. While specific defects or failure modes might occur in the gearbox, we theorize that vibration data can serve as a robust indicator of operational conditions. In our experiments, we designed a two-stage gearbox system with interchangeable gears, enabling us to simulate specific fault cases by substituting gears that bear the corresponding faults. Figure 4-4 displays the schematic and photograph of the designed gear system. The gear speed was manipulated by a motor, and torque was supplied through an adjustable magnetic brake. The first stage encompasses a 32-tooth pinion coupled with an 80-tooth gear, and the second stage features a 48-tooth pinion and a 64-tooth gear. A tachometer measured the input shaft speed, while an accelerometer captured gear vibration signals. The dSPACE system (DS1006 processor board, dSPACE Inc.) recorded the signals at a sampling frequency of 20KHz.

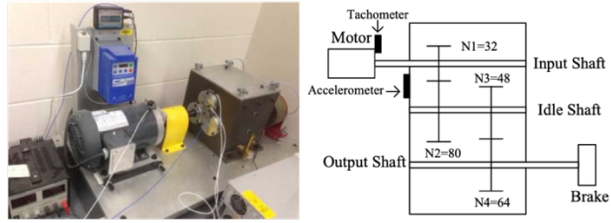

**Figure 4-4.** Schematic of gear system.

We adopted the Time Synchronous Averaging (TSA) approach [212]. In this method, time-aligned signals, resampled based on the shaft speed measured by the tachometer, are averaged in the angular domain. This is crucial as time-domain responses typically are not time-periodic due to speed variations, even when the gearbox system records at a fixed sampling rate. Employing the TSA approach allows us to address this non-stationarity issue and minimize the impact of uncertainty. We simulated nine different gear conditions for the input shaft's pinion, which included a healthy condition, missing teeth, root cracks, spalling, and chipping tips at five different severities, as presented in Figure 4-5. For each condition, we collected vibration data over 416 gear revolutions, with each revolution comprising 900 data points. We hypothesized that at least one revolution could elucidate the divergence between the varying gear conditions. Consequently, the entire data set for each condition was divided into 416 slices, with each slice treated as a single sample for model training. Ultimately, we generated 3744 training samples, evenly distributed across nine classes. Each sample represents a 1D time-series data containing 900 data points.

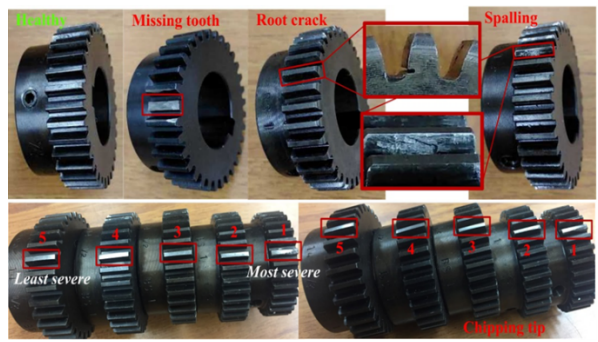

**Figure 4-5.** Gear conditions for classification.

### 4.3.2. Implementation

Each of the collected data points is segmented into 416 samples for each gear condition. Our aim is to input each of these samples into the model for training. However, the 1D time-series data is inherently incompatible with the 2D convolutional layers. As a resolution, we map each data point within a sample to corresponding image pixel values, thus reshaping the one-column data point arrays into 30x30 matrices. This allows the samples to be inputted into the model in an image format, as visualized in Figure 4-6.

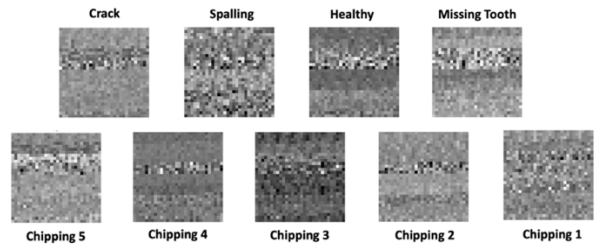

**Figure 4-6.** Raw data representation for 9 conditions.

In conventional methods, one might opt to plot these data points within each sample and then rescale these plots to make them compatible with the model input [213]. However, such an approach tends to lead to a loss of relevant information. As we know, plotting data traditionally serves to visualize inherent relationships within a dataset, transforming raw data points into observable physical metrics such as distance and amplitude. When it comes to deep learning model training, rescaling the plotted images is generally necessary, but the reduction in resolution can severely compromise the representation of the images. Variables such as the width of the plot curve and image resolution could significantly impact the representation of class features, complicating the interpretability. These uncertainties associated with data slice plotting render this method less reliable and operationally feasible.

Therefore, we prioritize preserving as much raw information as possible. This is achieved by directly transforming the data points into image representations. From visual perspective, differentiating between categories merely by observing these images is a challenging task, as any distinctions between them can be subtle and may not be universally applicable across all samples. Nonetheless, the inherent characteristics of neural networks enable our model to scrutinize these images on a pixel-by-pixel basis, extracting critical features that can effectively interpret the divergence between different classes. By adhering to this approach, we retain the maximum original data fidelity while also fully harnessing the power of deep learning.

In tandem with the image-based representation of raw data points, our proposed parallel deep learning model also requires wavelet scalograms to feed into its second branch. This dual-pronged input strategy necessitates that both the training and validation sections of our dataset be transformed into image-based formats. While the raw 1D data can be directly reshaped into a 30 × 30 image matrix, the process for wavelet scalograms is more complex. It begins with performing a CWT on each data slice, which results in 2D coefficient arrays. Each of these coefficients is then mapped to pixel values, akin to the method used for the raw data. Finally, the 2D coefficient arrays are reshaped into a 30 ×30 image matrix, maintaining compatibility with the feature map merge layers of our model. An illustration of a wavelet scalogram prepared for model training is presented in Figure 4-7. It's important to note that the wavelet scalograms can provide a more detailed view of the time-frequency characteristics of the data, capturing subtleties that might be lost in the raw data representation. Therefore, the incorporation of this additional layer of information in our model helps in reinforcing the robustness and versatility of the parallel deep learning model, enhancing its ability to discern between different gear conditions.

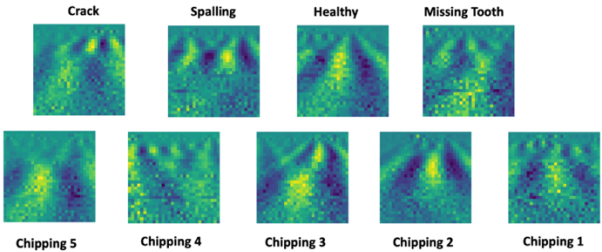

**Figure 4-7.** Wavelet scalogram representation for 9 conditions.

### 4.3.3. Model Selection

In our case study focused on gearbox fault detection, we performed a detailed model selection analysis to thoroughly evaluate the performance, stability, and efficiency of various machine learning models on our dataset. The proposed model in this study uniquely processes dual inputs, incorporating both 2D raw data and 2D CWT data. In contrast, established state-of-the-art (SOTA) methods typically handle either 1D raw data or 2D raw data. Regardless of the data format used, all models were trained on the same dataset, with 80% allocated for training and validation, and the remaining 20% for testing. The computational work was carried out using the PyTorch framework [214] on a CUDA NVIDIA A4000 GPU. Time measurements were systematically recorded to ensure accuracy and consistency, and all models underwent 100 epochs of training. The metrics reported, mean accuracy, standard deviation, and training time, are based on the average outcomes of five training trials. With this setup clearly outlined, we are now equipped to delve into the model selection, guided by the results summarized in Table 4-2.

Table 4-2. Performance comparison for model selection

| Model | Mean Acc | Best Acc | Std | Training Time |
|---|---|---|---|---|
| **1D- Input** | | | | |
| 1D-CNN | 0.9699 | 0.9760 | 0.0060 | 42.16 |
| RNN | 0.9821 | 0.9933 | 0.0093 | 83.00 |
| Transformer | 0.9981 | 1.0000 | 0.0016 | 147.35 |
| SVM | 0.9013 | 0.9053 | 0.0027 | 8.19 |
| **2D- Input** | | | | |
| ResNet | 0.9980 | 1.0000 | 0.0014 | 124.14 |
| CNN | 0.9548 | 0.9573 | 0.0019 | 38.33 |
| **Dual** | | | | |
| Proposed | 0.9979 | 1.0000 | 0.0011 | 40.42 |

Given the sequential nature of our data, it is appropriate to evaluate both state-of-the-art (SOTA) and established time series classification models to address this case. These include RNNs, Transformers, and 1D CNNs, alongside conventional machine learning methods, which are well-suited for simpler sequence problems. Our results confirm the robust capabilities of established sequence models. RNNs and Transformers, in particular, perform impressively in terms of mean accuracy. Notably, the Transformer model stands out for its stability, evidenced by a standard deviation of 0.0016. However, its comprehensive information extraction capabilities and complex structure make it the most time-consuming option in terms of training. In contrast, the 1D CNN proves to be quite efficient, although its accuracy does not reach the highest levels observed in other models. The SVM, representing traditional machine learning methods, is extremely fast but oversimplifies the problem, making it inadequate for complex sequence tasks.

Turning our attention to image-based models, they too show excellent performance. ResNet's results are on par with the best in our study, though it also suffers from lower efficiency. A standard 2D CNN, however, consistently demonstrates high efficiency and stability, as reflected in its standard deviation, making it a compelling choice for image-based modeling. This efficacy and stability support our decision to employ a dual 2D input design in our proposed model. Overall, the Transformer, RNN, and our proposed model can achieve perfect accuracy under optimal conditions, with comparable mean accuracies. However, our proposed model distinguishes itself by being significantly more efficient and stable than the others. This model selection process not only highlights the relative strengths and weaknesses of each model but also suggests potential trade-offs between accuracy and efficiency that could guide future selections depending on specific requirements.

In addition to the tabular data previously discussed, Figure 4-8 visually represents the comparative analysis of each model's performance in terms of mean accuracy and training time. This butterfly plot provides a clear graphical representation of the balance each model strikes between accuracy and computational efficiency. This plot arranges models along the y-axis with mean accuracy extended to the right and training time to the left of a central zero line. Each bar is capped with an error bar, representing the standard deviation of the model's accuracy, offering a visual gauge of each model's reliability alongside its performance metrics. The butterfly plot underscores the efficiency of the 1D-CNN and Proposed models, which exhibit not only high accuracy but also shorter training times, indicating a robust performance with less computational demand. Contrastingly, the Transformer model, while achieving near-perfect accuracy, requires significantly more training time, as depicted by the lengthy bar extending to the left. This visualization highlights the practical trade-offs involved when deploying these models in real-world applications.

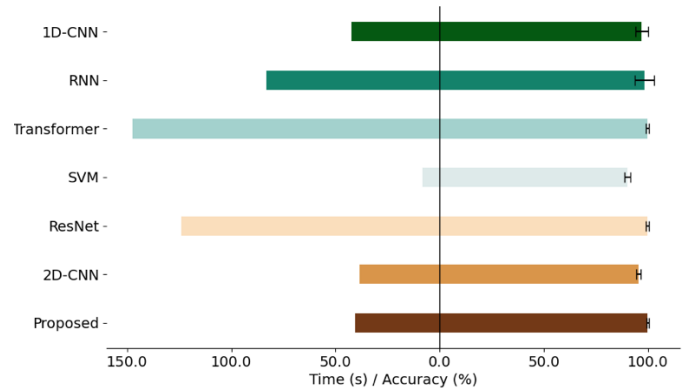

**Figure 4-8.** Performance and efficiency of model

## 4.4. Results and Discussion

### 4.4.1. Enhancement through Parallelism

This study centers on the aim of maximizing information extraction and feature detection from a limited dataset through our proposed parallel model structure. Moreover, by incorporating empirical and physical features into our methodology, we hope to enhance the model's inherent interpretability. Additionally, this approach enables the selection of diverse, yet information-rich, features to augment model performance. To examine whether our parallel model architecture can indeed surmount the challenges posed by limited data, we conducted model training across varying dataset sizes. It's crucial to understand how our model fares against other deep learning methodologies. As such, we have also trained a CNN, replicating the structure of a single branch of our parallel model, using the image-based representation of raw data. This comparison allows us to evaluate the intrinsic superiority of our proposed structure in the context of this case study. We posit that feeding the same data repeatedly is not an efficient way to enhance feature extraction and may contribute to overfitting. Therefore, we deployed traditional geometric transformation-based data augmentation techniques, leading to a fourfold increase in data size. All simulations involving these models and techniques were conducted on a consistent platform, and their performances were assessed using a uniform set of data for test.

For a comprehensive evaluation of our model, we executed training multiple times for each case, and subsequently averaged the test results for a balanced performance comparison. Upon assessing these results, as illustrated in Table 4-3, the merits of our proposed parallel model become evident. It consistently outperforms the single-branch CNN across diverse data sizes. When data volume is ample, the performance improvement due to the parallel structure is rather marginal, especially when non-generative data augmentation techniques are deployed. An exemplar of this can be seen where the CNN model achieved a classification rate of 0.9864 using data augmented four times. However, as the data volume dwindles, the performance enhancements facilitated by our parallel model become progressively more pronounced, maintaining a classification accuracy above 90% even with just 10% of the original dataset size. Notably, even when considering the CNN's performance with only 20 samples per class, the enhancement provided by our parallel model remains significant, despite an accuracy of 0.8185 that might not meet all expectations.

**Table 4-3.** Test results of different architectures

| MODEL | CNN | CNN (4 ×) | PARALLEL |
|---|---|---|---|
| 2970 (80%) | 0.9548 | 0.9864 | 0.9979 |
| 2520 (70%) | 0.8265 | 0.9026 | 0.9617 |
| 1440 (40%) | 0.7032 | 0.8372 | 0.9464 |
| 360 (10%) | 0.5948 | 0.6037 | 0.9226 |
| 180 (5%) | 0.4263 | 0.3591 | 0.8185 |

The superior performance of our parallel model, depicted in Figure 4-9, is an intuitive result. The multi-branch structure allows for twice as many features and data to be incorporated into the model's training, suggesting a richer information pool. Furthermore, the inclusion of empirical features deemed potentially relevant can enhance the interpretability of our model's learning process. The CNN's performance with data augmented four times offers a fresh perspective on the difference between the two models under similar data augmentation circumstances. The repetitive feeding of the same data does not necessarily mean the model will learn more relevant patterns, particularly when the limited data inadequately captures the wide distribution of features under different class conditions. As such, non-generative data augmentation might exacerbate the overfitting problem when performance is significantly below the baseline with limited data.

This is illustrated by the lower classification rate of the CNN with data augmentation training on just 20 samples per class, compared to that of the CNN without data augmentation.

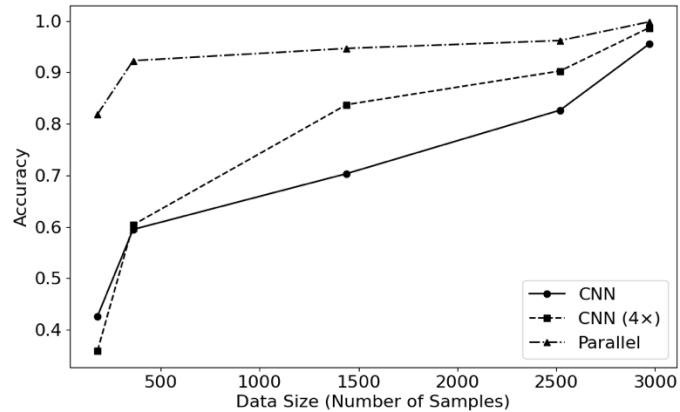

**Figure 4-9.** Classification rate comparison.

In our study on enhancing model performance through a dual-branch architecture, we employ t-SNE to visualize the effectiveness of feature extraction across different branches of our model. Specifically, we extract features from the convolutional layers of two distinct branches: Branch 1, which processes raw data, and Branch 2, which handles data transformed via CWT. Additionally, we examine features from the convolutional layer immediately following the merging of these branches. Figure 4-10 provides a 2D visualization of the learned features using t-SNE, laid out as follows: (a) represents the raw data branch, (b) illustrates the CWT data branch, and (c) depicts the merged layer's output. These visualizations are pivotal in understanding the interaction and complementary nature of the features processed by each branch.

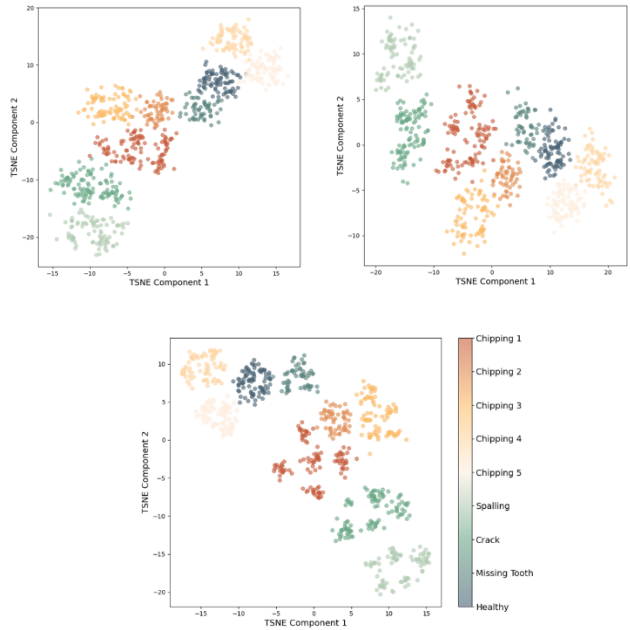

**Figure 4-10.** 2D visualization result of learned features.

Observing the clusters in the t-SNE plots, it's apparent that each branch contributes uniquely to the model's ability to differentiate between conditions. The raw data branch (a) shows distinct groupings, indicating effective initial separation of features relevant to various fault conditions. The CWT data branch (b) further refines these groupings, suggesting that the wavelet transform enhances the model's sensitivity to subtle, frequency-based patterns inherent in the data. The merged layer (c), expectedly, combines these perspectives, potentially offering a more integrated and robust feature set for classification tasks. The clusters in this plot are noticeably tighter and more distinct than those in the individual branches, underscoring the synergistic effect of combining raw and transformed data, which could lead to improved diagnostic accuracy and reliability in practical applications.

Building on the insights from our feature visualization, which highlighted the critical features distinguishing different classes, we are prompted to further investigate the model's performance across these classes as dataset size varies. In our analysis of the sensitivity of the proposed model to different labels as the dataset size decreases, we observe significant insights through the confusion matrices in Figure 4-11 obtained from testing results for models trained with varying amounts of data (80%, 70%, 40%, and 10%).

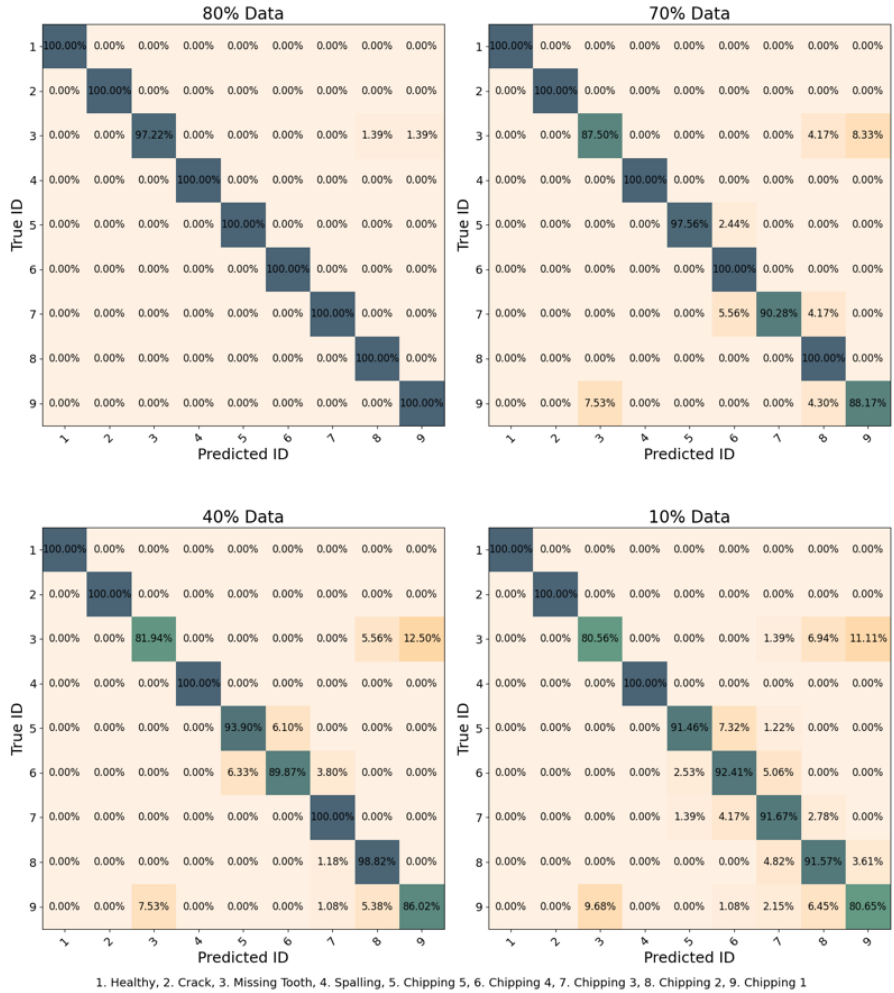

**Figure 4-11.** Confusion matrix of proposed model.

Firstly, it is notable that the confusion between 'Missing Tooth' (Label 3) and 'Chipping 1' (Label 9) is persistent across all training sizes. This observation is understandable considering the close similarity between severe chipping and a missing tooth in terms of their impact on the gearbox's vibration signatures. Both conditions produce distinct anomalies in the mechanical system that can be challenging to differentiate without sufficient data. However, as the dataset size decreases, the model's ability to accurately distinguish between the different severity levels of chipping deteriorates noticeably. For instance, in the 80% data training scenario, there is a moderate confusion between the chipping categories, but the model largely retains its discriminatory power. Contrastingly, in the 10% data scenario, the accuracy drops significantly, particularly for distinguishing between 'Chipping 1', which is the most severe, and other less severe chipping conditions like 'Chipping 5' (Label 5). This degradation in performance is illustrated by increased misclassifications among the chipping categories as data scarcity exacerbates the model's challenges in learning nuanced differences between similar fault types.

This trend underscores the model's dependency on data volume to maintain high precision in fault identification, particularly in distinguishing conditions with subtle differences. The reduction in data not only affects the overall accuracy but also impacts the model's ability to leverage detailed feature distinctions necessary for precise fault classification. These observations suggest the necessity of sufficient data for training robust diagnostic models, especially when dealing with a range of fault severities that exhibit closely related symptoms. Further, it highlights the importance of considering data volume and quality in the deployment of fault diagnostic systems in practical settings, ensuring that they are equipped with enough examples to understand and differentiate between various fault conditions effectively.

### 4.4.2. Post-Hoc Analysis for Interpretability and Representation Selection

In this part of research, we undertake an exploration of model interpretability through a mix of ad-hoc and post-hoc methodologies. Interpretability of a model is a complex yet crucial facet of machine learning, particularly in the realm of deep learning where the inner workings of a model often feel akin to a 'black box'. By making use of ad-hoc techniques, we endeavor to understand the model in situ, evaluating its various characteristics and behaviors during the development and training phases. However, post-hoc methods allow us to gather deeper insights retrospectively, once the model has been trained and validated.

Our approach adopts a unique standpoint, leaning heavily on post-hoc validation, where we experiment with a range of physics-informed data representations. By trialing an array of signal processing techniques like Empirical Mode Decomposition (EMD) [215], Variational Mode Decomposition (VMD) [216], and Fast Fourier Transform (FFT) [217] among others, we aim to glean more insight into the contribution of specific features in the model's decision-making process. In our analysis, we applied specific parameter settings for time-frequency decomposition: EMD was conducted using a standard stopping criterion with the standard deviation of the residue less than 0.3%; VMD was configured with 5 modes, an Alpha of 2000, and a Tau of 0.000001; CWT utilized 30 logarithmically spaced scales from 1 to 901 with a Morlet wavelet; and FFT was performed on the full 900-point signal with a Hanning window to mitigate spectral leakage. For methods other than FFT, the resulting coefficients were uniformly resized to $30 \times 30$ matrices using TensorFlow's resize function to standardize input dimensions for neural network processing. The intention here is to ascertain which physics-based features carry the most weight and are thus more pivotal in the diagnostic procedures. These techniques, coupled with our physics-informed merge layer, allow us to dig deeper into the layers of our model's learning mechanism, enabling a much more nuanced understanding of the model's operation.

Moreover, our physics-informed merge layer serves as a further tool for interpretability. It plays a key role in balancing the feature information from different branches in the model, which in turn can help us understand the model's reliance on the feature inputs from each branch. By adjusting the merge factors, we can observe how the model's performance varies with the emphasis placed on each branch, adding an additional layer of transparency to the model's operation. Through such post-hoc analysis, we strive to decode the 'black box', facilitating a comprehensive understanding of the model's operational dynamics and thus enhancing our ability to interpret its diagnostic reasoning. This is particularly vital in the field of fault diagnosis, where the interpretability of a model can significantly contribute to its usability in applications.

In our investigation, we adhere to a rigorous experimental protocol whereby each testing scenario is replicated five times, with results derived from an average of these individual trials in Table 4-4. This ensures the reliability of our performance metrics and the robustness of our findings. The architectural design across all experiments remains uniform, utilizing the dual-branch parallel structure, where one branch is fed with raw data representation. Notably, the introduction of the CWT as an additional source of data representation presents a compelling advantage in our model's performance, as we can observe in Table IV. When we compare the results from the model's operation with CWT against its counterpart using raw data, it becomes evident that the model's performance with CWT consistently surpasses that of the raw data across varying dataset sizes, which is evident in the Figure 4-12. This excellent performance is not merely an outcome of the parallel structure's enhancement, but also a testament to the value added by the CWT as a physics-informed data representation. The CWT imbues the model with additional, potentially crucial, physical information, further enriching the feature space used for classification. Consequently, our findings suggest that the choice of a physically insightful data representation method such as CWT, in conjunction with an appropriately designed model architecture, can significantly boost the performance of machine learning models in fault diagnosis tasks.

**Table 4-4.** Test results of different data representation

| REPRESENTATION | CWT | EMD | VMD | FFT | RAW DATA |
|---|---|---|---|---|---|
| 2970 (80%) | 0.9979 | 0.9764 | 0.9874 | 0.9822 | 0.9621 |
| 2520 (70%) | 0.9617 | 0.9352 | 0.9453 | 0.9217 | 0.8913 |
| 1440 (40%) | 0.9464 | 0.8497 | 0.8965 | 0.8564 | 0.7752 |
| 360 (10%) | 0.9226 | 0.7876 | 0.7944 | 0.7396 | 0.6279 |
| 180 (5%) | 0.8185 | 0.6745 | 0.6501 | 0.5937 | 0.5025 |

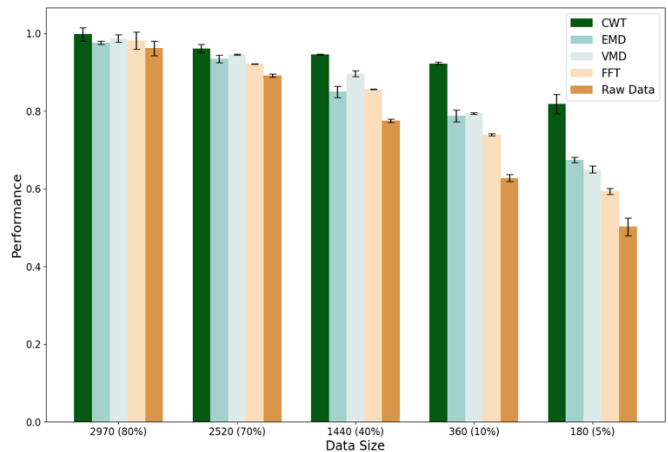

**Figure 4-12.** Model performances on varying data.

The performances of the models when using EMD, VMD, and FFT as data representations were also examined and compared. These methods, like the CWT, offer physics-informed insights that can potentially improve the performance of our model. In terms of the comparison, it was observed that the performance of the model using VMD was marginally superior to EMD and FFT. This subtle edge might suggest that VMD provides a slightly more informative or relevant feature set for our specific fault diagnosis. However, the performance margins among these three methods (EMD, VMD, FFT) were quite minimal, particularly when the data was abundant. This may imply that the extra information provided by these methods, over the raw data, is somewhat comparable in its contribution to the model's performance.

However, when we shift our focus to scenarios with limited data availability, the picture changes somewhat. While the performance of the models using EMD, VMD, or FFT, was significantly better than that using raw data, these methods fell noticeably short of the performance achieved with the CWT. This might suggest that, under data-constrained conditions, the CWT captures and provides the model with a more relevant or comprehensive set of features than EMD, VMD, or FFT. Consequently, the CWT appears to be more resilient and effective in handling data sparsity, thus proving its value in such situations. CWT is renowned for its ability to provide a time-frequency representation of signals, where both high and low frequencies can be analyzed simultaneously. This means that CWT can offer better resolution for capturing abrupt changes or non-stationary behaviors in signals, which are typical in faulty gearbox vibration data. EMD and VMD are also capable of handling non-stationary signals, but their decompositions can sometimes introduce artifacts or miss subtle features. FFT, in contrast, assumes signal stationarity, which can lead to inaccuracies when dealing with non-stationary signals. Additionally, CWT can be more robust to noise in comparison to other methods.

The hypotheses put forth here regarding the superiority of CWT for gearbox fault diagnosis are fundamentally grounded in performance validation and comparative analysis. They provide a lens through which we can begin to interpret the model's classification decisions and understand the influence of different physics-informed data representations on performance. However, it's crucial to emphasize that these findings do not provide a one-size-fits-all answer for the choice of data representation in every task. Instead, they highlight the importance of task-specific investigation and careful problem formulation. Moreover, these results underscore the importance of interpretability in model design and evaluation. By comparing different representations, we gain insights not just into their relative performance, but also into the characteristics they highlight in the data and how these are leveraged by the model. This has implications for both improving model performance and understanding the underlying processes.

The presented work thus offers a valuable framework for model interpretation and fine-tuning. It not only allows us to interpret the model's behavior post hoc by examining the influence of different representations but also provides a mechanism to guide the model's learning process a priori by selecting and tuning appropriate data representations. This dual capability supports both the enhancement of model performance and the advancement of our understanding of the classification task.

### 4.4.3. Robustness & Generalizability of Model

In order to further analyze the robustness and generalizability of the model, here in this part of research we employ a dataset sourced from the Case Western Reserve University (CWRU) Bearing Data Center, a well-recognized repository for its extensive collection of bearing fault data. This bearing system is considerably more complex. The dataset captures intricately single-point defects on bearings at various sizes, 0.007 inches, 0.014 inches, and 0.021 inches, located on different components such as the ball element, inner ring, and outer ring. Each defect is meticulously recorded through vibration signals in the time domain, providing a rich basis for analysis. To tailor the dataset for our analysis, we adopted a precise approach to data segmentation. Given the extensive raw signal length of approximately 120,000 data points per condition, we segmented the data using a window size of 512 points, ensuring that each sample comprised 1024 points. This segmentation strategy allowed us to generate 200 representative samples from each condition, facilitating a thorough examination of the bearing's operational health under varying fault conditions.

This CWRU bearing dataset provides a comprehensive array of fault diagnosis scenarios that test the versatility and adaptability of fault detection models. Recent research in this area employs advanced deep learning techniques to enhance diagnostic accuracy and model robustness. Notable among these are the CNN-BiLSTM with residual modules [218] which leverages spatial and temporal feature extraction, and the FaultFormer [219], which utilizes transformer architectures to adapt to varying fault conditions with minimal labeled data. Additionally, the integration of CNN with STFT by Yoo et al [220] aims to streamline computational processes, albeit at the potential cost of losing significant data detail. These methods highlight the ongoing evolution toward more complex but potentially computationally intensive solutions. The hybrid CNN-Transformer approach proposed by Kim et al [221], for example, attempts to balance local feature extraction with global context awareness, reflecting a trend towards models that aim for both high interpretability and accuracy. Each of these approaches brings forth advancements in handling the dynamic nature of bearing fault data, yet they also underscore the challenges of maintaining model efficiency and interpretability under diverse operational conditions.

In this context, our proposed model introduces a binary parallel convolutional architecture that efficiently processes both raw time-series data and scalogram representations, tailored specifically to harness the detailed fault characteristics inherent in the CWRU dataset. This dual-path strategy not only preserves the integrity and richness of the original data but also enhances interpretability and adaptability, allowing for a more nuanced understanding of fault dynamics. The subsequent sections will detail the robustness and generalizability of our model, demonstrating its superior performance across varied fault conditions and its ability to integrate and interpret complex datasets effectively.

To investigate the robustness and generalizability of our model, we conducted tests under varying load conditions and rotational speeds, as illustrated in Table 4-5. These tests were designed to assess the model's performance across a spectrum of operational states, from no load to a load of 3 HP, and at speeds ranging from 1730 rpm to 1797 rpm. Additionally, to explore the impact of sensor placement on model performance, data were collected from sensors located at both the drive end and the fan end of the test bearings. This dual-sensor setup allowed us to evaluate whether sensor displacement influences the model's diagnostic capabilities, providing insights into the adaptability of the model to different sensor configurations.

Table 4-5. Model performances on varying conditions

| Load | Sensor | Mean Acc | Best Acc | Std |
|---|---|---|---|---|
| 0 HP 1797 rpm | Drive | 0.9948 | 1 | 0.0033 |
| | Fan | 0.9990 | 1 | 0.0013 |
| 1 HP | Drive | 0.9854 | 0.9870 | 0.0013 |
| | Fan | 0.9990 | 1 | 0.0013 |
| 1772 rpm | | | | |
| 2 HP 1750 rpm | Drive | 0.9995 | 1 | 0.0010 |
| | Fan | 0.9995 | 1 | 0.0010 |
| 3 HP 1730 rpm | Drive | 1 | 1 | 0 |
| | Fan | 0.9984 | 1 | 0.0021 |

The results presented in Table 4-5 demonstrate that the proposed model maintains high mean accuracy across different sensor placements and operational conditions, underscoring its robustness. Notably, the model achieves near-perfect best accuracies and exhibits low standard deviations in almost all tested scenarios, indicating its reliability and generalizability. Whether tested under no load or full load, with data from the drive or fan end, the model consistently identifies bearing faults with high precision, showcasing its potential for real-world industrial applications.

Additionally, the specific distribution of five training trials on the bearing fault data across different operational conditions is revealed in Figure 4-13. The model exhibits a trend where higher loads generally correlate with improved accuracy. This pattern is particularly evident at 2 HP and 3 HP conditions, where the median accuracies are not only higher but also show less variability as indicated by the narrower interquartile ranges. This could be attributed to more pronounced fault signatures under higher mechanical stresses, which are easier for the model to detect and classify accurately. Notably, at 3 HP using the Drive End sensor, the model achieves perfect accuracy, which underscores the optimal alignment of sensor data capture with fault manifestation under heavy operational loads. Contrary to typical expectations and in contrast to the performance under higher loads, the Fan End sensor data demonstrate superior performance at lighter loads, specifically at 0 HP and 1 HP. This observation suggests that the Fan End sensors might be better suited for capturing fault data under conditions of minimal mechanical stress.

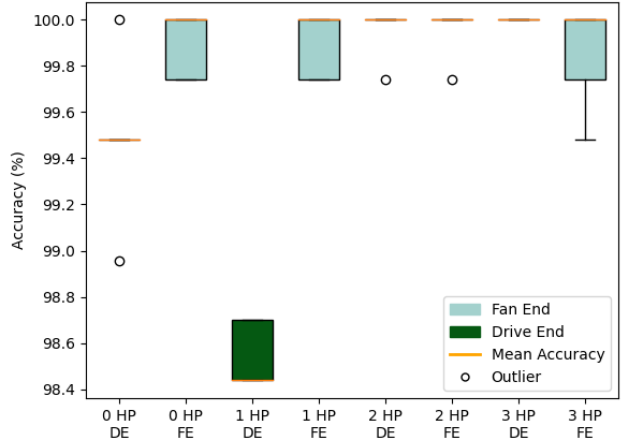

**Figure 4-13.** Prediction distribution across varying conditions.

While the insights derived from the analysis offer promising guidance for sensor placement and model optimization, it is important to acknowledge the presence of unknown disturbances that obscure the clear correlation between fault data and fault conditions. The outliers observed in the data, particularly under certain operational conditions, suggest that there may be external factors or intrinsic model uncertainties that could undermine the robustness of our conclusions. In practical terms, the findings emphasize the need for careful consideration of sensor placement in industrial applications, where the accessibility and cost implications of sensor installations are significant. Tailoring the model to effectively utilize data from variably placed sensors could significantly enhance its practical deployment and fault diagnostic capabilities.

### 4.4.4. Investigation of Merge Factor Impact

While it is crucial to identify and incorporate physics-informed data representations, it's equally important to effectively control and manage their contributions in a model. The inclusion of these representations indeed brings valuable insights into the model; however, their impact can vary depending on the specific problem context or data at hand. Thus, gaining a clear understanding of their individual contributions and adjusting them accordingly is essential to ensure the model's optimal performance. To this end, we delve into an investigation of the merge factors' impact. This analysis will allow us to control the influence of each representation and garner a better understanding of their roles within our parallel deep learning architecture.

By constraining the sum of merge factors to be 1, we ensure the model output remains consistent regardless of the specific values of $\gamma_1$ and $\gamma_2$. This also prevents overemphasis or underemphasis on one branch over the other due to the values of the merge factors. In this study, we seek to uncover the optimal balance between the contributions of our two different physics-informed data representations within our parallel deep learning architecture. To achieve this, we employ a systematic approach called a grid search, which systematically examines a range of possible combinations of merge factors to identify the optimal pairing.

The grid search results in Table 4-6 provide insights into the impact of merge factors on the performance of the model with different data sizes. $\gamma_1$ controls the weight given to the raw data branch, while $\gamma_2$ controls the weight given to the CWT branch. When the data size is sufficient, the results show that the performance is relatively consistent across different combinations of $\gamma_1$ and $\gamma_2$. Additionally, when comparing Trial 1 and Trial 11, we observe that the performance is worse in the case where only the CWT coefficients are utilized. This suggests that the contribution of the CWT branch is limited compared to the raw data branch, indicating that the CWT might lose some information compared to the raw data representation. However, it is interesting to note that relying solely on the CWT still achieves slightly better accuracy than relying solely on the raw data representation when the data is extremely limited. This indicates that even with limited data, the CWT is able to capture some critical physics features that contribute to the improved performance.

**Table 4-6.** Grid search on the merge factors

| Trial | $\gamma_1$ | $\gamma_2$ | 2970 (80%) | 1440 (40%) | 180 (5%) |
|---|---|---|---|---|---|
| 1 | 0.0 | 1.0 | 0.9384 | 0.7238 | 0.6457 |
| 2 | 0.1 | 0.9 | 0.9320 | 0.7542 | 0.7018 |
| 3 | 0.2 | 0.8 | 0.9439 | 0.8331 | **0.8185** |
| 4 | 0.3 | 0.7 | 0.9748 | 0.8823 | 0.7599 |
| 5 | 0.4 | 0.6 | 0.9735 | **0.9464** | 0.7474 |
| 6 | 0.5 | 0.5 | 0.9881 | 0.9284 | 0.7082 |
| 7 | 0.6 | 0.4 | 0.9854 | 0.9396 | 0.6325 |
| 8 | 0.7 | 0.3 | **0.9979** | 0.8954 | 0.5604 |
| 9 | 0.8 | 0.2 | 0.9946 | 0.8054 | 0.5252 |
| 10 | 0.9 | 0.1 | 0.9689 | 0.7377 | 0.4587 |
| 11 | 1.0 | 0.0 | 0.9548 | 0.7032 | 0.4263 |

We finally observe the heat map shown in Figure 4-14. As the data size decreases, the optimal combination of $\gamma_1$ and $\gamma_2$, highlighted by the yellow points starts to shift towards higher values of $\gamma_2$. This implies that a larger proportion of the CWT representation becomes more important in the absence of sufficient raw data. It suggests that increasing the emphasis on the CWT branch can be a good strategy when dealing with limited data, as it helps to capture relevant physics-informed features. Overall, these results highlight the importance of finding the right balance between the raw data and CWT representations based on the available data size. While the raw data contains more comprehensive information, the CWT can provide valuable insights, especially when the data is limited. By fine-tuning the merge factors, we can effectively leverage the benefits of both representations and enhance the performance of the model in gearbox fault diagnosis tasks.

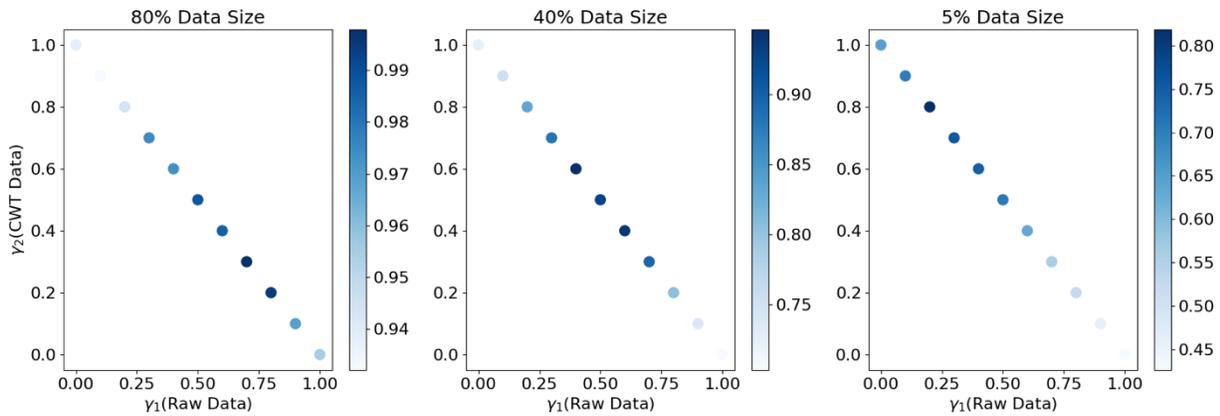

**Figure 4-14.** Heat map of grid search on merge factors.

### 4.5. Conclusion

In this study, we present an IoT-enhanced PHM system utilizing an interpretable parallel deep learning architecture that excels in the fault diagnosis of machinery systems. By capitalizing on empirical, physics-informed data representations of vibration signals, our system can reduce the reliance on data augmentation, even in scenarios of limited data. The architecture's design features a multi-pronged branch structure that converges into a physics-informed merge layer, meticulously devised to generate comprehensive feature maps throughout the training phase. Moreover, the integration of a self-attention mechanism within the LSTM layers empowers the model to judiciously weigh the significance of each sequential feature, thereby refining the feature extraction process and boosting predictive accuracy. Our primary evaluations, conducted on vibration data collected under nine distinct fault conditions of a laboratory gearbox, confirm the model's excellent performance over conventional single-branch convolutional neural networks and other models reliant on data augmentation, irrespective of dataset size. Our detailed post hoc analysis and targeted grid search to optimize the merge factor enrich the model's interpretability and training efficacy. Furthermore, the accuracy and robustness of the proposed algorithm are examined by a second case study employing a publicly accessible bearing dataset, which also illustrates the generalizability and scalability. Leveraging the advantages of our model, edge computing can be employed in future to facilitate efficient, localized diagnosis, minimizing latency and maximizing computational efficiency.

# Chapter 5. Spatially-Informed Online Prediction of Milling Surface Deformation Using Multiphysics-Infused Graph Neural Network for Digital Twinning

Milling is essential for achieving precise surface flatness and durability, yet residual stress-induced deformation in high-strength alloys, coupled with the computational intensity of traditional finite element methods, poses significant challenges for rapid prediction within the digital twin framework. Efficient and accurate surrogate modeling technique is needed to facilitate process design. An emerging concept is to leverage graph neural networks (GNNs), which map each mesh node to a graph vertex and reflect finite element connectivity as graph edges, to directly capture both local and global interactions. The limited, existing attempts, however, employ an entire mesh, yielding very large graphs and undermining iterative process assessment. In this work, we propose a physics-infused Graph Neural Network (PhyGNN) and apply it to machining process for the first time. Our approach uses a concise, dynamically evolving node graph composed exclusively of critical measurement points, ensuring an effective while accurate surrogate model. To enhance expressiveness, we introduce an attention-based dynamic graph mechanism that adaptively learns strong interactions between these selected nodes. By embedding physics-informed constraints derived from the Johnson–Cook damage model, the proposed approach accurately predicts post-milling deformation aligned with experimental measurement protocols. This will enable rapid surface quality prediction within a digital twin environment, supporting data-driven process planning. Extensive ablation studies on A2024 aluminum alloy demonstrate that integrating node graph features with a physics-informed loss function significantly enhances model performance and stability, unleashing the full potential of physics-driven surrogate modeling in machining applications.

## 5.1. Introduction

Milling operations are important machining processes in manufacturing for shaping materials into precise geometries, requiring rigorous control to mitigate quality loss due to residual stresses [222,223]. Surface quality is essential, as it directly impacts workpiece durability and performance. Accurate surface deviation characterization is key for guiding post-machining decisions, enabling proactive quality assessments and determining the necessity of further machining to adhere to strict standards [224,225]. This predictive ability enhances workflow efficiency and reduces production costs by minimizing excessive rework, ensuring products meet stringent requirements. Consequently, understanding and accurately predicting surface quality have emerged as significant research areas in advanced manufacturing. Traditionally, direct measurement methods have been the standard for assessing surface flatness [226,227]. These methods utilize precision instruments such as coordinate measuring machines (CMMs) [228,229], laser scanners [230], and profilometers [231,232] to capture detailed topographical data of machined surfaces. While these techniques provide accurate measurements, they are often time-consuming and costly, both in terms of the required equipment and the potential production downtime. Moreover, direct measurement methods are limited to assessing already-machined products and are thus inadequate for large-scale adaptive process adjustments.

Given these challenges, there has been a need towards model-based methods for analyzing surface quality, particularly using finite element analyses [233-235]. Finite element analyses allow for proactive predictions of machining process outcomes by simulating the dynamics of milling and the mechanisms of material removal. This leads to the possibility of predicting surface deviations and subsequently adjusting the machining parameters [236,237] before the actual cutting process begins. For instance, Yao et al [238] used a finite element model to predict blade deformation by considering residual stress effects. Similarly, Liu et al [239] optimized the curvilinear micro-groove structure to improve cutting performance using a three-dimensional finite element analysis. While seemingly promising, finite element analyses require substantial computational resources. As model complexity increases with the inclusion of multi-physics effects such as the thermo-mechanical coupling [240,241] and high mesh density to ensure convergence and accuracy, the computational cost increases significantly, making real-time applications challenging. Strategies to reduce computation time include simplifying the model from three-dimensional to two or excluding certain interactions, such as thermo-mechanical coupling. For example, Taraphdar et al [242] used two-dimensional (2D) full geometry models to study weld-induced residual stresses. Shiomi et al [243] developed a 2D model to examine the effects of melting and solidification on the thermal history of laser rapid prototyping with metallic powders. However, these simplifications may lead to reduced accuracy and potentially diminish the predictive quality of the simulations while the reduced computational costs may still be too much for conducting process assessment.

Recognizing the limitations of finite element simulations, particularly their impracticality for real-time output generation following changes in input parameters, it becomes evident that finite element model alone cannot serve as the online computational engine within a digital twin framework. Indeed, both process optimization and uncertainty propagation analysis for process assurance will require mass sampling to assess milling outcome across multiple variables. To enable feasible grid search and iterative design adjustments that satisfy surface quality criteria, rapid and reliable online prediction becomes indispensable, prompting an increasing focus on data-driven techniques [244,245] and the development of lightweight surrogate models [246,247]. Surrogate modeling involves creating simplified representations of the intricate physical processes, utilizing comprehensive datasets derived from previous simulation results. By leveraging deep learning algorithms [248,249], surrogate models can reliably predict the effects of various machining parameters on surface flatness, requiring only a fraction of the computational resources compared to direct finite element simulations. To facilitate this numerically, an automated high-quality data generation pipeline can be established, which involves sampling from specified ranges of input parameters and integrating these values into the finite element model. The resulting finite element analysis data, extracted from specific nodes within the model by reading the output database files generated during the simulations, provides a robust dataset for training the surrogate model.

Nevertheless, most existing surrogate modeling approaches lack the capacity to capture the vital spatial dependencies inherent in computational mechanics. For instance, Heydari and Tai [250] developed a convolutional neural network (CNN)-based framework for rapid prediction of temperature distributions in three-dimensional moving source problems in manufacturing. While their approach effectively accelerates thermal simulations, it does not explicitly address the spatial interactions at the node level crucial for detailed finite element analyses. Similarly, Oddiraju et al [251] proposed a differentiable physics-informed machine learning model to simulate laser-based micro-manufacturing processes. Their model adeptly integrates machine learning and physical models to predict process dynamics. However, it primarily focuses on temporal evolution and lacks explicit modeling of spatial dependencies among mesh nodes. Although these methods bring speed gains, they are unable to capture the nuanced node-to-node interactions crucial for high-fidelity finite element simulations. Graph neural networks (GNNs) have recently gained considerable attention due to their capability to capture nuanced node-level graph information. However, they are predominantly applied in areas such as fault diagnosis and prognosis [252], generative design [253], and manufacturing optimization problems [254,255]. Although effective, these studies primarily structure data graphically rather than capturing the intricate spatial dependencies necessary for physics-driven engineering problems, such as those encountered in finite element-based simulations. In a way, these applications have not yet fully unleashed GNNs' potential for capturing rich and interconnected spatial node dependencies, a characteristic that makes them ideally suited for surrogate modeling in finite element analysis. On the other hand, it has recently been suggested that GNNs may align with finite element models by mapping each mesh node to a graph vertex and directly translating finite element connectivity into graph edges, enabling both local and global interactions to be captured in a topologically consistent manner. Although promising, explorations along this direction remain at preliminary stage. Related investigations [256,257] generally adopt a straightforward approach by including every node in the finite element mesh. This may result in very large graph sizes, severely constraining mesh density and thus limiting both model accuracy and practical feasibility for iterative assessment processes.

Aiming at advancing the state-of-the-art, in this research we propose a node graph strategy that avoids relying on a fine-mesh finite element model yet retains essential spatial dependencies. Rather than incorporating all mesh nodes, our model forms a concise yet dynamically evolving graph of only the few key nodes most relevant to deformation prediction or measurement points. We enhance this reduced graph with a dynamic attention mechanism, allowing graph connectivity to adapt based on learned feature relationships rather than relying on a rigid, pre-defined adjacency. Consequently, the graph isolates the strongest node interactions without incurring the high computational cost of a full mesh. By coupling this attention-based graph structure with physics-guided constraints, we achieve a balance between computational efficiency and predictive accuracy, making it ideal for rapid, iterative process planning in machining applications. Meanwhile, while physics-informed GNNs have shown promise in various domains, their application to machining processes remains largely unexplored. For instance, Dalton et al [258] leveraged physics-informed GNNs for soft-tissue mechanics, incorporating energy-based constraints, and Liao et al [259] combined thermal physics laws with neural networks for additive manufacturing. However, machining-specific dynamics, particularly damage evolution, have yet to be fully addressed in a physics-informed GNN framework. In this work, we introduce one of the first physics-infused GNN architectures explicitly tailored for machining processes. Our approach uniquely integrates the Johnson–Cook (J-C) damage model through a unified, physics-informed loss function that combines damage consistency constraints and graph structural regularization. By focusing on essential nodes and employing a dynamic attention mechanism, we balance computational efficiency and accuracy, supporting rapid iterative machining process assessment under varying parameters including feed rate and material property values. The key contributions of this work thus include:

- Concise and Adaptive Node Graph: We address the typical graph-size explosion by constructing a minimal but highly expressive graph from only essential measurement nodes. A dynamic, attention-based connectivity mechanism preserves critical spatial dependencies without resorting to a full mesh.

- Physics-Infused GNN for Machining Analysis: To ensure the accuracy of the surrogate model for machining, we propose to establish a physics-infused GNN framework, integrating the J-C damage model to ensure physically consistent deformation predictions.
- Unified Physics-Informed Loss Function: We incorporate a grand loss formulation combining J-C model consistency and graph regularization, balancing computational efficiency with physics fidelity.

While our physics-infused GNN addresses the computational bottlenecks of predicting milling-induced deformation, deploying the model alone may overlook the operational drift that routinely occurs on the shop floor. In practice, even with fixed design parameters, the machining state evolves across batches due to material variability, ambient temperature changes, tool wear, and fixture stiffness. Because the residual-stress–driven deformation is small in scale, these seemingly minor shifts can accumulate and push predictions off target unless the model is continuously updated. In our intended use, periodic CMM validation provides fresh measurements that are fed back to fine-tune the surrogate, closing the loop and correcting bias in near real time. This is precisely the role of a digital twin: to couple fast predictive modeling with on-line sensing and targeted simulation updates so the surrogate remains calibrated to the current process. Embedding PhyGNN in such a framework elevates it from a stand-alone predictor to a self-correcting, deployable tool, supporting real-time response, hybrid-fidelity integration, and sustained accuracy under realistic aerospace manufacturing conditions.

Indeed, digital twins in manufacturing are not merely high-fidelity simulations, but dynamic, cyber-physical systems that continuously integrate simulation, sensing, and data-driven modeling to support real-time decision-making. Technically, most digital twin solutions fall into two broad categories: reduced-order models, which simplify physics-based equations for faster evaluation, and data-driven surrogates, which learn mappings from input conditions to outcomes based on historical simulations or measurements. While both approaches aim to achieve near real-time response, their applicability becomes increasingly constrained as the underlying process physics grow more complex, particularly in high-stakes domains like aerospace manufacturing, where both speed and physical fidelity are critical. Our motivation stems from collaboration with an authentic aerospace manufacturer specializing in precision milling of thin-walled

A2024 components. Here, traditional finite element simulations require over 20 hours per configuration, making them infeasible for on-machine feedback or batch-level design adjustments. This industrial need reveals key pain points in practical digital twin deployment: how to accelerate prediction, how to remain physically consistent, and how to enable continuous model refinement based on new measurements. The proposed digital-twin framework integrates high-fidelity finite element analysis and a lightweight GNN surrogate, each playing a distinct and complementary role. Finite element analysis offers deterministic, physics-based solutions with high accuracy and is executed offline in large batches to cover a wide range of tunable machining parameters, including cutting speed, feed, depth of cut, material grade, and coolant condition. This simulation campaign is automated and continuously expands the database in the background. Whenever a non-tunable design feature, such as CAD geometry, meshing strategy, or tool path, is modified, a new finite element baseline is generated to keep the underlying physics current. These high-fidelity results are then distilled into a GNN surrogate, which learns the governing trends and delivers predictions in milliseconds. This makes the GNN ideal for fast parametric analysis during process design and monitoring, enabling users to explore variations, obtain surface-quality predictions, and identify optimal settings without rerunning expensive simulations.

Figure 5-1 illustrates the proposed cyber-physical coordination loop that operationalizes our digital twin framework for adaptive milling process control. The architecture fuses offline high-fidelity finite element simulation with online, low-latency inference via a lightweight physics-infused Graph Neural Network (PhyGNN). Offline, an automated parametric simulation campaign is executed across the design space and translated into compact graph representations via structured node sampling. This provides the GNN with a physics-anchored prior that captures both local deformation dynamics and global material interactions under complex multiphysics. Once deployed, the GNN enables millisecond-level forward inference at the edge, supporting on-machine, event-driven predictions with latency budgets suitable for modern in-line metrology systems. Crucially, when deviations are detected, typically via CMM scans or embedded sensor alerts, the framework initiates a loop-closure protocol wherein:

(1) Targeted re-simulation is triggered only at the discrepancy setting using a condition-aware triggering function;

(2) The resulting simulation and measurement data are fused using a multi-fidelity domain adaptation strategy, where discrepancies are weighted based on confidence-aware uncertainty metrics;

(3) The PhyGNN is incrementally fine-tuned using gradient-preserving low-rank adaptation (LoRA) techniques or selective replay buffers to preserve prior knowledge while minimizing catastrophic forgetting. The result is a self-updating, self-validating digital twin that generalizes across product designs and supports reliable, efficient process optimisation.

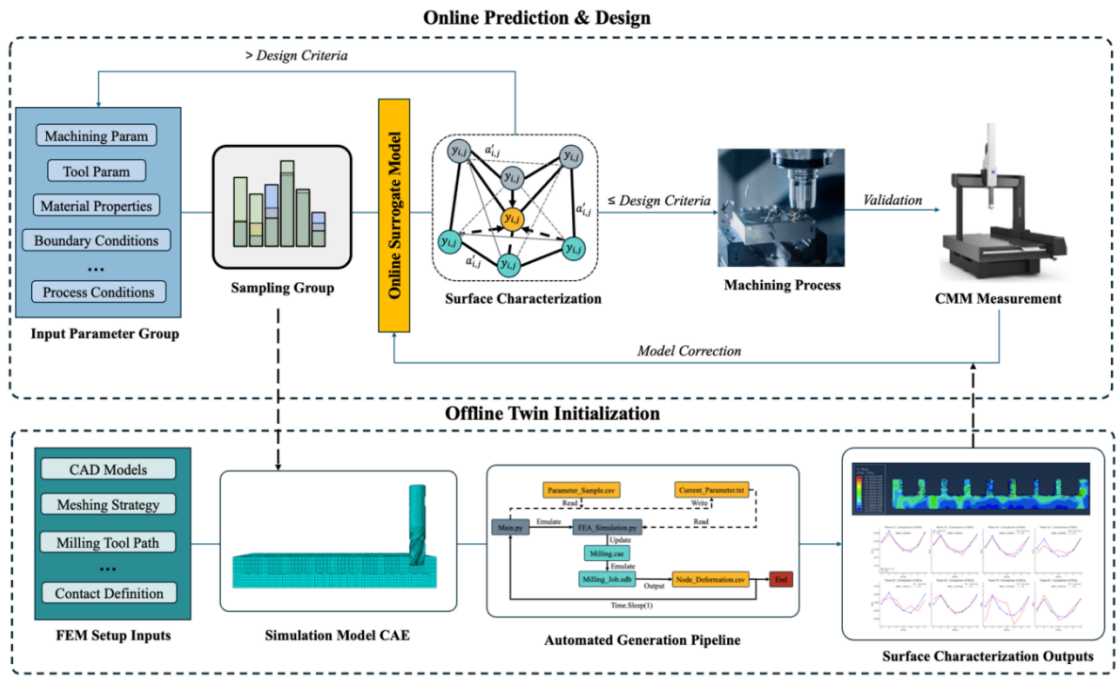

**Figure 5-1.** Digital twin-driven framework for intelligent surface quality assessment.

This asynchronous online–offline orchestration enables real-time adaptation without full retraining, and ensures the surrogate evolves continuously with the changing machining environment. The embedded Johnson–Cook consistency loss and attention-based graph dynamics reinforce physics-constrained generalization and semantic interpretability, both critical for explainability and traceability in aerospace-grade digital twins. Unlike generic surrogates, this framework is explicitly designed for runtime feedback integration, hybrid fidelity alignment, and modular retraining triggers, enabling practical deployment in closed-loop, high-consequence manufacturing settings. The rest of the paper is organized as follows: Section 5.2 describes the data generation pipeline, which employs a comprehensive finite element model to simulate the milling process. Section 5.3 presents the spatially aware, physics-infused graph neural network (PhyGNN) framework designed for assessing surface quality. In Section 5.4, we evaluate the effectiveness of the proposed approach and compare it with benchmark models. Section 5.5 provides concluding remarks and discusses the implications of our findings.

## 5.2. Physics-based Simulation of Milling Processes for Data Generation

This section details a first-principles based simulation of milling processes, implemented with finite element discretization. We also introduce an automated data generation pipeline to facilitate the efficient generation of training data for the surrogate model.

### 5.2.1. Physics-based modeling

Here we outline the finite element modeling of the milling process. Specifically, we first describe the J-C model employed, followed by the meshing scheme of the milling operation involved in this research. Finally, we examine node spatial relationships and deformation distribution to support the subsequent spatially-aware framework.

**5.2.1.1. Johnson-Cook material and damage models**

To model the stress-strain response of A2024 aluminum under the thermo-mechanical coupling present in milling, the J-C constitutive model is employed. The standard J-C model is mathematically expressed as

$$\sigma = (A + B\varepsilon^n)(1 + Cln\hat{\varepsilon})(1 - T^m) \qquad (5.1)$$

where $\sigma$ represents the flow stress, $\varepsilon$ is the plastic strain, $A$, $B$, $C$, $n$ and $m$ are material constants, $\hat{\varepsilon}$ is the dimensionless strain rate, and $T$ is the homologous temperature. This relation effectively captures the material behavior under varying strain rates, temperatures, and mechanical loads. To simulate chip formation and material failure, the J-C damage model is incorporated to describe damage initiation and evolution. In this model, the equivalent strain at failure follows [260]:

$$\varepsilon_f = (D_1 + D_2 e^{D_3 \sigma^*})(1 + D_4 ln\dot{\varepsilon}^*)(1 + D_5 T^*) \tag{5.2}$$

where $\sigma^*$ is the stress triaxiality, defined as the ratio of hydrostatic stress to equivalent von Mises stress. The parameters $D_1$, $D_2$, $D_3$, $D_4$ and $D_5$ are material-dependent constants governing damage initiation under different loading conditions. Among the damage parameters, $D_2$ is selected as a process input parameter as it directly influences strain-to-failure under varying stress triaxialities. This parameter affects the transition from plastic deformation to failure without introducing excessive complexity from the J-C damage model. Adjusting $D_2$ allows for controlled investigation of its impact on chip formation and material failure in the milling process. The specific values of the parameters used for A2024 aluminum are presented in Table 5-1.

**Table 5-1.** Johnson-Cook Parameters for A2024 Aluminum [261]

| Visco plastic Parameters | | | | |
|---|---|---|---|---|
| $A$ (Mpa) | $B$ (Mpa) | $C$ | $n$ | $m$ |
| 352 | 440 | 0.083 | 0.42 | 1 |
| **Damage Parameters** | | | | |
| $D_1$ | $D_2$ | $D_3$ | $D_4$ | $D_5$ |
| 0.13 | 0.13 | -1.5 | 0.011 | 0 |

### 5.2.1.2. Three-dimensional milling finite element modeling

A 3-dimensional finite element model is constructed using ABAQUS/Explicit to simulate the milling process. It is tailored to offer a comprehensive depiction of the milling process, utilizing the J-C model [262,263] to represent material behaviors under various machining conditions. The focus of the simulation is on A2024 aluminum [264,265], a material frequently used in manufacturing due to its pertinent properties. The specific attributes of A2024 aluminum are outlined in Table 5-2 [266]. In the finite element model, we systematically configure the workpiece and milling cutter geometries with specific element types to strike a balance between accuracy and computational efficiency. The workpiece geometry utilizes C3D8R hexahedral elements with reduced integration, while the intricate geometry of the milling cutter is captured using C3D4 tetrahedral elements

Table 5-2. Physical properties of workpiece material (A2024)

| Property | Value |
|---|---|
| Density | 2,780 kg / m$^3$ |
| Modulus of Elasticity | 73.1 GPa |
| Poisson ratio | 0.33 |
| Specific Heat Capacity | 875 J/kg·°C |
| Thermal Conductivity | 121 W/m·K |

As depicted in Figure 5-2, the milling groove is treated as a distinct mesh instance and is detailed with finer element sizing. This method enhances the accuracy of the milling process simulation and manages computational demands effectively due to the 3D complexity. The mesh consists of 225,815 elements for the workpiece, contributing 721,440 degrees of freedom (DOFs), and 25,744 elements for the endmill, with 16,590 DOFs in a 3D structural formulation. This extensive detailing necessitates substantial computational resources, highlighting the importance of developing an efficient and reliable surrogate model to reduce the computational load. Additionally, the endmill, specified as a Supermill SR4F500R, is made of carbon steel. The geometric specifications and milling parameters are detailed in Table 5-3. The design of the milling process, as outlined in Table 5-3, is engineered to explicitly control the surface quality of the final product. The total depth of cut for each milling groove is set at 0.25 inches, with each pass incrementally achieving half this depth at 0.125 inches, requiring two passes for completion. The endmill operates at a spindle speed of 3000 RPM and a feed rate of 30 inches per minute, following a programmed zig-zag trajectory. This specific path promotes consistent material removal and efficient chip evacuation, both of which are vital for extending tool life and ensuring the quality of the machined part.

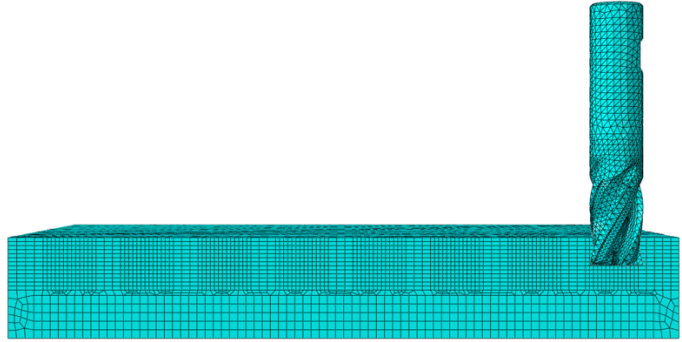

**Figure 5-2.** Finite element model of milling process.

**Table 5-3.** Specifications of endmill and milling process

| Geometric specs | |
|---|---|
| Cutting Diameter | 12.70 mm (0.5 inch) |
| Shank Diameter | 12.70 mm (0.5 inch) |
| Length of Cut | 31.75 mm (1.25 inch) |
| Overall Length | 76.20 mm (3 inches) |
| Corner Radius | 0.76 mm (0.03 inch) |
| **Milling process specs** | |
| Depth of Cut (each pass) | 3.18 mm (0.125 inch) |
| Number of Pass | 2 |
| RPM | 3000 |
| Feed Rate | 762 mm/min (30 inch/min) |

The milling process simulation depends on key parameters that influence material deformation, heat generation, and residual stress. The J-C model describes the material's response under high strain rates and thermal effects, shaping chip formation and surface characteristics. The friction coefficient at the tool-workpiece interface affects heat generation and material flow, influencing cutting forces, tool wear, and temperature distribution. Heat transfer coefficients determine how heat dissipates within the tool and workpiece, impacting thermal expansion and residual stresses. These parameters, drawn from Tables 5-1–3, represent key aspects of the milling process, ranging from damage modeling to material response and process settings that collectively influence product quality. Including them allows us to assess the surrogate's ability to support both process optimization and uncertainty-aware quality evaluation, thereby providing more comprehensive guidance for improving machined surface performance.

### 5.2.1.3. Representative node selection for efficient quality assessment

In the finite-element simulation of the machined workpiece, each groove is discretized with fine spatial resolution to capture detailed surface deformation. Three planar surfaces are extracted per groove for quality inspection, resulting in a total of 24 inspection planes across the eight grooves. Each plane is meshed using a 14 × 96 grid, yielding 1,344 nodes per plane and a cumulative 32,256 nodes for the entire part, as indicated in Figure 5-3. While this level of granularity is desirable for accuracy, it poses a significant challenge for online implementation. Specifically, feeding such high-dimensional inputs into a surrogate model, or validating predictions using CMM measurements across all nodes, is computationally prohibitive and time-inefficient in real-time settings.

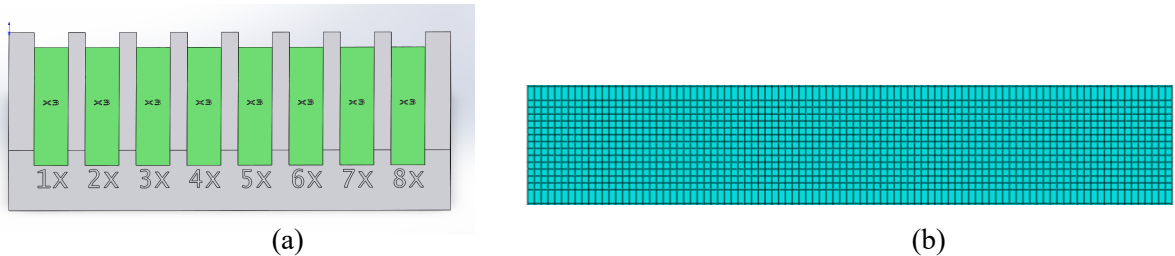

**Figure 5-3.** Finite element mesh density for surface deviation.

To address this issue, we adopt a node selection strategy aimed at identifying a small set of representative points that can preserve the essential deformation characteristics of each plane. Instead of processing over a thousand nodes per surface, we focus on chosen nodes that retain spatial coverage and sensitivity to surface shape while ensuring real-time feasibility. As illustrated in Figure 5-4, four representative planes are visualized using surface plots with magnified differences to enhance interpretability. Despite their individual variations, these surfaces all exhibit a smooth and structured deformation trend along the Y-direction, which corresponds to the long axis of the groove. This consistent spatial behavior suggests that surface deviation can be effectively characterized without exhaustive node-wise evaluation. The assumption of spatial regularity, supported by both physical insight and simulation results, forms the basis for the reduced-order sampling strategy described in the next section. To enable fast and reliable surface deviation prediction with limited sensing, we propose a three-stage node selection procedure that combines physical insight, data-driven refinement, and statistical control. The goal is to identify a reduced subset of representative nodes, $N^*$, that preserves the essential spatial deformation profile while satisfying real-time constraints on computation and measurement. The selection problem can be formulated as: $\min_{N \subset N_0} RMSE(\hat{U}(N), U)$, subject to $|N| \leq N_{\max}$, and pairwise correlation $\rho_{ij} < \rho_{\max}$, where $\hat{U}(N)$ is the predicted displacement using only the selected nodes, and $U$ is the full reference field. The total node pool per plane is $N_0$ with $|N_0|=1344$, and we target $N_{\max}=7$ representative nodes per plane. The upper bound on correlation, $\rho_{\max}=0.90$, prevents redundant nodes from being selected. To determine this value, we performed a parameter sweep from 0.80 to 0.99, using a lightweight multi-layer perceptron-based surrogate model to reconstruct the full-field deformation cross 1344 nodes from selected nodes.

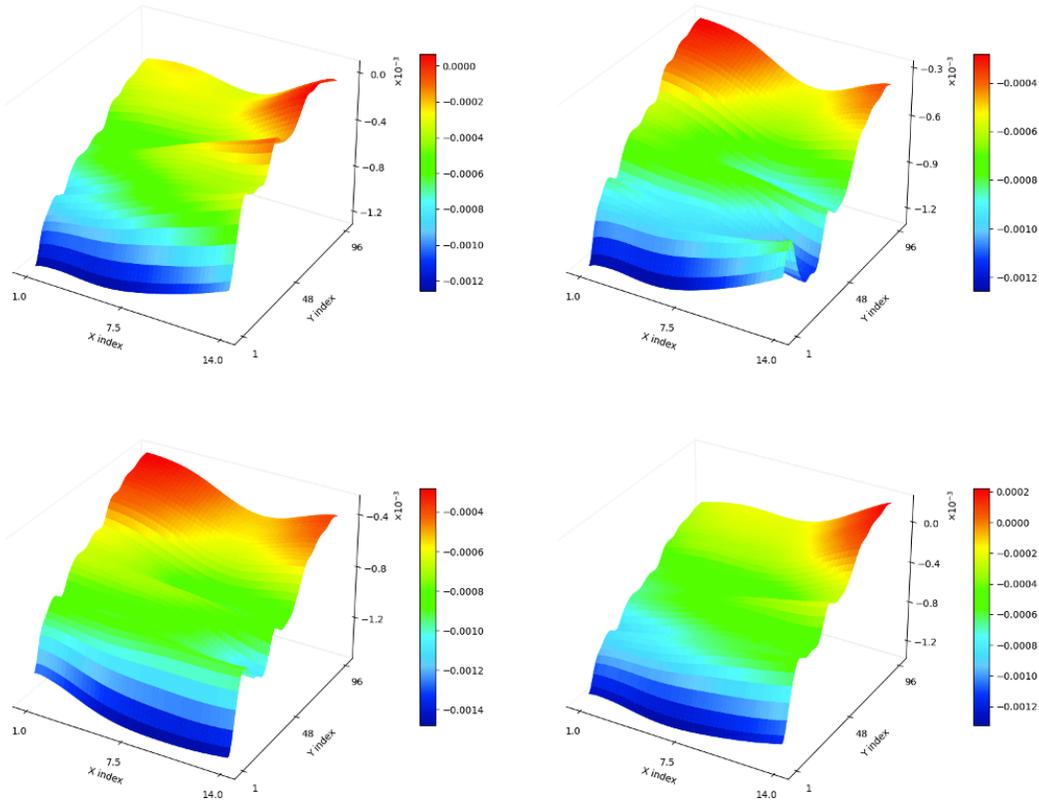

**Figure 5-4.** Deformation contour surfaces of four representative planes.

Our goal is to balance diversity and prediction accuracy, a larger $\rho_{max}$ retains more nodes but risks redundancy, while a smaller $\rho_{max}$ promotes diversity but may underrepresent key deformation patterns. We observed that when $\rho_{max}$ dropped below 0.85, the reconstruction accuracy began to degrade noticeably. In contrast, values between 0.88 and 0.93 consistently delivered stable performance, with less than 2% variation in five-fold cross-validated RMSE. We therefore chose 0.90 as a secure and representative threshold that filters highly correlated nodes while preserving predictive accuracy.

To reduce the search space from the full mesh, we first compute the time-averaged displacement gradient for each node on the groove surface across the entire FE simulation. Only nodes in the top 15% of displacement-gradient magnitude are retained as candidates. This filtering yields a physically meaningful node pool of approximately 200 points per plane, denoted as $N_{cand}$, which we empirically found to cover all major regions of deformation activity while eliminating over 85% of low-sensitivity nodes. Next, we construct the final node subset using a greedy forward-selection strategy. Starting with an empty set, nodes from $N_{cand}$ are iteratively added based on their contribution to reducing the cross-validated reconstruction error of a lightweight surrogate model. At each step, the node that yields the largest drop in five-fold CV-RMSE is added to the set. The process stops either when the number of selected nodes reaches $N_{max}$, or when the marginal improvement falls below 1%. To further prevent redundancy, we evaluate the pairwise Pearson correlation coefficient between all selected nodes. Any node whose displacement history shows a correlation exceeding $\rho_{max} = 0.90$ with another is removed, and the next-best candidate is inserted. This final pass ensures the selected set retains sensitivity without overlapping signal characteristics.

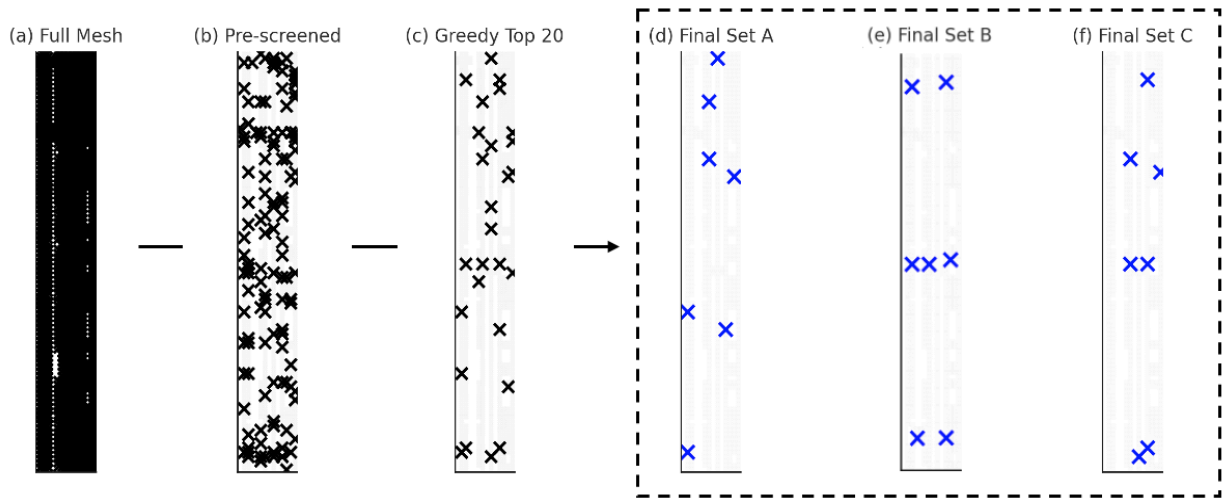

**Figure 5-5.** Node selection workflow illustrated through a stage refinement process.

To balance predictive accuracy with computational efficiency, the node selection process first identifies a greedy top-20 candidate pool per plane based on each node's sensitivity to deformation variation across different input conditions. These candidates are selected not only for their contribution to reconstructing the global displacement field but also for capturing meaningful spatial correlations and ensuring coverage across the surface domain. From this refined pool, multiple feasible final sets, such as the examples shown in Figure 5-5(d–f), can be constructed depending on the target number of nodes. In this study, we constrain the final configuration to 7 nodes per plane, prioritizing those that jointly capture the dominant variance modes and provide spatially uncorrelated, domain-representative information. Among the variants, Set B is adopted for subsequent model training and visualization, as it reflects a balanced and physically interpretable sampling pattern well-suited for efficient surrogate modeling. This strategy yields a total of 168 sampled points per workpiece, offering a quick yet effective assessment of the surface deviation and its compliance with the design criteria before proceeding with further process decision-making. The corresponding deviation distributions across all sampled planes and grooves are visualized in Figure 5-6. For example, PLN12 represents the right-side surface (plane 2) of groove 1. Observing the deformation trends across grooves, consistent patterns emerge that are critical for the surrogate model to learn and generalize effectively for rapid surface characterization.

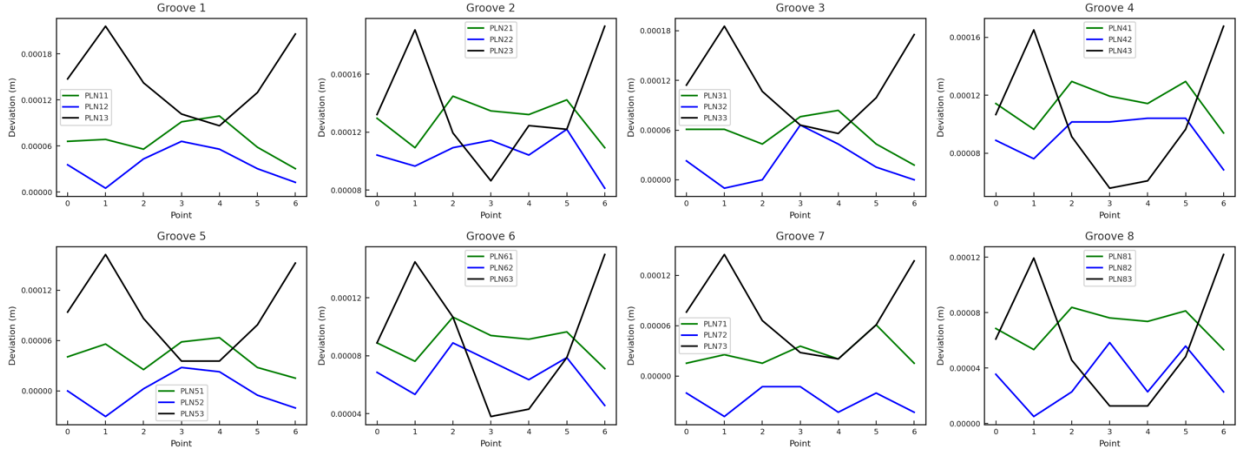

**Figure 5-6.** Distribution patterns of residual deformation across selected node samples.

This structured approach simulates surface deformation by capturing displacement distributions at key nodes, providing insight into the milling process. However, the high-fidelity finite element model requires over 10 hours per simulation on a desktop computer (Dell Precision 5820, Intel Xeon W-2155 processor, 10 cores), creating challenges for large-scale data generation. To address this, we simplified the model by reducing the workpiece length and focusing on a single groove, eliminating the need to simulate tool adjustments and inter-groove movements. This modification lowers the simulation time to about 20 minutes per run while preserving the essential deformation patterns, enabling more efficient data generation and supporting an automated pipeline.

### 5.2.2. Streamlined data generation pipeline

Here we implement a fully automated data generation pipeline to streamline surrogate modeling. This autonomous system eliminates manual oversight and maintains a seamless transition from data generation to modeling. We balance thorough statistical coverage with computational feasibility by focusing on four key input parameters of the milling process listed in Table 5-4. They are representative of process control and material property uncertainties, which will provide guidance to process optimization and quality assurance. To systematically generate simulation samples across these parameters, we adopted a computer-based Design of Experiments (DoE) approach using multivariate uniform random sampling. This method randomly and independently samples each parameter within its specified range, thus ensuring comprehensive and unbiased statistical coverage across the defined parameter space. The choice of multivariate uniform sampling was motivated by the need for adequate representation and even exploration of the parameter space without predefined bias or correlation, which helps prevent unintended clustering or gaps. A total of 1,000 simulation cases was determined by balancing the computational feasibility against the required resolution to sufficiently characterize process behaviors. Increasing the sample size beyond this point resulted in diminishing returns on surrogate model accuracy relative to computational effort. Thus, this sampling strategy and scale provided robust coverage of essential physical behaviors while remaining computationally practical for surrogate model training.

**Table 5-4.** Process input parameter ranges

| Input Parameters | Range |
| --- | --- |
| J-C Damage Model Parameter | 0.1 – 0.3 |
| Friction Coefficient | 0.1 – 0.5 |
| Heat Transfer Conduction | 1e5 – 1.5e5 W/m$^2$/°C |
| Feed Rate | 20 – 60 inch/min |

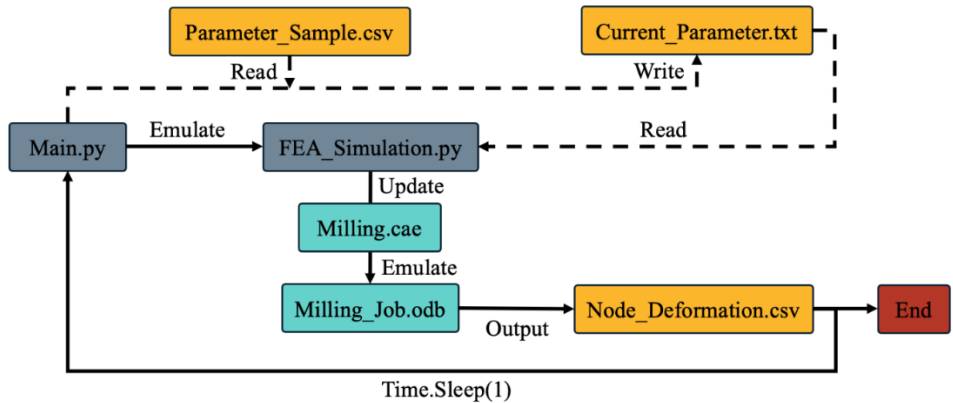

**Figure 5-7.** Automated data generation pipeline.

The automated data generation process is managed in Python, serving as a central interface between Abaqus and external systems. As illustrated in Figure 5-7, multiple Python scripts and file types enable finite element simulations and data retrieval from ODB files. The CAE file defines model parameters, while ODB files store simulation outputs. Python scripts oversee Abaqus CAE operations, extract data from ODB files, and organize the input parameters and output data for subsequent surrogate modeling. As detailed in Figure 5-7, the automation of the data generation process is depicted through a series of Python scripts, represented by light green blocks in the diagram. These scripts load and record the input parameters for each sample in a text file, then update the finite element model within the CAE file to launch the simulation. Upon completion, the ODB file is processed to extract node displacement data, which is saved as a CSV file. To maintain data integrity, the system pauses briefly after each run before moving on to the next simulation. Despite this automated setup, not every simulation finishes successfully. Out of 1,000 samples, 85 were discarded due to simulation failures linked to numerical instability or element distortion. Numerical instability often arises from complex interactions or input parameters that do not converge, while element distortion can occur in high-strain or detailed-contact scenarios, causing the solver to halt. These issues highlight the balance required in finite element simulations, where detailed physics-based models must also align with computational constraints and solver capabilities.

## 5.3. Physics-Infused GNN Framework for Milling Predictions

The proposed spatially-informed predictive framework leverages a Physics-Infused Graph Neural Network (PhyGNN) to effectively capture and utilize node structure information, ensuring precise handling of spatial dependencies and relationships essential for accurate predictions. We enhance the model's accuracy by customizing its loss function to include physics-based insights from the J-C model. This integration of physical principles significantly deepens the framework's understanding of the milling process, yielding a robust, physics-informed solution for predictive modeling [267,268].

### 5.3.1. Graph convolutional network

The Graph Convolutional Network (GCN) emerges as an innovative neural network architecture, ideally suited for datasets with inherent graph structures. This capability makes the GCN [269] particularly
136136136136136136136136136136136136136136136136136136136136136136136136136136136136136136136136136136136136136136136136136136

effective in addressing the complex challenges associated with the interconnected and spatially-oriented data typical in finite element modeling for physical manufacturing processes. A hallmark of the GCN is its ability to harness the connectivity patterns and feature correlations present in graph-structured data, providing a detailed and nuanced representation of the data landscape. A key feature of the GCN is its layer-wise propagation rule [270,271], which enhances the update of node features through the aggregation and transformation of information from adjacent nodes. This approach enables the incorporation of local neighborhood data at each node while allowing deeper network layers to assimilate more global information. Additionally, GCNs excel at integrating edge weights, which facilitates distinguishing the importance of different connections within the graph. This flexibility is essential for making precise predictions where the impact of node positions is critically variable.

In the domain of GCN, the process of aggregating structure features alongside other input variables is pivotal for synthesizing the complex interplay of node relationships and features within a graph. The mechanism of graph convolution, especially within a graph $G = (V, E)$, where $V$ and $E$ represent the sets of vertices and edges respectively, is defined through a series of operations that combine the node features with the graph topology to produce a richer representation of the data. The matrix $X \in \mathbb{R}^{d \times |V|}$ representing the finite element model process parameters as the input features, where $d$ is the number of features per node, serves as the foundation for feature propagation. The aggregation and update of node features through the layers of GCN can be formalized as follows:

$$m_{N(u)}^{(k)} = AGGRE^{(k)}\left(\left\{h_v^{(k-1)}, \forall v \in N(u)\right\}\right) \tag{5.3}$$

In Equation 3, the aggregation step involves collecting the features of neighboring nodes to a target node $u$, denoted by $N(u)$, to form an aggregated message $m_{N(u)}^{(k)}$ at the $k$-th layer. Additionally, $h_v^{(k-1)}$ represents the features of neighbor $v$ at layer $k-1$. Subsequently, the feature update step recalculates the feature vector for node $u$ at layer $k$ by integrating $u$'s own previous features with the aggregated message from its neighbors, which can be described as

$$h_u^{(k)} = \sigma(W_u^{(k)} \cdot h_u^{(k-1)} \oplus W_{N(u)}^{(k)} \cdot m_{N(u)}^{(k)} + b^{(k)}) \tag{5.4}$$

where $\sigma$ is non-linear activation function, and $W_u^{(k)}$, $W_{N(u)}^{(k)}$, and $b^{(k)}$ are trainable parameters specific to layer $k$. After combining the above equations, then the feature update equation becomes

$$h_u^{(k)} = \sigma(MERGE^{(k)}(W_u^{(k)} \cdot h_u^{(k-1)}, W_{N(u)}^{(k)} \cdot m_{N(u)}^{(k)}) + b^{(k)}) \tag{5.5}$$

In these equations, *AGGRE* and *MERGE* functions play decisive roles in defining how the features of neighboring nodes are combined and how the updated features of a node are derived from its own and those of its neighbors. These operations are fundamental for ensuring that the GCN captures the spatial dependencies and relationships inherent in the graph structure. In the context of predicting deformation values in a milling process, this sophisticated mechanism of feature aggregation and update allows the GCN to integrate spatial dependencies and physical interactions between different points on the workpiece.

### 5.3.2. Dynamic attention-based graph mechanism

In our proposed framework, we first construct a concise graph using only a finite set of essential nodes, such as measurement points aligned with CMM data. Initially, we capture general spatial dependencies between these nodes using a fixed geometric location matrix, where edge weights and adjacency relationships are established based on node positions. However, this fixed representation alone may not adequately capture the subtle and evolving node-to-node correlations inherent in complex machining processes, such as localized stress interactions or damage progression. Therefore, we further incorporate a dynamic attention-based mechanism that refines node connectivity during the learning process. Specifically, we dynamically update edge weights according to learned correlations between node features, thus augmenting the initial geometric relationships. Formally, the adaptive edge weight $a_{ij}$ between nodes $i$ and $j$ is computed as:

$$a_{ij} = \frac{\exp\left(\sigma\left(W^T\left[h_i \| h_j \| l_{ij}\right]\right)\right)}{\sum_{k \in N(i)} \exp\left(\sigma\left(W^T\left[h_i \| h_k \| l_{ik}\right]\right)\right)} \tag{5.6}$$

where $h_i$ and $h_j$ represent the learned feature vectors at nodes $i$ and $j$, sigma is a nonlinear activation, and $N(i)$ denotes the neighbor nodes connected to node $i$.

Through this mechanism, the model adaptively highlights the most influential interactions among nodes during training, combining both geometric and learned feature relationships. Consequently, the dynamic attention-based graph mechanism effectively balances computational efficiency with expressive modeling power, which is illustrated by three key stages of the dynamic graph construction in Figure 5-8. In the initial stage, critical nodes are positioned based on geometric location information $l_{i,j}$, forming initial spatial dependencies represented by dashed lines. In the intermediate stage, node feature embeddings $h_{i,j}$ are introduced, and the attention mechanism dynamically updates edge weights $a_{i,j}$, highlighting strong correlations while diminishing weaker ones. In the final predictive stage, refined correlations $a_{i,j}$ guide accurate prediction of node displacements $y_{i,j}$, ensuring physically meaningful predictions based on dynamically learned relationships.

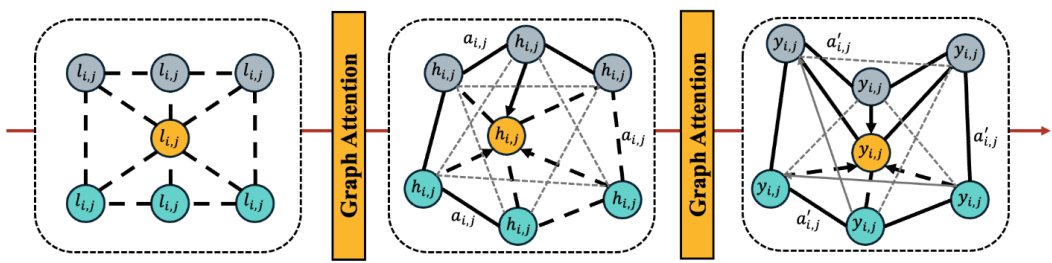

**Figure 5-8.** Dynamic attention-based node graph.

### 5.3.3. Unified physics-infused loss function

To ensure our model delivers both high predictive accuracy and physical consistency, we introduce a unified physics-infused loss function, combining a J-C physics consistency loss, a parameter estimation loss, and a graph structural regularization loss. Each component targets a distinct aspect of our modeling framework to improve overall predictive performance. First, the J-C physics consistency loss enforces that the deformation predictions align with fundamental physical principles governing machining-induced stress and damage evolution. Specifically, node-level strains are first computed from the predicted node displacements; these computed strains are then substituted into the J-C constitutive model as described in Equation (5.1) to derive corresponding predicted stresses. These stress predictions are directly compared to ground truth stress values from finite element simulations, resulting in the J-C consistency loss:

$$L_{JC} = \frac{1}{N} \sum_{i=1}^{N} (\sigma_{JC}(i) - \sigma_T(i))^2 \tag{5.7}$$

This term ensures that our deformation predictions remain physically valid and consistent with the known constitutive behavior. Furthermore, due to our concise yet dynamically adaptive node graph, we incorporate a graph structural regularization term. This regularization stabilizes node interactions, prevents overfitting, and preserves physically meaningful correlations by penalizing large feature variations among strongly connected nodes. Guided by dynamically learned attention-based edge weights $a_{ij}$, it is expressed as:

$$L_{reg} = \sum_{(i,j) \in \varepsilon} a_{ij} \| h_{pred}(i) - h_{pred}(j) \|^2 + \lambda \|W\|^2 \tag{5.8}$$

where $h_{pred}$ represents predicted node-level deformation, and $W$ represents network parameters. This regularization ensures spatial coherence and smoothness in predictions, effectively mitigating physically implausible variations.

These two loss terms, physics consistency and graph regularization, are combined with the primary data-driven deformation prediction loss $L_{pred}$ into the unified physics-infused loss function:

$$L_{total} = L_{pred}(Y_t, Y_p) + \beta_1 L_{JC} + \beta_2 L_{reg} \tag{5.9}$$

Here, $Y_p$ represents the displacement predictions generated by the predictive model and $Y_t$ denotes the corresponding finite element-based ground truth displacements. In this comprehensive formulation, weighting parameters $\beta_1$ and $\beta_2$ balance the contributions of physics-infused J-C loss and the graph structural regularization, respectively, with displacement prediction loss. This constructed loss function guides our model towards a balanced solution that is physically consistent, computationally efficient, and highly predictive, effectively supporting rapid iterative predictions and virtual process assessment.

Figure 5-9 illustrates the GCN framework, where machining parameters are first transformed into node-level features through linear transformations. A concise node graph, formed from selected critical measurement nodes, is then dynamically updated through a graph attention mechanism to capture evolving spatial correlations. While previous methods such as Mou et al [272] constructed GNN-driven digital twins for predicting stress in turbine blades, and Park and Kang [273] employed Bayesian optimization to adapt mesh densities in a GNN framework, both approaches rely on extensive node sets or dense mesh inputs, making them challenging to deploy for frequent, high-fidelity machining simulations. In contrast, our method focuses on a compact node selection strategy and dynamically learned attention to handle machining-specific damage evolution efficiently. Once the dynamic graph connectivity is established, graph convolutional layers propagate updated node embeddings, which feed into a fully connected layer to predict node displacements. These predicted displacements yield nodal stresses via the J-C model, subsequently validated against finite element references. A unified physics-infused loss function integrates deformation prediction accuracy, J-C-based stress consistency, and a graph structural regularization term. This integrated approach ensures physically meaningful, computationally efficient, and highly accurate deformation predictions, ideal for rapid iterative prediction and decision support in machining processes.

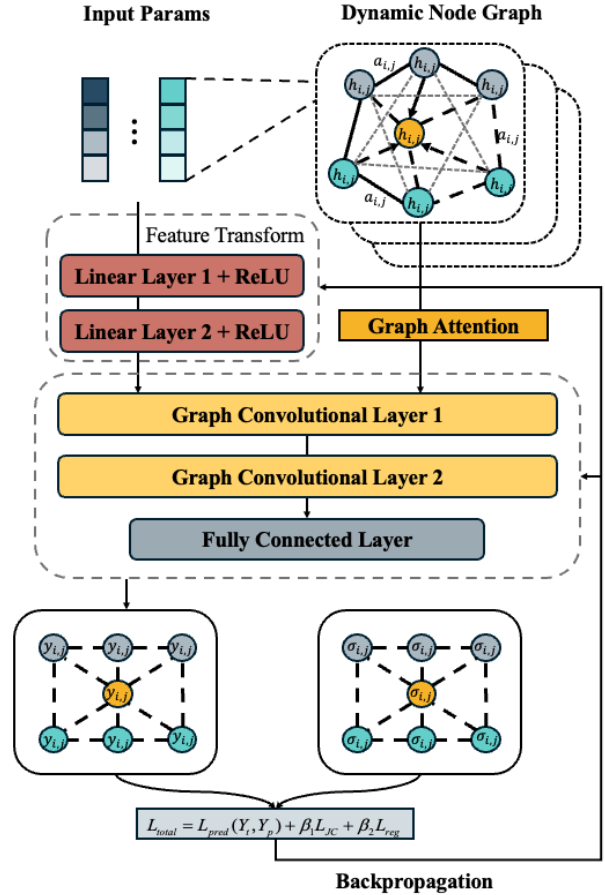

**Figure 5-9.** Architecture of PhyGNN featuring dynamic attention-based node graph.

### 5.4. Case Studies and Discussions

In this section, we assess the performance of the proposed physics-infused Graph Convolutional Network (PhyGNN) through a detailed comparison with other models. An ablation study is performed to evaluate the contribution of relevant components to the accuracy, including the integration of physics-based insights and the node graph representation. These analyses underscore the advantages of incorporating domain-specific physics, resulting in enhanced model performance and robustness.

#### 5.4.1. Performance comparison

We conduct an in-depth analysis of the surface deviation data collected from the seven selected nodes. Figure 5-10 visualizes these distributions using probabilistic histograms, where each subplot corresponds to a specific node. Across all nodes, the histograms exhibit approximately symmetric, bell-shaped patterns centered near zero, indicating a balanced distribution of tensile and compressive residual deformations. This symmetry suggests that, after milling and stress relaxation, the local surfaces tend to return close to their original baseline states, supporting the feasibility of lightweight deformation characterization for rapid online assessment.

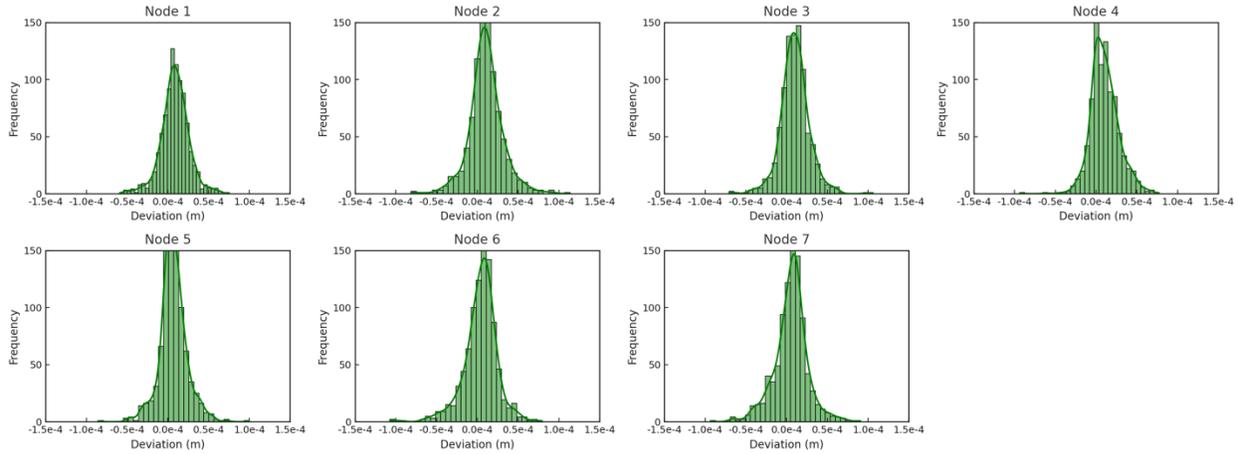

**Figure 5-10.** Distribution of residual deformation at each selected node after relaxation.

The residual deformation values are represented in the finite element output as the displacement of elements or nodes in the axis perpendicular to the milling surface, which we refer to as deformation magnitude. This displacement, measured in meters, typically falls within the range of $10^{-5}$ to $10^{-6}$. To enhance the learning, we normalize the deformation magnitudes before training. This normalization scales the values, making the data more conducive to the predictive model's learning process. The combined visual and quantitative analysis provided by this histogram offers a deeper insight into the material behaviors, ensuring that the model accurately captures the subtle variations present in the empirical data. Our research addresses a complex multi-output regression problem, focusing on predicting deformation across multiple nodes. To explore a variety of potential methods, we evaluate both traditional machine learning techniques, such as Multivariate Linear Regression (MLR) [274,275], Random Forest (RF) [276,277], and Multivariate XGBoost (MXGBoost) [278], as well as deep learning models, including 1D Convolutional Neural Networks (1DCNN) [279,280], Multi-Layer Perceptron (MLP) [281], and Physics-Based Graph Convolutional Networks (PhyGNN). Additionally, we included Multi-response Gaussian Processes (MRGP), a well-established method in machine learning for capturing correlations between multiple outputs.

For traditional machine learning models, we apply specific modifications to adapt them to the multi-output regression task. Multivariate Linear Regression is used to handle multiple dependent variables simultaneously. Similarly, Random Forest, which inherently supports multi-output tasks, is employed to predict multiple targets by constructing several decision trees for each output. For XGBoost, we leverage its multi-output variant, which allows it to handle multiple dependent variables within a gradient-boosting framework. The MRGP is chosen due to its capability of modeling correlated responses, where the outputs are linked, rather than treating them as independent predictions. This method uses Gaussian processes to predict multiple outputs simultaneously, accounting for the correlations between them. We select these methods for their ability to model multi-output relationships, each offering a different approach to handling the complexity of the data. The traditional machine learning models provide practical and efficient baselines, offering simplicity and relatively fast training times. On the other hand, deep learning models were included due to their capacity to learn complex input-output mappings, which could be beneficial in capturing intricate patterns in the data. To evaluate the performance of each model, we used both Root Mean Squared Error (RMSE) and Mean Absolute Error (MAE). RMSE is useful for identifying large prediction errors, while MAE provides a clearer measure of the average error, allowing for a more comprehensive assessment of model performance. Through this analysis, we aim to determine the most effective and efficient method for predicting deformation, balancing model complexity and practical applicability. This ensures that our findings consider not only predictive accuracy but also the computational and practical constraints of different modeling scenarios. The performance and stability of all baseline models are summarized in Table 5-5.

**Table 5-5.** Performance and stability comparison across predictive models

| Model | MAE (*meter*) | RMSE (*meter*) |
|---|---|---|
| MLR | 1.0661e-5±0.0000e-5 | 1.1402e-5±0.0000e-5 |
| RF | 7.8084e-6±0.0152e-6 | 8.2661e-6±0.0136e-6 |
| MXGBoost | 5.5810e-6±0.0000e-6 | 5.8834e-6±0.0000e-6 |
| MRGP | 7.2958e-7±0.0263e-7 | 9.3734e-7±0.0391e-7 |
| MLP | 2.5713e-6±0.1429e-6 | 2.9516e-6±0.1290e-6 |
| 1DCNN | 3.4709e-6±0.2705e-6 | 3.7152e-6±0.2835e-6 |
| PhyGNN | 3.2094e-7±0.0359e-7 | 5.0117e-7±0.0477e-7 |

Table 5-5 compares the performance and stability of various predictive models in terms of their Mean Absolute Error (MAE) and Root Mean Squared Error (RMSE), along with the standard deviations across 10 experimental runs. To facilitate a fair comparison, the RMSE calculations are performed after re-scaling the model predictions to match the scale of the actual observed data. This re-scaling ensures that the predictions are directly comparable to the ground truth, with any variability between models reflecting differences in their predictive capabilities rather than differences in scale. Each model is evaluated over 10 runs with fixed random seeds to expose both performance (as measured by MAE and RMSE) and stability (as indicated by the standard deviation of these metrics across runs). This setup accounts for inherent randomness in training and provides a robust measure of both the accuracy and consistency of each model.

Overall, the machine learning models (MLR, RF, and MXGBoost) tend to exhibit more stable performance, as evidenced by their smaller standard deviations, especially when compared to the deep learning models (MLP, 1DCNN, and PhyGNN). This greater stability is expected due to the more deterministic nature of traditional machine learning algorithms. For example, MLR and MXGBoost exhibit no standard deviation for either MAE or RMSE, with values such as 1.0661e-5 ± 0.0000 (MLR MAE) and 5.5810e-6 ± 0.0000 (MXGBoost MAE), indicating that these models produced identical predictions across all 10 runs. This is consistent with the behavior of deterministic methods, where fixed seeds result in no variability in outcomes. Deep learning models exhibit higher standard deviations across runs due to the stochastic components of neural network training, such as random initialization and dropout layers. MLP, for example, has an MAE of 2.5713e-6 ± 0.1429e-6 and an RMSE of 2.9516e-6 ± 0.1290e-6, indicating relatively higher variability compared to machine learning models. 1DCNN shows even greater variability, with a standard deviation of 0.2705e-6 for MAE and 0.2835e-6 for RMSE. This suggests that while deep models can capture complex patterns, they are also more sensitive to random initialization and model configuration. Despite this general trend, PhyGNN stands out as an exception among the deep learning models. It exhibits both strong performance and relatively low variability, with an MAE of 3.2094e-7 ± 0.0359e-7 and an RMSE of 5.0117e-7 ± 0.0477e-7, demonstrating better stability compared to other deep models. This lower standard deviation may be attributed to the inclusion of graph-based learning and physics-informed components, which help guide the model towards predictions that adhere to underlying physical principles. These constraints likely reduce the variability in predictions by preventing extreme outliers, thereby improving both performance and stability, as will be demonstrated in further validation.

Among all the models tested, PhyGNN and MRGP emerge as top performers. PhyGNN achieves the lowest MAE (3.2094e-7) and RMSE (5.0117e-7), combining strong predictive accuracy with computational scalability. MRGP also shows good numerical accuracy (MAE: 7.2958e-7 ± 0.0263e-7, RMSE: 9.3734e-7 ± 0.0391e-7), notably better than other benchmarks. However, considering the extremely small scale of deformation magnitudes (around 1e-5 to 1e-6 meters, as seen in Figure 10), prediction errors around the 1e-7 scale, despite appearing numerically small, can still represent substantial relative inaccuracies, leading to potentially incorrect deformation patterns. Additionally, MRGP's Gaussian-process structure becomes computationally prohibitive as the dataset grows larger, whereas PhyGNN remains effective due to its physics-informed deep-learning architecture, making it more suitable for practical digital twin applications.

Moreover, the ability of a model to predict consistently well across all noses is fundamental, especially when the data is structured in a way that individual nodes might exhibit different behaviors or responses. To address this, we selected four of the best-performing models from the overall analysis, MRGP, MLP, 1DCNN, and PhyGNN, and examined their performance on each node, as shown in Figure 5-11. The log10(MSE) values for each model across seven nodes, which represent the average results across multiple runs, can provide insight into how consistently each model performs. Both PhyGNN and MRGP demonstrate relatively stable performance across all nodes, indicated by the minimal variation in bar heights. This consistency suggests that these models are effective in capturing underlying patterns and correlations across different nodes, with PhyGNN showing the lowest overall error. PhyGNN's ability to integrate graph-based relationships and physics-informed components likely helps maintain consistent performance across nodes, making it especially suitable for applications requiring uniform prediction quality. MRGP also performs stably, though slightly less accurately, likely due to its limitations with scalability on larger datasets. On the other hand, MLP and 1DCNN show more variability across nodes, with noticeable differences in performance from node to node. For example, their predictions for Node 3 are relatively poor, while their performance on Node 6 is much better. This variability highlights their challenges in maintaining consistent prediction accuracy across all nodes. While these models can capture non-linear patterns, their lack of explicit mechanisms to model node-to-node dependencies, such as the graph-based structure in PhyGNN, may lead to inconsistent performance.

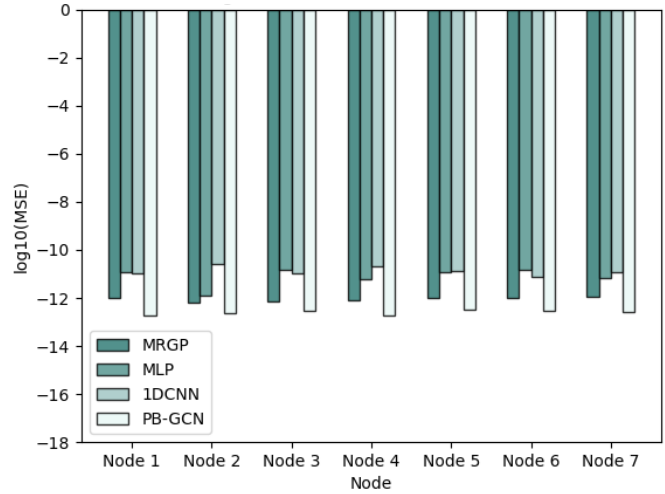

**Figure 5-11.** MSE of models across 7 nodes.

While earlier metrics such as MSE and MAE provided overall performance insights, we are trying to investigate how each model handles the specific deformation distribution at each node in Figure 5-12, where we plotted the predicted vs. actual deformation for every node using a single test run for each model. We can observe how deformation patterns differ across nodes, presenting challenges for models to predict each node accurately. Consistent performance across nodes is significant because variation in deformation distributions can lead to fluctuating model accuracy. PhyGNN and MRGP exhibit strong consistency in their predictions, with points clustering tightly around the diagonal line for most nodes. In contrast, MLP and 1DCNN show slightly broader distributions of predictions, particularly at higher or lower deformation values, which suggests that these models may struggle more with extreme cases. Node 7, in particular, poses challenges for MLP and 1DCNN, with predictions more spread out, indicating potential difficulties in handling complex deformation patterns unique to that node. Conversely, for Node 1, all models perform relatively well, with predictions tightly clustered around the actual values. This consistency across nodes with simpler deformation patterns reinforces the importance of robust model performance across a variety of node behaviors. Ultimately, the ability to consistently handle the diverse deformation conditions across all nodes is a key indicator of a model's effectiveness.

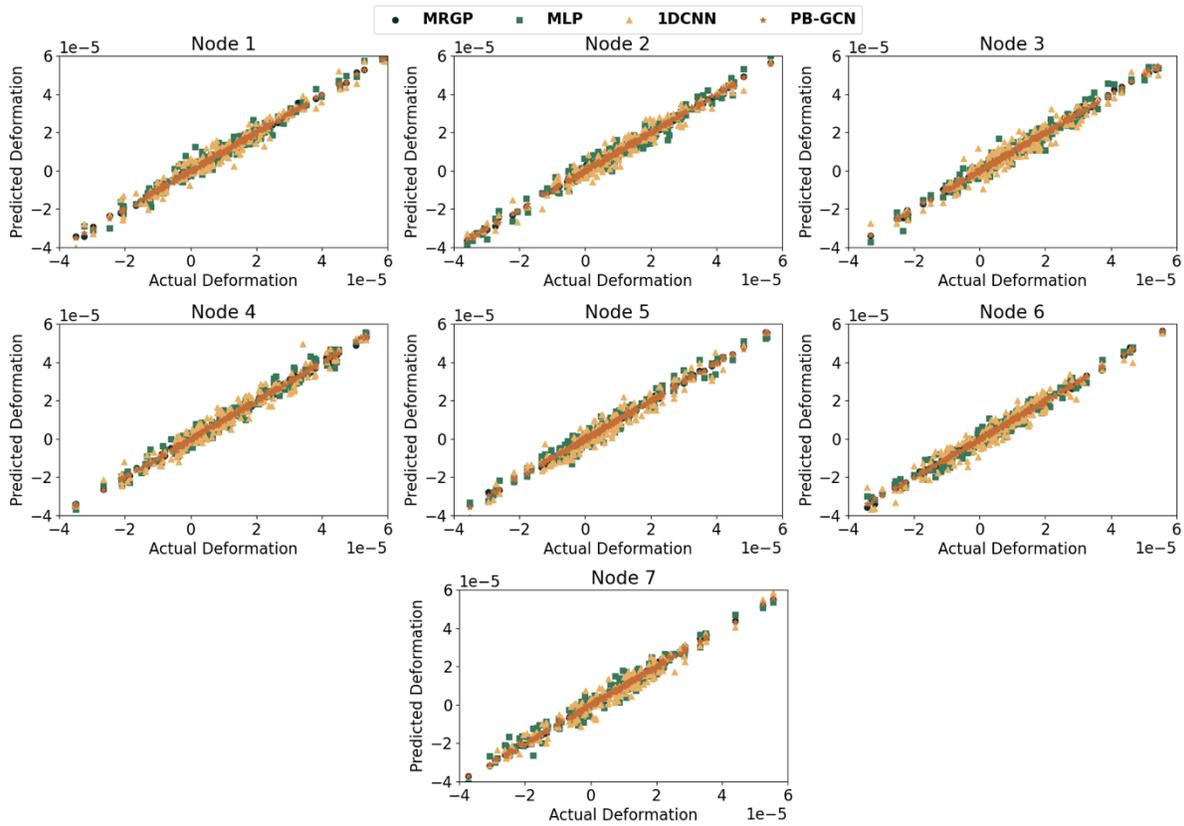

**Figure 5-12.** Testing accuracy of deformation at different nodes with multiple well-performing.

Beyond surface displacement, the finite-element database exposes additional outputs, von Mises stress, plastic strain, heat generation, reaction forces, etc., that are equally valuable for milling-process optimization. By replacing the target field and retraining on the same seven-node layout, PhyGNN can predict these variables without altering its architecture. Figure 5-13 illustrates this ability for three representatives: von Mises stress, equivalent plastic strain (PEEQ) and logarithmic strain (LE). For stress and PEEQ, the predicted points cluster tightly around the 45 ° reference line, confirming that the model captures their smooth, spatially coherent patterns with limited data. Accuracy is lower for LE, whose rapid local gradients are more difficult to approximate, resulting in a wider scatter. The comparison highlights that the physics-guided graph framework generalizes well to additional responses. Moreover, the performance naturally depends on the physical complexity and spatial variability of the quantity being modelled. Future work will extend this capability by refining feature normalization, adapting node-sampling density in highly stressed regions, and introducing variable-specific loss terms, with the goal of furnishing a single surrogate that delivers reliable predictions for the full suite of process-critical responses needed in closed-loop machining optimization.

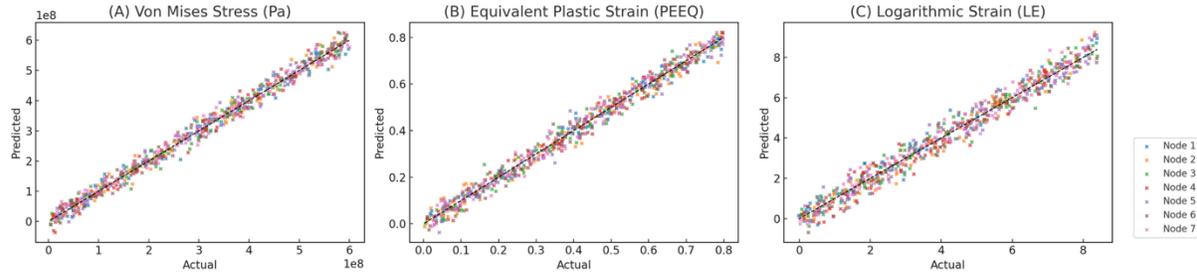

**Figure 5-13.** Testing performance of PhyGNN on multiple response variables.

The node-level analyses above confirm that the proposed pipeline delivers stable accuracy across a wide range of deformation magnitudes, lending confidence in its effectiveness for the present aerospace workpiece. Moving from effectiveness to domain generalizability, the same accuracy must be reproduced when the underlying material model, fixturing, or tool path changes. In our framework these factors are encoded only in the high-fidelity finite element batch that feeds the surrogate; all downstream modules, automatic meshing, stress-gradient pre-screening, greedy node selection, and GNN training, remain unchanged. Consequently, transferring the method to a new alloy or substrate simply requires re-running the FE solver with updated constitutive data and boundary conditions, after which the surrogate can be retrained on the regenerated data set. To streamline this step, we have modularized the workflow and are developing scripted material libraries plus transfer-learning routines that reuse learned graph weights when physics remains similar. These additions will extend the demonstrated performance to a broader class of parts and machining scenarios without altering the core methodology.

### 5.4.2. Ablation study

The above comparison of performance reveals the power of the deep learning methods and our proposed model, PhyGNN, both in terms of overall performance and in the detailed, node-level predictions. By examining not only the aggregate performance but also the model's predictions for each data point at each node, we gain a more comprehensive view of how well the models perform. However, while these results are promising, they still do not provide definitive evidence for the specific enhancements made to the model architecture. To address this, we perform an ablation study aimed at investigating the effects of two key components: the Node Graph Feature (NGF) and the Physics-Based Loss (PBL).

### 5.4.2.1. Performance contribution analysis

In this study, w/o NGF refers to removing the node graph feature during training. For the GCN architecture, this involves omitting the graph input, which is typically provided as a node adjacency matrix or edge list that defines the connections between nodes. By removing the NGF, we effectively treat the input as if the nodes are independent, thereby eliminating the spatial relationships that GCN relies on to propagate information between connected nodes. This allows us to assess the contribution of the node graph feature to the model's performance. Similarly, w/o PBL refers to removing the physics-based loss from the optimization process. In this case, the model is trained solely on the standard loss function (mean squared error), without the additional physics-informed constraints that enforce physical consistency. This comparison allows us to isolate the effect of the physics-based loss on prediction accuracy and consistency. By systematically removing these features, we aim to draw conclusions about which enhancements, NGF or PBL, contributes more to the overall performance improvement. Additionally, recognizing the unique architecture of the GCN, we also incorporate GraphSAGE and GAT as benchmark graph-based models in our ablation study. These models can also take graph input and node features, mapping spatial relationships to the final output. By applying the same ablation process to GraphSAGE [282] and GAT [283], we can validate whether the benefits of NGF and PBL are generalizable across different graph-based architectures. This will help us better understand the contribution of these modifications and whether they are specific to the GCN or beneficial for a broader range of graph neural networks.

The RMSE values in Table 5-6 are averaged over 10 runs to ensure that the performance metrics reflect both accuracy and stability across various random seed initializations. From the results, it is clear that the enhancements we introduced, namely the Node Graph Feature (NGF) and the Physics-Based Loss (PBL), are generalizable to all models tested (PhyGNN, GraphSAGE, and GAT), bringing significant improvements in their performance. NGF consistently has a more pronounced impact on performance than PBL, particularly for PhyGNN and GAT. The most noticeable improvement is seen in GAT, which suffers a significant degradation of 31.18% when NGF is removed, highlighting that GAT heavily relies on global node correlations. As GAT utilizes an attention mechanism to assign weights to different nodes, removing the node graph feature compromises its ability to capture these relationships, leading to the steep drop in performance. PhyGNN also experiences degradation (26.09%) without NGF, reflecting its strong dependence on node-to-node correlations, which the NGF provides. This behavior is intuitive since GCNs are designed to leverage the structural information in graphs, and removing NGF disrupts this functionality.

**Table 5-6.** Performance comparison of ablation studies

| Method | Ablation Study | RMSE (*meter*) | Degradation Deg. |
|---|---|---|---|
| PhyGNN | w/o NGF | 6.3193e-7±0.0944e-7 | -26.09% |
|  | w/o PBL | 5.4978e-7±0.0850e-7 | -9.70% |
|  | Baseline | 5.0117e-7±0.0477e-7 | N/A |
| GraphSAGE | w/o NGF | 7.0026e-7±0.0621e-7 | -17.18% |
|  | w/o PBL | 6.8974e-7±0.0401e-7 | -15.42% |
|  | Baseline | 5.9759e-7±0.0268e-7 | N/A |
| GAT | w/o NGF | 7.3233e-7±0.1036e-7 | -31.18% |
|  | w/o PBL | 6.1425e-7±0.0926e-7 | -10.03% |
|  | Baseline | 5.5825e-7±0.0837e-7 | N/A |

While the impact of PBL is somewhat less than NGF, it still contributes around 10% performance improvement across all models. This highlights that physics-based constraints serve as a valuable regularization mechanism, guiding the models toward more physically plausible predictions, even when NGF is present. PBL's ability to refine predictions further by incorporating domain-specific knowledge makes it a complementary enhancement. GraphSAGE stands out as an exception where the removal of NGF results in only a 17.18% degradation, which is smaller compared to PhyGNN and GAT. This suggests that GraphSAGE relies less on the explicit graph features (neighboring nodes), possibly because it samples a subset of neighbors and aggregates information differently from GCN or GAT. In contrast, the removal of PBL results in a relatively higher degradation (15.42%), implying that GraphSAGE benefits more from physics-informed constraints than from the global node correlations provided by NGF. These observations offer heuristic insights into the types of enhancements that should be emphasized depending on the model. Models like PhyGNN and GAT, which rely heavily on global node correlations, should prioritize incorporating NGF for optimal performance. In contrast, models like GraphSAGE, which are less dependent on such correlations, may benefit more from incorporating physics-based losses, as seen from its comparatively smaller degradation without NGF and higher sensitivity to PBL.

When examining the spread of multiple runs in Figure 5-14, it becomes apparent that GraphSAGE is remarkably stable across all settings. This could be attributed to its neighbor-sampling aggregation method, which likely makes it more robust to fluctuations in input data and randomness in training, compared to models that rely heavily on graph-wide correlations (like GCN and GAT). By using a more localized sampling approach, GraphSAGE seems to generalize better with less variance across runs. Moreover, the enhancements (NGF and PBL) contribute not only to performance improvements but also to the stability of the predictions. The additional information provided by NGF and PBL ensures that predictions adhere to known physical principles and global node relationships, which helps constrain the output values, preventing large deviations from expected results. This makes predictions more stable across different random initializations. Therefore, PhyGNN stands out as the most effective model for our case, largely due to its tailored modifications that effectively capture the simpler node correlations and leverage available physical constraints to guide the prediction process. These enhancements make it particularly well-suited to our dataset and problem requirements. However, both GraphSAGE and GAT remain strong alternatives, with GraphSAGE offering greater stability in contexts that prioritize local node relationships, and GAT excelling in capturing intricate node interactions through its attention mechanism. Ultimately, the choice of model should be guided by the dataset's characteristics, such as node complexity and the presence of physical constraints.

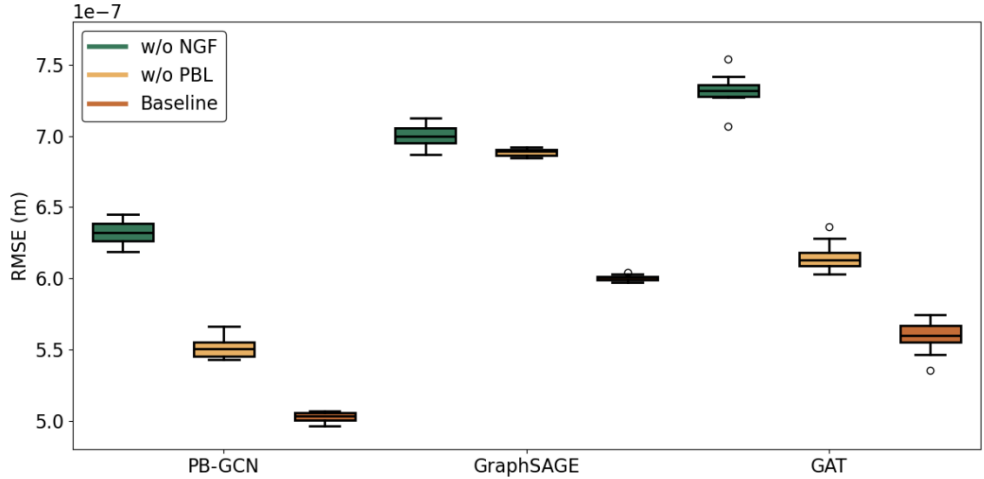

**Figure 5-14.** RMSE distribution of ablation studies among all models.

### 5.4.2.2. Data efficiency contribution

To complement the performance ablation, we also investigated how the integration of the physics-based loss (PBL) affects the model's sample size requirements, an essential consideration for enabling fast surrogate updates in online digital twin applications. For this study, we prepared a total of 900 simulation samples, with 100 samples held out as a fixed test set to evaluate all models under the same conditions. The remaining 800 samples were used to train models under different data availability scenarios: 100 %, 75 %, 50 %, 25 %, and 10 % of the training pool. To ensure a fair comparison across varying sample sizes, we performed hyperparameter tuning for each configuration to identify optimal training settings, and averaged performance over 10 independent runs. The experiments were conducted for both the PhyGNN (with PBL) and the baseline GNN (without PBL) to isolate the effect of physics integration on data efficiency.

The results in Table 5-7 highlight the significant data-efficiency advantage gained by embedding physics into the graph neural network framework. At full data availability (800 training samples), both PhyGNN and the baseline GNN perform reasonably well, with PhyGNN achieving slightly better accuracy (RMSE = $5.01 \times 10^{-7}$ m vs. $6.32 \times 10^{-7}$ m) and stronger agreement in both $R^2$ and Pearson correlation. As the training data are reduced, however, the differences become much more pronounced. When the training size drops to 75 % and 50 %, PhyGNN demonstrates a relatively stable performance, with only moderate increases in RMSE and minor reductions in correlation and R² score. In contrast, the baseline GNN shows noticeably faster degradation, especially at 50 % training, where its RMSE climbs to $8.50 \times 10^{-7}$ m (a 35 % increase from full data), and R² falls to 0.79. This contrast indicates that the physics-based loss in PhyGNN acts as a form of inductive bias, guiding the model to learn physically plausible patterns even when data is limited.

**Table 5-7.** Effect of training size on predictive performance

| Training Ratio | Model | Evaluation Metrics | | | |
| --- | --- | --- | --- | --- | --- |
| | | RMSE(1e-7) | MAE(1e-7) | $R^2$ Score | Pearson Corr. |
| 100% | PhyGNN | 5.01±0.05 | 3.21 ± 0.04 | 0.96 ± 0.02 | 0.98 ± 0.01 |
| | GNN | 6.32±0.10 | 5.05 ± 0.08 | 0.91 ± 0.05 | 0.95 ± 0.03 |
| 75% | PhyGNN | 5.28 ± 0.06 | 4.15 ± 0.05 | 0.95 ± 0.03 | 0.97 ± 0.02 |
| | GNN | 7.04 ± 0.12 | 5.68 ± 0.10 | 0.87 ± 0.06 | 0.90 ± 0.05 |
| 50% | PhyGNN | 5.64 ± 0.07 | 4.45 ± 0.06 | 0.93 ± 0.04 | 0.96 ± 0.02 |
| | GNN | 8.50 ± 0.18 | 6.80 ± 0.15 | 0.79 ± 0.08 | 0.85 ± 0.07 |
| 25% | PhyGNN | 6.26 ± 0.09 | 4.95 ± 0.07 | 0.90 ± 0.05 | 0.94 ± 0.03 |
| | GNN | 10.59 ± 0.30 | 8.47 ± 0.24 | 0.64 ± 0.12 | 0.73 ± 0.09 |
| 10% | PhyGNN | 9.10 ± 0.33 | 8.63 ± 0.20 | 0.62 ± 0.09 | 0.70 ± 0.08 |
| | GNN | 14.10 ± 0.45 | 11.20 ± 0.37 | 0.38 ± 0.15 | 0.55 ± 0.11 |

As summarized in Table 5-7, the benefits of this physics integration become even more evident at 25 % training (200 samples), where PhyGNN still explains 90 % of the variance and maintains a strong correlation with the true values ($\rho = 0.94$). In comparison, the baseline GNN's performance deteriorates rapidly, with R² dropping to 0.64 and Pearson correlation to 0.73. These results suggest that PhyGNN can reliably function with just 200 samples, roughly a quarter of the original data, without substantial compromise in predictive accuracy. This has practical implications in realistic digital twin applications, where the cost of generating finite element data is nontrivial and data availability is often constrained.

At the lowest data level (10 %, or 80 samples), both models experience a clear drop in stability. However, PhyGNN still retains moderate predictive capacity ($R^2 = 0.62$, $\rho = 0.70$), while the baseline model collapses ($R^2 = 0.38$, $\rho = 0.55$), reflecting overfitting and loss of generalization. The relatively smaller standard deviations in PhyGNN across all settings further indicate that physics-based constraints improve not only mean accuracy but also robustness across different training splits. In summary, these results demonstrate that the integration of physics, via the Johnson–Cook-based loss, significantly enhances data efficiency. PhyGNN can match or exceed baseline performance using 50 % fewer samples, making it more suitable for rapid updates and deployment in practical, resource-limited scenarios.

**5.4.2.3. Node correlation analysis**

Building on the results of our ablation study, we now turn our attention to the node graph information. Specifically, we aim to explore how nodes are related to each other based on their predicted outputs, providing a statistical view of the node interactions. By investigating these correlations, we can gain deeper insights into how the model captures node-to-node relationships. To assess the relationships between nodes, we construct a Node Graph Correlation Matrix, which reveals how strongly the predictions from each node are associated with others. For this analysis, we use the Pearson correlation coefficient, a widely used measure of linear correlation. It effectively captures the degree to which changes in one node's predictions are linearly related to changes in another node's predictions, offering a clear statistical understanding of their interaction. The Pearson correlation coefficient [284] is defined as

$$\gamma_{xy} = \frac{\sum_{i=1}^{n}(x_i - \bar{x})(y_i - \bar{y})}{\sqrt{\sum_{i=1}^{n}(x_i - \bar{x})^2}\sqrt{\sum_{i=1}^{n}(y_i - \bar{y})^2}} \quad (5.10)$$

The variables $x_i$ and $y_i$ represent the predicted outputs for two different nodes, where $x_i$ corresponds to node $x$ and $y_i$ to node $y$. The terms $\bar{x}$ and $\bar{y}$ denote the mean values of the predictions for each node across all samples. The coefficient $\gamma_{xy}$ known as the Pearson correlation coefficient, quantifies the linear relationship between two nodes. A higher value of $\gamma_{xy}$ indicates a stronger positive correlation, suggesting that the predictions of the two nodes move together, while a lower or negative value implies weaker or inverse relationships between the nodes.

By calculating these correlations for each pair of nodes, we create the Node Graph Correlation Matrix, which visualizes the degree of similarity between nodes' predictions in Figure 5-15. Strong positive correlations indicate that two nodes have closely aligned predictions, while weaker or negative correlations suggest less similarity or inverse relationships. This analysis helps us better understand how the model's output reflects the underlying relationships within the node graph structure. In examining the node graph information, we observe patterns in the correlation matrix that reflect spatial relationships among nodes. For instance, the moderate correlations between Node 1 and Node 3 (0.378), and between Node 4 and Node 5 (0.484), indicate possible shared influences, likely due to their proximity or symmetrical placement within the structure. On the other hand, the weaker or negative correlations involving Node 7, such as with Node 1 (-0.063) and Node 2 (0.057), suggest that certain nodes may be subject to different conditions, leading to more independent predictions. These variations in correlation, even across distant nodes, demonstrate that spatial dependencies are being captured by the model in some form. This analysis provides a perspective that spatial relationships may play a role in shaping the predictions of individual nodes. The correlation patterns offer insight into how the node graph could be encoding these spatial interactions, reinforcing the idea that incorporating spatial information is important in this context.

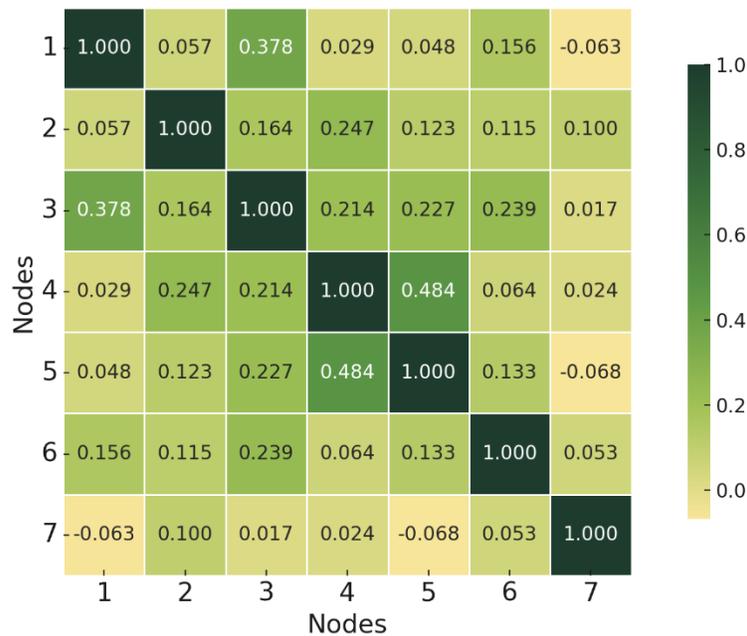

**Figure 5-15.** Node graph correlation matrix on deformation.

**5.4.2.4. Ablation study on graph granularity and efficiency trade-offs**

While the upper bound on node correlation ensures representativeness and avoids redundancy, the number of selected nodes directly affects the downstream model's computational load and predictive accuracy. This raises an important question: how does the performance evolve as we progressively reduce the number of nodes? Because our target setting is flexible manufacturing with a continuously calibrated digital twin, we first establish what is achievable on the full graph before exploring reductions. To this end, we introduce a full-graph reference configuration, 1344 (BM), using a moderately larger PhyGNN (6 layers, 128-dimensional hidden features, multi-head attention with residual connections and normalization) and a slightly longer training schedule with early stopping. This benchmark model provides an upper-bound on accuracy and a cost baseline for later comparisons in computational time, performance, and node selection.

With this upper bound in place, we then conduct the ablation on node granularity. To evaluate the impact of node reduction on surrogate model performance and computational efficiency, we systematically analyze multiple representative node subsets selected at various stages of our refined node-selection pipeline (outlined in Figure 5-5). Initially, the full finite-element mesh comprises 1,344 nodes per plane, representing the most detailed spatial resolution. To reduce computational complexity while preserving the critical deformation information, we first employed a displacement-gradient–based filtering method to retain the top 15% (approximately 200 nodes per plane) exhibiting the highest sensitivity. From these 200 pre-screened nodes, we further conducted a greedy forward-selection process that identified a top-20 candidate subset per plane based on its contribution to reducing cross-validated reconstruction error. Finally, informed by manufacturing constraints and practical measurement accessibility (e.g., CMM locations), we selected the most representative 7 nodes per plane. This compact setting is not prescriptive; seven nodes align with the available CMM layout and serve as a practical low-cardinality case for validation. For a more comprehensive comparison, we also include an intermediate subset of 400 nodes per plane, corresponding to the top 30% ranked by displacement-gradient magnitude.

For this ablation study, we evaluate five distinct node subsets: the full set (1,344 nodes), intermediate-scale subsets (400 and 200 nodes), a candidate subset (20 nodes), and the final selected subset (7 nodes). Each case is assessed using Mean Absolute Error (MAE), Root Mean Square Error (RMSE), model training time, and inference speed. We further consider computational cost through FLOPs (floating-point operations), quantifying complexity and runtime efficiency at both training and inference stages. Given the permutation-invariant and topology-adaptive nature of PhyGNN, varying node cardinality requires dynamic adjustment of input and hidden-layer dimensions; the input embeddings and graph convolutional layers automatically reconfigure to accommodate changing feature sizes and adjacency, ensuring architectural consistency across resolutions. Hereafter we quantify these trade-offs in terms of FLOPs, runtime, and memory usage to assess the scalability of our approach for real-time digital twin deployment.

**Table 5-8.** Performance report on graph granularity and model efficiency

| Node No. | Evaluation Metrics | | | | |
|---|---|---|---|---|---|
| | RMSE(1e-7) | MAE(1e-7) | Runtime (min) | Inf. Time (ms) | FLOPs (G) |
| 1344 (BM) | 5.12 ± 0.24 | 3.42 ± 0.13 | 165.7 | 385.4 | 5.14 |
| 1344 | 26.85 ± 1.83 | 24.40 ± 1.57 | 98.2 | 298.1 | 2.40 |
| 400 | 9.73 ± 0.52 | 8.96 ± 0.44 | 35.7 | 119.3 | 0.91 |
| 200 | 8.45 ± 0.21 | 6.52 ± 0.17 | 17.1 | 58.7 | 0.48 |
| 20 | 4.96 ± 0.09 | 3.13 ± 0.07 | 2.2 | 10.1 | 0.09 |
| 7 | 5.01 ± 0.05 | 3.21 ± 0.04 | 0.96 | 5.2 | 0.04 |

The ablation study summarized in Table 5-8 illustrates the relationship between the number of representative nodes and the predictive accuracy of the consistent PhyGNN model. A key observation is that increasing the number of nodes does not necessarily improve predictive performance. Specifically, when scaling from the baseline of 7 nodes to 20 nodes, the RMSE and MAE remain relatively stable or slightly improved (RMSE: 5.01 to 4.96, MAE: 3.21 to 3.13). This indicates that a moderate increase in representative nodes can slightly enhance the representativeness without introducing excessive model complexity, maintaining an efficient and accurate prediction. However, as the number of nodes continues to increase significantly, the model accuracy sharply deteriorates, RMSE values increase substantially at 200 nodes (8.45), 400 nodes (9.73), and dramatically at the full set of 1344 nodes (26.85). This trend underscores that while more nodes provide theoretically greater spatial representativeness, the increased complexity and expanded input-output mapping task overwhelm the model's lightweight architecture. The larger node set leads to a more complex relational mapping between node-level input parameters and deformation outputs, resulting in an underfitting scenario where the model cannot adequately capture subtle node interactions and spatial dependencies. Hence, an optimal node selection lies in a balanced range, here exemplified effectively by around 7–20 nodes, where the model remains computationally feasible, yet sufficiently representative. To contextualize the attainable accuracy on the full graph, we also report a 1344 (BM) benchmark that employs a moderately larger architecture; this configuration achieves accuracy comparable to the best small-node cases but requires substantially higher compute and longer training, confirming that strong performance on large graphs is feasible albeit at a much higher cost.

As shown in Table 5-8, increasing the number of nodes leads to higher training time, inference latency, and FLOPs, but not necessarily better accuracy. With a lightweight GNN architecture, larger graphs introduce more complex input-output mappings that can overwhelm the model. For instance, the 1,344-node case results in significantly worse RMSE despite the highest computational cost, while the 20-node case achieves slightly better performance than the 7-node baseline with only modest overhead. This highlights that the trade-off between computation and performance is not straightforward. More nodes can improve representativeness up to a point, but beyond that, they reduce efficiency without meaningful gains. Even though inference time differences are small (e.g., 5 ms vs. 10 ms), computational efficiency remains crucial, especially for training and updating. Our framework is designed to support rapid iteration between offline finite element data generation and online model refinement. As new CMM or simulation data become available, the surrogate must be retrained or fine-tuned. A compact model enables fast updates, low resource demand, and real-time deployment on edge or cloud platforms, avoiding the need for prolonged retraining cycles. Thus, model efficiency directly supports the responsiveness and scalability of the digital twin system. Taken together, these results indicate that GNNs are sufficiently powerful to model the full surface when resources permit, and they can also be tailored into lightweight variants for rapid on-site updates; in practice, the choice between full-graph and compact designs can be adapted to the application's accuracy, latency, resource requirements with appropriate architectural and training adjustments.

### 5.4.3. Validation with respect to CMM experimental data

In this section, we focus on the task of validating our proposed model using experimental data. The detailed surface deformation data from four samples are measured using Coordinate Measuring Machines (CMMs). These data provide insights into the actual deformations across seven selected nodes, offering a thorough understanding of the surface profile after machining.

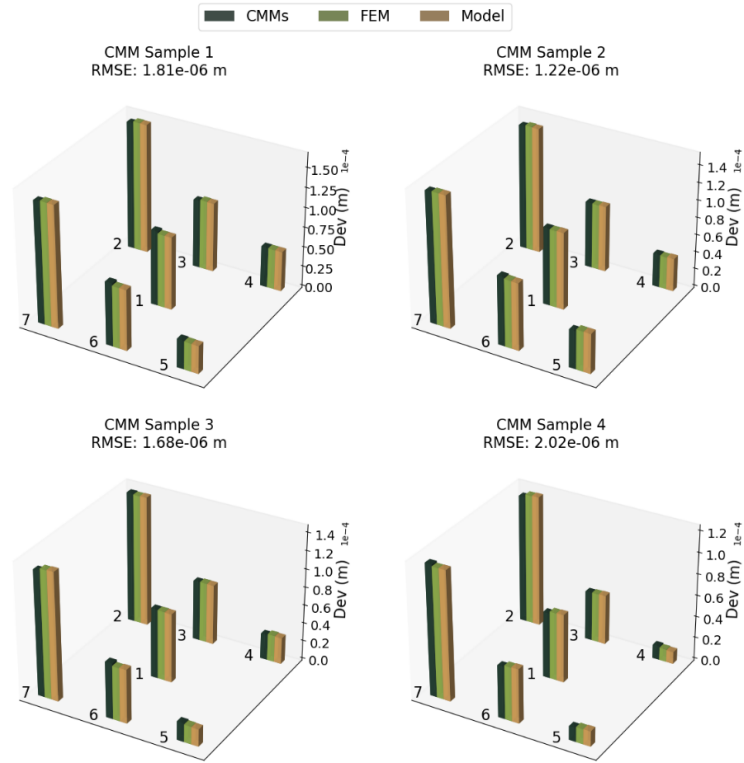

**Figure 5-16.** Milling surface deviation sample from CMM, finite element model, proposed model.

The comparison between the experimental CMM data and our model's predictions is visualized in Figure 5-16. The bar plots for each node are placed at their corresponding locations on the *xy*-platform, accurately representing the spatial distribution of deformations. This layout offers a clear visualization of how deformation varies across the sample surface. Each node has three bars: one for the CMM measurements, another for the finite element model results, and a third for the model's predictions. The RMSE values, shown in the plot titles, quantify the error between the model predictions and the CMM data. While the RMSE values appear larger, particularly for some nodes, this is partly due to the inherent limitations of the finite element model in capturing fine-grained surface details. Finite element models typically approximate surface deformations based on assumed material behavior and boundary conditions, which may contribute to deviations from experimental measurements, given the complexity of simulating machining processes with perfect accuracy. One consistent observation in Figure 5-16 is that the left side of the samples exhibits higher deformation compared to the right. This pattern likely stems from the specific dynamics of the milling process, such as the tool's entry and exit points and the varying stresses applied during the cut. As the tool interacts more intensively with certain regions, it results in uneven material removal and residual stresses, causing higher deformation on the left side. Such spatial deviations are common in machining and highlight the complexity of accurately predicting deformation distribution across an entire surface.

From a practical standpoint, this surrogate model can serve as a rapid assessment tool for surface deformations and enabling informed process adjustments. By visualizing the surface profile after machining, the model allows for iterative back-and-forth predictions, enabling manufacturers to fine-tune process parameters for optimal outcomes. This capability is particularly useful for reducing production time, improving surface quality, and minimizing material waste.

## 5.5. Conclusion

This study addresses the challenge of post-milling deformations in A2024 aluminum workpieces, a material prone to residual stresses that compromise both flatness and surface integrity. Traditional finite element models, while valuable, suffer from high computational costs, when directly employed in process assessment. To overcome these limitations, we introduce a spatially-aware predictive framework using a physics-based graph neural network. By integrating spatial interconnections among nodes with physics-driven principles from the Johnson-Cook model, this framework enables accurate predictions of post-milling surface deviation. Our study also included an ablation analysis, demonstrating the notable contributions of node graph features and physics-based loss functions in enhancing predictive performance and model stability. Furthermore, validation with industrial experimental data from Coordinate Measuring Machines reinforced the model's effectiveness in practical machining scenarios. Despite some deviation between the finite element model and CMM data, the GNN-based model provided a reliable overview of deformation distribution, capturing key trends across the machined surfaces. The results confirm the PhyGNN's high predictive proficiency, particularly when compared to other established models, and its exceptional stability across all nodes. This robustness, driven by the integration of physical laws into deep learning, proves essential for developing adaptive manufacturing strategies. Such strategies can dynamically adjust for stress-induced deformations, ensuring that workpieces meet stringent manufacturing tolerances while maintaining high quality throughout the machining process. Future work will explore reconstructing full-surface deformation from predicted representative nodes, as well as developing surrogate models for direct full-field prediction using more expressive architectures.

# Chapter 6. Edge-Aware Physics-Informed Concurrent Multiplexed Transformer for High-Fidelity Signal Compression, Reconstruction, and Diagnosis/Prognosis

Prognostics & Health Management (PHM) systems collect high-rate, multi-sensor signals that must traverse bandwidth-limited networks and run on resource-constrained edge devices. Uniform, ratio-driven signal compression often erases weak yet diagnostic structures and degrades decision quality under operating drift. Many learned codecs store dense floating-point latent codes that are ill-suited to edge budgets, require per-sensor or per-fault specialization, and rarely embed physics. Under drift, these methods smear the sidebands and harmonics that carry critical evidence of degradation in vibration and acoustic signals. We introduce the Edge-aware Physics-informed Concurrent Multiplexed Transformer (EPicMT), a signal-processing framework, which couples selective transmission with physics-guided signal reconstruction within a single architecture spanning compression, reconstruction, and decision tasks. EPicMT implements content-aware token selection (sampling) to control the number of transmitted coefficients, and structure-aware, non-uniform quantization that assigns finer steps to harmonic/envelope bands and coarser steps to broadband background, achieving bit-budgeted signal compression rather than mere dimensionality reduction. On the decoder, a physics-guided Transformer with time–frequency priors, multi-scale envelope consistency and a physically-constrained diffuser, regularizes reconstruction so that spectral energy, phase relations, and envelope trends are preserved. Concurrent reflects the joint optimization of diagnostic and prognostic objectives within the same network, while multiplexed denotes parameter sharing across sensors and tasks through a common backbone and token interface, enabling unified training and signal-adaptive bit allocation under varying operating conditions. Case studies on an in-house gearbox testbed, the CWRU bearing dataset, and NASA CMAPSS demonstrate high-fidelity signal reconstructions at compression ratios up to 64:1 while maintaining or improving fault-classification accuracy and remaining useful life prediction relative to representative autoencoder and Transformer baselines. The learned gating reduces transmitted bits and on-device compute without hand-tuned heuristics, and an embedded deployment confirms feasibility on resource-constrained hardware.

## 6.1. Introduction

Modern Prognostics & Health Management (PHM) systems [285,286] have increasingly been deployed in aerospace platforms [287], energy systems [288], precision instrumentation [289], and advanced manufacturing. These systems couple dense sensor arrays with host workstations, robotic cells, network switches, decision modules, and edge–cloud compute. They generate heterogeneous, high-rate data signals that must traverse constrained links while meeting strict latency and reliability requirements. Achieving this level of responsiveness requires not only efficient communication between networked units but also capturing evolving signal signatures that decisions depend on (e.g., harmonics, sidebands, envelope modulations). Bandwidth ceilings, however, force aggressive signal reduction, and conventional compressors too often discard precisely those structures. The result is a persistent mismatch: transmission budgets are optimized for bytes, whereas PHM accuracy is governed by physics-salient signal information. High-fidelity signal reconstruction [290] is therefore necessary but not sufficient; reconstructions must retain decision-relevant content so that fault classification, remaining useful life (RUL) estimation [291] remain stable after compression. As a result, the field needs signal compression–reconstruction mechanisms that are explicitly information-preserving with respect to downstream tasks and that operate within tight edge budgets [292].

Early work emphasized transform-based signal processing methods, such as Fourier [293], Discrete Cosine [294], and Wavelet analysis, that exploit sparsity in fixed, human-designed bases. These methods are attractive for stationary signals and deliver competitive ratios with modest compute, but the premise of a fixed basis is fragile when signals are non-stationary, multi-source, or exhibit condition-dependent modulation. Compressive sensing (CS) [295] alleviated sampling burdens by assuming transform sparsity and solving an inverse problem, yet still focuses on hand-picked models, iterative optimization, and tuning that is brittle under domain shift. Both routes externalize intelligence to manual design choices and task-specific thresholds; when operating conditions evolve, re-tuning is often unavoidable, and guarantees degrade in exactly the regimes where PHM pipelines must remain reliable.

Machine-learning-based signal representation approaches partially relaxed these constraints. Principal component analysis (PCA) [296], independent component analysis (ICA), sparse coding, dictionary learning, random projections, and manifold methods such as kernel PCA [297], ISOMAP [298], and locally linear embedding (LLE) [299] replaced fixed analytic bases with data-adapted representations. These techniques improved compactness on curated regimes yet remained fundamentally limited for high-rate industrial signals as linearity assumptions cap representational power, shallow architectures cannot express the hierarchical, cross-scale interactions that encode degradation, and performance is sensitive to hyperparameters that seldom transfer cleanly across assets, sensors, or environments. For continuous PHM with evolving faults and variable speed/load, models must preserve subtle harmonics and envelope modulations while operating at edge latencies, requirements that shallow learners satisfy only in narrow, hand-tuned settings.

Autoencoders (AE) [300,301] overcame many of these signal compression limits by learning non-linear latent representations; convolutional variants captured local correlations, and recurrent variants [302] injected sequence information for time-series monitoring. However, AE latent spaces are typically dense floating-point vectors. Dimensionality reduction does not equate to bit-rate reduction when each latent coefficient consumes high precision; the hidden bit-budget of the code becomes the bottleneck. This is precisely where AE-based signal compression underperforms for edge deployment, uniform encoding of all latent features forces either conservatively high bit-rates to preserve fidelity or unacceptable distortion when aggressively quantized. Attempts, such as fault division autoencoder multiplexing (FDAM) [303] that trains a separate AE per fault class, reduce within-class error but sacrifice scalability and generalizability, as it is impractical to maintain a model for every possible condition. These challenges reveal that conventional AE-based compression strategies are not truly edge-aware, nor adaptive enough for dynamic and continuous monitoring scenarios.

Transformers [304-306] introduce a different lever as attention naturally reallocates representational capacity toward salient tokens, and the tokenization mechanism admits computational and bit-rate control through pruning, gating, and selective forwarding. In contrast to AEs that pack all information into dense codes, Transformers can condition their compute and I/O footprint on signal content, enabling targeted preservation of health-critical structures while discarding redundancy. Despite this structural advantage, Transformer-based signal compression and reconstruction remain underdeveloped in PHM. Existing efforts, such as FusionOpt-Net [307] integrating FISTA and Transformers for improved compressive image, DR-TransNet [308] using unrolling, TTS-GAN [309] for structural response synthesis, demonstrate potential but do not address the joint requirement of extreme signal compression, physics-credible signal reconstruction, and decision-level invariance on resource-constrained edge devices. The open technical gap is thus concrete: a single architecture must (i) adapt transmission to health state and operating condition, (ii) constrain reconstruction by physically meaningful priors that protect fault harmonics and envelopes, and (iii) sustain diagnostic/RUL performance under drift, all within realistic bandwidth and compute budgets.

To close this gap, we introduce the Edge-Aware Physics-Informed Concurrent Multiplexed Transformer (EPicMT), a unified system that couples selective, quantized signal encoding at the edge with physics-guided signal reconstruction and multi-task inference. Here, concurrent refers to the fact that compression, reconstruction, classification, and RUL prediction are carried out in parallel within the same architecture, while multiplexed indicates parameter sharing across sensors and tasks. In this formulation, a single Transformer backbone ingests per-channel tokens, applies cross-channel attention to capture inter-sensor dependencies, and connects to lightweight task-specific heads. The resulting framework integrates functions that are often handled by separate models into a single network that distributes computation and learning efficiently across the monitoring stack.

Edge-side signal compression in EPicMT is health-aware by construction. A harmonic-preserving quantizer (HP-Quantizer) performs discretization while explicitly protecting spectral components that carry diagnostic information, maintaining energy and phase relations among characteristic harmonics and sidebands that index fault severity and kinematics. In parallel, signal-adaptive token skimming (SATS) modulates the subset of tokens admitted to transmission using indicators derived from the incoming signal and operating condition; when the system is stationary and healthy, SATS reduces the token budget aggressively, and when evolving signatures emerge, it allocates additional tokens to the regions that dominate health-relevant attention. The encoder thus converts the traditional static, ratio-driven pipeline into closed-loop signal compression where the bit-rate is a controlled variable tied to estimated decision impact rather than a fixed hyperparameter.

On the reconstruction side, EPicMT incorporates a physically-constrained diffuser (PC-Diffuser) then reconstructs signals under physics-inspired priors, reinforced by a multi-scale envelope consistency (MEC) constraint that preserves fault-relevant patterns across temporal and spectral scales. Together, PC-Diffuser and MEC ensure that the decoder's objective is not generic denoising but targeted preservation of the temporal–spectral signal signatures on which fault classifiers and RUL regressors depend. Because EPicMT is multiplexed, the same backbone that reconstructs signals also conditions the classification and RUL heads; this tight coupling enables the network to learn task-coherent representations in which the features preserved by compression are precisely those validated by decision loss. The main contributions of this work can be summarized as follows:

- **Sensor-status-aware compression**: A novel combination of harmonic-preserving quantization and signal-adaptive token skimming enables adaptive, edge-friendly compression under strict bandwidth budgets.
- **Physics-guided reconstruction**: A physically-constrained diffuser with multi-scale envelope consistency preserves fault-relevant temporal–spectral patterns during reconstruction.
- **Unified multi-task framework**: EPicMT jointly supports compression, reconstruction, fault diagnosis, and RUL prediction in a single scalable Transformer architecture.

The remainder of this paper is organized as follows. Section 6.2 introduces the architecture of the EPicMT framework, highlighting how domain knowledge is embedded through physics-guided modules and constraints. Section 6.3 describes the experimental setup, including the in-house gearbox testbed and benchmark datasets, and outlines how the framework enables efficient edge-to-cloud transmission and reconstruction. Section 6.4 presents the results, where the reconstructed signals are validated through diagnostic and prognostic tasks such as fault classification and RUL estimation. Finally, Section 6.5 summarizes the key findings and discusses conclusions, offering insights into the broader implications of this work for real-time, bandwidth-limited PHM applications.

## 6.2. Edge-Cloud Framework for Compression and Reconstruction

This section details the EPicMT, beginning with the baseline Transformer architecture as the foundation for sequence modeling, followed by the novel compression design that integrates harmonic-preserving quantization and signal-adaptive token skimming, and finally the reconstruction stage that incorporates a continuous-condition router, a physically-constrained diffuser, and multi-scale envelope consistency to achieve high-fidelity signal restoration and predictive integrity.

### 6.2.1. Signal-processing under bit-budget constraints

In edge transmission, the governing resource is the payload in bits, not merely the length of a latent vector. Dimensionality is at best a proxy: a short code stored in high-precision floats can still be expensive to transmit, whereas a longer code with low precision may be cheap but distortive. For vibration signals used in PHM, the risk is that bit-starved encoders erase line spectra, sidebands, and envelope-AM patterns that carry diagnostic meaning. We therefore make the bit budget explicit at the window level and pose compression as joint control of token admission and precision allocation. Let a selector $\delta$ choose the tokens to transmit and a quantizer assign per-coefficient bits. The total payload is then expressed as follows:

$$B = \sum_{i \in \delta} b_i = \sum_{i \in \delta} \sum_{m=1}^{d_i} q_{i,m} \leq B_{\max}, \tag{6.1}$$

where $K = |\delta|$ denotes the number of transmitted tokens, $d_i$ the number of coefficients within token $i$, and $q_{i,m}$ the bit-width assigned to the $m$-th coefficient. In the uniform case, with equal token length $d$ and identical bit-width $q$, this simplifies to

$$B = Kdq. \tag{6.2}$$

This formulation reveals that efficiency can be improved in two complementary ways. One is through selective token transmission: many tokens often carry redundant or background information, and excluding these allows the system to reduce K while preserving those tokens that carry distinctive spectral or temporal cues. In vibration signals, the informative regions tend to be harmonics, sidebands, or envelope variations that are strongly linked to fault progression. By suppressing tokens that add little diagnostic value, the overall payload decreases without undermining decision quality. In addition, efficiency can be enhanced through differential precision allocation. Instead of quantizing all coefficients uniformly, more bits can be concentrated on harmonic and modulation components that encode machine health, while coarser quantization suffices for broadband background. This uneven allocation maintains the fidelity of diagnostic structures even at aggressive compression ratios.

Within our framework, these two perspectives are instantiated by content-aware token reduction together with structure-aware quantization. In practice, this means that fewer tokens are transmitted, and the ones that remain are represented with higher bit-widths precisely in those spectral bands and temporal envelopes where physics indicates diagnostic salience. Such a formulation makes the bit budget not only an accounting constraint but also a principled signal-processing tool that directly links compression behavior with the preservation of health-relevant features. This mechanism is schematically illustrated in Figure 6-1. On the left, the original tokens are first scored and only a subset is retained, reflecting the content-aware selection principle. On the right, the surviving tokens are quantized with non-uniform bit allocation, so that harmonics and envelope-related structures receive finer resolution while less informative background components are coarsely represented. Together, the two levers, token selection and bit allocation, highlight how bit budgeting can be turned into a signal-centric design tool rather than a purely numerical constraint.

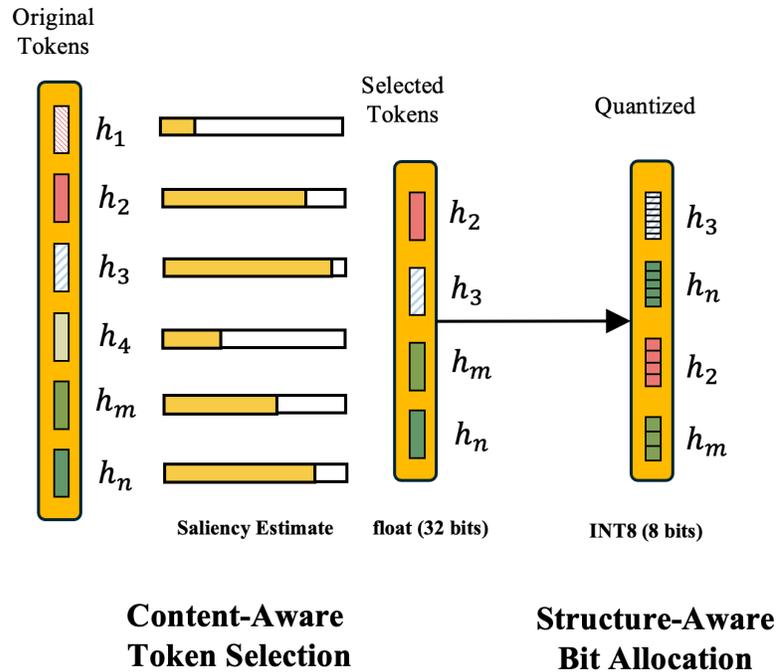

**Figure 6-1.** Signal compression under bit budgets by selecting tokens and allocating bits adaptively.

### 6.2.2. Baseline transformer architecture

The Transformer architecture has gained prominence for sequence modeling tasks by overcoming two fundamental limitations of conventional autoencoder-based approaches: their reliance on fixed latent representations and their inability to capture long-range dependencies. Built around self-attention, the Transformer can selectively weight input tokens, allowing it to prioritize fault-sensitive regions of sensor signals while suppressing redundant background patterns. This selective capacity is especially appealing for data compression in PHM, where the diagnostic content of the signal is often sparse yet distributed across extended time horizons. Moreover, the Transformer's modular structure, comprising embedding layers, attention blocks, and residual normalization, provides natural entry points for introducing quantization, token pruning, and physics-guided constraints, making it an ideal foundation for adaptive compression and reconstruction.

In its canonical form, the Transformer encoder distills sequential inputs into compact latent embeddings, while the decoder reconstructs the original signal from these compressed representations. This setup ensures that the reconstructed signals are not only numerically accurate but also capable of retaining diagnostic and prognostic features critical for downstream tasks such as fault classification and RUL prediction. Although Transformers are often viewed as computationally demanding, their inherent flexibility offers a unique advantage: they leave room for innovative adaptations, such as sensor-status-aware quantization, condition-adaptive routing, and physics-constrained diffusion, that transform the baseline model into an edge-aware, domain-informed solution for real-time PHM.

**6.2.2.1. Attention-based context modeling**

The self-attention mechanism [310] is the cornerstone of the Transformer's capacity to model long-range dependencies in sequential data. Unlike conventional convolutional or recurrent operations, which capture local neighborhoods or propagate information step by step, self-attention allows the model to compute pairwise relationships across all time points simultaneously. This capability enables the Transformer to globally contextualize each token, making it possible to recognize subtle patterns that may be dispersed over long sensor sequences, a property particularly valuable in structural health monitoring and prognostics. Formally, the mechanism is implemented through three learned transformations: Query (Q), Key (K), and Value (V), all derived from the input sequence $X$:

$$\begin{aligned} Q &= X \cdot W_q \\ K &= X \cdot W_k \\ V &= X \cdot W_v \end{aligned} \qquad (6.3)$$

The attention weights are then obtained by measuring the similarity between queries and keys, normalized by the key dimension:

$$\text{Attention}(Q, K, V) = \text{softmax}\left(\frac{Q \cdot K^T}{\sqrt{D(K)}}\right) \qquad (6.4)$$

This operation produces a weighted aggregation of values V, where the contribution of each token is adaptively determined by its relevance to the others. As a result, the Transformer can highlight diagnostically important subsequences, such as localized transients caused by faults, while downplaying redundant or steady-state regions. From the perspective of compression and reconstruction, this dynamic weighting is crucial: it effectively acts as a built-in selection mechanism, concentrating representation capacity on information-rich segments. However, the quadratic complexity of self-attention with respect to sequence length introduces non-trivial computational cost, particularly for high-frequency sensor signals with long time windows. This trade-off between global contextual power and computational efficiency motivates the exploration of token-level adaptivity and condition-aware routing, directions that our proposed framework develops further.

### 6.2.2.2. Embedding and positional encoding

In the Transformer architecture, raw sensor signals $X \in \mathbb{R}^{T \times d_{in}}$ must first be mapped into a latent space that is suitable for attention-based operations. This is accomplished by a linear embedding layer [311], which projects each time step into a $d$-dimensional representation:

$$E = X \cdot W_e + b_e \tag{6.5}$$

where $W_e \in \mathbb{R}^{d_{in} \times d_{model}}$ is the embedding weight matrix and $b_e$ is a bias term. The resulting matrix $E \in \mathbb{R}^{T \times d_{model}}$ provides a compact but expressive representation of the input, serving as the bias for attention mechanisms. For continuous sensor data streams, such embeddings are particularly important for balancing compression efficiency with the retention of diagnostic features.

However, embeddings alone do not encode temporal ordering, which is critical in health monitoring where degradation evolves over time. To resolve this, positional encodings are added to $E$, injecting information about the relative or absolute position of each token. A common formulation uses sinusoidal functions of varying frequencies:

$$PE(t, 2i) = \sin\left(\frac{t}{10000^{2i/d_{model}}}\right), \quad PE(t, 2i+1) = \cos\left(\frac{t}{10000^{2i/d_{model}}}\right) \tag{6.6a,b}$$

where $t$ denotes the time index and $i$ is the dimension index. The positional encoding [312] is added element-wise to the embedding, producing the final input to the Transformer. This combination enables the Transformer to preserve both signal content and temporal structure, ensuring that localized anomalies (e.g., fault-related transients) remain properly contextualized. From the perspective of compression and reconstruction, embedding and positional encoding also open opportunities for further adaptation: embeddings can be discretized through quantization, while positional signals can guide token selection. These mechanisms thus form not just preliminary steps, but natural precursors to the edge-aware and physics-guided innovations developed in our framework.

**6.2.2.3. Residual normalization layers**

Within the baseline Transformer architecture, residual connections and layer normalization [312,314] play a central role in stabilizing training and enabling deep network structures. After each sub-layer, whether attention or feedforward, the output is added to its input through a residual shortcut, and the combined result is normalized before being passed to the next stage. This design mitigates vanishing and exploding gradients, while ensuring that the learning dynamics remain stable even as the network depth increases. Formally, given an input $x$ to a sublayer $F(\cdot)$, the residual pathway can be written as:

$$y = \text{LayerNorm}(x + F(x)) \tag{6.7}$$

This mechanism preserves the identity of the input while allowing the network to learn complex transformations, effectively combining robustness with flexibility.

Layer normalization itself standardizes activations within a single training instance, operating across features rather than across the batch as in batch normalization. For an input vector $x = (x_1, x_2, ..., x_d)$, the normalized output is:

$$\text{LayerNorm}(x_i) = \gamma \cdot \frac{x_i - \mu(x)}{\sqrt{\sigma^2(x) + \varepsilon}} + \beta \tag{6.8}$$

where $\mu(x)$ and $\sigma^2(x)$ are the mean and variance of the features, and $\gamma$, $\beta$ are learnable scale and shift parameters. By ensuring consistent distribution of activations across layers, LayerNorm reduces training variance and accelerates convergence, which is crucial when working with noisy or heterogeneous sensor signals.

#### 6.2.2.4. Signal reconstruction decoder

Decoder remains a fundamental component of the original Transformer design, particularly suited for generative or reconstruction tasks. In the context of sensor data compression, this role becomes indispensable: once the encoder produces a compact latent representation, the data must be transmitted and subsequently expanded back into the signal domain. This pipeline naturally restores the importance of the decoder, allowing it to reconstruct signals with high fidelity and ensuring that diagnostic features are preserved after compression. Structurally, the decoder mirrors the encoder: it applies attention mechanisms to contextualize latent tokens, followed by feedforward transformations and residual normalization to refine the reconstruction. This symmetry allows the compressed latent $z$ to be progressively expanded into a high-dimensional signal approximation $\hat{X}$. Formally, given a compressed representation $z \in \mathbb{R}^{T \times d_{model}}$, the decoder computes:

$$H = \text{Attention}(z, z, z) \tag{6.9}$$

$$\hat{x} = FFN(h)W_d + b_d \tag{6.10}$$

where $W_d$ and $b_d$ project the decoder output back into the raw signal domain. This design ensures that reconstructed signals are not only numerically accurate but also capable of retaining fault-relevant features critical for downstream tasks such as fault classification and RUL prediction. In practice, this structure provides a natural foundation for our later enhancements, including physics-constrained diffusion and multi-scale envelope consistency, which further strengthen the diagnostic fidelity of the reconstructed signals.

### 6.2.3. Edge-aware compression module

The use of Transformer architectures for data compression remains relatively underexplored, as most prior work has focused on their strength in sequence modeling and representation learning. In fact, the vanilla Transformer encoder has no inherent mechanism for compression: it preserves the input length and dimensionality, only reshaping representations through attention and feedforward layers. This limitation makes it unsuitable for direct deployment in bandwidth-constrained sensor networks. In the context of edge computing, a rapidly emerging paradigm in the industrial IoT and prognostics, compression efficiency becomes as critical as predictive accuracy. Edge devices deployed in PHM systems are inherently constrained by computation, storage, and energy budgets, while being tasked with handling continuous high-frequency streams. To be practically deployable, a compression model must therefore deliver compact, efficient representations that can be transmitted under stringent hardware limits, yet still preserve sufficient fidelity for diagnostics and prognostics.

Traditional autoencoder-based schemes highlight why such adaptations are necessary. Their latent representations are typically dense floating-point vectors devoid of explicit physical meaning, forcing networks to retain many coefficients to reconstruct signals with diagnostic fidelity. This indiscriminate treatment of input tokens results in heavy storage and computational overhead, undermining their suitability for real-time edge applications. Attempts to alleviate this burden, such as handcrafted architectures tailored to specific fault classes, lack scalability and generalizability. To address these shortcomings, our framework introduces two complementary mechanisms: a Harmonic-Preserving Quantizer (HP-Quantizer), which reduces bit usage per token while safeguarding fault-relevant frequency harmonics, and a Signal-Adaptive Token Skimming (SATS) module, which adaptively discards non-informative tokens to achieve controllable compression ratios. By combining quantization and adaptive token reduction, these modules directly embed compression into the Transformer pipeline, turning it into a lightweight yet diagnostic-preserving compressor suitable for real-world PHM deployments.

#### 6.2.3.1. Harmonic-preserving quantizer

In diagnostics and prognostics, many critical fault indicators manifest in the frequency domain, such as harmonic peaks corresponding to bearing fault frequencies or gear meshing frequencies. Preserving these harmonics during compression is essential, as even slight distortions may obscure fault signatures and compromise diagnostic reliability. Traditional autoencoder-based compression typically encodes signals into dense floating-point latent vectors, which lack explicit physical meaning and often distribute information uniformly across coefficients. This indiscriminate representation not only leads to excessive storage requirements but also risks erasing subtle harmonic components under aggressive compression. To address this, we introduce the Harmonic-Preserving Quantizer, a module designed to explicitly safeguard frequency-domain content during the quantization process, ensuring that diagnostic harmonics remain intact even at high compression ratios.

At its core, quantization reduces continuous values into discrete levels. The simplest case is a uniform step size,

$$q_i = \Delta \cdot round\left(\frac{x_i}{\Delta}\right) \qquad (6.11)$$

where $\Delta$ controls how finely features are represented. A smaller $\Delta$ means higher fidelity but more bits, while a larger $\Delta$ saves space at the cost of accuracy. Our key change is to let this step size depend on frequency, so that bands containing harmonics get finer resolution, while background regions are coarsely represented:

$$\Delta(f) = \Delta_o \cdot \frac{1}{1+\alpha \cdot I(f \in H)} \qquad (6.12)$$

where $H$ denotes fault-related harmonic bands. This simple adjustment makes the quantizer physics-aware, as it saves bits where the signal is uninformative, but spends them carefully where PHM requires precision. During reconstruction, the quantized coefficients are reassembled into the time domain:

$$\hat{x}(t) = \sum_k q_k \cdot \phi_k(t) \qquad (6.13)$$

where the preserved coefficients around harmonics ensure that envelope peaks and sideband spacings are still visible. In effect, the HP-Quantizer provides a controllable way to trade bit-rate for fidelity, while ensuring that the most sensitive information is never sacrificed.

Placed at the front end of our Transformer encoder, HP-Quantizer reduces the bit width of each token representation, directly improving edge-friendliness. Combined with the token reduction handled by SATS, it provides two orthogonal dials for compression: one controlling how many tokens are transmitted, and the other controlling how many bits are allocated per token. Together, they enable high compression ratios such as 32:1 or 64:1 while retaining fault-critical harmonics. As illustrated in Figure 2, the sensor signal carries not only dominant harmonic components, but also subtle sidebands closely tied to machine health and fault progression. Conventional compression schemes tend to indiscriminately attenuate or eliminate these diagnostically critical components. In contrast, our design explicitly targets the preservation of such frequency-domain signatures, ensuring that both harmonics and sidebands remain distinguishable after compression and reconstruction. This emphasis provides a clear link between the proposed edge-aware compression strategy and its utility for reliable PHM applications. The practical effectiveness of this strategy will be further validated in Section 6.4 through experimental evaluation and ablation analysis.

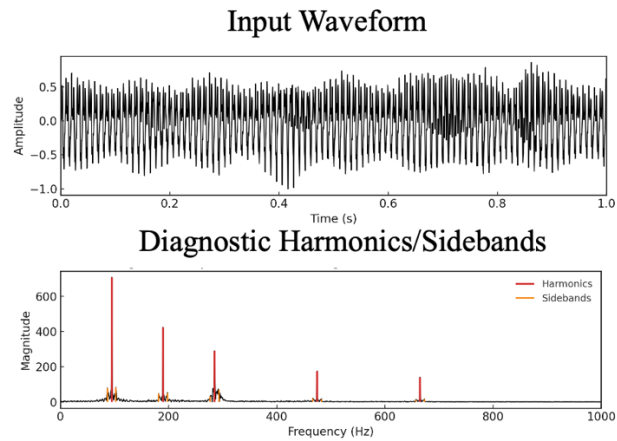

**Figure 6-2.** Sensor signal with highlighted harmonics and sidebands.

### 6.2.3.2. Signal-adaptive token skimming

Signal-Adaptive Token Skimming (SATS) serves as the core mechanism for achieving compression within our framework. While a standard Transformer encoder simply propagates all tokens through multiple layers, SATS introduces an adaptive selection strategy that regulates the number of tokens transmitted according to their diagnostic relevance. In practice, this means that tokens carrying fault-critical patterns, such as harmonics or sidebands, are preferentially retained, while tokens dominated by redundant or noise-like content are discarded. This design ensures that compression is not only aggressive but also sensitive to the health information embedded in sensor signals.

Formally, each token representation $h_i$ is first mapped to an importance score that quantifies how much information this token contributes to downstream diagnostics. The scoring function is defined as

$$s_i = \sigma\left(W_f \cdot h_i + b_f\right) \tag{6.14}$$

where $W_f$ and $b_f$ are trainable parameters learned during training. The linear projection can be understood as a learned filter that checks whether the token contains fault-relevant spectral or temporal patterns. The sigmoid function then compresses this projection into the range [0,1], allowing each token to be softly interpreted as "keep" ($s_i = 1$) or "discard" ($s_i = 0$). Intuitively, this step converts a continuous feature vector into a simple relevance score that is easier to use for compression decisions. Next, the scores are applied back to the token embeddings through element-wise gating:

$$\tilde{h}_i = s_i \odot h_i \tag{6.15}$$

where $\odot$ denotes element-wise multiplication. This means that tokens judged to be unimportant are directly suppressed in magnitude, while important tokens are transmitted almost intact. Conceptually, this operation functions as a token-level switch: rather than treating every token equally, the model learns to silence those that contribute little to identifying health states or degradation trends.

Finally, to ensure that the compression module respects a target compression ratio, SATS introduces a global constraint:

$$\sum_{i=1}^{N} s_i \approx \frac{N}{R} \tag{6.16}$$

where $N$ is the number of input tokens and $R$ is the target compression ratio. This equation means that, on average, only $N/R$ tokens are expected to survive the gating process. It acts like a regulator, forcing the model to make hard trade-offs, only tokens that truly matter for diagnostics are worth keeping under this budget. In practice, this constraint ensures that compression performance can be precisely controlled, which is critical for deployment on edge devices where bandwidth or memory limits are strict. To better illustrate the working mechanism of the proposed SATS module,

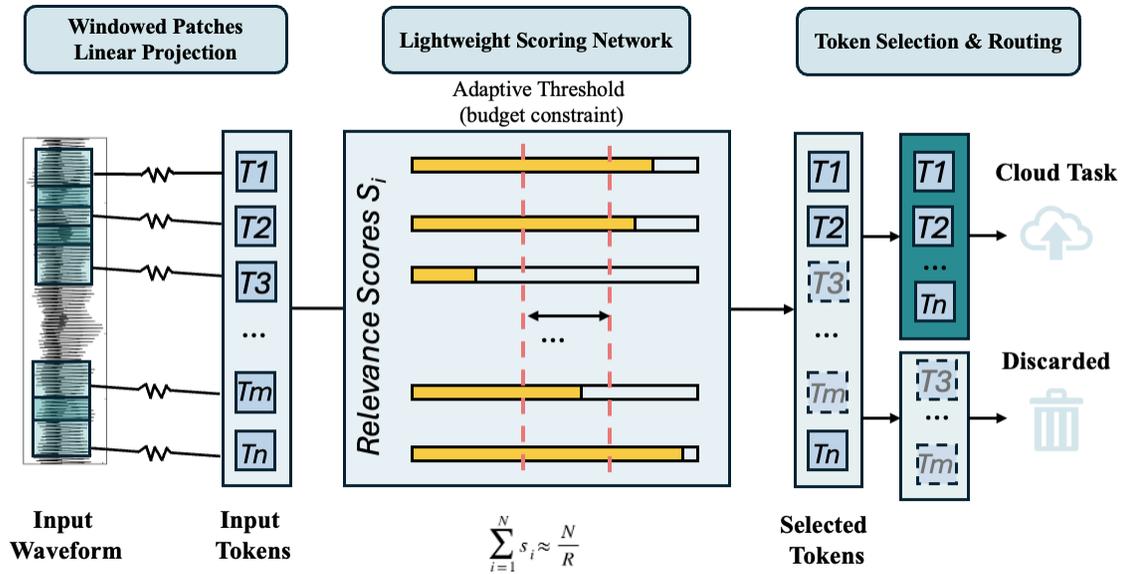

**Figure 6-3.** Illustration of signal-adaptive token skimming (SATS) framework

Figure 6-3 presents a schematic overview of the process. Raw input waveforms are first divided into windowed patches and projected into token embeddings. Each token is then assigned a relevance score through a lightweight scoring network, with an adaptive threshold enforcing the overall compression budget. Tokens surpassing the threshold are retained and routed to downstream tasks such as RUL prediction and fault classification, while less informative tokens are discarded. This design highlights the edge-aware nature of our framework: compression is no longer static but adaptively regulated by the diagnostic relevance of each token.

### 6.2.4. Condition-aware encoder backbone

The proposed framework integrates edge-aware compression through the HP-Quantizer and SATS modules, but a dedicated encoder backbone remains essential to extract meaningful features from the compressed tokens. In PHM applications, machinery signals are heavily influenced by operating conditions such as rotational speed, applied load, temperature, and degradation stage. These condition variables are routinely recorded in industrial monitoring systems or can be directly measured by auxiliary sensors. Explicitly introducing these condition inputs into the encoder design enhances robustness, ensuring that feature extraction accounts for the evolving state of the monitored system. Without condition-awareness, compressed representations risk conflating condition changes with fault signatures, undermining both diagnostic fidelity and RUL estimation accuracy.

To realize condition awareness, we embed operating condition vectors directly into token processing and attention computations. First, each token is augmented with a condition embedding:

$$h_i' = h_i + W_c \cdot c \tag{6.17}$$

where $h_i$ is the *i*-th token representation, $c \in \mathbb{R}^{d_c}$ is the condition vector, and $W_c \in \mathbb{R}^{d \times d}$ is a learnable mapping. This shifts latent representations into a condition-informed space, ensuring that compressed tokens reflect both local signal content and system state. Second, condition-awareness is integrated into the attention mechanism by adding a condition-dependent gating bias:

$$\text{Attention}(Q, K, V | c) = softmax\left(\frac{QK^T}{\sqrt{d_K}} + U \cdot c\right) V \tag{6.18}$$

where $U \in \mathbb{R}^{n \times d_c}$ projects the condition vector into a bias term aligned with the number of tokens n. This term selectively strengthens or weakens token interactions according to the operating condition, allowing the encoder to adaptively highlight harmonics and sidebands most relevant under specific regimes. Together, these two mechanisms enable the encoder backbone to act as a condition-aware representation learner, preserving prognostically critical features across varying operating states while remaining lightweight enough for edge deployment.

To concretize the above formulation, Table 6-1 outlines the detailed specification of the proposed encoder backbone. The design remains faithful to the Transformer encoder structure but incorporates optional condition embeddings and gated attention as described in Equations (6.12–13). By explicitly listing patch sizes, token budgets, embedding dimensions, and quantization settings, the table highlights how the encoder balances expressive power with edge feasibility. The relatively small depth (4 layers), moderate hidden dimensions (192 with 4 heads), and shallow feed-forward expansion ensure that the model remains computationally tractable on edge devices. Supplementary condition vectors, when available, can provide generic operating-state context, but are not central to the proposed contributions. Meanwhile, compression knobs (via SATS and HP-Quantizer) enable ratios like 64:1 without discarding harmonics or sidebands critical for diagnosis. This backbone thus forms the foundation of EPicMT, marrying lightweight architecture with adaptive feature extraction suitable for edge deployment.

Table 6-1. Condition-aware encoder backbone (EPicMT-Encoder)

| Stage | Key Settings | Values |
| --- | --- | --- |
| Input | Sampling rate/ Patch length/ Stride | 4096 Hz/ 64 / 32 |
| Tokens (per 1-s) | Before SATS → After SATS | 127 → 32 |
| Model Dim/ Heads | Token embedding/ Attention heads | 192/4 (48 per head) |
| FFN Dim | Expansion ratio | 384 |
| Encoder Depth | Stacked CA-Transformer layers | 4 |
| Condition Input | Vector dimension | 8 |
| Quantization | HP-Quantizer bit width | INT8 |
| Parameters | Encoder only | ~1.18M |
| Compute | MACs per layer (post-SATS) | ~0.18 GMAC |

### 6.2.5. Physics-informed reconstruction module

Aggressive edge-side compression raises the bar for the decoder: it must restore not only waveform fidelity but also the diagnostic essence, harmonics, sidebands, transients, and slowly evolving envelopes that drive PHM decisions. Optimizing solely with MSE often gravitates toward average solutions that wash out these details. We therefore adopt a composite, physics-informed objective that guides the decoder to reconstruct signals consistent with domain physics: (i) time–frequency fidelity with periodicity preservation, (ii) global trend consistency at the sequence scale, and (iii) multi-scale envelope preservation tied to degradation signatures. This design narrows the solution space to reconstructions that remain usable for fault diagnosis and RUL, even at high compression ratios.

#### 6.2.5.1. Time-frequency and autocorrelation composite

Unlike Fast Fourier Transform (FFT), Continuous Wavelet Transform (CWT) [315] provides time–frequency localization, making it better suited to PHM where faults appear as transient, evolving features (e.g., sidebands drifting with speed or load). Let $W_x$ and $W_{\hat{x}}$ denote the CWT coefficient tensors of the original and reconstructed signals with the same mother wavelet and matched scales. We penalize their mismatch:

$$L_{CWT} = \frac{1}{N}\sum i = 1^N \left\| W_{x^{(i)}} - W_{\hat{x}^{(i)}} \right\|_2^2 \tag{6.19}$$

To reinforce physical resonance and naturally suppress uncorrelated noise, we add an autocorrelation alignment term:

$$L_{AC} = \frac{1}{N}\sum i = 1^N \left\| r_{x^{(i)} x^{(i)}} - W_{\hat{x}^{(i)} \hat{x}^{(i)}} \right\|_2^2 \tag{6.20}$$

Together, $L_{CWT}$ preserves where energy lives over time–frequency, and $L_{AC}$ preserves how it repeats, two complementary constraints that stabilize reconstruction under condition drift.

### 6.2.5.2. Global dependency alignment

While frequency-domain fidelity ensures that harmonics and sidebands are preserved, an equally important aspect of sensor signals lies in their global temporal dependencies. In practice, many PHM signals, especially those related to degradation, do not only fluctuate locally but also exhibit overarching trends, such as gradual amplitude increases, baseline shifts, or correlated phase patterns across cycles. These long-range dependencies are vital for RUL prediction, because it is the trend of the signal that encodes how a system is progressing toward failure. To enforce this property, we adopt a correlation-based alignment loss. Let $x$ and $\hat{x}$ denote the original and reconstructed signals within a time window. Their Pearson correlation coefficient is defined as:

$$\rho(x,\hat{x}) = \frac{\sum_t (x_t - \bar{x})(\hat{x}_t - \hat{x})}{\sqrt{\sum_t (x_t - \bar{x})^2}\sqrt{\sum_t (\hat{x}_t - \hat{x})^2}} \tag{21}$$

The associated loss is expressed as:

$$L_{GDA} = 1 - \rho(x,\hat{x}) \tag{22}$$

By maximizing correlation, the reconstruction is compelled to move in lockstep with the true signal trajectory, not just match values point by point. This encourages the decoder to reproduce monotonic drifts, global envelopes, and sequence-level alignments that are central to degradation monitoring. Unlike pure MSE, which can be minimized even when the overall trend is distorted, correlation loss guards against reconstructions that look right locally but mislead higher-level prognostic models.

### 6.2.5.3. Multi-scale envelop consistency

Another defining characteristic of machinery fault signals is their amplitude modulation: localized defects modulate carrier frequencies, producing sidebands whose amplitudes evolve as the fault progresses. These modulations are captured most naturally by the signal envelope. If compression and reconstruction fail to preserve envelope dynamics, critical markers of fault severity may be lost, leading to false-healthy conclusions. To address this, we compute analytic envelopes using the Hilbert transform across multiple frequency scales. Let $e_s(x)$ and $e_s(\hat{x})$ denote the envelopes of the original and reconstructed signals at scale $s$. The MEC loss is defined as:

$$L_{MEC} = \frac{1}{S}\sum s = 1^S \|e_s(x) - e_s(\hat{x})\|_2^2 \tag{6.23}$$

This term explicitly enforces cross-scale envelope fidelity, ensuring that modulation depth, beat frequencies, and slow amplitude drifts remain consistent across the reconstructed signals. In PHM contexts, this means that the model can still distinguish between early-stage micro-cracks as small modulation depth and severe spalling or looseness as large, unstable envelopes. Importantly, by operating at multiple scales, MEC guards against aliasing effects or scale-specific blind spots, yielding reconstructions that retain both fine-grained and coarse-grained envelope features.

---

**Algorithm 1:** Physics-Informed Reconstruction Module (Training)

**Input:** Mini-batches $x$; compressed integer codes $z_q$; explicit condition vector $c$; task labels $(y_{cls}, y_{rul})$; loss weights $\lambda$; learning rate $\eta$.
**Result:** Trained decoder $D_\theta$ with PC-Diffuser; shared condition pathway $H_\phi$.
Initialize decoder parameters $\theta$ and condition pathway $H_\phi$.
**for** each mini-batch $x$ **do**
**Dequantize & embed:** $z \leftarrow \text{Dequantize}(z_q), h_0 \leftarrow \text{TokenEmbed}(z) + PE_{dec}$;

**Condition features:** $h_c \leftarrow H_\phi(c)$.
**Cross-decoding backbone (2 blocks):**
  $h \leftarrow \text{CrossAttn}(q = h_0, k = h_c, v = h_c) \rightarrow \text{FFN} \rightarrow \text{Residual} + \text{LayerNorm}$ (repeat twice)
**Synthesis projection:** $x_0 \leftarrow \text{OverlapAdd/Deconv}(h)$.
**PC-Diffuser refinement (K steps):**
  **for** $k = 1..K$ **do**
    $g \leftarrow \text{DecoderRefine}(x)$
    $p \leftarrow \text{PhysicsProjector}(x; B(c))$
    $x \leftarrow x + \eta_k [g + \alpha(p - x)]$
  **end for**
  $\hat{x} \leftarrow x$
   **Auxiliary task heads** (concurrent, multiplexed):
    $\hat{y}_{cls} \leftarrow \text{Head}_{cls}(h), \hat{y}_{rul} \leftarrow \text{Head}_{rul}(h)$
**Compute Losses:**
  $\hat{x} = D(z)$
  $L_{CE} = \text{CrossEntropy}(y_{cls}, \hat{y}_{cls})$
  $L_{RUL} = \|y_{rul} - \hat{y}_{rul}\|^2$
**Total objective:** $J = \lambda_{MSE} L_{MSE} + \lambda_{TF} L_{TF} + \lambda_{GDA} L_{GDA} + \lambda_{MEC} L_{MEC} + \lambda_{CE} L_{CE} + \lambda_{RUL} L_{RUL}$.
**Update:** $(\theta, \phi) \leftarrow \text{Adam}(\nabla_{\theta,\phi} J; \eta)$
  **end for**

### 6.2.6. Edge-cloud workflow of the proposed framework

The overall workflow of the proposed EPicMT is presented in Figure 6-4. The design follows a two-tier edge–cloud paradigm that explicitly addresses the dual challenge of bandwidth efficiency and diagnostic fidelity in real-time PHM. On the edge side, raw gearbox sensor signals are first segmented and embedded into tokens. A Transformer encoder contextualizes these tokens, after which two compression mechanisms are applied. The SATS module selectively retains only diagnostically relevant tokens, while the HP-Quantizer encodes retained tokens with frequency-dependent step sizes, ensuring that critical harmonics are preserved under quantization. This design converts raw high-frequency signals into compact, edge-friendly packets that minimize transmission cost while safeguarding health-related content. The compressed packets are then transmitted to the cloud side, where the decoder reconstructs signals under physics-guided priors. A dequantization and token expansion stage first restores the input dimensionality. Subsequently, a condition-aware Transformer decoder integrates encoder features with auxiliary operating information through cross-attention, while the Physics-Constrained Diffuser enforces temporal–spectral consistency. A multi-objective loss module further aligns reconstruction with both numerical fidelity (low error) and diagnostic fidelity (preservation of harmonics, sidebands, and envelopes).

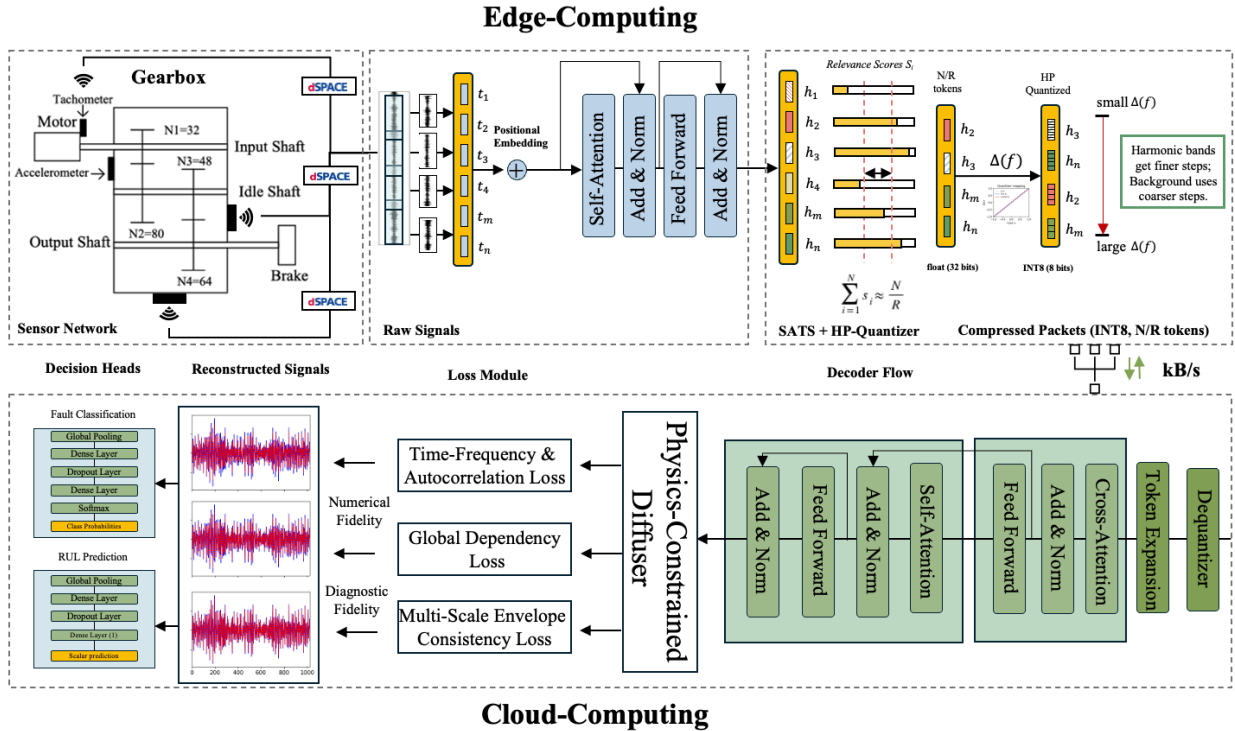

**Figure 6-4.** Schematic workflow of the EPicMT

Beyond reconstruction, EPicMT incorporates multiplexed decision heads for fault classification and RUL prediction. These lightweight heads are trained jointly with the backbone, and their task-specific losses feed back into the compression–reconstruction process. In effect, the classification head informs the network which frequency bands and temporal patterns are indispensable for diagnostic accuracy, while the RUL regression head highlights evolving features that are predictive of system degradation. By propagating these supervisory signals backward, the multi-task learning setup reinforces the preservation of health-relevant content during compression and guides the reconstruction toward prognostic integrity.

The proposed workflow embodies the design philosophy of EPicMT: compression at the edge is no longer treated as a static reduction process but as an adaptive mechanism that selectively emphasizes health-relevant tokens and harmonics within stringent bit budgets. On the cloud side, reconstruction is not limited to data fitting but is reinforced by physics-guided priors and informed by task-level supervision. By unifying adaptivity at the edge, physics-infusion in reconstruction, and task-aware feedback through multiplexed heads, EPicMT establishes a scalable and resource-rational framework for real-time PHM under severe bandwidth limitations.

## 6.3. Case Study Setup and Evaluation Protocol

Reliable evaluation of compression–reconstruction models requires diverse datasets and clear experimental setups. In PHM, sensor signals vary with load, speed, degradation stage, and acquisition settings. Relying on a single dataset would risk overstating the performance. We therefore employ both in-house measurements and widely used public datasets, covering classification tasks such as fault diagnosis and regression tasks like RUL prediction. This ensures that the proposed model is tested across a broad range of conditions and objectives. Beyond accuracy metrics, diagnostic signatures, harmonics, sidebands, and envelope modulations, must be preserved. Visualizing these patterns highlights the true challenge of compression: whether critical cues survive reduction and reconstruction. Such evidence complements quantitative results with direct interpretability. We also present the specific training configurations, including windowing strategy, normalization, optimization parameters, and learning schedule. These details provide a complete picture of how the case studies are conducted. Our goal is to facilitate an transparent demonstration of the new framework proposed.

### 6.3.1. Datasets analyzed

#### 6.3.1.1. UConn-DSCL gearbox dataset

The UConn-DSCL (Dynamics, Sensing and Controls Laboratory at University of Connecticut) gearbox dataset [316-318] was generated using a customized two-stage gearbox test rig designed to capture realistic gear fault behaviors under controlled laboratory settings, as shown in Figure 6-4(a). The rig employed a motor to regulate shaft speed and an adjustable magnetic brake to apply varying torque loads, producing operating conditions representative of practical machinery. The gearbox itself consisted of two stages: a 32-tooth pinion meshing with an 80-tooth gear in the first stage, and a 48-tooth pinion paired with a 64-tooth gear in the second stage.

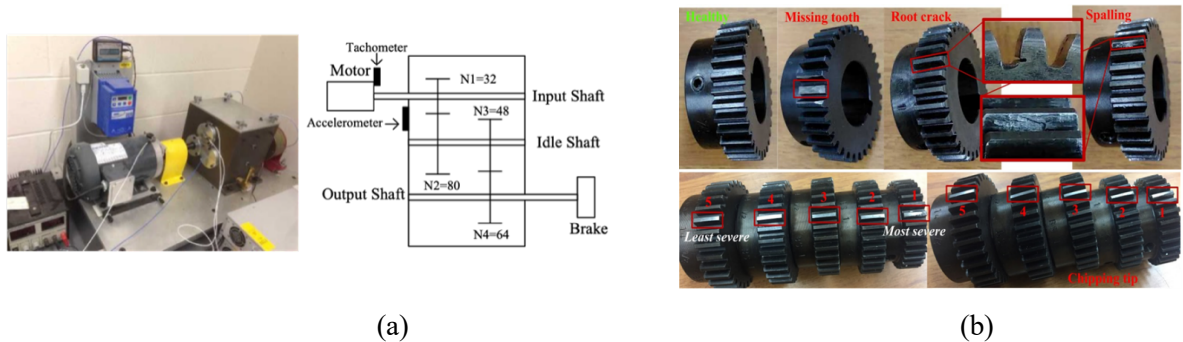

(a)                                      (b)

**Figure 6-5.** (a) Schematic of gear system; (b) 9 gear conditions.

High-fidelity vibration signals were collected using an accelerometer mounted near the gear meshes, with shaft speed simultaneously recorded via a tachometer. Data were sampled at 20 kHz using a dSPACE DS1006 processor board, and Time Synchronous Averaging (TSA) was applied to mitigate non-stationarity and align the signals with shaft rotation. This preprocessing step yields stationary-like signals that are more suitable for fine-grained diagnostic analysis. As shown in Figure 5(b), nine distinct gear conditions were simulated, including healthy operation, missing teeth, root cracks, spalling, and multiple levels of chipping tip severity. This diversity captures both clearly distinct fault modes and subtle intra-class variations in fault severity. Such nuance makes the dataset particularly challenging, distinguishing between similar defect types under continuously varying conditions closely mirrors the realities of industrial PHM systems.

### 6.3.1.2. CWRU bearing dataset

The bearing fault dataset from Case Western Reserve University (CWRU) [319,320] provides a widely used benchmark for evaluating signal compression and reconstruction methods. Accelerometer vibration data were collected while operating a 2-horsepower (HP) motor under 0-, 1-, 2-, and 3-HP loads. In addition to normal operating conditions, seeded defects were introduced in the drive-end bearing, including ball faults, inner race faults, and outer race faults with crack sizes of 0.007 inch, 0.014 inch, and 0.021 inch. For outer race faults, defects could occur at the 90º, 180º, and 0º positions. In this study, data from the 180º position were used. All signals were sampled at 12 kHz, and both drive-end and fan-end accelerometer measurements are available.

The dataset covers 40 unique operating configurations when considering the combinations of load, fault type, and fault severity. This diversity, particularly the inclusion of multiple load conditions, makes the CWRU dataset valuable for testing compression schemes under realistic variability. While the faults themselves are relatively simple to classify with modern deep learning models, the high sampling rate and varied operating conditions create a challenging environment for compression algorithms, which must retain diagnostically relevant signatures while reducing data volume.

### 6.3.1.3. NASA CMAPSS turbofan dataset

The NASA C-MAPSS (Commercial Modular Aero-Propulsion System Simulation) dataset [321,322] provides simulated run-to-failure trajectories of turbofan engines, designed to capture progressive degradation under varying operational conditions. Each engine instance is monitored across cycles until failure, with every cycle recording 26 channels: engine ID, cycle number, three operational settings, and 21 sensor signals including temperatures, pressures, and shaft speeds. Noise is injected to mimic realistic sensor variability. The dataset is organized into four subsets (FD001–FD004), with increasing complexity in terms of fault modes and operating conditions. For example, FD001 contains a single fault type under a fixed condition, while FD004 involves two fault types under six different operating conditions. Depending on the subset, hundreds of trajectories are available, with each trajectory ending at the point of simulated failure. Evaluating on C-MAPSS allows us to test whether the proposed compression and reconstruction framework can preserve these gradual trends across diverse operating conditions, making it an effective benchmark for assessing the generalizability of our model beyond conventional diagnostic tasks. By working with continuous degradation signals rather than categorical labels, the dataset highlights the importance of retaining prognostically relevant features after compression, ensuring that reconstructed signals remain suitable for accurate RUL estimation and decision-making.

**Table 6-2.** Summary of datasets and experimental tasks

| Dataset | Equipment | Task Type | Fault Types |
|---|---|---|---|
| UConn-DSCL | Gearbox | Classification | Healthy, Missing teeth, Root crack, Spalling, Chipping tips (5 severities) |
| CWRU | Bearing | Classification | Healthy, Ball defect (3 severities), Inner ring defect (3 severities), Outer ring defect (3 severities) |
| NASA C-MAPSS | Turbofan | RUL | FD001–FD004 subsets, up to 249 RUL trajectories under 6 conditions |

### 6.3.2. Fault-relevant signal features

In addition to describing the datasets, we present visualizations of characteristic signal patterns observed under different operating and fault conditions. The motivation for this step is to better understand what types of information may be embedded in the raw vibration data and how these features can inspire the design of lightweight yet accurate compression and reconstruction methods. By examining both obvious and subtle structures in the time–frequency representations, we aim to highlight features that could serve as useful cues for subsequent diagnostic and prognostic tasks, as illustrated in Figure 6-6, while also considering the need for generalizable representations across varying conditions.

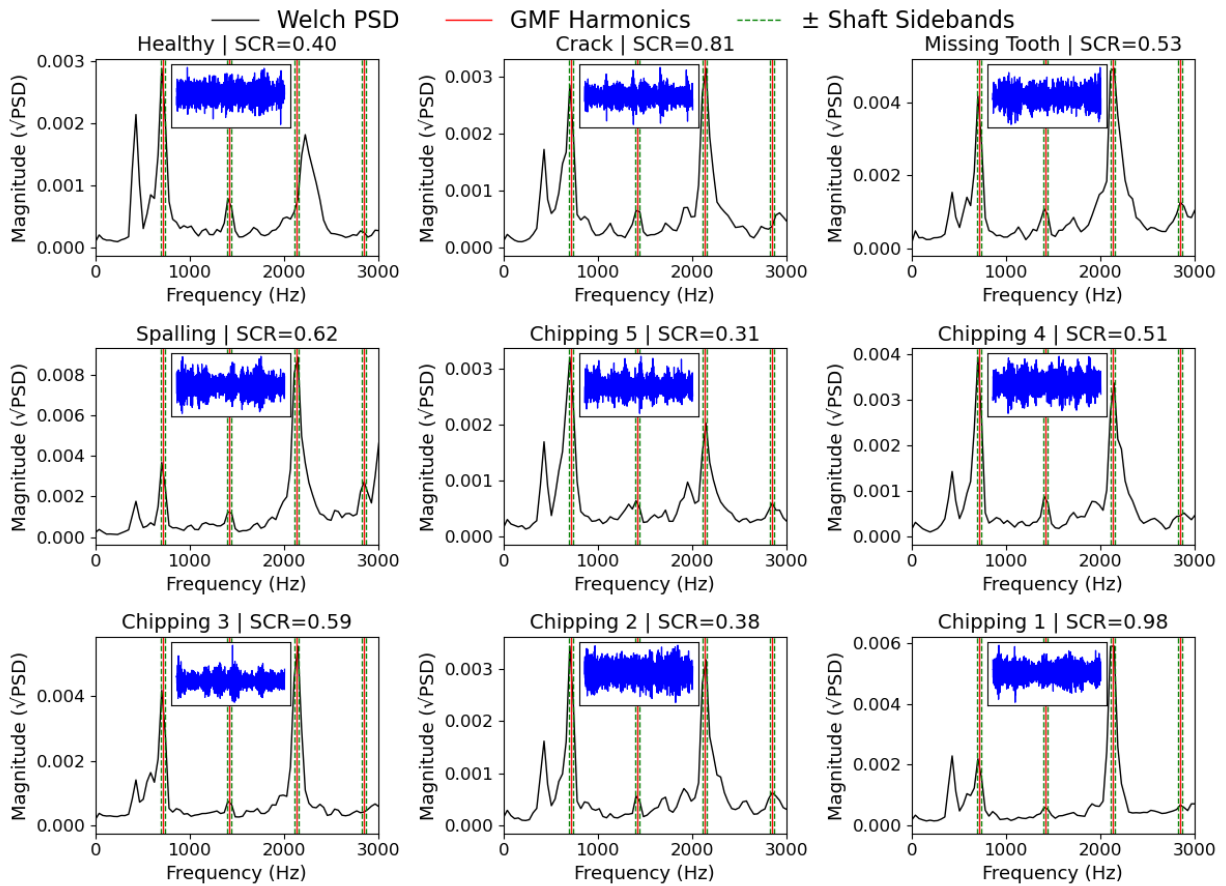

**Figure 6-6.** Spectral signatures of gearbox conditions with harmonics, shaft sidebands, and SAR values.

To verify that such spectral features indeed provide discriminative information about different operating conditions, we first examine the raw vibration signals collected from the UConn-DSCL gearbox dataset. These signals inherently contain gear mesh frequency (GMF) harmonics and their sidebands [423], which are widely recognized in PHM as indicators of fault progression. By visualizing these spectral structures, we aim to highlight both their presence and their variability across fault categories, motivating their preservation in downstream compression and reconstruction. To illustrate this, Figure 6-5 presents the spectral features of nine gearbox conditions, including healthy, missing tooth, crack, spalling, and five severities of chipping. The power spectral density (PSD) [324] of each condition is computed using Welch's method [325], with the corresponding raw time-domain signal inset in blue. Red vertical lines mark the GMF harmonics, while green dashed lines denote shaft sidebands. These characteristic frequencies are not arbitrary; they are derived analytically from the shaft speed (~1333 RPM, determined from 3600 samples per 4 revolutions at 20 kHz) and gear ratio parameters, and are found to align precisely with observable spectral peaks. This agreement confirms that the annotated frequencies correspond to physically meaningful features rather than spurious spectral artifacts.

Within each subplot, the amplitude variations at these fixed harmonic and sideband locations differ markedly between fault types and severities. For example, certain defects amplify higher-order harmonics, while others primarily modulate the sideband amplitudes, producing distinct patterns of spectral energy distribution. To quantify this, we compute the Spectral Consistency Ratio (SCR), which measures the relative energy retained at annotated diagnostic frequencies compared to the broadband spectrum. The SCR values, shown in each subplot, provide a compact metric for comparing how consistently each condition emphasizes fault-relevant frequencies. The SCR, together with the harmonic and sideband annotations, forms the basis for integrating physics-aware signal preservation into our compression module, ensuring that the encoded representations retain diagnostically critical content.

### 6.3.3. Training configurations

All models are implemented using the PyTorch framework with high-level orchestration via PyTorch Lightning. Training runs are executed on a Windows workstation equipped with an Intel Xeon Silver 4210R CPU (20 cores, 2.4 GHz base frequency), 128 GB of RAM, and an NVIDIA RTX A4000 GPU with 16 GB of memory. Mixed-precision training with 16-bit tensors is enabled throughout to accelerate computations and reduce GPU memory consumption. Each configuration, defined as a combination of model variant and compression ratio, is repeated five times with different random seeds. This repetition strategy ensures that data shuffling and weight initialization vary across trials, allowing a more robust assessment of variability while keeping all other factors fixed. Compression ratios of 4:1, 8:1, 16:1, 32:1, and 64:1 are investigated, corresponding to latent sequence lengths ranging from half to one sixty-fourth of the original input size. This range provides a balanced evaluation of fidelity versus efficiency, especially under the hardware constraints of edge computing scenarios.

For all training sessions, the Adam optimizer is employed with an initial learning rate of 0.001. A batch size of 64 is used unless restricted by GPU memory, and early stopping with a patience of 10 epochs is applied to prevent overfitting. The maximum number of training epochs is set to 100, and learning rate decay by a factor of 0.5 is triggered when the validation loss plateaus. Gradient clipping with a threshold of 1.0 is further adopted to stabilize optimization. These hyperparameters are consistently applied across datasets and compression settings to maintain comparability.

The primary metric for reconstruction quality is the signal-to-noise ratio (SNR) [326], defined as:

$$SNR = 10\log_{10}\left(\frac{P_{signal}}{P_{noise}}\right) \tag{6.24}$$

where $P_{signal}$ denotes the power of the original signal and $P_{noise}$ represents the power of the reconstruction error. This metric contextualizes the distortion introduced by compression relative to the strength of the original signal, providing a direct measure of how well fault-relevant features are preserved after reconstruction.

## 6.4. Implementation Results and Discussions

To evaluate the effectiveness of the proposed framework, we include both classical compression methods and modern deep-learning approaches as baselines. Classical signal processing schemes such as Discrete Cosine Transform (DCT), Compressed Sensing (CS), Principal Component Analysis (PCA) and Time-domain Component Analysis (TCA) provide lightweight references grounded in established theory. Traditional autoencoder families including deep convolutional autoencoder (DCAE), and variational autoencoder (VAE) serve as neural compression benchmarks. More recent architectures such as vector-quantized VAE (VQ-VAE) [327] and a vanilla Transformer autoencoder (TAE) [328] further extend the comparison to advanced, sequence-oriented models. Together these eight baselines span a spectrum from hand-crafted to data-driven, against which the effectiveness of our proposed framework can be fairly assessed.

The evaluation employs multiple criteria that together capture reconstruction quality, diagnostic fidelity, and deployment feasibility. Reconstruction fidelity is assessed using Percent Root-mean-square Difference (PRD) and SNR, with the latter providing a scale-invariant measure of distortion. To better reflect the information content relevant to fault diagnosis, we additionally compute the sideband-to-harmonic ratio (SCR), which explicitly quantify how well harmonic and sideband components are preserved. Beyond reconstruction, we evaluate task performance, classification accuracy for fault identification and RUL prediction error for prognostics, as well as efficiency metrics such as compression ratio, bit-width per token, retained-token percentage, and latency on edge hardware.

### 6.4.1. Classification case study

In this part, we investigate two vibration datasets to evaluate the classification performance of the proposed framework under different perspectives. The UConn gearbox dataset emphasizes fault type diversity and severity differentiation, while the CWRU bearing dataset focuses on robustness across varying load conditions. Together, they provide complementary viewpoints that allow us to assess the framework's ability to preserve diagnostic features across diverse operating scenarios.

**6.4.1.1. UConn gearbox: compression fidelity and diagnostic feature preservation**

We first evaluate EPicMT on the UConn gearbox dataset, focusing on whether aggressive compression can still retain both overall signal fidelity and diagnostically critical features. This case study serves as a baseline demonstration of reconstruction performance before extending to other datasets and tasks.

To assess compression fidelity, we benchmark EPicMT against both classical and neural compression approaches, including DCT, Wavelet, PCA, Compressive Sensing (OMP), DCAE, VQ-VAE, and Tiny-Transformer AEs, under compression ratios of $\{4:1, 8:1, 16:1, 32:1, 64:1\}$. Fidelity is reported using Signal-to-Noise Ratio (SNR) and Percent Root-mean-square Difference (PRD). While SNR provides an absolute reconstruction quality measure, PRD offers a normalized error metric that facilitates cross-dataset comparisons. PRD is defined as:

$$PRD(\%) = 100 \cdot \frac{\|x - \hat{x}\|_2}{\|x\|_2} \qquad (6.25)$$

where $x$ denotes the original signal and $\hat{x}$ the reconstructed counterpart. We emphasize PRD as it captures distortion relative to the signal energy, thereby serving as a more informative alternative to raw MSE. The resulting performance across methods and ratios is summarized in Table 6-3.

Table 6-3. Comparative performance of baseline and proposed methods across compression ratios

| Model | Compression Ratio (SNR in dB and PRD in %) | | | | | | | | | |
|---|---|---|---|---|---|---|---|---|---|---|
| | 4 | | 8 | | 16 | | 32 | | 64 | |
| | SNR | PRD | SNR | PRD | SNR | PRD | SNR | PRD | SNR | PRD |
| CS | 4.09 | 62.45 | 2.84 | 72.11 | 0.73 | 91.94 | -0.66 | 107.89 | -1.07 | 113.11 |
| DCT | 13.57 | 20.97 | 9.59 | 33.15 | 7.22 | 43.55 | 5.83 | 51.11 | 3.83 | 64.34 |
| PCA | 14.52 | 18.36 | 10.74 | 28.30 | 8.89 | 36.34 | 6.28 | 48.53 | 4.03 | 62.88 |
| TCA | 13.74 | 20.67 | 9.93 | 31.45 | 7.21 | 43.60 | 5.71 | 51.82 | 3.61 | 65.99 |
| DCAE | 18.12 | 12.42 | 12.57 | 23.06 | 9.68 | 32.81 | 7.19 | 43.70 | 5.52 | 52.97 |
| VAE | 17.56 | 13.24 | 11.81 | 24.69 | 9.12 | 34.99 | 7.04 | 44.46 | 5.05 | 55.91 |
| VQVAE | 22.48 | 7.52 | **18.48** | 12.44 | 12.16 | 24.72 | 10.45 | 30.03 | 6.89 | 45.76 |
| TAE | 21.77 | 8.16 | 16.19 | 15.51 | 11.09 | 28.15 | 8.74 | 36.56 | 6.58 | 47.16 |
| EPicMT | **24.38** | 6.31 | 18.24 | 12.76 | **15.97** | 15.90 | **12.22** | 25.13 | **10.09** | 31.30 |

At moderate compression levels such as 4:1, most approaches remain within a usable range, and the relative gaps between methods are not pronounced, as shown in Table 3. Classical schemes like DCT, PCA, and TCA deliver tolerable reconstructions, but they already begin to trail behind neural autoencoders that adapt their latent space more flexibly. VQVAE and TAE stand out in this regime, consistently offering high SNR and low PRD, which shows that discrete codebooks and self-attention provide tangible benefits in compact representations. EPicMT also lies in the top tier, performing on par with or only slightly behind the strongest baselines at 8:1. The slight advantage of VQVAE at this point likely reflects its discrete latent quantization, which tends to preserve sharp structures well under moderate compression. Still, EPicMT's values remain stable and close to the leader, suggesting that its design does not sacrifice compression quality.

As the compression ratio increases to 16:1 and beyond, the divergence across methods becomes more revealing. Traditional techniques such as CS and DCT degrade quickly, highlighting the limitation of fixed basis functions once redundancy is aggressively removed; by 32:1, their reconstructions approach unusable levels. Neural baselines like AE and DCAE also begin to falter, with PRD rising steeply, while VAE and VQVAE, despite their strong start, show sharper drops than expected, suggesting that their latent bottlenecks become less robust under extreme constraints. By contrast, EPicMT maintains a smoother degradation curve, remaining competitive at 16:1 and still delivering usable fidelity even at 64:1, where most other models collapse. This endurance implies that the condition-aware design and token adaptation allow EPicMT to retain signal features that other frameworks lose once the latent dimension becomes too tight. Taken together, the comparison indicates that while several advanced methods excel under mild compression, EPicMT demonstrates stronger resilience in the high-compression regime, which is often the more practical operating point in bandwidth-limited monitoring.

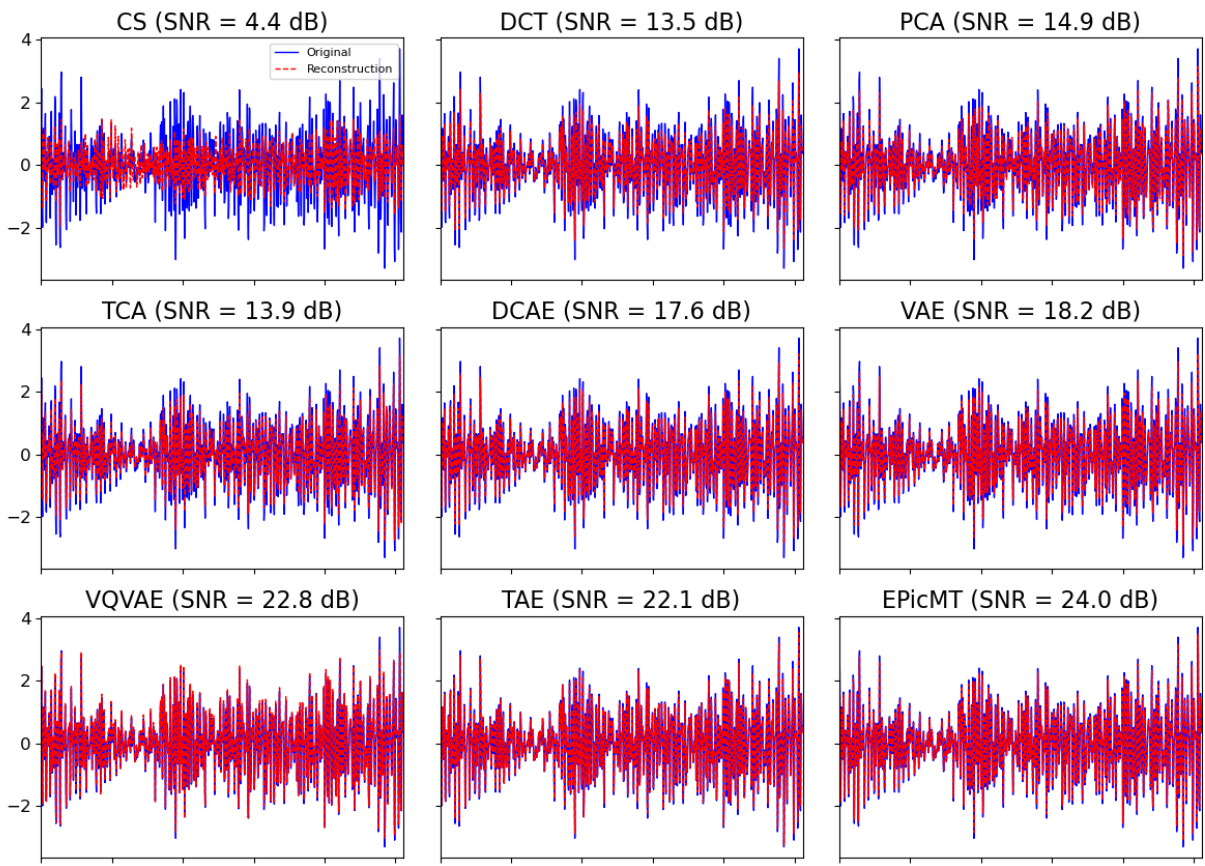

**Figure 6-7.** Reconstruction of Spalling sample at 4:1 compression.

Building on the aggregate evaluation in Table 6-3, we further examined how different methods behave on a concrete example. Here, we randomly selected a single Spalling sample for closer inspection. The input length was set to 1024 points, which is slightly longer than the 900-point revolution window and thus better preserves a complete fault pattern within each segment. This choice ensures that the sample retains representative information about both periodic shaft behavior and fault-related irregularities. We visualized the reconstructions at two contrasting compression ratios: 4:1 in Figure 6-7, representing the low-compression regime where most methods still preserve recognizable patterns, and 32:1, representing a high-compression regime where differences between methods become more pronounced. At 4:1, the reconstructed signals across methods remain close to the original waveform, with only subtle variations in smoothness or detail retention. In contrast, at 32:1 as shown in Figure 8, certain traditional schemes (e.g., CS, DCT, PCA) show clear loss of fidelity, while more advanced autoencoder-based or transformer-based approaches manage to preserve key oscillatory patterns with varying degrees of distortion.

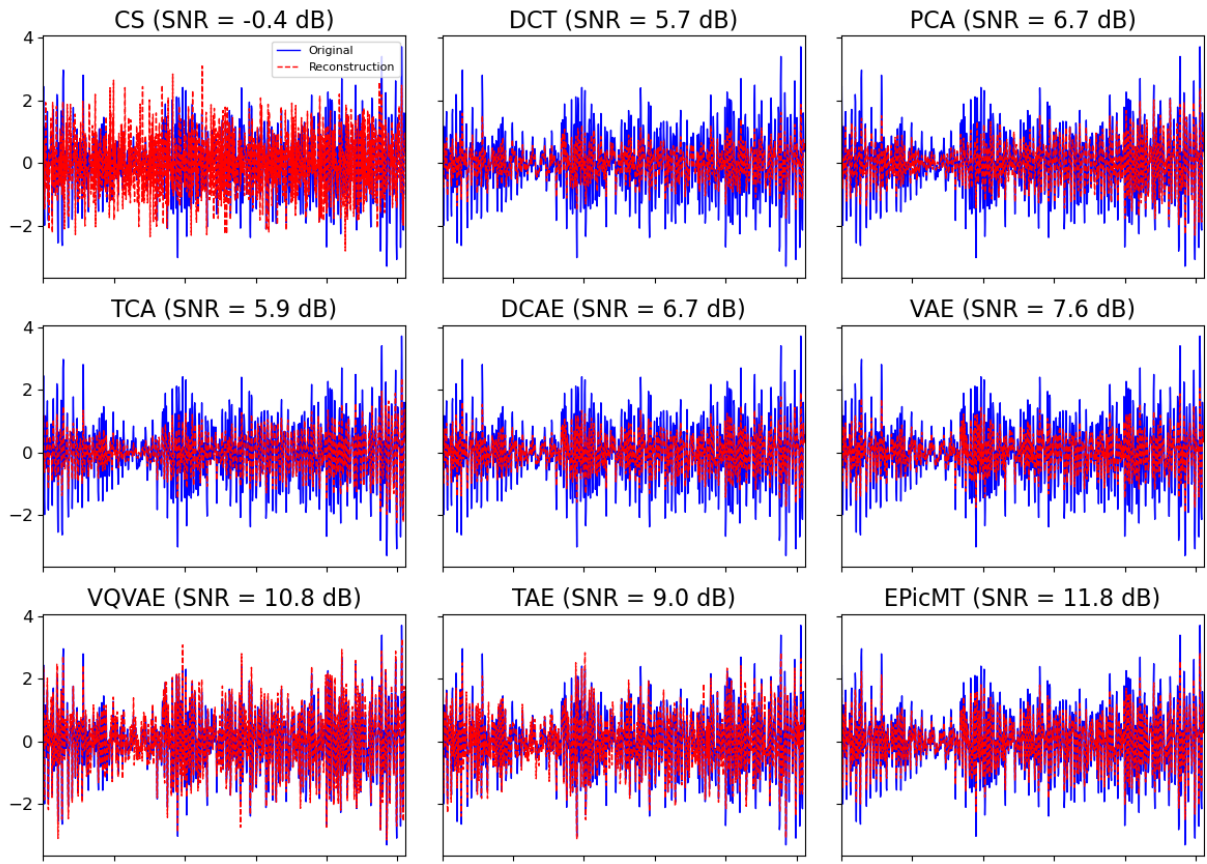

**Figure 6-8.** Reconstruction of Spalling sample at 32:1 compression.

While fidelity metrics quantify reconstruction accuracy, they do not by themselves guarantee that health-relevant structures are preserved. In practice, the value of compression lies not only in reducing error but also in maintaining the spectral and temporal signatures that encode machine condition. This motivates a closer look at whether compressed representations still carry the diagnostic content required for reliable monitoring. Our framework is designed with mechanisms that inherently aim to retain such features during the compression and reconstruction process. Therefore, a dedicated investigation is warranted to examine the fidelity of harmonics and sidebands in the reconstructed signals. Preserving these features ensures that the reconstructed data remain meaningful for downstream diagnostic tasks, rather than being limited to mere waveform similarity. In this subsection, we visualize and compare how different methods retain these diagnostic features under varying compression ratios and damage severities.

To operationalize this analysis, we compare the Welch PSD of the original and reconstructed signals under the strictest compression (CR = 64). We report four representative operating conditions, Healthy (Figure 6-9), Chipping 5 (Figure 6-10), Chipping 3 (Figure 6-11), and Chipping 1 (Figure 6-12), covering a progression from no damage to the highest severity. For each condition, gray dashed lines mark harmonic references to guide interpretation at diagnostically relevant frequencies. This setup stresses the methods at a challenging compression level and asks a practical question: how much of the diagnostically important frequency content can each method still retain across severities?

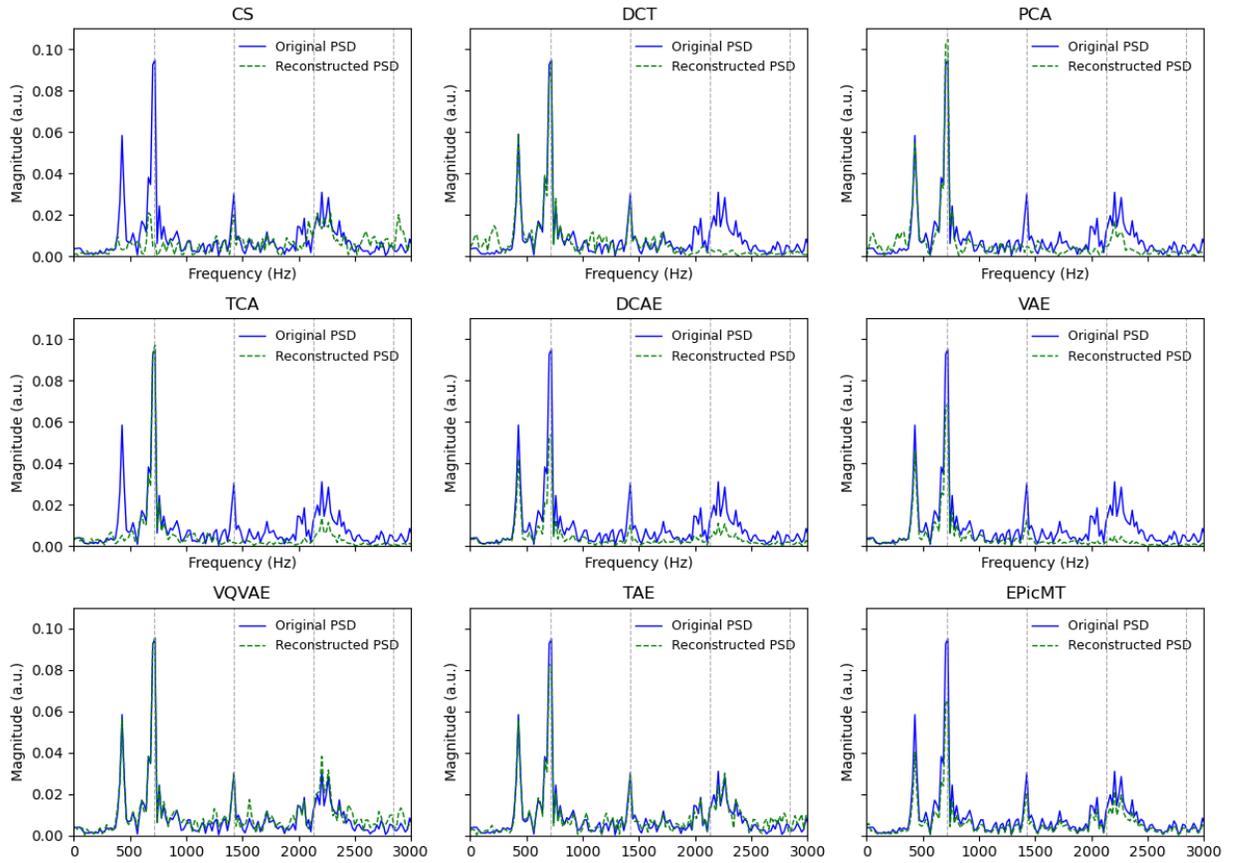

**Figure 6-9.** Welch PSD comparison between original and reconstructed signals under the Healthy condition (CR=64).

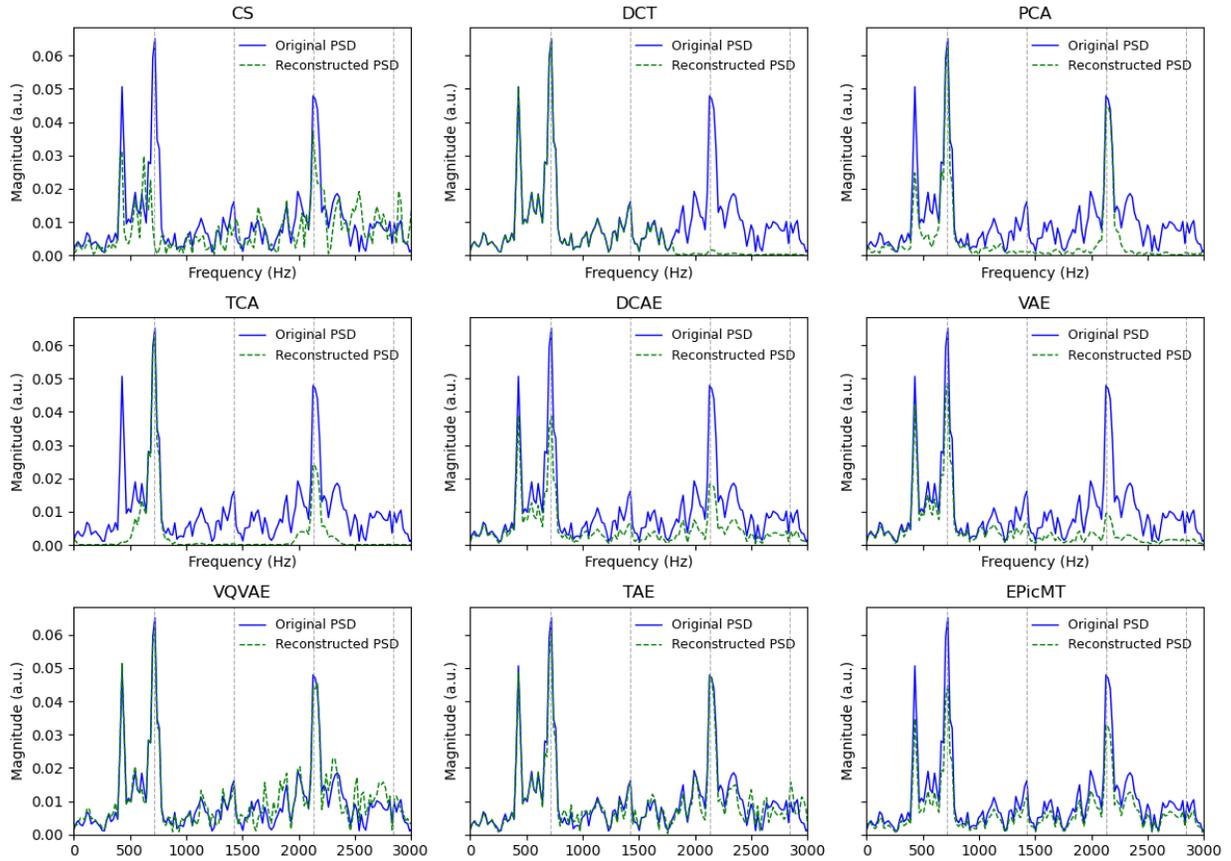

**Figure 6-10.** Welch PSD comparison between original and reconstructed signals under the Chipping 5 condition (CR=64).

As can be observed in Figures 6-8—11, diagnostic harmonics are weak and the spectra are relatively clean in the Healthy case, and most methods appear visually close to the original, though DCT already shows its tendency to under-represent higher frequencies despite nearly perfect low-frequency agreement. As severity increases (Fig. 6-8 and Fig. 6-9), harmonic content and sideband-like structures become more pronounced. Here, the divergence among methods is clearer: DCT retains low-frequency fidelity but progressively degrades at mid-to-high bands, while VQVAE, TAE, and EPicMT provide more stable preservation across the spectrum. At the highest severity (Fig. 6-10, Chipping 1), diagnostic peaks dominate the spectrum, and TAE in particular achieves the closest reconstruction of harmonic magnitudes, whereas EPicMT demonstrates steady and reliable tracking of spectral energy distribution across bands.

Overall, these comparisons highlight that methods differ not only in average reconstruction accuracy but also in how they preserve fault-related frequency features across severities. DCT is characterized by strong low-frequency but weak high-frequency retention. VQVAE and TAE maintain balanced performance, with TAE especially excelling under severe faults. EPicMT stands out for its cross-condition stability, consistently retaining the relative spectral trends from Healthy through to Chipping 1. The results emphasize that evaluating feature retention at diagnostically relevant harmonics provides a more nuanced view of how compression and reconstruction impact diagnostic usability.

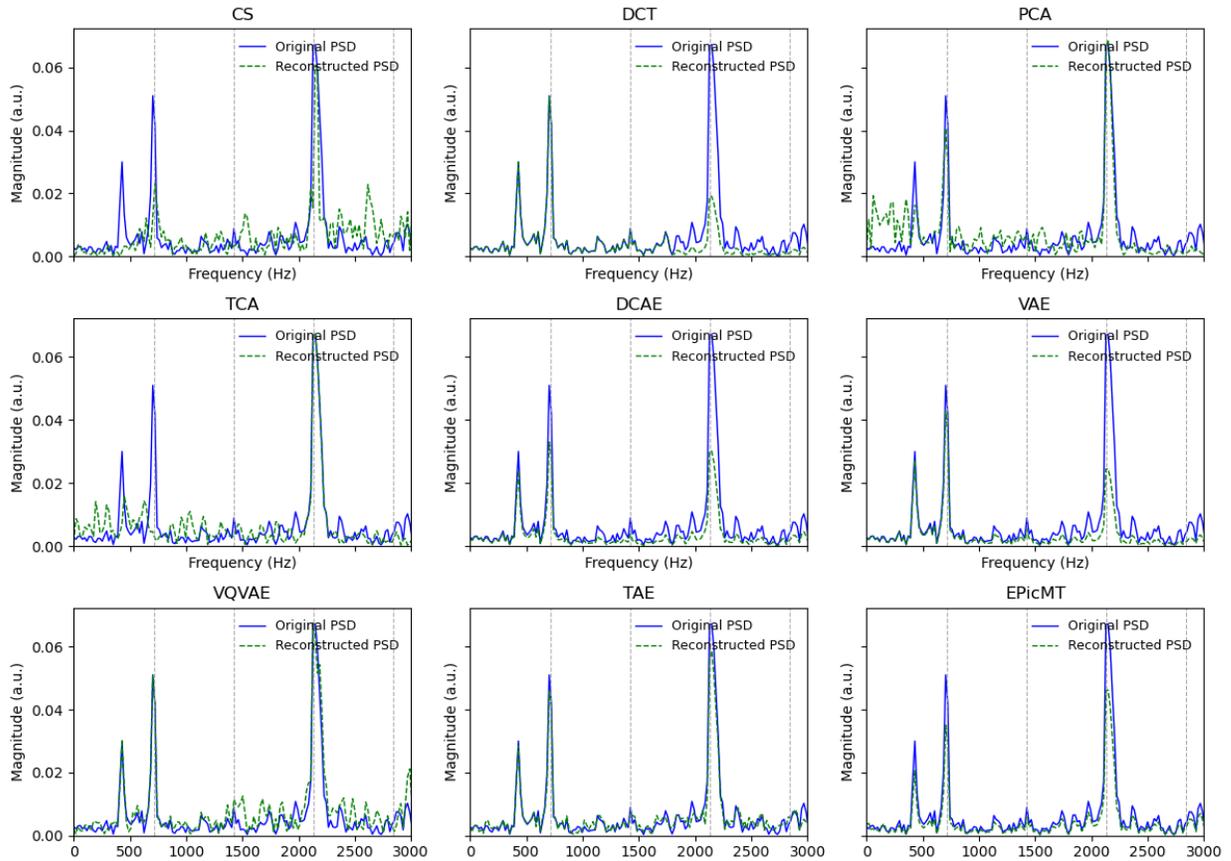

**Figure 6-11.** Welch PSD comparison between original and reconstructed signals under the Chipping 3 condition (CR=64).

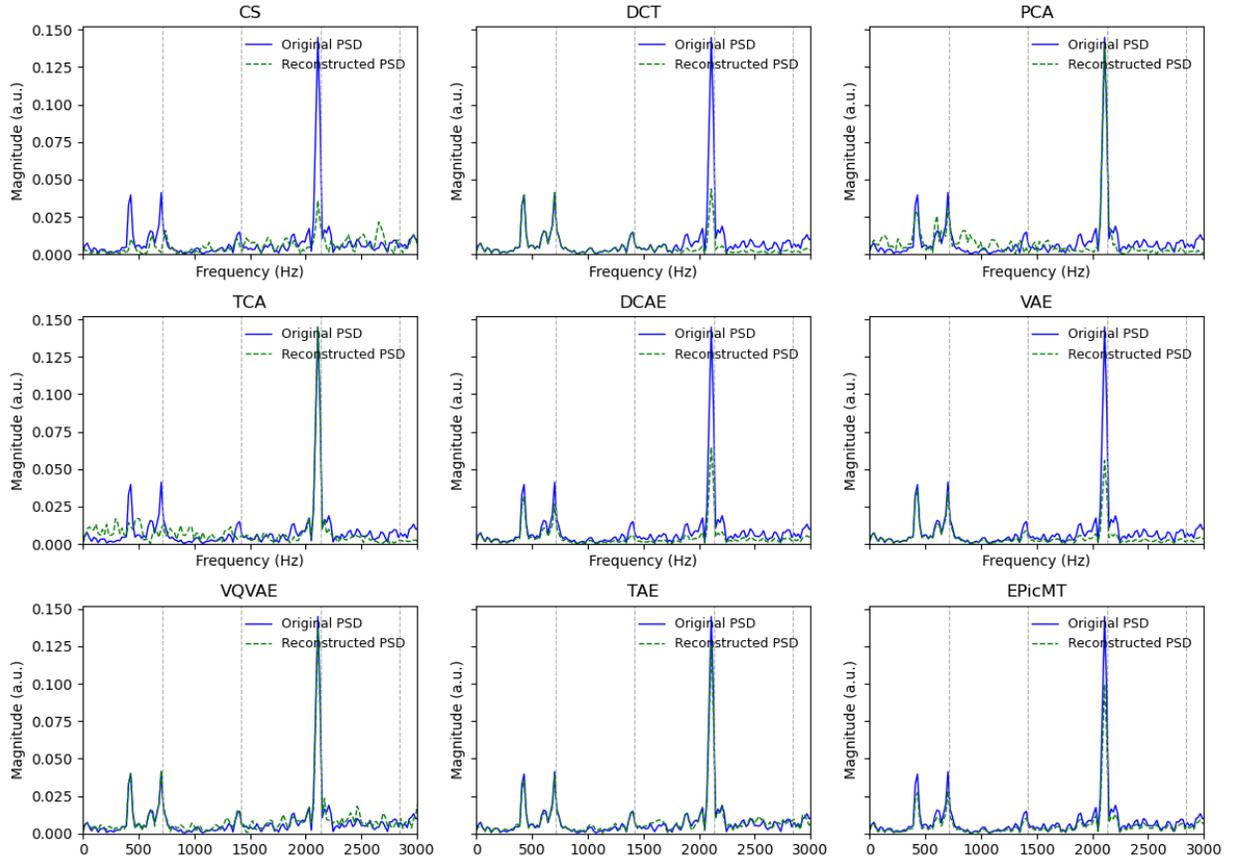

**Figure 6-12.** Welch PSD comparison between original and reconstructed signals under the Chipping 1 condition (CR=64).

#### 6.4.1.2. CWRU Bearings: post-reconstruction accuracy and edge feasibility

To further assess the quality and diagnostic usability of the reconstructed signals on the Bearing dataset, we perform a comprehensive evaluation with a particular emphasis on post-reconstruction accuracy. Four complementary metrics are employed: SNR, PRD, classification accuracy, and the sideband-to-carrier ratio (SCR). PRD quantifies the relative reconstruction error, classification accuracy directly reflects diagnostic performance after reconstruction. For classification accuracy, we adopt a baseline diagnostic classifier based on a lightweight 1D convolutional neural network (1D-CNN). This model has been widely validated in prior work [ref], and we use it here solely as a consistent benchmark across all reconstruction methods. SCR measures the preservation of fault-related harmonic content by comparing the relative energy between sidebands and the carrier at the gear mesh frequency, defined as

$$SCR = \frac{E_{sidebands}}{E_{carrier}} \quad (6.25)$$

To facilitate comparison across methods and compression ratios, each metric is presented in the form of a heatmap, with rows representing methods and columns compression ratios. All reported values are averaged over five independent runs to ensure robustness and reduce randomness.

Across the examined compression ratios, the behavior of different reconstruction methods reveals distinct quantitative patterns, as reported in Figure 6-13. For conventional transforms, DCT achieves an SNR of 10.8 dB at 4:1 compression and maintains 10.0 dB at 8:1, but its performance deteriorates to 4.5 dB at 64:1. Correspondingly, PRD increases from 28.8% to 64.3% across this range. PCA shows a similar progression, with SNR decreasing from 9.2 dB to 3.8 dB and PRD rising from 34.7% to 64.6%. TCA follows the same trajectory, dropping from 9.5 dB to 4.0 dB and PRD increasing from 33.5% to 63.1%. These trends indicate that while such transforms can retain dominant low-frequency components under mild compression, their ability to preserve overall fidelity diminishes substantially at higher ratios. Learning-based approaches display greater robustness. VQVAE sustains an SNR of 14.0 dB at 4:1 and still maintains 7.6 dB at 64:1, with PRD rising only from 20.0% to 41.7%. TAE exhibits a similar pattern, with SNR reducing from 13.8 dB to 7.2 dB and PRD increasing from 20.4% to 43.6%. EPicMT shows the most stable profile, maintaining 15.0 dB at 4:1 and 8.2 dB at 64:1, while PRD increases moderately from 17.8% to 38.9%. This stability across all compression ratios underscores the resilience of these approaches in preserving essential spectral content.

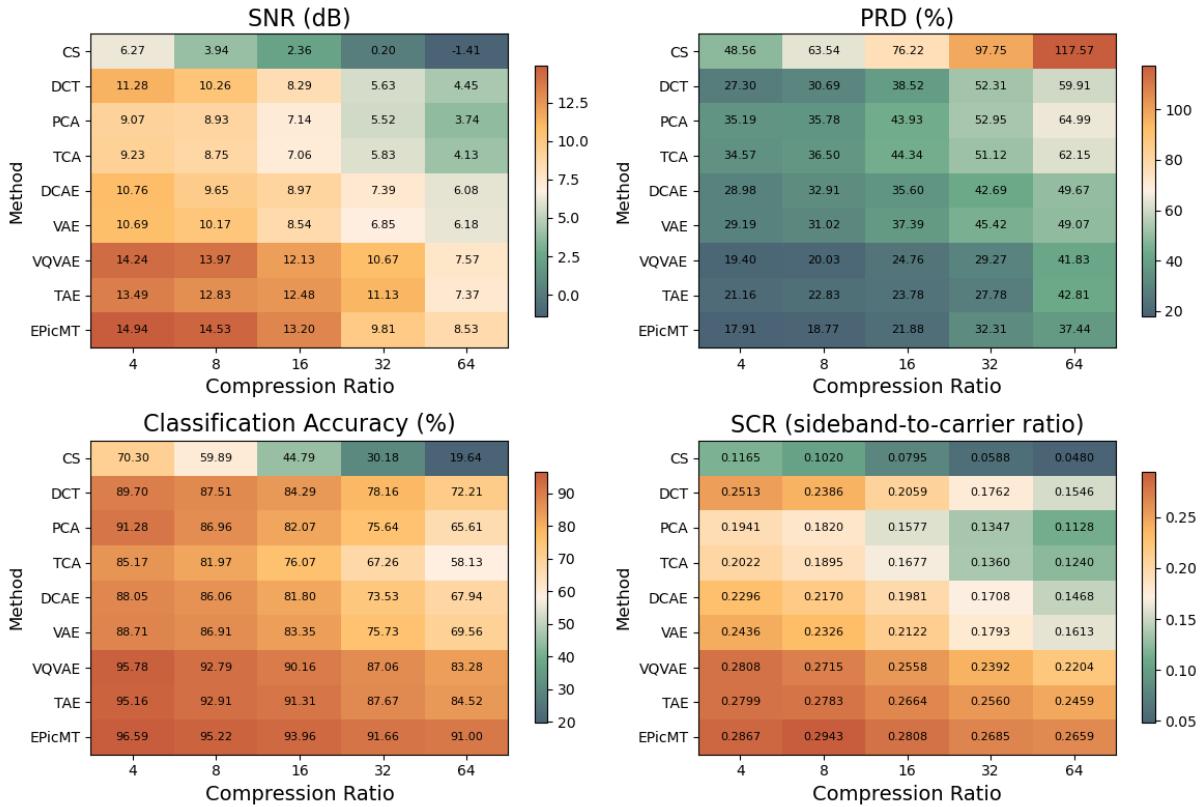

**Figure 6-13.** Post-reconstruction performance metrics across compression ratios.

Classification accuracy further highlights differences in diagnostic relevance. For DCT, accuracy remains relatively high at 90% for 4:1 compression but drops to 72% at 64:1. PCA decreases from 91% to 66%, and TCA from 85% to 58% across the same range. In contrast, VQVAE, TAE, and EPicMT consistently retain higher levels: VQVAE decreases from 96% to 83%, TAE from 95% to 85%, and EPicMT sustains the highest stability, ranging from 97% at 4:1 to 91% at 64:1. These results indicate that, while conventional methods can achieve competitive fidelity under mild compression, they are less effective at retaining discriminative features necessary for accurate fault classification. By comparison, EPicMT consistently preserves the dominant diagnostic features, enabling reliable classification across all tested scenarios. Overall, the combination of fidelity metrics (SNR, PRD) and application-oriented evaluation (classification accuracy) demonstrates that reconstruction stability is not solely dependent on waveform preservation but on selectively retaining features most critical for diagnostic tasks.

Although the bearing dataset experiments confirm the strong performance of all evaluated models, reconstruction accuracy alone is not the sole objective. In practical edge–cloud deployments, the encoder is executed at the edge device under limited resources, while the decoder resides on the cloud. This design emphasizes that token selection at the encoder side must strike a balance: the most critical diagnostic features should be preserved, but with minimal computational overhead. The goal is not to exhaustively search for the optimal representation at high cost, but rather to ensure that edge-aware compression retains essential fault-related signatures efficiently. To validate this principle, we assess edge awareness by examining whether transient and high-frequency components are preserved after compression, and computational efficiency by quantifying the encoder-side complexity in terms of parameter count, FLOPs, and inference time.

**Table 6-4.** Efficiency comparison at CR = 16 and CR = 64.

| Model | Compression Ratio | | | | | | | | | |
|---|---|---|---|---|---|---|---|---|---|---|
| | 16 | | | | | 64 | | | | |
| | Param | MACs | FLOPs | Runtime | Bits | Param | MACs | FLOPs | Runtime | Bits |
| Unit | (M) | (MMAC) | (MFLOP) | ms | | (M) | (MMAC) | (MFLOP) | ms | |
| CS | N/A | 0.0627 | 0.1254 | 0.0013 | 2048 | N/A | 0.0164 | 0.0328 | 0.0003 | 512 |
| DCT | N/A | 0.0146 | 0.0292 | 0.0002 | 2048 | N/A | 0.0125 | 0.0250 | 0.0002 | 512 |
| PCA | N/A | 0.0615 | 0.1230 | 0.0011 | 2048 | N/A | 0.0131 | 0.0262 | 0.0007 | 512 |
| TCA | N/A | 0.0628 | 0.1256 | 0.0024 | 2048 | N/A | 0.0166 | 0.0332 | 0.0005 | 512 |
| DCAE | 0.5952 | 124.93 | 249.86 | 2.5564 | 2048 | 0.5952 | 125.09 | 250.18 | 2.4960 | 512 |
| VAE | 0.5154 | 100.09 | 200.18 | 1.9556 | 2048 | 0.5154 | 99.94 | 199.88 | 2.0676 | 512 |
| VQVAE | 0.8635 | 214.98 | 429.96 | 4.3035 | 512 | 0.8635 | 130.08 | 260.16 | 2.6316 | 128 |
| TAE | 1.0839 | 194.91 | 389.82 | 3.9609 | 512 | 1.0839 | 119.96 | 239.92 | 2.3121 | 128 |
| EPicMT | 1.2152 | 159.99 | 319.98 | 3.2532 | 512 | 1.2152 | 89.97 | 179.94 | 1.8169 | 128 |

To ensure a fair and interpretable comparison across different encoder designs, we report efficiency-related metrics at two representative compression ratios, CR = 16 and CR = 64, in Table 6-4. All runtime values are averaged over five runs. Parameters are reported only for neural encoders, as traditional linear transforms (CS, DCT, PCA, TCA) contain no learnable weights. The bit budget denotes the number of bits required to represent compressed features, reflecting communication or storage overhead rather than computational cost. This measure is essential because compression is not only about reducing raw signal size but also about transmitting sufficient feature information to enable reliable reconstruction and subsequent diagnostics.

Across the methods, distinct efficiency trends can be observed. Linear baselines such as CS, PCA, and TCA show reductions in runtime and FLOPs that are nearly proportional to the compression ratio, which is consistent with their direct dependence on the number of retained coefficients. In contrast, DCAE and VAE exhibit only marginal variation across different ratios. A plausible explanation is that their latent representations remain dense floating-point vectors, so the computational cost is dominated by encoder–decoder operations rather than latent dimensionality. VQVAE and TAE fall in between: while quantization reduces the bit budget, their runtime remains largely dictated by token-level processing.

EPicMT demonstrates a more favorable scaling profile. Owing to the joint effect of signal-adaptive token skimming and harmonic-preserving quantization, the bit budget decreases more aggressively and runtime drops more markedly as compression becomes stronger. At $CR = 16$, runtime is comparable to VAE due to the attention layers, but at $CR = 64$ both runtime and FLOPs decline more sharply, indicating improved scalability under stringent compression. This suggests that EPicMT not only reduces transmission overhead but also achieves more consistent efficiency gains, providing a practical advantage for deployment in bandwidth-limited PHM scenarios.

The joint visualization of runtime and SNR in Figure 6-14 highlights the trade-off between computational efficiency and signal fidelity across compression ratios. Linear methods (CS, DCT, PCA, TCA) incur negligible runtime but yield limited SNR. Neural encoders (DCAE, VAE) improve SNR at higher computational cost. Quantization-based approaches (VQVAE, TAE, EPicMT) achieve a more balanced profile. Importantly, EPicMT becomes increasingly efficient at higher compression ratios, sustaining competitive runtime while preserving SNR. This highlights its suitability for edge-aware deployment under constrained bit budgets.

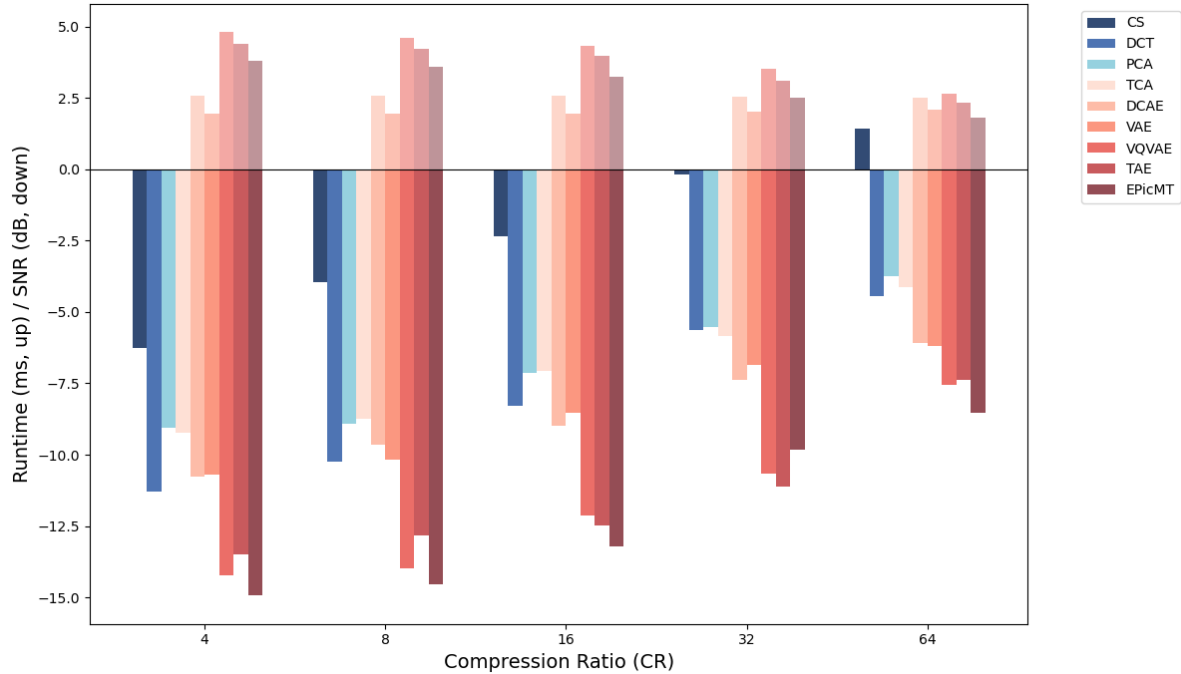

**Figure 6-14.** Comparison of runtime (upward bars, ms) and reconstruction quality measured by SNR (downward bars, dB) across compression ratios for all evaluated methods.

### 6.4.2. RUL prediction case study

Having established the comparative performance of EPicMT across fidelity and diagnostic tasks, the focus now turns to an ablation study aimed at disentangling the contributions of individual design modules. Rather than revisiting comparisons with external baselines, this analysis isolates the effect of removing specific components, using the full EPicMT as the internal reference. In this way, the ablation study provides a controlled view of how compression, reconstruction, and physics-informed constraints jointly sustain prognostic accuracy in the RUL prediction setting.

We further extend our evaluation to the NASA C-MAPSS dataset for turbofan engine RUL estimation, a regression task characterized by continuous degradation trajectories rather than discrete fault classes. The study focuses on ablation: testing the contribution of architectural and training designs, including the loss formulation and the token-selection mechanisms, under controlled compression. We report standard regression metrics (RMSE, and $R^2$) to quantify predictive fidelity.

This ablation isolates the contribution of the three physics-informed loss terms (TF, GDA, MEC) and two architectural mechanisms (HPQ, SATS), as elaborated in Table 6-5. The full EPicMT integrates all five components, while each ablation disables one element under otherwise identical training and inference settings. Loss weights are fixed, with re-normalized coefficients when one term is removed. Slice length is standardized at 1024 points, and tokenization yields 6 tokens per slice ($\approx$170 points/token), corresponding to a 48-bit budget at CR=32. By holding computational cost constant, any change in regression metrics directly reflects the diagnostic value of the removed component.

Table 6-5. Ablation study design for EPicMT on C-MAPSS dataset

| Model | Modification | Loss weights |
| --- | --- | --- |
| EPicMT | N/A | $\lambda_{TF}=0.6, \lambda_{GDA}=0.2, \lambda_{MEC}=0.2$ |
| EPicMT-w/o TF | Remove time–frequency & autocorrelation | $\lambda_{TF}=0, \lambda_{GDA}=0.5, \lambda_{MEC}=0.5$ |
| EPicMT-w/o GDA | Remove global correlation | $\lambda_{TF}=0.6, \lambda_{GDA}=0, \lambda_{MEC}=0.4$ |
| EPicMT-w/o MEC | Remove multi-scale envelope | $\lambda_{TF}=0.6, \lambda_{GDA}=0.4, \lambda_{MEC}=0$ |
| EPicMT-w/o HPQ | Replace HPQ with uniform quantizer | $\lambda_{TF}=0.6, \lambda_{GDA}=0.2, \lambda_{MEC}=0.2$ |
| EPicMT-w/o SATS | Replace adaptive token selection | $\lambda_{TF}=0.6, \lambda_{GDA}=0.2, \lambda_{MEC}=0.2$ |

With the ablation design fixed (full EPicMT vs. five single-factor removals), we evaluate the framework on the most challenging FD004 subset of NASA C-MAPSS. FD004 combines multiple fault modes and operating conditions, with 21 sensor channels and ~250 train/test units, making it a demanding testbed for compression and reconstruction. Results are reported as the average of five runs in Figure 6-14, with error bars denoting standard deviation. We consider four complementary metrics: RMSE and $R^2$ for regression accuracy, and SNR and SCR for signal fidelity.

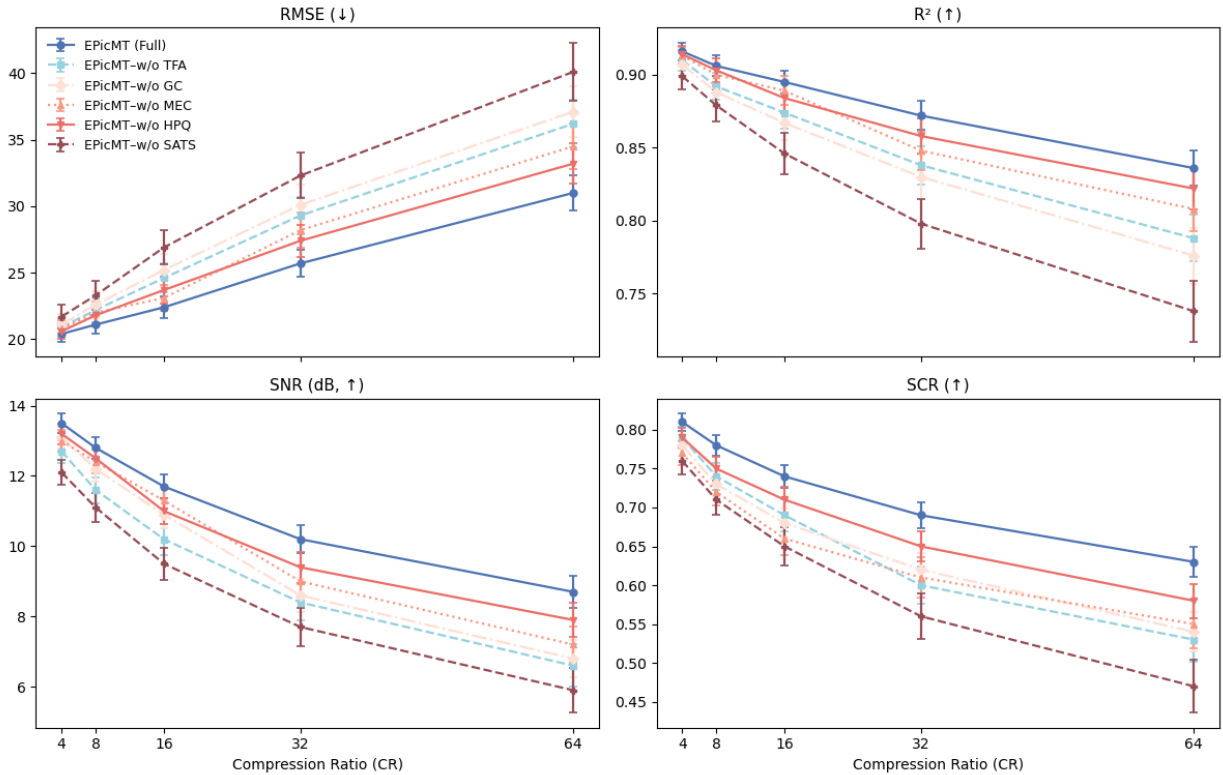

**Figure 6-15.** Performance of EPicMT and its ablated variants on FD004 subset of C-MAPSS.

Across the four metrics in Figure 6-15, removing different components yields distinct performance degradations. Eliminating SATS produces the largest deterioration at higher compression ratios: RMSE rises sharply and $R^2$ declines accordingly, indicating that token adaptivity is central for preserving regression accuracy under aggressive compression. Removing GDA also worsens long-horizon alignment, visible in elevated RMSE and reduced $R^2$, while the associated error bars widen, reflecting reduced stability across runs. In contrast, HPQ has only moderate impact on RMSE and $R^2$, but its absence shows clearer differences in SNR and SCR, especially at CR $\geqslant$ 32, suggesting that harmonic-aware quantization contributes more directly to signal fidelity than regression precision. The other two terms, TF and MEC, show complementary influences. Without TF, the curves in SNR and SCR drop faster, consistent with weakened time–frequency localization of fault signatures. Without MEC, the degradation of RMSE and $R^2$ is moderate, but SCR decreases more noticeably, highlighting its role in maintaining envelope-related modulation depth. The error bars also reveal differences in stability: MEC removal results in larger variance in SCR, whereas TF removal yields more variability in SNR. Notably, at intermediate compression ratios some curves partially overlap or even cross, such as w/o MEC and w/o HPQ in SCR, reflecting that different mechanisms dominate different aspects of signal preservation.

## 6.5. Conclusion

This study sets out to design an edge-aware compression–reconstruction framework capable of reducing communication and computation costs without discarding the diagnostic features needed for PHM. By integrating token adaptivity, harmonic-preserving quantization, and physics-informed loss terms, the framework was validated across diverse datasets ranging from rotating machinery to turbofan prognostics. Quantitative evaluations confirm that this design maintains both efficiency and diagnostic fidelity. On rotating machinery datasets, classification accuracy remained above 90% under compression ratios as high as 64, with runtime contained within ~2 ms per 1 k-sample slice and FLOPs reduced by over 30% relative to standard autoencoders. For turbofan prognostics (C-MAPSS FD004), the framework sustained RMSE below 27 cycles and $R^2$ above 0.85 at CR = 32, outperforming ablated variants by 5–10 cycles. Signal fidelity metrics further highlight the advantage that SNR improved by 1–2 dB and SCR stayed above 0.70 even at high compression, ensuring that harmonics and envelopes remained interpretable for diagnosis. Beyond confirming the feasibility of transmitting fewer bits without sacrificing critical harmonics, envelopes, and long-term trends, the findings suggest strong potential for deployment in scenarios such as real-time edge monitoring, wireless sensor networks, and digital-twin pipelines, where both efficiency and fidelity are indispensable. Limitations remain: the loss weights were heuristically set and may benefit from more systematic tuning, and future work will extend evaluation to noisy, in-field datasets. Nonetheless, the results establish that combining token selection, quantization, and physics-informed reconstruction provides a principled and practical path to high compression ratios with sustained diagnostic utility.

# Chapter 7. Few-Shot Visual Reasoning for Wind Turbine Damage Detection via RAG with Vision-Language Model

Wind turbine blade inspection using drone-based imaging has emerged as a highly promising solution for scalable, low-cost monitoring of large wind farms. However, the majority of existing visual inspection methods still rely heavily on either handcrafted features or large-scale labeled datasets to train traditional deep learning models. These approaches face significant challenges in terms of adaptability, requiring time-consuming retraining or fine-tuning whenever new defect patterns, lighting conditions, or inspection angles arise. In this work, we propose a novel few-shot visual reasoning pipeline based on visual -text Retrieval-Augmented Generation (RAG) integrated with a pre-trained Vision-Language Model (VLM), designed to reduce reliance on manual labeling and enhance adaptability across blade inspection scenarios. Unlike conventional pipelines, our system does not require task-specific fine-tuning. Instead, it performs in-context few-shot reasoning by retrieving relevant visual-textual examples from a structured knowledge base and prompting the language model to reason over these alongside the current image's description. To demonstrate its practical potential, we construct a domain-specific knowledge base including structured textual files and optional curated image-caption examples. We show that our visual-text RAG-VLM framework is able to reason about damage type and severity in a flexible, interpretable, and data-efficient manner.

## 7.1. Introduction

Wind energy plays a key role in the global transition to sustainable power. As wind farms grow in scale, efficient inspection and maintenance of turbine blades has become increasingly important [329]. Blades are prone to defects such as cracks, delamination, erosion, and impact damage, which can compromise performance if left undetected. While various techniques, like acoustic sensing, ultrasonics, and thermography, have been explored, machine vision–based methods offer clear advantages: they are non-contact, high-resolution, and capable of full-surface coverage. Combined with drone technology, which enables automated and flexible image capture, these systems provide a scalable, safe, and non-intrusive solution for large-scale turbine blade inspection.

Over the past decade, extensive research has focused on developing vision-based models for automated wind turbine blade damage detection. With the rise of deep learning, many data-driven approaches have been proposed, including CNNs for crack detection [330], semantic segmentation for multi-class classification [331], and transformer-based models for high-resolution defect localization [332]. While these models perform well on curated datasets, they typically require large amounts of labeled data under controlled conditions. In real-world deployments, however, collecting balanced and representative samples is difficult due to environmental variability, seasonal effects, and inspection inconsistencies. As a result, datasets are often dominated by common defects like stains, while critical types such as cracks or burns remain underrepresented and inconsistent over time. Some studies have explored generative models like GANs to augment rare classes, but these require sufficient real examples and incur high computational costs. Despite advances, current deep learning pipelines still struggle with scalability and robustness in practical, data-scarce scenarios.

Few-shot learning [333] addresses the limitations of traditional supervised learning by enabling accurate predictions from only a few examples. This approach has gained traction through large language models (LLMs) like GPT and T5 [334], which generalize across tasks via in-context learning without retraining. Extending this to vision, vision-language models (VLMs) [335] offer powerful image understanding through natural language, capable of captioning, visual Q&A, and pattern recognition with minimal task-specific data. However, directly applying VLMs to industrial tasks like turbine blade inspection remains difficult, as domain-specific knowledge, e.g., damage types or severity scales, is rarely covered in pretraining. Fine-tuning VLMs is also resource-intensive. To address this, retrieval-augmented generation (RAG) [336] allows models to access external, task-specific knowledge at inference time, combining general reasoning capabilities with contextual precision without additional training.

Vision-language inspection has gained traction. Lei & Qi [337] fine-tune a VLM with dual-state prompts for rail-car surface anomalies; Zhang et al [338] adapt CLIP for zero-shot defect detection using a textual-domain bridge; Tan et al. [339] add multi-level contrastive training to transfer a VLM to steel-surface defects. These approaches validate the promise of VLMs but still require task-specific training and lack an explicit mechanism to inject domain knowledge. Moreover, they address scenes with relatively uniform backgrounds. Wind-turbine blades pose tougher challenges, large texture variation, changing perspectives, and stringent safety demands. To bridge this gap, we keep the pre-trained VLM intact and augment it at inference time with a retrieval-based knowledge base, enabling data-scarce, auditable, and adaptable blade-damage assessment.

In this work, we propose a novel few-shot visual reasoning pipeline that integrates a pre-trained VLM with a lightweight visual-text RAG framework for wind turbine blade damage assessment. The system begins by generating a natural language caption from a blade image using a VLM, which acts as a compact semantic representation of the visual content. This caption is then embedded and used to retrieve semantically relevant entries from a domain-specific knowledge base that includes both curated textual files, covering defect definitions, severity guidelines, and inspection protocols, and image-text exemplars that serve as visual analogs. What sets our framework apart is its hybrid knowledge base design and retrieval-enhanced reasoning strategy. We combine dual-modal embeddings (for both text and image-caption pairs) with a semantic reranking mechanism that selects contextually concise and domain-relevant entries. These are used to construct a dynamic prompt for a second-stage language model (e.g., Qwen-Plus), which generates an interpretable diagnostic output, including damage classification, severity level, and suggested actions. This framework is lightweight, adaptable, and requires no retraining, making it well-suited for blade inspection tasks where annotated data is limited and inspection conditions vary. Furthermore, the use of structured retrieval allows the model's decisions to be traceable and easily updated by modifying knowledge files, enabling transparent, few-shot reasoning grounded in real domain knowledge. While we focus on wind turbine blades in this study, the overall approach is broadly applicable to other engineering inspection scenarios, such as aircraft structures, rail surfaces, or manufacturing components, where visual variability, limited labels, and interpretability are critical.

The remainder of this paper is organized as follows. Section 7.2 introduces the proposed system architecture, detailing each module including image captioning, knowledge retrieval, and language-guided reasoning. Section 7.3 describes the experimental setup, along with the qualitative and quantitative evaluations based on public wind turbine dataset. Finally, Section 7.4 concludes the paper and outlines future directions.

## 7.2. Methodology

In this section, we detail the modular components that comprise our pipeline, including image caption generation, semantic retrieval from a domain-specific knowledge base, and contextual reasoning using a language model. Additionally, we describe the preparation of textual and visual knowledge sources that support few-shot inference within the system.

### 7.2.1. Image Captioning via Vision-Language Model

VLMs are multimodal architectures that learn joint visual-textual representations from large-scale image-text pairs. They excel in tasks like image captioning, visual question answering, and multimodal reasoning, often under few- or zero-shot settings. In our system, the VLM serves as a front-end to automatically generate captions for turbine blade images, converting complex visual input into structured natural language. These captions act as semantic queries, enabling downstream modules to interact with textual knowledge purely in the language domain, rather than relying on traditional visual features or handcrafted descriptors.

The architecture of the VLM used in our system, Qwen-VL, follows a transformer-based encoder-decoder framework. The vision encoder, often instantiated as a Vision Transformer (ViT), processes the input image and produces a sequence of visual embeddings that encode both spatial and semantic content. These embeddings are then passed to a transformer decoder, which generates a natural language caption in an autoregressive fashion. Formally, given an input image $I \in \mathbb{R}^{H \times W \times 3}$, the encoder produces a latent sequence of embeddings $z = f_v(I)$. The decoder then generates a caption $C = \{w_1, w_2, \ldots, w_T\}$, where each token is predicted as:

$$P(C|I) = \prod_{t=1}^{T} P(w_t | w_1, w_2, \ldots, w_{t-1}, z) \tag{7.1}$$

### 7.2.2. Visual-Text Retrieval-Augmented Generation

RAG is a powerful framework that enhances LLMs by incorporating external knowledge at inference time. Unlike traditional LLMs that rely solely on static, pre-trained parameters, RAG retrieves relevant content from a dynamic knowledge base, enabling domain adaptability and improved factual accuracy without retraining. It follows a three-step pipeline: embedding the input query, retrieving semantically similar documents from a vector store, and generating a response conditioned on both the query and the retrieved context.

Originally designed for text-only tasks like open-domain question answering, textual RAG operates entirely within the language domain, using a language encoder, a text-based knowledge base, and an LLM as generator. It supports zero-shot generalization and offers modularity by allowing knowledge updates without modifying the model itself. The components of textual RAG include:

- Query Encoder: Converts the input text into an embedding vector.

- Knowledge Base: A set of textual documents or chunks embedded in advance.

- Retriever: A vector search mechanism that finds the most similar documents to the input.

- LLM Generator: Combines the query and retrieved context to generate a final answer.

While traditional RAG handles text-based queries, many real-world applications, especially in engineering, require reasoning over multimodal inputs like images. To address this, RAG has been extended to the visual domain via integration with vision-language models (VLMs), which translate images into natural language captions. These captions serve as textual queries in a standard RAG pipeline, enabling retrieval of semantically relevant knowledge and generation of informed responses. We adopt this visual-text RAG framework for wind turbine blade inspection using drone imagery, where challenges like data scarcity, class imbalance, and environmental variability hinder the scalability of supervised models. Unlike conventional textual RAG, our knowledge base includes both structured domain-specific documents (e.g., defect definitions, severity criteria) and expert-authored image-caption exemplars, enabling few-shot reasoning through visual-text analogy.

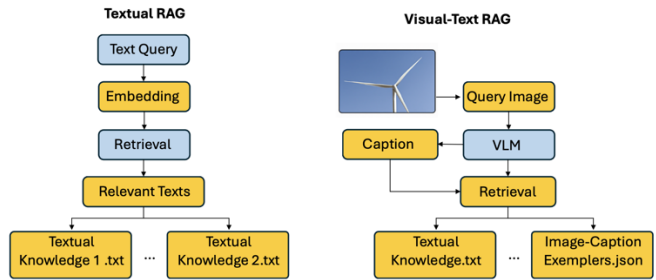

**Figure 7-1.** Workflows of textual RAG and visual-text RAG.

As shown in Figure 7-1, in visual-text RAG, an image is first processed by a VLM to produce a caption. This caption serves as the query for retrieving both textual knowledge and image-caption exemplars. The retrieved content is then used to generate a domain-aware diagnostic report, enabling few-shot reasoning from visual input. In our implementation, the caption is first embedded using the Qwen-Plus text encoder, which shares a common embedding space with all knowledge base entries. A FAISS index then performs cosine similarity search to retrieve the top-k most relevant items. To refine retrieval quality, a lightweight re-ranking step boosts entries containing key domain terms such as damage types or blade zones, ensuring that the retrieved context is both semantically aligned and inspection-specific.

### 7.2.3. Domain-Specific Knowledge Base Construction

A central component of our visual-text RAG pipeline is the curated domain-specific knowledge base, which provides the external information necessary for retrieval-augmented reasoning. To support few-shot, context-aware reasoning, our knowledge base is constructed in a hybrid format containing core textual knowledge and visual-text exemplars. Each category plays a distinct role in enriching the model's understanding of domain-specific concepts and decision criteria.

**Table 7-1.** Structure of the Domain-Specific Knowledge Base

| File Entry | Description |
|---|---|
| **Core Knowledge** | |
| damage_types.txt | Definitions and visual characteristics of common blade defects |
| severity_grading.txt | Criteria for grading defect severity based on size, shape, and location |
| maintenance_threshold.txt | Safety thresholds for triggering inspection or shutdown decisions |
| blade_zones.txt | Functional segmentation of the blade into zones (tip, mid-span, root) and inspection priorities |
| **Visual-Text Exemplars** | |
| tip_crack_case.json | {"id": "tip_crack_case", "caption": "A 15 cm longitudinal crack runs along the upper tip edge."} |
| leading_edge_erosion.json | {"id": "leading_edge_erosion", "caption": "Erosion marks along leading edge from persistent wind load."} |
| healthy_blade_sample.json | {"id": "healthy_blade", "caption": "No visual defects observed; surface smooth and uniformly colored."} |
| salt_damage_pattern.json | {"id": "salt_damage", "caption": "Surface pitting and discoloration from long-term salt exposure."} |

Table 7-1 outlines the structure of our domain-specific knowledge base, which supports the retrieval component of the RAG framework. It consists of two layers: core knowledge files and visual-text exemplars. The core entries provide standardized definitions, severity grading criteria, and operational thresholds essential for consistent inspection interpretation. In parallel, the visual-text exemplars comprise curated JSON files pairing descriptive captions with image identifiers. These serve as analogical references that help the model contextualize current inputs by comparing them to past visual cases, reinforcing few-shot reasoning during inference.

### 7.2.4. System Overview

To unify vision-language modeling, domain-specific retrieval, and language-based reasoning, we propose a modular pipeline for wind turbine blade inspection. The system enables few-shot, interpretable damage assessment from drone imagery without model fine-tuning. It begins with a pre-trained VLM generating a caption that summarizes key visual features, such as damage type, location, or surface condition, which serves as the query for retrieval. The caption is embedded and used to search a hybrid knowledge base containing both textual entries (e.g., defect definitions, severity criteria) and image-caption exemplars stored in JSON format. The top-k semantically relevant items are retrieved and combined with the original caption to form the input for a language model, which generates a structured diagnostic report. This includes damage classification, severity estimation, and reasoning supported by retrieved domain knowledge. As shown in Figure 7-2, the pipeline bridges visual perception and symbolic inference, enabling scalable, label-efficient inspection under real-world conditions. Since the retrieval process operates independently of model weights, the system can accommodate larger or more diverse knowledge bases through simple re-indexing, without requiring any retraining.

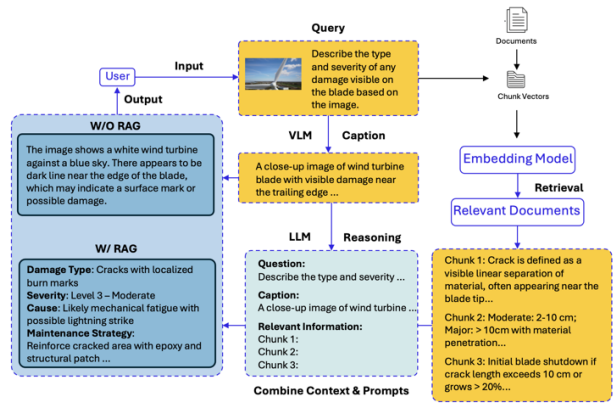

Figure 2.

**Figure 7-2.** System Overview.

## 7.3. Experimental Setup

### 7.3.1. Dataset

We evaluate our method on the publicly available DTU wind turbine blade dataset [340], which contains high-resolution drone images captured under varied environmental and positional conditions. While extensive, the dataset lacks consistent labeling and balanced class distribution, limiting its suitability for supervised classification. To support few-shot reasoning in our RAG framework, we curated a subset of representative images to build a visual-text knowledge base.

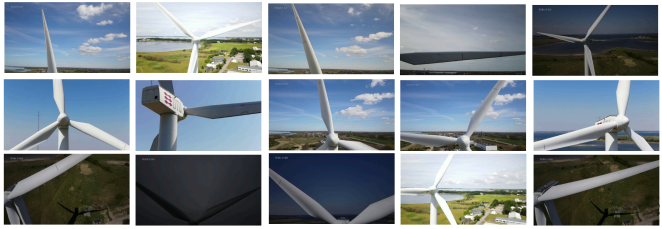

**Figure 7-3.** Image samples in visual-text knowledge base.

As shown in Figure 7-3, we selected 15 drone-captured images, each paired with a human-written caption describing visible damage, location (e.g., tip, trailing edge), and contextual details. These exemplars, encoded into a shared vector space, serve as semantic anchors for retrieval, enabling the system to match incoming image queries with structurally similar cases. Despite the small sample size, this curated set enables effective few-shot learning without requiring model fine-tuning or large annotated datasets.

### 7.3.2. Implementation Details

Our visual-text RAG system is built on the Qwen-VL-Max model [341], a 32.5B-parameter VLM with 64 transformer layers and a 7,500-token context window, enabling rich image-text interaction. For embedding and retrieval, we use Qwen-Plus, based on the Qwen2.5 architecture, supporting up to 131,072 tokens for efficient processing of large textual inputs.

To benchmark performance, we compare against standard image-based classifiers: CNN, VGG16, ResNet50 [342], ViT [343], and ConvNext [344]. All models are trained with a learning rate of 1e-4 for 50 epochs using the Adam optimizer, with data augmentation [345] applied for generalization. Experiments are conducted on a workstation with an Intel Xeon Silver 4210R CPU, NVIDIA RTX A4000 GPU, and 128 GB RAM using PyTorch, ensuring fair comparison in training and inference efficiency.

### 7.3.3. Evaluation Results

The original DTU wind turbine blade dataset does not come with predefined labels, making it necessary to manually curate a well-structured labeled dataset for fair benchmarking. To ensure fair comparison with deep learning baselines, we limited classification to four well-represented classes: Healthy, Erosion, Crack, and Stain, while rarer classes (e.g., ice, burn marks) were excluded from training but remain detectable by our method through the knowledge base. For each class, we selected 15 representative images and applied geometric data augmentation (flipping, cropping, scaling, rotation) to generate 60 training samples. An additional 15 diverse samples per class were selected for testing. This setup provides a balanced dataset for benchmarking both conventional and few-shot RAG-based approaches. We report overall accuracy and macro-averaged F1 across the four classes. Accuracy is the fraction of correctly classified test images: $Accuracy = \frac{\sum_c TP_c}{N}$, where N is the total test set size. For each class $F1_c = \frac{2TP_c}{2TP_c+FP_c+FN_c}$, and the reported score is the macro average over all classes.

**Table 7-2.** Performance Comparison

| Model | Accuracy | F1-Score | Inf. Time(s) |
|---|---|---|---|
| CNN | 0.7333 | 0.7149 | 0.014 |
| VGG16 | 0.7667 | 0.7336 | 0.025 |
| ResNet50 | 0.7833 | 0.7757 | 0.030 |
| ViT | 0.8500 | 0.8449 | 0.041 |
| ConvNext | 0.8167 | 0.8237 | 0.045 |
| Ours | 0.9833 | 0.9821 | 16.178 |

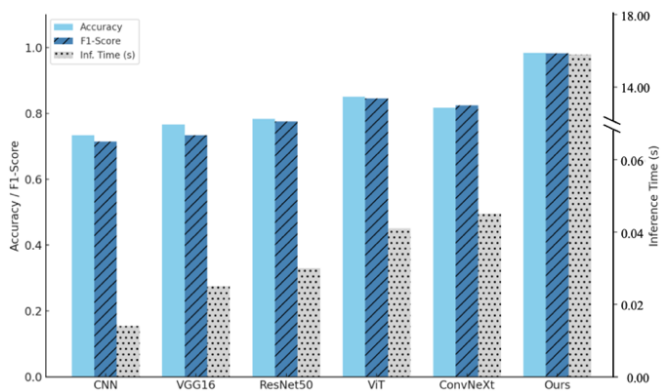

**Figure 7-4.** Model Performance Comparison.

The results in Table 7-2 demonstrate that our few-shot learning approach using a visual-text RAG system significantly outperforms traditional deep learning models, including the state-of-the-art ViT, in both accuracy and F1-score. This superior performance can be attributed to the inherent limitations of conventional supervised models under constrained data regimes. Despite applying standard data augmentation techniques, such as flipping, cropping, and scaling, the overall diversity and informational richness of the training set remain limited. This is primarily because the images were extracted from videos captured under similar conditions and time frames, leading to substantial overlap in content across training samples. As a result, the models trained via supervised learning tend to overfit on redundant patterns, whereas the proposed method excels by leveraging external domain-specific knowledge to enrich its reasoning capabilities.

However, this improved performance comes with increased inference time, as illustrated in Figure 7-4. Unlike conventional feed-forward CNNs or vision transformers, our RAG-enabled pipeline includes additional steps, image captioning, semantic retrieval, and generative reasoning, which contribute to higher latency. To clearly present this disparity, we apply a split y-axis to highlight the significantly longer inference time of our method while preserving comparability across other models. Despite this overhead, the trade-off yields richer outputs: in addition to classification, the system provides severity estimates, diagnostic reasoning, and maintenance suggestions grounded in expert knowledge (see Figure 7-5). Notably, inference time can be significantly reduced in future deployments by hosting the model locally, eliminating current network-induced delays from cloud-based API calls. Additional optimization strategies, such as model distillation, hardware acceleration, and embedding caching, may further reduce latency and improve deployment efficiency in edge or real-time scenarios.

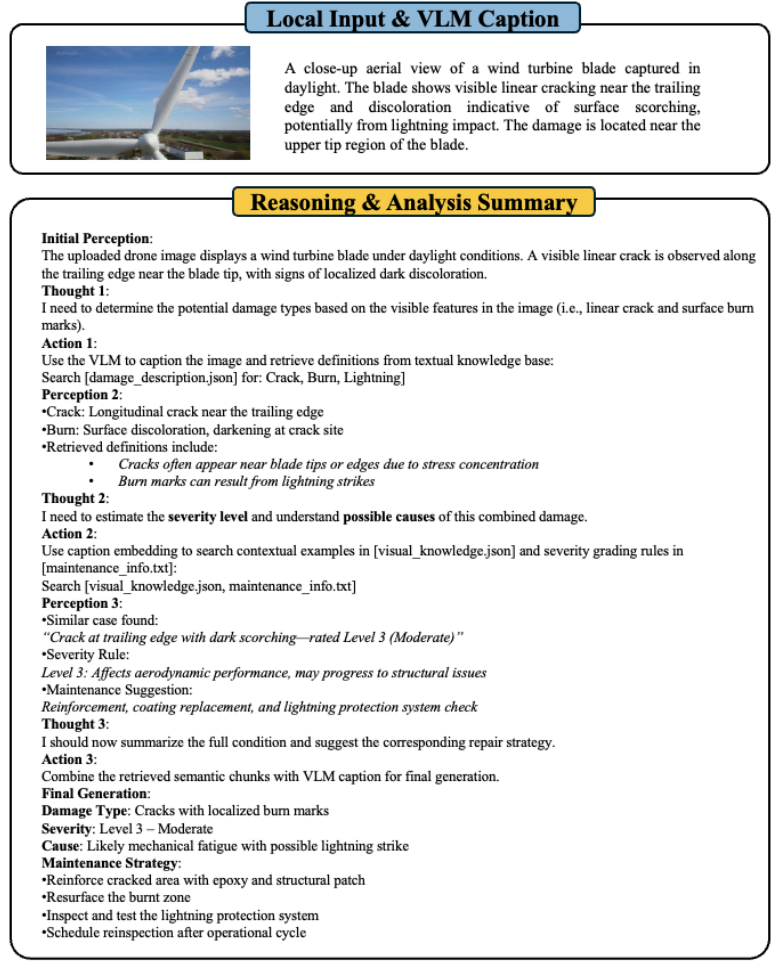

**Figure 7-5.** Model reasoning and analysis summary.

To further evaluate the classification performance of our proposed RAG-based few-shot learning system, we compare it against the baseline VLM without retrieval assistance. Figure 7-6 presents the normalized confusion matrices for the four-class classification task (Healthy, Erosion, Crack, Stain) under both configurations: without RAG (w/o RAG) and with RAG (w/ RAG). Each matrix cell represents the proportion of test samples classified into each category, averaged over 15 samples per class.

The results reveal a clear performance gap between the two settings. The w/o RAG model misclassifies a notable portion of Erosion and Stain samples, likely due to visual similarities and the lack of external context to disambiguate subtle damage types. In contrast, the w/ RAG system achieves near-perfect classification, with only a single error where a Stain sample is misinterpreted as Erosion. This highlights the strength of the RAG framework in grounding the VLM-generated image caption with domain-specific knowledge, improving recognition accuracy, especially for visually ambiguous or underrepresented categories.

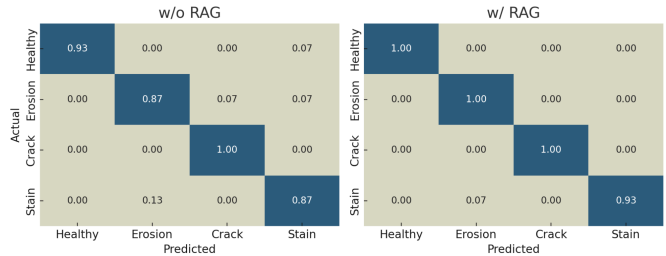

**Figure 7-6.** Result Comparison between with and without RAG knowledge base.

In this study, we present a novel visual-text Retrieval-Augmented Generation (RAG) framework tailored for wind turbine blade surface inspection using drone imagery. Unlike conventional deep learning approaches that rely on large volumes of labeled data and often struggle with underrepresented or ambiguous classes, our proposed method integrates a pre-trained vision-language model (VLM) with a domain-specific knowledge base to support few-shot learning in real-world conditions. By transforming image inputs into descriptive captions and enriching them through semantic retrieval of textual and visual knowledge, our system enables robust, interpretable, and scalable damage classification. Experimental results on the DTU wind turbine dataset demonstrate that the RAG-enhanced model significantly outperforms traditional image-based classifiers in both accuracy and interpretability, achieving up to 98.33% classification accuracy with a minimal number of training samples. Remaining challenges are higher inference latency than pure vision models and the absence of pixel-level localization. We will address these by optimizing and locally hosting the pipeline, caching embeddings, and adding lightweight grounding or hybrid backbones to return spatial cues. Future work will also explore temporal video information, richer multimodal knowledge, and on-line adaptation to evolving inspection conditions.

## 7.4. Conclusions

In this study, we present a novel visual-text Retrieval-Augmented Generation (RAG) framework tailored for wind turbine blade surface inspection using drone imagery. Unlike conventional deep learning approaches that rely on large volumes of labeled data and often struggle with underrepresented or ambiguous classes, our proposed method integrates a pre-trained vision-language model (VLM) with a domain-specific knowledge base to support few-shot learning in real-world conditions. By transforming image inputs into descriptive captions and enriching them through semantic retrieval of textual and visual knowledge, our system enables robust, interpretable, and scalable damage classification. Experimental results on the DTU wind turbine dataset demonstrate that the RAG-enhanced model significantly outperforms traditional image-based classifiers in both accuracy and interpretability, achieving up to 98.33% classification accuracy with a minimal number of training samples. Remaining challenges are higher inference latency than pure vision models and the absence of pixel-level localization. We will address these by optimizing and locally hosting the pipeline, caching embeddings, and adding lightweight grounding or hybrid backbones to return spatial cues. Future work will also explore temporal video information, richer multimodal knowledge, and on-line adaptation to evolving inspection conditions.

# Chapter 8.  Concluding Remarks

This dissertation set out to narrow the gap between laboratory-grade learning systems and the realities of industrial sensing, where computation is limited, bandwidth is intermittent, and labeled data are scarce. Across multiple threads, semi-supervised learning for label-lean regimes, physics-guided representations, and edge-aware compression/reconstruction, we developed a coherent toolkit aimed at edge-ready, resource-constrained deployment without sacrificing diagnostic fidelity. The studies span analytical formulations, algorithmic design, and cross-domain case studies on gearboxes, bearings, and milling processes, offering both principled methods and empirical evidence. Our main contributions are as follows.

(1) Edge-ready signal intelligence under tight resource budgets.

We designed an end-to-end edge–cloud pipeline that compresses high-rate multi-sensor signals on device and reconstructs them in the cloud while preserving fault-critical content. On the edge, the encoder performs content-aware token skimming and structure-aware quantization guided by spectral priors (e.g., gear-mesh harmonics and sidebands) so that the bit budget scales with diagnostic relevance rather than with raw sampling rate. On the cloud, a physics-constrained decoder, regularized by time–frequency and autocorrelation consistency, restores waveforms and maintains task-level discriminability. Across compression ratios up to 64:1, this framework consistently retained spectral signatures required for downstream fault classification and RUL surrogates, while reducing runtime, memory footprint, and transmission cost to levels compatible with embedded controllers. The resulting family of encoders and decoders provides a transferable recipe for edge deployment: isolate the health-bearing portions of a signal, assign them disproportionate representational precision, and reconstruct under physics-consistent losses to prevent diagnostic drift.

(2) Learning in the small-data regime: feasibility, methods, and guarantees of utility.

We examined label scarcity from complementary angles and showed that performance can be recovered—and sometimes improved—without exhaustive annotation. First, we formalized a mixed-training strategy in which pseudo-labels are injected with a time-varying confidence weight. The schedule begins conservatively, increases only as the model stabilizes, and yields measurable gains in both training stability and test accuracy. Second, we analyzed feature-preserving data representations, demonstrating that 2-D time–frequency embeddings (e.g., CWT/STFT images) paired with parallel branches capture complementary structures and mitigate overfitting relative to single-stream baselines. Third, we integrated physics-guided priors—for example, harmonic and sideband placements and Johnson–Cook style constraints—into loss design and architecture selection to regularize learning when data are sparse. The collective outcome is a set of small-data feasible recipes that combine modest augmentation, semi-supervision with controlled trust, and physics-aware inductive bias to achieve competitive generalization with orders-of-magnitude fewer labels than conventional pipelines.

(3) Physics-guided representation and inference for resource-conscious diagnostics.

Beyond generic architectures, we articulated a viewpoint in which task physics defines the representational budget. In gearboxes, the allocation of bits and network attention tracks known excitation lines; in milling, graph-structured predictors respect spatial coupling of nodes along the toolpath and are constrained by constitutive relations. We operationalized this by: (i) embedding spectral anchors into the encoder's token importance ranking and the decoder's consistency penalties; (ii) coupling graph attention with process-aware node selection so that computation is spent on mechanically sensitive locations; and (iii) multiplexing decision heads (fault class, proxy-RUL) to feed back task signals that shape what the compression stage must preserve. This physics-guided, resource-aware perspective yielded models that are not only accurate but auditable, their decisions can be traced to physically meaningful components of the signal or structure, thereby improving trust and transfer across operating conditions.

Collectively, the contributions above were validated through a suite of case studies that mirror realistic deployment: variable-speed gearbox diagnostics from vibration, bearing health under aggressive compression, and milling surface deformation prediction with graph-based surrogates. In each scenario, we quantified reconstruction fidelity (SNR/PRD), preservation of diagnostic content (spectral-consistency metrics), downstream utility (classification accuracy, proxy-RUL), and system cost (parameters, FLOPs, runtime, and bit budgets). The empirical evidence consistently supports three claims: (i) compression can be health-aware rather than uniformly lossy; (ii) physics-guided priors are indispensable in small-data regimes; and (iii) models that are designed from the outset for edge constraints transfer more reliably than those retrofitted for deployment. While the proposed methods improve robustness and efficiency, several fronts merit continued work: formal sample-complexity results for the mixed-training schedule; automated discovery of physics anchors when first-principles knowledge is partial; joint optimization of sensing policy, compression, and inference; and hardware-in-the-loop evaluation across heterogeneous MCUs and industrial gateways. Extending the pipeline to multi-modal streams (acoustic emission, current, thermography) and to closed-loop prognostics will further stress-test the approach.

In summary, this dissertation advances a cohesive agenda for resource-conscious, edge-ready intelligence in manufacturing systems. By aligning representation, learning, and deployment with the physics of the process and the realities of the platform, it demonstrates that practical constraints need not be barriers to high-fidelity diagnostics; correctly leveraged, they can be design signals that lead to compact, trustworthy, and field-viable solutions.

[325] Villwock, S., & Pacas, M. (2008). Application of the Welch-method for the identification of two-and three-mass-systems. IEEE Transactions on Industrial Electronics, 55(1), 457-466.

[326] Lacy, F., Ruiz-Reyes, A., & Brescia, A. (2024). Machine learning for low signal-to-noise ratio detection. Pattern Recognition Letters, 179, 115-122.

[327] Rodriguez, A., & Kokalj-Filipovic, S. (2024). VQalAttent: a Transparent Speech Generation Pipeline based on Transformer-learned VQ-VAE Latent Space. arXiv preprint arXiv:2411.14642.

[328] Wang, H., Wang, S., Sun, W., & Xiang, J. (2024). Multi-sensor signal fusion for tool wear condition monitoring using denoising transformer auto-encoder Resnet. Journal of Manufacturing Processes, 124, 1054-1064.

[329] Memari, M., Shakya, P., Shekaramiz, M., Seibi, A. C., & Masoum, M. A. (2024). Review on the advancements in wind turbine blade inspection: Integrating drone and deep learning technologies for enhanced defect detection. IEEE Access.

[330] Guo, J., Liu, C., Cao, J., & Jiang, D. (2021). Damage identification of wind turbine blades with deep convolutional neural networks. Renewable energy, 174, 122-133.

[331] Li, W., Pan, Z., Zhu, Q., & Du, Y. (2024). Wind turbine blade defect detection and measurement technology based on improved SegFormer and pixel matching. Optics & Laser Technology, 179, 111381.

[332] Wei, C., Han, H., Wu, Z., Xia, Y., & Ji, Z. (2024). Transformer-based multi-scale reconstruction network for defect detection of infrared images. IEEE Transactions on Instrumentation and Measurement.

[333] Zajec, P., Rožanec, J. M., Theodoropoulos, S., Fontul, M., Koehorst, E., Fortuna, B., & Mladenić, D. (2024). Few-shot learning for defect detection in manufacturing. International Journal of Production Research, 62(19), 6979-6998.

[334] Liang, Z., Xu, Y., Hong, Y., Shang, P., Wang, Q., Fu, Q., & Liu, K. (2024, January). A Survey of Multimodel Large Language Models. In Proceedings of the 3rd International Conference on Computer, Artificial Intelligence and Control Engineering (pp. 405-409).

[335] Zhang, J., Huang, J., Jin, S., & Lu, S. (2024). Vision-language models for vision tasks: A survey. IEEE Transactions on Pattern Analysis and Machine Intelligence.

[336] Chen, J., Lin, H., Han, X., & Sun, L. (2024, March). Benchmarking large language models in retrieval-augmented generation. In Proceedings of the AAAI Conference on Artificial Intelligence (Vol. 38, No. 16, pp. 17754-17762).
252